\documentclass[journal]{IEEEtran}
\usepackage{amssymb}
\usepackage{latexsym}

\usepackage{url}
\usepackage{xcolor}

\usepackage{hyperref}

\usepackage{titlesec}
\usepackage[utf8]{inputenc}
\usepackage{amssymb}
\usepackage{latexsym}
\usepackage{siunitx}
\usepackage{xcolor}
\usepackage{colortbl}
\usepackage{mathtools}
\usepackage{nicefrac}
\usepackage{subcaption}
\usepackage{hyperref}
\usepackage{array}
\usepackage{caption}
\usepackage{footnote}   % footnote in tables
\usepackage{tablefootnote}
\usepackage{sidecap}   % caption 
\usepackage{placeins}  % Floatbarrier
\usepackage[ruled,vlined]{algorithm2e}
\usepackage{dblfloatfix} % fix for bottom-placement of figure
\usepackage[square,sort,comma,numbers]{natbib}  %% citep pacakge

\SetKwInput{KwInput}{Input}                % Set the Input
\SetKwInput{KwOutput}{Output}              % set the Output

\newcolumntype{C}[1]{>{\centering\arraybackslash}p{#1}}
\newcolumntype{R}[1]{>{\raggedright\arraybackslash}p{#1}}
\newcolumntype{L}[1]{>{\raggedleft\arraybackslash}p{#1}}

\usepackage{arydshln}
\usepackage{array,makecell}
\usepackage{booktabs}
\usepackage{tikz}
\usetikzlibrary{calc}
\usepackage{flushend}  % [keeplastbox]

\titleformat{\paragraph}
{\normalfont\normalsize\itshape}{\theparagraph}{1em}{}
\titlespacing*{\paragraph}
{0pt}{3.25ex plus 1ex minus .2ex}{1.5ex plus .2ex}
\definecolor{newcolor}{rgb}{.8,.349,.1}

\begin{document}
	
\title{Autoencoding Low-Resolution MRI for Semantically Smooth Interpolation of Anisotropic MRI}%

\author{Jörg~Sander,
		Bob D.~de Vos
			and~Ivana~Išgum% <-this % stops a space
		\thanks{Jörg Sander, Bob D. de Vos and Ivana Išgum are with the Department of Biomedical Engineering and Physics, Amsterdam University Medical Centers - location AMC, University of Amsterdam, The Netherlands e-mail: \mbox{j.sander1@amsterdamumc.nl.}}
		\thanks{Bob .D. de Vos and Ivana Išgum are with Amsterdam Cardiovascular Sciences, Amsterdam University Medical Centers - location AMC, University of Amsterdam, The Netherlands}
		\thanks{Jörg Sander, Bob D. de Vos and Ivana Išgum are also with Informatics Institute, University of Amsterdam, The Netherlands}
		\thanks{Ivana Išgum is with Department of Radiology and Nuclear Medicine, Amsterdam University Medical Centers - location AMC, University of Amsterdam, The Netherlands}% <-this % stops a space
		}
	
		\maketitle
	
		\begin{abstract}
			%%%
			High-resolution medical images are beneficial for analysis but their acquisition may not always be feasible. Alternatively, high-resolution images can be created from low-resolution acquisitions using conventional upsampling methods, but such methods cannot exploit high-level contextual information contained in the images. Recently, better performing deep-learning based super-resolution methods have been introduced. However, these methods are limited by their supervised character, i.e. they require high-resolution examples for training. Instead, we propose an unsupervised deep learning semantic interpolation approach that synthesizes new intermediate slices from encoded low-resolution examples. To achieve semantically smooth interpolation in through-plane direction, the method exploits the latent space generated by autoencoders. To generate new intermediate slices, latent space encodings of two spatially adjacent slices are combined using their convex combination. Subsequently, the combined encoding is decoded to an intermediate slice. To constrain the model, a notion of semantic similarity is defined for a given dataset. For this, a new loss is introduced that exploits the spatial relationship between slices of the same volume. During training, an existing \textit{in-between} slice is generated using a convex combination of its neighboring slice encodings. The method was trained and evaluated using publicly available cardiac cine, neonatal brain and adult brain MRI scans. In all evaluations, the new method produces significantly better results in terms of Structural Similarity Index Measure and Peak Signal-to-Noise Ratio ($p< 0.001$ using one-sided Wilcoxon signed-rank test) than a cubic B-spline interpolation approach. Given the unsupervised nature of the method, high-resolution training data is not required and hence, the method can be readily applied in clinical settings.
			%%%%
		\end{abstract}

		\begin{IEEEkeywords}
			Image Super-Resolution, Image Synthesis, Semantic Interpolation, Autoencoder, Latent Space Interpolation, Deep Learning, Unsupervised, Cardiac MRI, adult Brain MRI, neonatal Brain MRI
		\end{IEEEkeywords}

	\flushbottom
	
	\thispagestyle{empty}
	%\linenumbers
	
	\section{Introduction}
	
	High spatial resolution of medical images is considered a key quality component for accurate disease diagnosis and prognosis. However, in case of magnetic resonance imaging (MRI) acquiring high-resolution images comes at the cost of reduced signal-to-noise ratio (SNR) or decreased temporal resolution. Higher image resolution can be achieved by increasing acquisition time. However, in clinical practice, fast scanning is often required to mitigate the risk for motion artifacts and to sustain patient comfort. 
	
	\begin{figure}
		\captionsetup[subfigure]{justification=centering}
		\centering
		\includegraphics[width=0.45\textwidth]{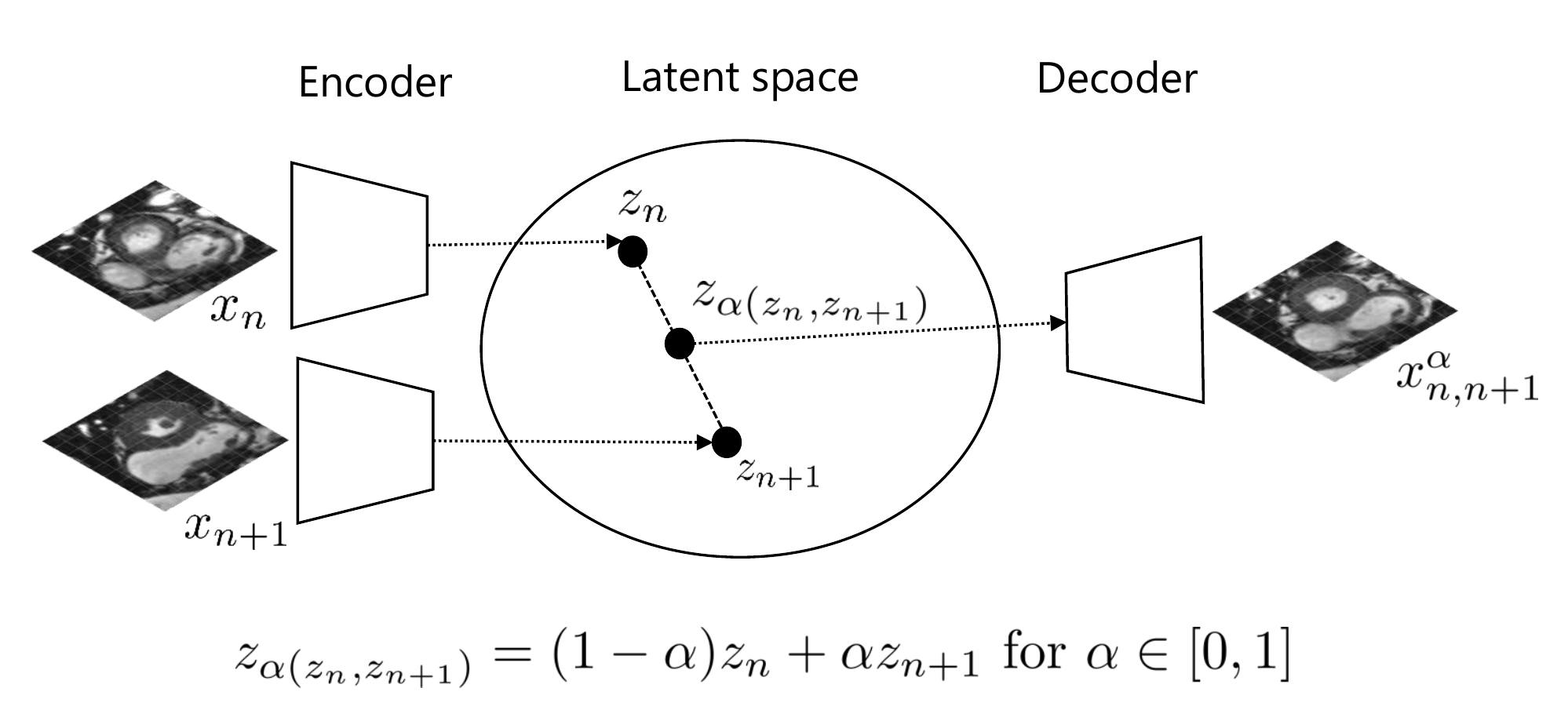}  
		\caption{Visualization of proposed method. To perform upsampling in anisotropic medical images we exploit the ability of autoencoders to interpolate in latent space. The trained autoencoder is used to project two spatially adjacent slices onto a latent space. Thereafter, latent space encodings $z_n$ and $z_{n+1}$ are combined using their convex combination. Increasing $\alpha$ from \num{0} to \num{1} results in a sequence of new
		slices where each subsequent slice is progressively less semantically similar to $x_n$ and more semantically similar to $x_{n+1}$. Anatomy in the obtained stack of slices appears semantically smooth in the direction that is perpendicular to the imaging plane.}
		
		\label{fig_aesr_approach} 
	\end{figure}
	% This trade-off is especially challenging for magnetic resonance imaging (MRI) where SNR loss incurred by reduced voxel size cannot be recovered by averaging adjacent pixels after acquisition (\cite{edelstein1986intrinsic, parker1990signal}). Moreover, in clinical practice, fast scanning is often required to mitigate the risk for motion artifacts e.g. due to breathing. 
	%As a result, MRI scans with high temporal resolution are often highly anisotropic, which may hamper accurate analysis. 
	As a result, MRI scans with high temporal resolution are often highly anisotropic, which may hamper accurate analysis.
	%As a result, MRI scans with high temporal resolution are often highly anisotropic. Low through-plane resolution in these scans is caused by large slice thickness and/or spacing. Consequently, accurate analysis of the scans may be hampered.
	There is ample research on methods that enable faster acquisition while maintaining high SNR and high spatial resolution such as compressed sensing (\cite{lustig2007sparse}) and parallel MRI \citep{heidemann2003brief}. However, these methods are mainly available in a research setting and are not available for the majority of MRIs obtained in clinical practice.
	
	Conventional interpolation methods like Linear, B-spline, and Lanczos resampling \citep{duchon1979lanczos} are often used to increase through-plane resolution of anisotropic MRIs at post-processing. The possibility to apply such methods retrospectively, i.e. after image acquisition, is often advantageous because they do not require raw image data. Conventional interpolation methods are easily and broadly applicable. Nevertheless, they cannot exploit high-level contextual information contained in the images. Furthermore, to upsample low-resolution images these methods quintessentially compute weighted intensity averages using existing image voxel values. % Therefore, these methods cannot correct for aliasing that is often present in subsampled i.e. low-resolution medical images.
	
	More sophisticated super-resolution methods have been developed that either perform denoising, deblurring, anti-aliasing, upsampling, or a combination thereof, aiming to recover a high quality of medical images from their degraded versions. In the context of this work, super-resolution for medical images refers to the process of recovering information that was lost or degraded during the low-resolution sampling process. Hence, such methods can recover anatomical structures that are finer than the sampling grid. As a result of this process anatomical structures also appear smooth and plausible in through-plane direction.	%to restore a high-resolution image from a single or several degraded, aliased images. Super-resolution refers to the process of } 
	% to recover the high-resolution information that was lost during image acquisition.
	% https://link.springer.com/content/pdf/10.1007/s10851-005-2028-5.pdf
	
	Super-resolution methods can be broadly divided into two categories. First, approaches that combine several low-resolution images to estimate, or reconstruct, the high-resolution image (\cite{greenspan2009super, gholipour2010robust}). Typically, such methods require registering the low-resolution scans with each other. Therefore, performance of these approaches depends on the quality of image alignment. Given that alignment of non-moving organs can be achieved easier than for moving organs like the heart, early super-resolution approaches in the medical imaging domain were first developed for brain MRI (e.g. \cite{peled2001superresolution, peeters2004use}). Second, to extend the applicability of super-resolution methods to moving organs approaches were developed that learn a non-linear mapping between (paired) low-resolution (LR) and high-resolution (HR) image patches \citep{manjon2010non, rueda2013single, shi2013cardiac, bhatia2014super, alexander2014image}. 
	Recently, these methods were superseded by deep-learning based super-resolution approaches using convolutional neural networks (CNN) \citep{dong2015image, kim2016deeply, oktay2016multi, ledig2017photo, lim2017enhanced, oktay2017anatomically, timofte2018ntire, chen2018brain, shi2018super, basty2018super, chen2018efficient, pham2019multiscale, mahapatra2019image, xuan2019reconstruction, masutani2020deep}. % to mitigate this limitation supervised super-resolution 
	
	To learn the non-linear mapping between low and high-resolution images the aforementioned methods require high-resolution training examples. However, in clinical practice such images are impractical or even impossible to acquire. Hence, to circumvent this short-coming, several unsupervised methods \citep{jog2016self, zhao2018self, zhao2018deep, zhao2020smore, dalca2018medical, sander2021unsupervised, xia2021super} have been proposed to increase resolution of images without using high-resolution training data. These approaches take advantage of anisotropic images and exploit the high in-plane resolution to increase the low through-plane resolution. \cite{jog2016self} proposed a Fourier-based method \citep{delbracio2015removing} that combines multiple augmentations of a single \num{3}D low-resolution input image to estimate Fourier coefficients of higher frequency ranges absent in the anisotropic images. To create required image augmentations the method learns a regression function between blurred low-resolution slices and their corresponding high-resolution slices extracted from the high-resolution in-plane direction. Later \cite{zhao2018self} proposed to alter the method of \cite{jog2016self} by replacing the regression approach with the deep-learning super-resolution approach by \cite{lim2017enhanced}. Both methods were evaluated on brain MRI with relatively mild anisotropy. Subsequently, \cite{zhao2018deep, zhao2020smore} extended their approach \citep{zhao2018self} to suppress aliasing artifacts. In parallel, \cite{dalca2018medical} proposed a Gaussian Mixture Model that learns to encode anatomical similarities extracted from a large collection of anisotropic brain MRIs. Using the learned latent structure, low-resolution scans are upsampled by imputing missing slices.
	% using a second deep-learning model based on approach of \cite{lim2017enhanced}
	% a generative image synthesis approach that increases through-plane resolution by decreasing slice spacing in highly anisotropic brain MRIs. Their model learns to capture fine-scale anatomical similarities across volumes of the training set to recover high-frequency information in through-plane direction  

	We propose a deep learning semantic interpolation approach that synthesizes new intermediate slices from encoded low-resolution examples. Specifically, in our preliminary study \citep{sander2021unsupervised} we trained an autoencoder to compress and reconstruct high-resolution slices taken from highly anisotropic volumes. Using the latent space of the trained autoencoder new intermediate slices are synthesized by mixing the encodings of two adjacent slices. The process is depicted in Figure~\ref{fig_aesr_approach}. Our method can synthesize an arbitrary number of intermediate slices exploiting the mixing coefficient of the neighboring slice's latent vectors. Note that the method effectively imputes image slices that are of similar appearance, i.e. slice thickness is retained, but slice spacing will be decreased, resulting in anatomically and semantically smooth transitions in through-plane direction. 
	The method was evaluated on cardiac cine MRI. In parallel to our study, \cite{xia2021super} presented a super-resolution approach for cardiac cine MRI that employs a conditional generative adversarial network (GAN) that takes two spatially adjacent cardiac MRI (CMRI) slices as input to synthesize a slice in-between the input slices. To guide adversarial training the approach generates an auxiliary image using previously developed optical \citep{jiang2018super} and depth-aware \citep{bao2019depth} flow-based interpolation approaches. The approach increases anisotropic resolution with upsampling-factor of two, synthesizing one slice in through-plane direction. By recursively applying the method, higher upsampling-factors of two can be achieved. 
	
	Building further on our preliminary method using convolutional autoencoders, we introduce an additional training loss function that exploits the spatial relationship between neighboring slices in \num{3}D images. This enables us to define a notion of semantic similarity for a given dataset. As a result, the autoencoder is encouraged to generate new slices that provide a semantically smooth morphing between two input images. Furthermore, we provide evidence that our extended approach leads to improved upsampling performance when compared to \citep{sander2021unsupervised}. Unlike \cite{xia2021super}, our approach can be applied with any desired upsampling factor, i.e. it can synthesize an arbitrary number of slices between two given slices in a straightforward fashion. Moreover, our approach uses only a single encoder-decoder structure and does not rely on auxiliary networks. Compared to \cite{dalca2018medical} our approach does not require a common atlas space to operate in. Moreover, our method is easy to implement and requires little GPU memory.
	
	Like in our preliminary work \citep{sander2021unsupervised}, we evaluate performance on cardiac cine MRI. Moreover, to show that our approach generalizes to other anatomies the evaluation has been substantially extended. In the experiments, three publicly available MRI datasets were used: cardiac cine MRI from the MICCAI \num{2017} Automated Cardiac Diagnosis Challenge (ACDC) (\cite{bernard2018deep}); neonatal brain MRI of the developing Human Connectome Project (dHCP) (\cite{hughes2017developing}) and adult brain MRI from the OASIS project (\cite{marcus2007open}). Evaluation on neonatal and adult brain MRI enabled performance comparison with related unsupervised \citep{jog2016self, zhao2018self} and supervised \citep{pham2019simultaneous} super-resolution methods. Finally, to demonstrate that our model is invariant to MRI scanners and voxel intensity distributions, we apply a model trained on cardiac MRI scans from the ACDC dataset to cardiac MRIs of the Sunnybrook dataset \citep{radau2009evaluation}.

	%------------------------------  DATA -------------------------------------------
	
	\section{Data}\label{section_data}
	% The method has been trained on three and evaluated on four publicly available MRI datasets. 
	% The method has been trained and evaluated on three publicly available MRI datasets: cardiac cine MRI from the MICCAI \num{2017} Automated Cardiac Diagnosis Challenge (ACDC) (\cite{bernard2018deep}); neonatal brain MRI of the developing Human Connectome Project (dHCP) (\cite{hughes2017developing}) and adult brain MRI from the OASIS project (\cite{marcus2007open}). To demonstrate the ability of our proposed method to generalize to other datasets with the same modality and anatomy, cardiac cine MRI from the publicly available Sunnybrook Cardiac dataset was used for additional model evaluation.
	
	\subsection{Cardiac Cine MRI}\label{dataset_cardiac_mri}
	
	\subsubsection{Automated Cardiac Diagnosis Challenge}\label{dataset_cardiac_mri_acdc}
	
	Cardiac cine MRIs from the MICCAI \num{2017} Automated Cardiac Diagnosis Challenge (ACDC) (\cite{bernard2018deep}) were used. The dataset consists of short-axis cardiac cine MRIs from 100 patients uniformly distributed over normal cardiac function and four disease groups: dilated cardiomyopathy, hypertrophic cardiomyopathy, heart failure with infarction, and right ventricular abnormality. Detailed acquisition protocol is described by \cite{bernard2018deep}. %Briefly, short-axis CMRIs were acquired with two MRI scanners of different magnetic strengths (\num{1.5} and \num{3.0} T). Images were made during breath hold using a conventional steady-state free precession (SSFP) sequence.

	Briefly, MRIs have an in-plane resolution ranging from \num{1.37} to \SI{1.68}{\milli\meter} (average reconstruction matrix \num{243}$\times$\num{217} voxels) with slice thickness and spacing varying from \num{5} to \SI{10}{\milli\meter}. The ACDC dataset specifies slice spacing for each image volume while slice thickness is only specified as a range for the complete dataset. Per patient 28 to 40 time points cover the cardiac cycle. Each volume consists of on average ten slices including the heart. % Some volumes\footnote{\cite{bernard2018deep} do not provide precise information.} have an additional interslice gap of \SI{5}{\milli\meter} i.e. slice spacing is greater than slice thickness.
	To correct for intensity differences among scans, in the current work, image intensities of each volume were rescaled and clamped between [\num{0}, \num{1}] based on their \num{1}$^{st}$ and \num{99}$^{th}$ percentiles. Furthermore, to correct for differences in-plane voxel sizes, image slices were resampled to $\SI{1.4}\times\SI{1.4}{\milli\meter}^2$.
	
	\subsubsection{Sunnybrook Cardiac Data}\label{dataset_cardiac_mri_sunnybrook}
	
	To demonstrate the ability of our proposed method to generalize to other datasets with the same modality and anatomy, cardiac cine MRI from the publicly available Sunnybrook Cardiac dataset \citep{radau2009evaluation} was used for additional model evaluation. The dataset contains \num{45} short-axis cine MRI images distributed over four pathology categories: healthy subjects, patients with hypertrophy, patients with heart failure and infarction, and patients with heart failure without infarction. 
	
	Each scan contains \num{20} time points (i.e. volumes) encompassing the entire cardiac cycle, which results in \num{45}$\times$\num{20} volumes in total. All scans have a slice thickness and spacing of \SI{8}{\milli\meter} and an in-plane resolution of \num{1.25}$\times\SI{1.25}{\milli\meter}^2$. Scans are made with a \num{256}$\times$\num{256} reconstruction matrix and consist of about \num{10} slices. In this work, image intensities of each volume were rescaled and clamped between [\num{0}, \num{1}] based on their \num{1}$^{st}$ and \num{99}$^{th}$ percentiles.
	
	\subsection{Neonatal Brain MRI}\label{dataset_neonatal_brain_mri}
	
	In this study neonatal brain MRIs of the developing Human Connectome Project (dHCP) \citep{hughes2017developing} were used (second release). The dataset constists of \num{508} infants with gestational age at birth ranging from \num{24} to \num{42} weeks.  All infants were scanned without sedation in a scanner environment optimized for safe and comfortable neonatal imaging. A comprehensive description of the acquisition protocol can be found in \cite{hughes2017developing}.
	
	The T$_2$-weighted (T2w) images are provided with an isotropic resolution of \num{0.5}$\times$\num{0.5}$\times\SI{0.5}{\milli\meter}^3$. To reduce image size, in this work, volumes were cropped to cortical brain structures resulting in an axial reconstruction matrix of \num{256}$\times$\num{256} voxels for all images. Finally, to correct for intensity differences among scans, voxel intensities of each volume were scaled to the [\num{0}, \num{1}] range. 
	% The data was acquired at the Evelina Newborn Imaging Center at St. Thomas Hospital (London, UK) using a \num{3}T Philips Achieva system (Philips Medical Systems, Best, The Netherlands). All infants were scanned without sedation in a scanner environment optimized for safe and comfortable neonatal imaging.
	% To reduce the effects of motion, T$_2$-weighted (T2w) images were obtained using a Turbo Spin Echo (TSE) sequence, acquired in two stacks of 2D slices (in sagittal and axial planes), using parameters: field of view $\SI{145}\times\SI{145}\times\SI{108}{\milli\meter}^3$, TR=$\SI{12,000}{\milli\second}$, TE=$\SI{156}{\milli\second}$, SENSE factor 2.11 (axial) and 2.58 (sagittal) with overlapping slices. Motion corrected reconstruction techniques were employed by the data providers, resulting in isotropic volumes of resolution  $\SI{0.5}\times\SI{0.5}\times\SI{0.5}{\milli\meter}^3$. To reduce image size, volumes were cropped to cortical brain structures resulting in an axial reconstruction matrix of \num{256} $\times$ \num{256} voxels for all images. 
	
	\subsection{Adult Brain MRI}\label{dataset_adult_brain_mri}
	
	Brain MRIs of \num{416} subjects aged \num{18} to \num{96} from the OASIS project (\cite{marcus2007open}) were used.  Detailed information about the acquisition can be found in the paper by \cite{marcus2007open} and on the OASIS website\footnote{https://www.oasis-brains.org/}.
	
	Briefly, T$_1$-weighted brain MRIs were provided with an isotropic resolution of \num{1.0}$\times$\num{1.0}$\times\SI{1.0}{\milli\meter}^3$. To correct for intensity differences among scans, in this work, voxel intensities of each volume were scaled to the [\num{0}, \num{1}] range. % Images have an average axial reconstruction matrix of \num{208} $\times$ \num{176} voxels. $\SI{1.0}\times\SI{1.0}\times\SI{1.0}{\milli\meter}^3$
	
	% The T$_1$-weighted images were acquired on a Siemens Vision \num{1.5}T scanner in the sagittal plane with a voxel size of $\SI{1.0}\times\SI{1.0}\times\SI{1.25}{\milli\meter}^3$ and were subsequently resized to an isotropic resolution of $\SI{1.0}\times\SI{1.0}\times\SI{1.0}{\milli\meter}^3$. Detailed information about the acquisition can be found in the paper by \cite{marcus2007open} and on the OASIS website\footnote{https://www.oasis-brains.org/}. Images have an average axial reconstruction matrix of \num{208} $\times$ \num{176} voxels. To correct for intensity differences among scans, voxel intensities of each volume were scaled to the [\num{0}, \num{1}] range. 
	
	% The proposed method is trained and evaluated on publicly available cardiac cine, neonatal and adult brain MR images. Detailed description of the three datasets can be found in Section~\ref{exp_cardiac_acdc}, \ref{exp_neonatal_brain_dhcp} and \ref{exp_adult_brain_mri}.
	
	%MRI datasets: cardiac cine MRI from the MICCAI \num{2017} Automated Cardiac Diagnosis Challenge (ACDC) (\cite{bernard2018deep}); neonatal brain MRI (T$_2$-weighted) of the developing Human Connectome Project (dHCP) (\cite{hughes2017developing}) and adult brain MRI (T$_1$-weighted) from the OASIS project (\cite{marcus2007open}). 
	
	%------------------------------  METHOD -------------------------------------------
	
	\section{Method} \label{section_method}
	
	We propose a method to synthesize new slices in anisotropic \num{3}D medical images by using the ability of a trained autoencoder to interpolate in latent space. %\mycomments{During training the model learns to encode a shared anatomy extracted from the collection of training images. Each individual volume might capture only some partial aspect of the complete structure.} 
	The autoencoder is trained to compress and reconstruct high-resolution \num{2}D slices taken from volumes with low through-plane resolution. We postulate that the autoencoder learns to encode anatomy from a collection of training images. While an individual image may only capture a partial aspect of a complete anatomical structure, such as the heart, the autoencoder may infer the missing information from similarly appearing images that captured different aspect of the anatomical structure.
	
	After training, input slices can be \textit{reconstructed} with minimal information loss. More important, new slices are \textit{synthesized} through convex combinations of latent space encodings of the two adjacent slices, which is followed by decoding of the convex combinations to the new intermediate slices. Note that increasing the mixing coefficient from \num{0} to \num{1} results in a sequence of new slices where each subsequent slice is progressively less semantically similar to the first input slice and more semantically similar to the second input slice. Although thickness of synthesized slices will be similar to the input slices, slice spacing will decrease. As a result, anatomical structures appear smooth and plausible in through-plane direction. 
	% As a consequence, spatial resolution of volumes in through-plane direction is increased by reducing slice spacing while slice thickness remains unchanged.}
	% Hence, our method increases spatial resolution of anisotropic \num{3}D medical images by reducing slice spacing while slice thickness remains unchanged.}
	% The method exploits the fact that fine scale anatomical structures are shared between medical images of the same anatomy although each individual volume might capture only some partial aspect of the complete structure.
	% shared in a population of medical images, and each scan with sparse slices captures some partial aspect of this structure.
	% The approach is depicted in Figure~\ref{fig_aesr_approach}.
	
	\subsection{Autoencoder}
	
	An autoencoder (\cite{rumelhart1985learning}) is an unsupervised learning algorithm that aims to learn a lower-dimensional representation of the input. It consists of an encoder and decoder implemented as neural network. The encoder $f_{\theta}$ parametrized by $\theta$ compresses the input $x \in \mathbb{R}^{d_x}$ into a lower-dimensional space $z = f_{\theta}(x), z\in \mathbb{R}^{d_z}$, i.e. the latent space representation, which captures the most salient features of the input. Typically, this layer has the least amount of neurons and is also referred to as the bottleneck of an autoencoder. 
	
	The decoder $g_{\phi}$ parametrized by $\phi$ uses the latent space representation to generate an approximate reconstruction of the input, $\hat{x} = g_{\phi}(z)$. The network layers in encoder and decoder are fully connected. In general, training an autoencoder aims to minimize a loss function $\mathcal{L}(x, \hat{x})$ that quantifies dissimilarity between the input and the corresponding reconstruction. 
	
	This work uses a convolutional autoencoder\footnote{The terms autoencoder and convolutional autoencoder will be used interchangeably hereafter.} (CAE) (\cite{masci2011stacked})
	that has the same overall structure as a standard autoencoder but replaces the fully connected layers with convolutions. Latent space encodings generated by standard autoencoders are vectors with dimensionality equal to the size of the lower-dimensional space. In comparison, latent space representation of an input tensor generated by a convolutional autoencoder is a tensor with rank equal to the rank of the input tensor. The rank of an image is three (width, height, number of input channels). Throughout this work the number of input channels is one for gray scale images. 
	
	% The terms autoencoder and convolutional autoencoder will be used interchangeably hereafter.
	
	\subsection{Autoencoder Architecture} \label{subsection_ae_architecture}
	
	The architecture of the convolutional autoencoder used in this work is the same for all datasets and experiments. The architecture of the encoder consists of two blocks, each with two consecutive convolutional layers using a kernel size of \num{3}$\times$\num{3} voxels and zero-padding of size \num{1}, followed by batch normalization and \num{2}$\times$\num{2} voxels average pooling. The first and second block use \num{32} and \num{64} kernels, respectively. 
	The last block is followed by two additional convolutional layers of \num{128}, and \num{128} kernels for the final output layer. The output of the final convolutional layer is used as latent space representation of the input. All convolutional layers except for the final use a leaky ReLU nonlinearity. 
	The combination of two average pooling layers of size \num{2}$\times$\num{2} voxels and \num{128} kernels for the latent space representation results in an \textit{over-complete} autoencoder. In other words, the information that can potentially be stored in the latent space is larger than the information contained in the grayscale input image. %
	%As a result of the encoder architecture, the autoencoder is \textit{over-complete} i.e. the information that can potentially be stored in the latent space is larger than the information contained in the input image. % \cite{bengio2007greedy}
	
	The architecture of the decoder is reverse of the encoder. It consists of two blocks of two consecutive convolutional layers with leaky ReLU nonlinearities followed by batch normalization and \num{2}$\times$\num{2} voxels nearest neighbor upsampling. The number of kernels is halved after each upsampling layer. The last block is followed by two additional convolutional layers of \num{32} kernels, and \num{1} kernel for the last layer.  All convolutional layers of the autoencoder use a kernel size of \num{3}$\times$\num{3} voxels and zero-padding of size \num{1}. To ensure that output values are in the range of $[0, 1]$ the final layer uses the sigmoid function. Moreover, using two average pooling layers of size \num{2}$\times$\num{2} voxels in the encoder requires the width and height of the input image each to be divisible by four. Nevertheless, test images do not have to match the size of the training patches.

	\subsection{Autoencoding for Semantic Interpolation}\label{method_synthesize_images}
	
	Our interpolation approach projects two spatially adjacent slices ($x_n$, $x_{n+1}$) onto a latent space. It requires high in-plane resolution. Thereafter, latent representations $z_n=f_{\theta}(x_n)$ and $z_{n+1}=f_{\theta}(x_{n+1})$ are combined using a convex combination:
	
	\begin{equation}  \label{eq_convex_combination}
	z_{\alpha(z_n, z_{n+1})} = (1 - \alpha) z_n + \alpha z_{n+1} \; \text{for} \; \alpha \in [0, 1] \;.
	\end{equation}
	
	\noindent where $\alpha$ denotes the mixing coefficient. Finally, the decoder generates a new slice $x_{n, n+1}^{\alpha}=g_\phi(z_{\alpha(z_n, z_{n+1})})$ by decoding the mixture of latent codes. We refer to this process as \textit{synthesizing} slices for values of $\alpha \in (0,  1)$, as opposed to \textit{reconstructing} encoded input slices when $\alpha \in \{0, 1\}$ for slices $x_n$ and $x_{n+1}$, respectively. Increasing $\alpha$ from \num{0} to \num{1} results in a sequence of new slices where each subsequent slice is progressively less semantically similar to $x_n$ and more semantically similar to $x_{n+1}$. As a result, anatomy in the obtained stack of slices also appears semantically smooth in the direction from which the slices were extracted. We refer to this approach as Autoencoding for Semantic Interpolation (ASI).

	To attain upsampling of anisotropic images by factor $K$ in through-plane direction, $K-1$ slices need to be synthesized where the set of alpha values $\mathcal{A}$ is defined as follows:
	
	\begin{equation}  \label{eq_alpha_set}
	\begin{split}
	\mathcal{A} = \Big\{\alpha_i | \alpha_i = \frac{i}{K} \Big\}_{i=1}^{K-1} \;,\; &\text{where} \; K = \{m | m \in \mathbb{N}, m > 1 \} \\
	\text{and} \; |\mathcal{A}| = K - 1 \; ,
	\end{split}
	\end{equation}
	
	\noindent and $|.|$ denotes the cardinality of a set.

	\subsection{Loss Function}\label{method_combined_loss}
	
	The autoencoder is trained to compress and reconstruct high-resolution slices taken from an anisotropic \num{3}D medical imaging dataset. The model aims to minimize the reconstruction loss between the original $x$ and the reconstructed $\hat{x}$ slice. % $\mathcal{L}_{\text{reconstruction}}$
	
	To further constrain the model a notion of semantic similarity is defined for a given dataset. For this, the spatial relationship between slices of the same volume is exploited.
	% More specific, synthesized images are enforced to provide a semantically smooth morphing between two neighboring images. 
	% Moreover, to enforce synthesized images i.e. intermediate points along the interpolation, to provide a semantically smooth morphing between two images (interpolation endpoints) a notion of semantic similarity is defined for a given dataset. For this, the spatial relationship between neighboring slices in image volumes is exploited. 
	%To improve performance of the proposed method we exploit the spatial relationship between neighboring slices in image volumes. 
	During training an existing \textit{in-between} slice $x_{n}$ (where $n \in \mathbb{N}^{+}$) that has two neighboring slices, $x_{n-1}$ and $x_{n+1}$, is synthesized using a convex combination of the neighboring slice encodings where $\alpha$ (the mixing coefficient) of Equation~\ref{eq_convex_combination} is set to \num{0.5}. The new slice encoding is mapped through the decoder to an approximation $\hat{x}_{n}^{\alpha=0.5}$ of the original in-between slice $x_{n}$. 
	% $\hat{x}_{n}^{\alpha=0.5}=x_{n-1, n+1}^{\alpha=0.5}$
	
	Finally, a distance loss is computed between the original in-between $x_{n}$ slice and its approximation i.e. synthesized slice $\hat{x}_{n}^{\alpha=0.5}$ resulting in the following combined loss: 
	
	\begin{align}\label{eq_combined_loss}
	\begin{split}
	\mathcal{L} &= \underbrace{d(x_{n}, \hat{x}_n)}_{\mathcal{L}_{\text{reconstruction}}} + \lambda \; \underbrace{d(x_{n}, \hat{x}_{n}^{\alpha=0.5})}_{\mathcal{L}_{\text{synthesis}}} \; \text{where} \; \lambda \geq 0 ,\\
	\end{split}
	\end{align}
	
	\noindent $d$ denotes a distance function between two images and $\lambda$ is a hyperparameter weighting the contribution of the synthesis loss. Setting $\lambda$ to zero disables the synthesis loss during training.
	Minimizing the synthesis loss during training should encourage the autoencoder to linearize the latent space of images. Therefore, a convex combination of slice encodings should result in smooth nonlinear interpolations in image space. % Figure~\ref{fig_loss_synthesis} visualizes computation of the synthesis loss term.
	
	To compute the reconstruction loss, this work used the pixel-wise mean squared error between original $x_{n}$ and reconstructed $\hat{x}_{n}$ image. In addition, to compute the synthesis loss between reference $x_{n}$ and synthesized image $\hat{x}_{n}^{\alpha=0.5}$ the Learned Perceptual Image Patch Similarity (LPIPS) metric (\cite{zhang2018perceptual}) was used. The LPIPS metric is a perceptually-based pairwise image distance that is calculated as a weighted difference between the VGG-16 (\cite{simonyan2014very}) embedding of the reference and synthesized image. LPIPS uses the embeddings of VGG-16 layers \texttt{conv\_1} to \texttt{conv\_5}. The VGG-16 CNN is pretrained on ImageNet and the weights to compute the weighted difference were fit so that the metric agrees with human perceptual similarity judgments.

	% intermediate points provide a semantically smooth morphing between the endpoints. The latter characteristic is hard to enforce because it requires defining a notion of semantic similarity for a given dataset, which is often hard to explicitly codify.
	%It is important to note that these paths do not go through an “average” image of the dataset as the path interpolation between the 3 images of Figure 2 shows.
	
	\begin{figure*}
		\captionsetup[subfigure]{justification=centering, labelformat=empty}
		\setlength{\tabcolsep}{1pt}
		\renewcommand{\arraystretch}{0.6} 
		\begin{center}
			\subfloat[]{
				\begin{tabular}{c c c c c c}
					% row 1  References
					& \num{0}$^{\circ}$ & \num{10}$^{\circ}$ & \num{20}$^{\circ}$ & \num{30}$^{\circ}$ & \num{40}$^{\circ}$  \\
					%\begin{tabular}{@{}c@{}} Original \\ \phantom{x} \end{tabular} & \begin{tabular}{@{}c@{}} Original \\  \num{10}$^{\circ}$ \end{tabular}  & \begin{tabular}{@{}c@{}} Original  \\ \num{20}$^{\circ}$ \end{tabular}  &  \begin{tabular}{@{}c@{}} Original  \\ \num{30}$^{\circ}$ \end{tabular} & \begin{tabular}{@{}c@{}} Original  \\ \num{40}$^{\circ}$ \end{tabular} \\
					\begin{tabular}{@{}c@{}} Original  \\ images \\ \phantom{x} \\ \phantom{x} \end{tabular} &
					\subfloat[Reconstructed]{\includegraphics[width=.05\textwidth]{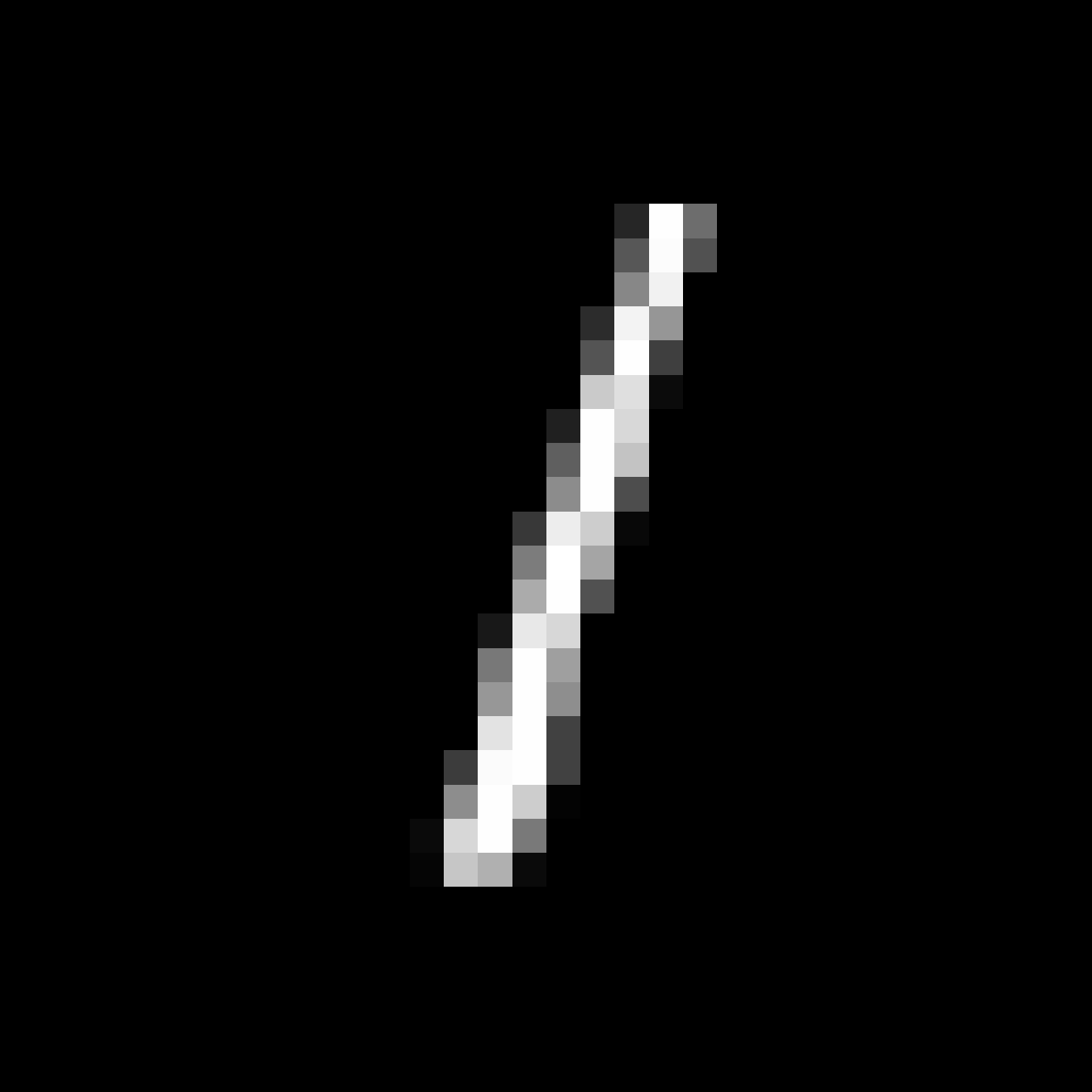} } &
					\subfloat[$\alpha=0.25$]{\includegraphics[width=.05\textwidth]{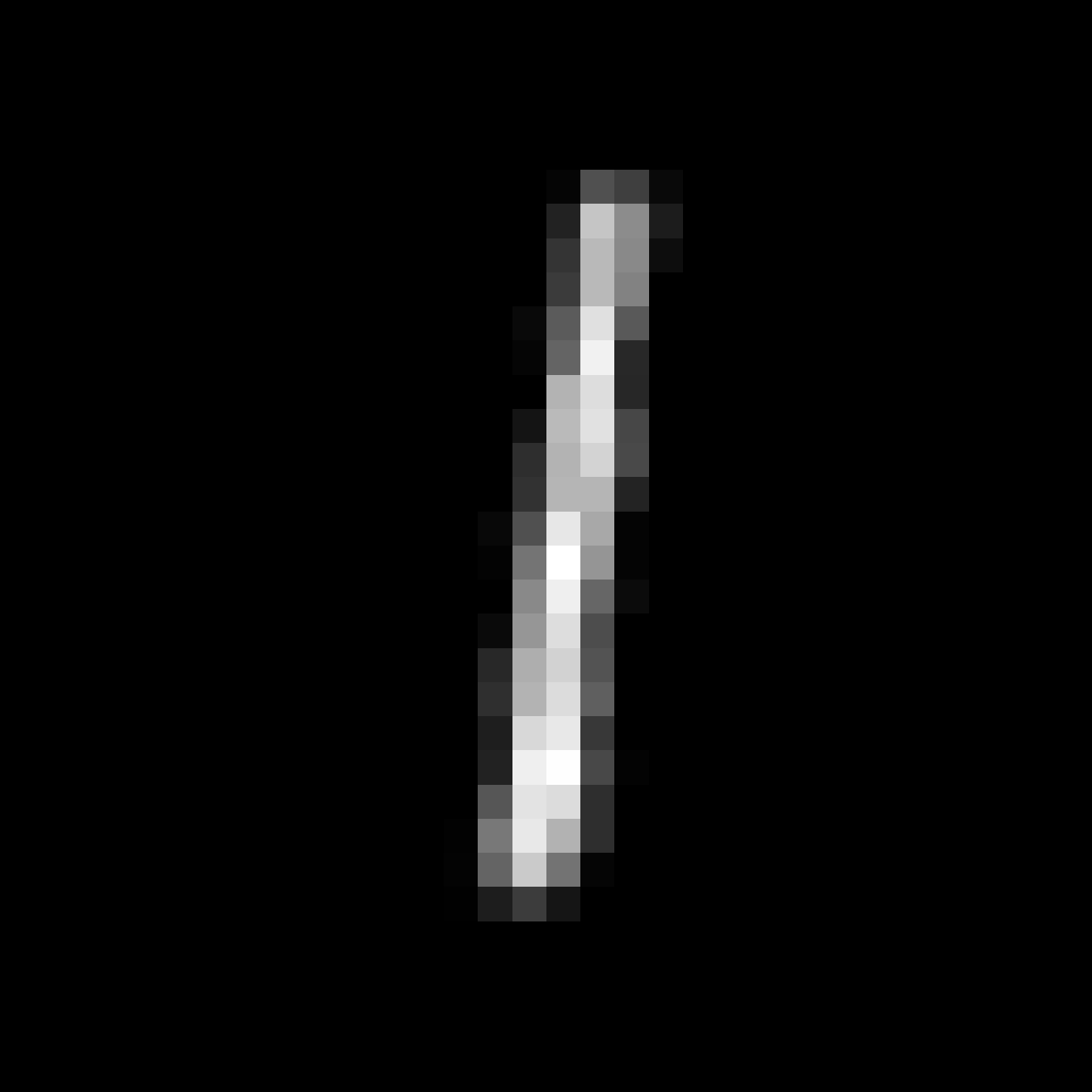} } &
					\subfloat[$\alpha=0.5$]{\includegraphics[width=.05\textwidth]{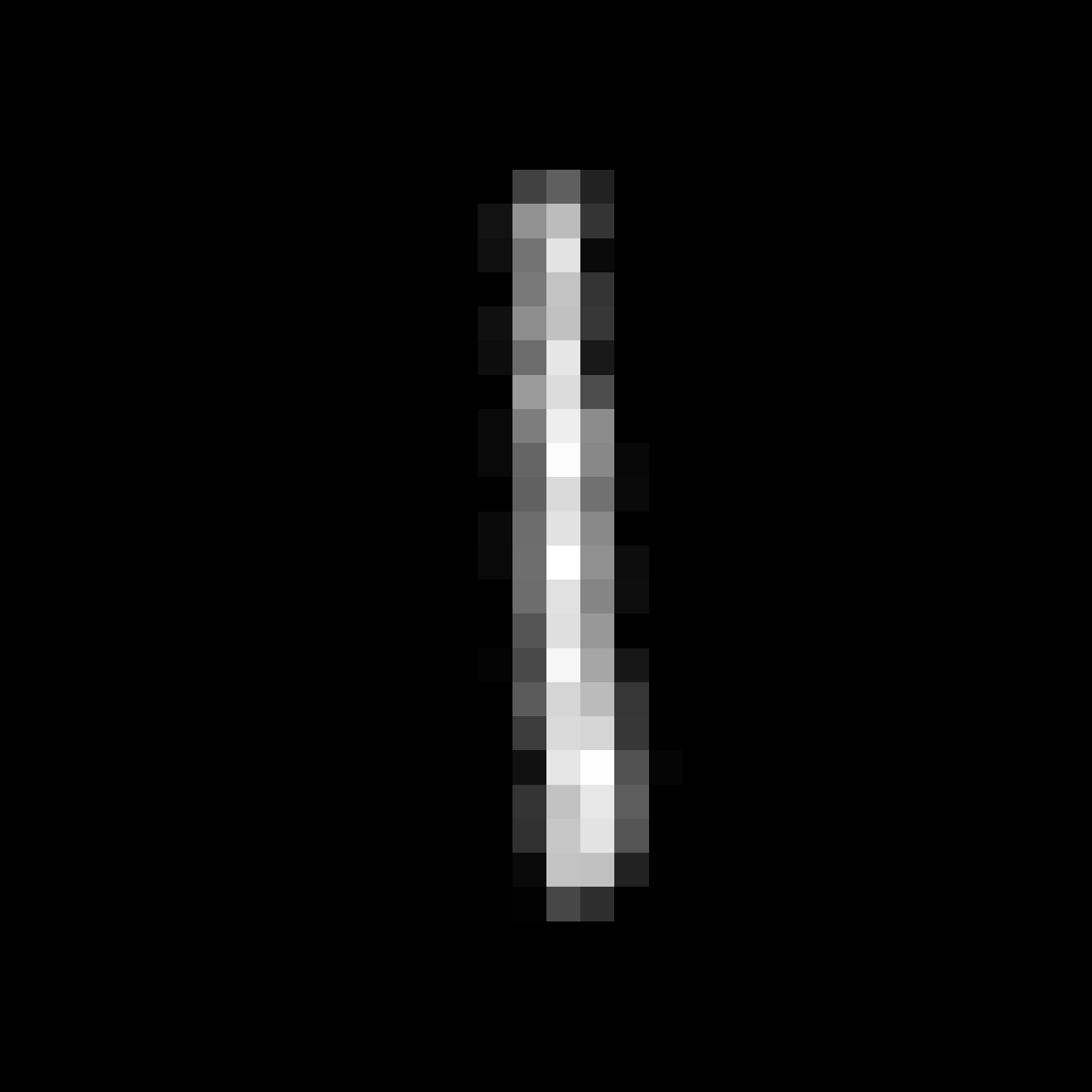} } &
					\subfloat[$\alpha=0.75$]{\includegraphics[width=.05\textwidth]{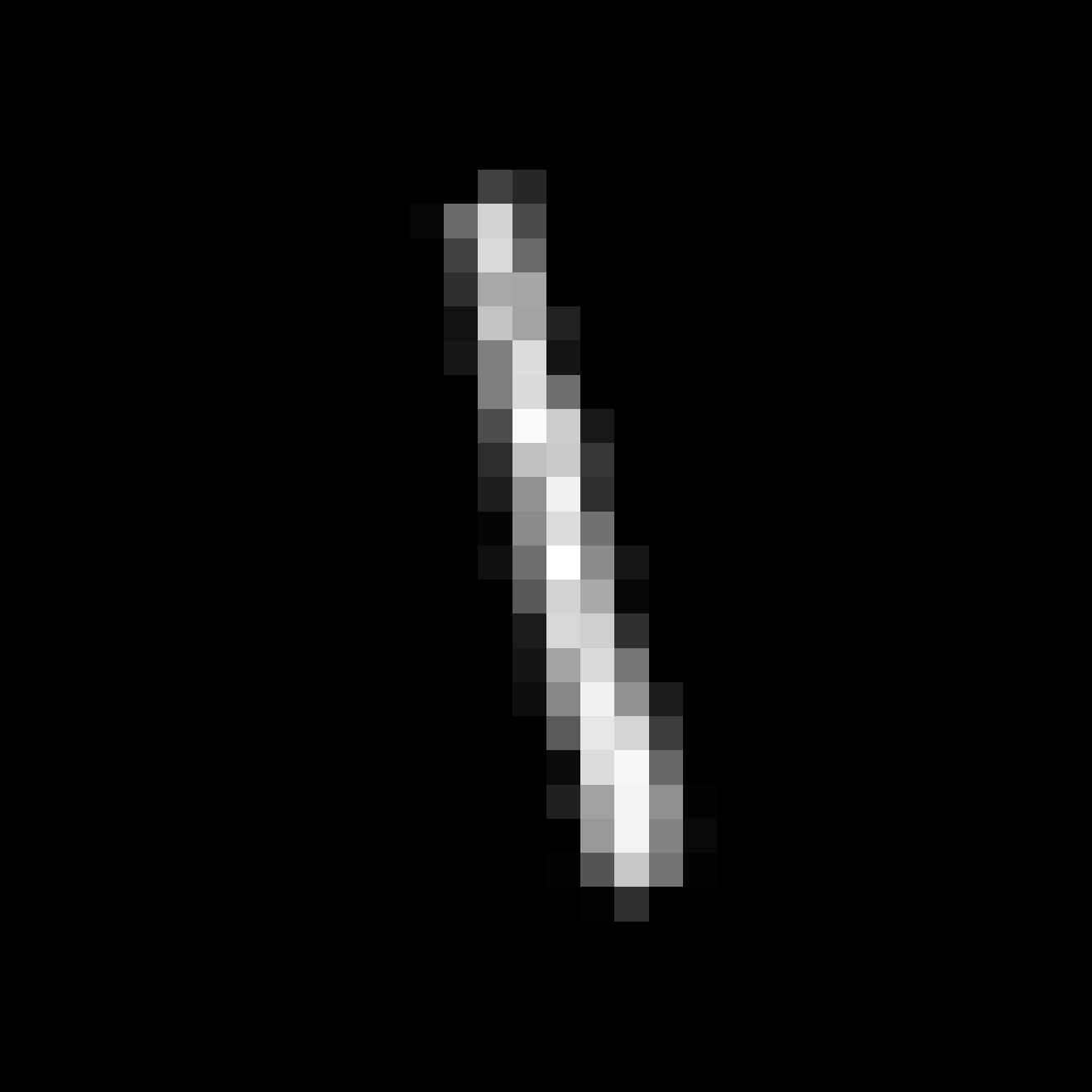} } &
					\subfloat[Reconstructed]{\includegraphics[width=.05\textwidth]{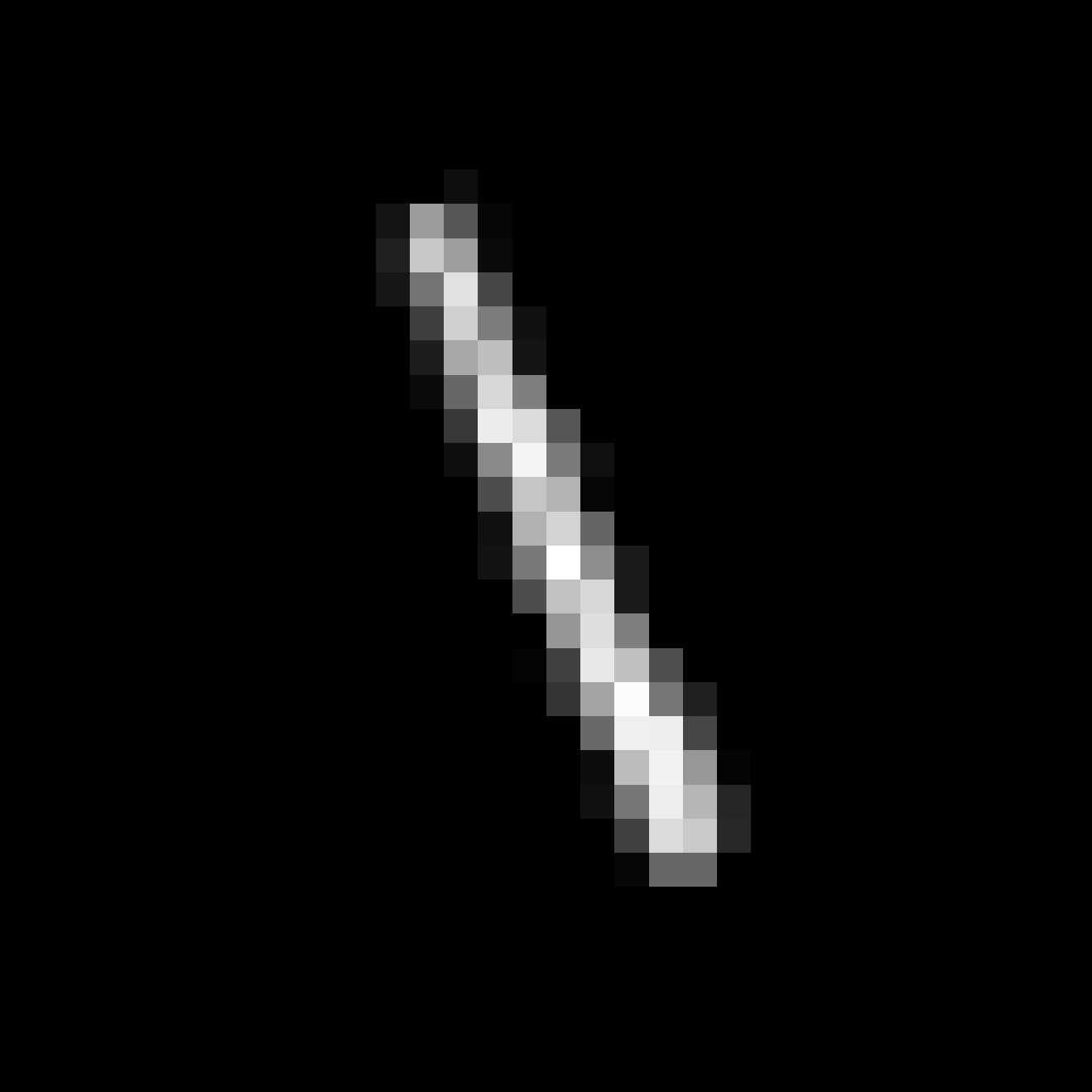} }  \\
					
					% row VAE	
					\begin{tabular}{@{}c@{}} VAE \\ \\    \phantom{x}  \end{tabular}  &
					\subfloat[]{\includegraphics[width=.05\textwidth]{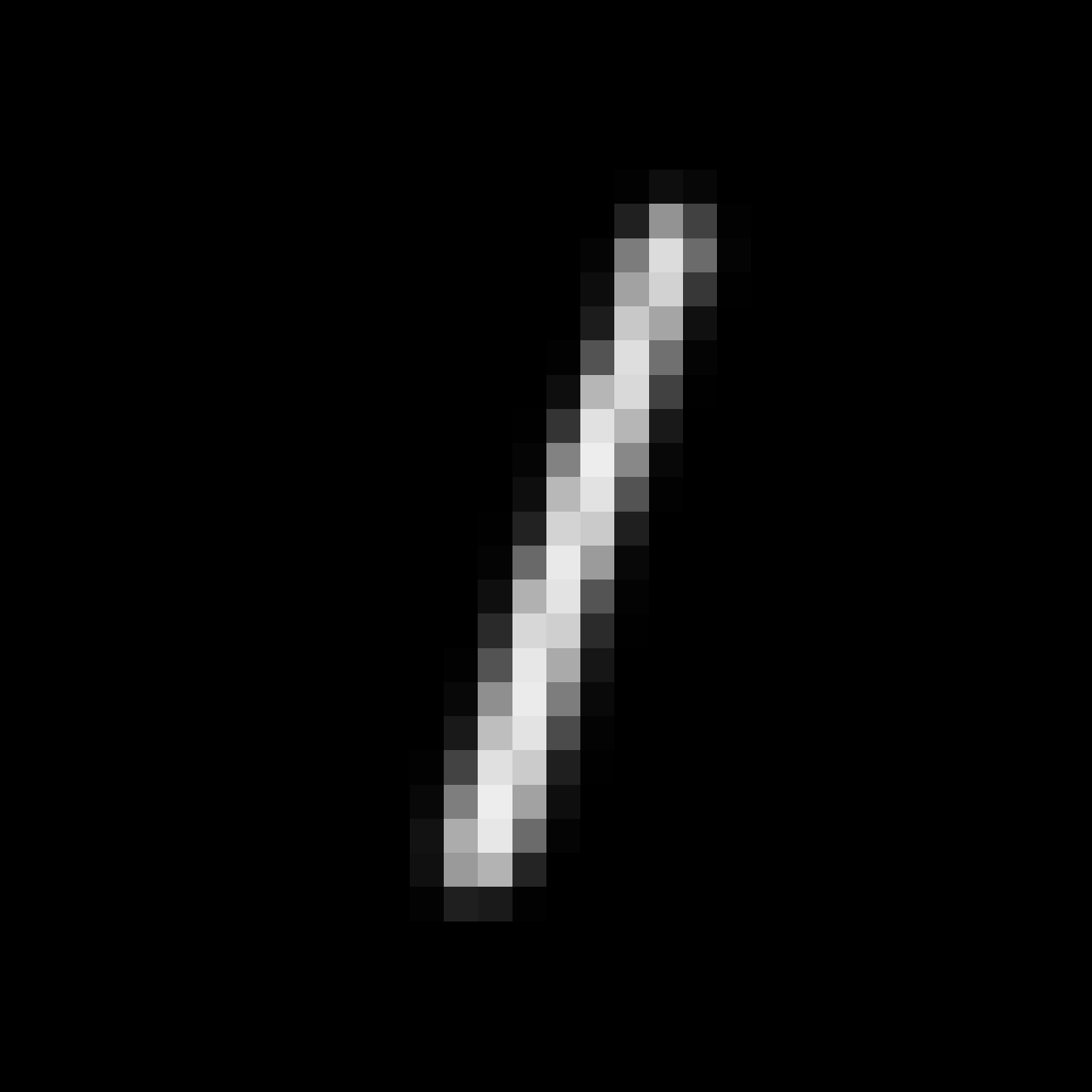} } &
					\subfloat[ ]{\includegraphics[width=.05\textwidth]{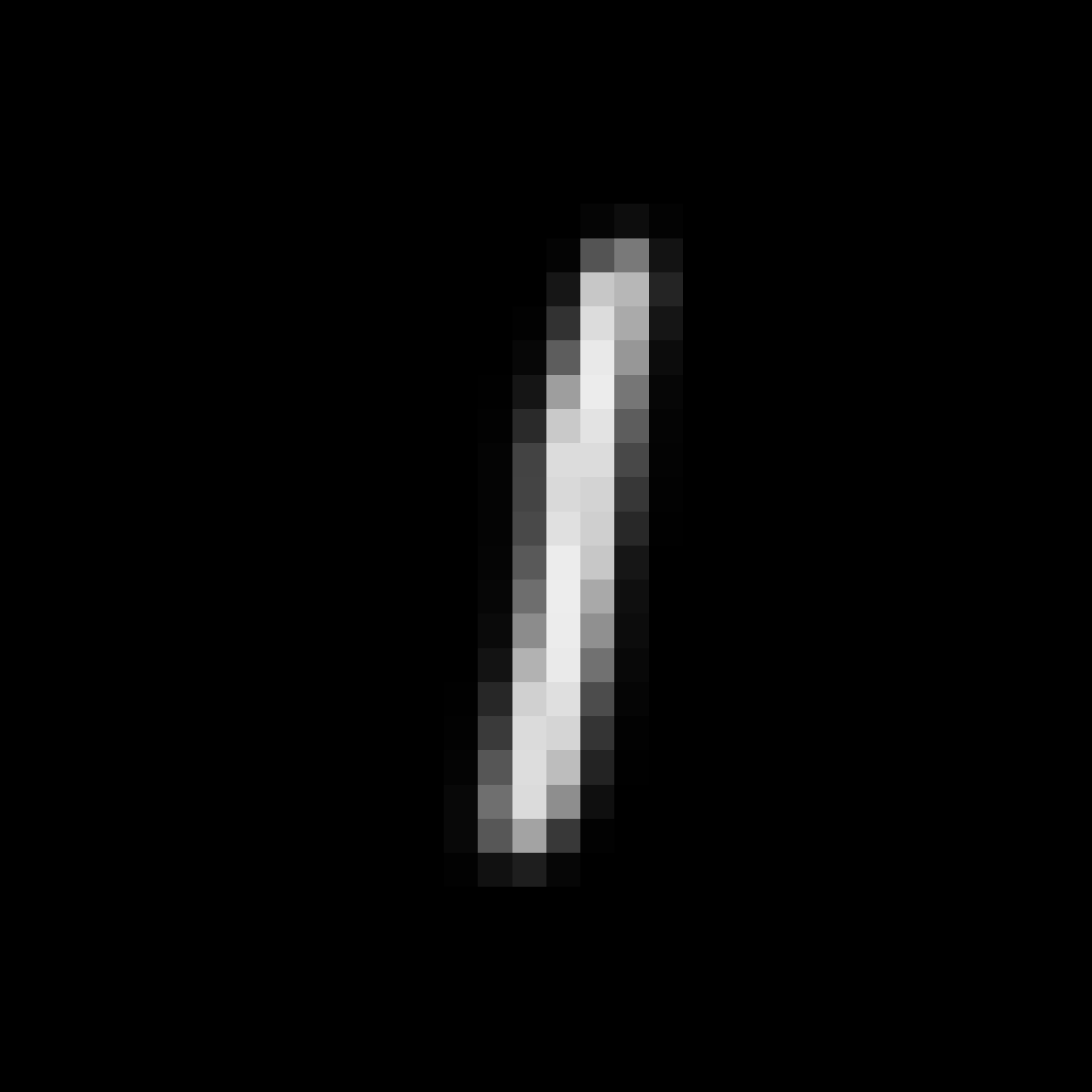} } &
					\subfloat[]{\includegraphics[width=.05\textwidth]{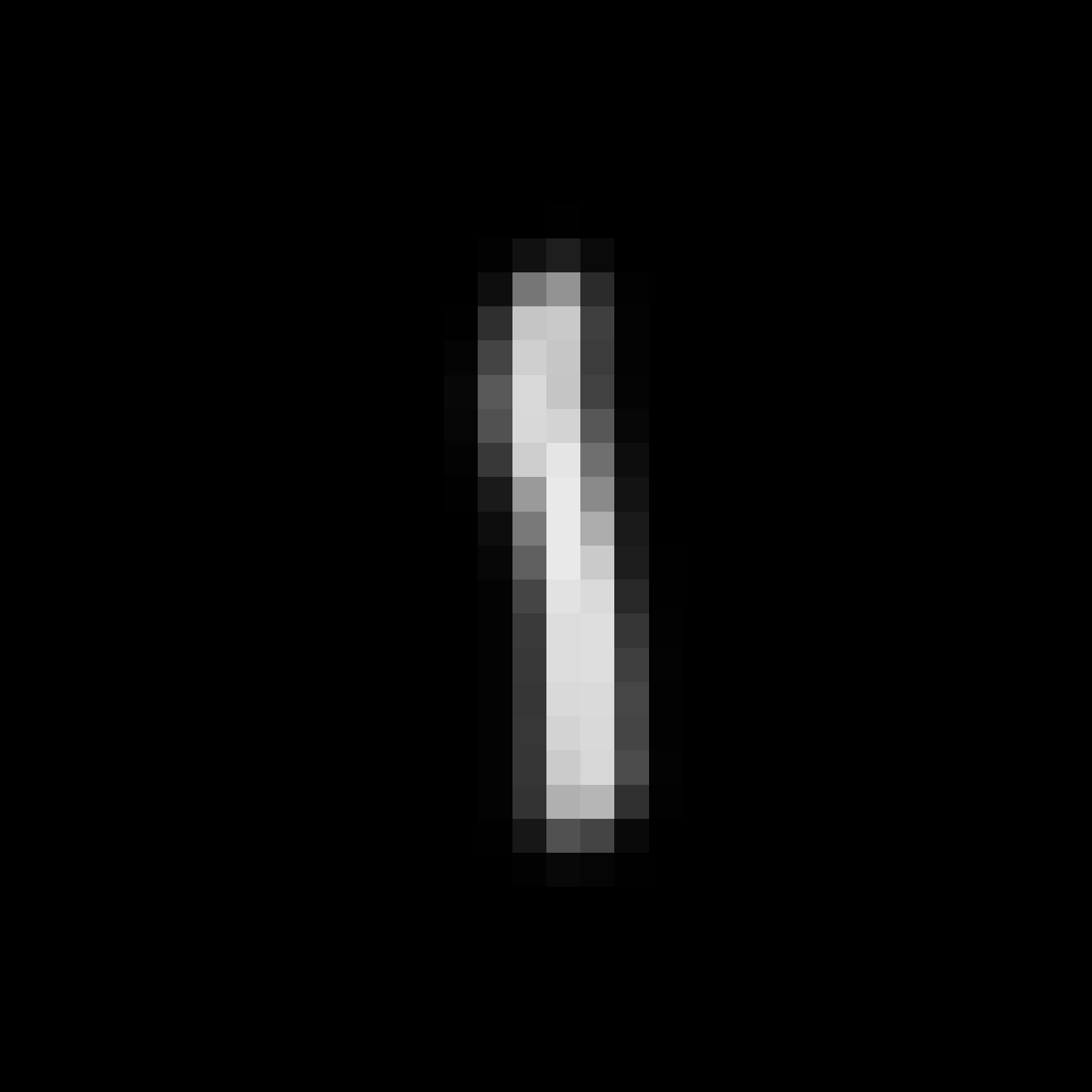} } &
					\subfloat[]{\includegraphics[width=.05\textwidth]{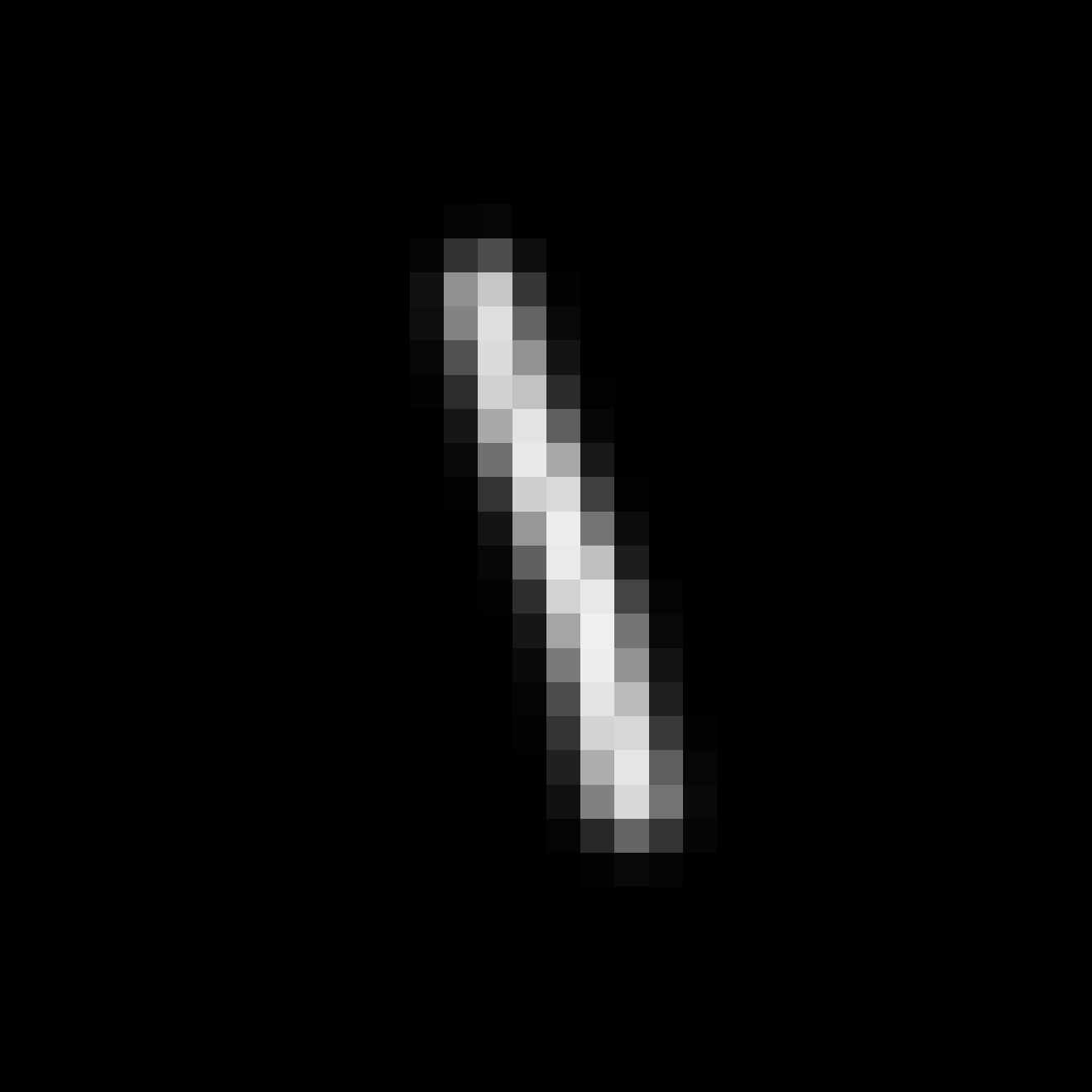} } & 
					\subfloat[]{\includegraphics[width=.05\textwidth]{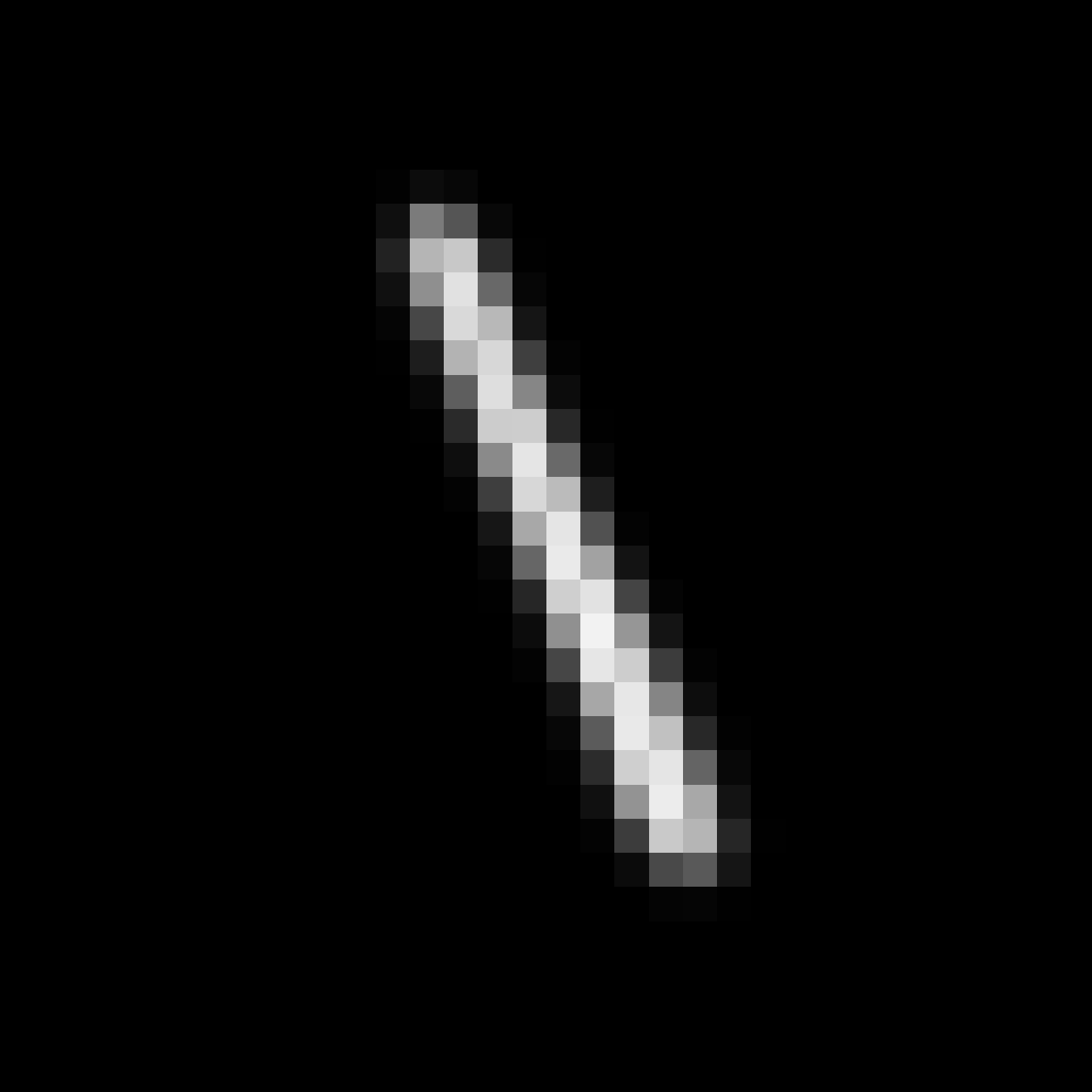} } \\
					
					% row ACAI	
					\begin{tabular}{@{}c@{}} ACAI   \\ \\   \phantom{x}  \end{tabular} & 
					\subfloat[]{\includegraphics[width=.05\textwidth]{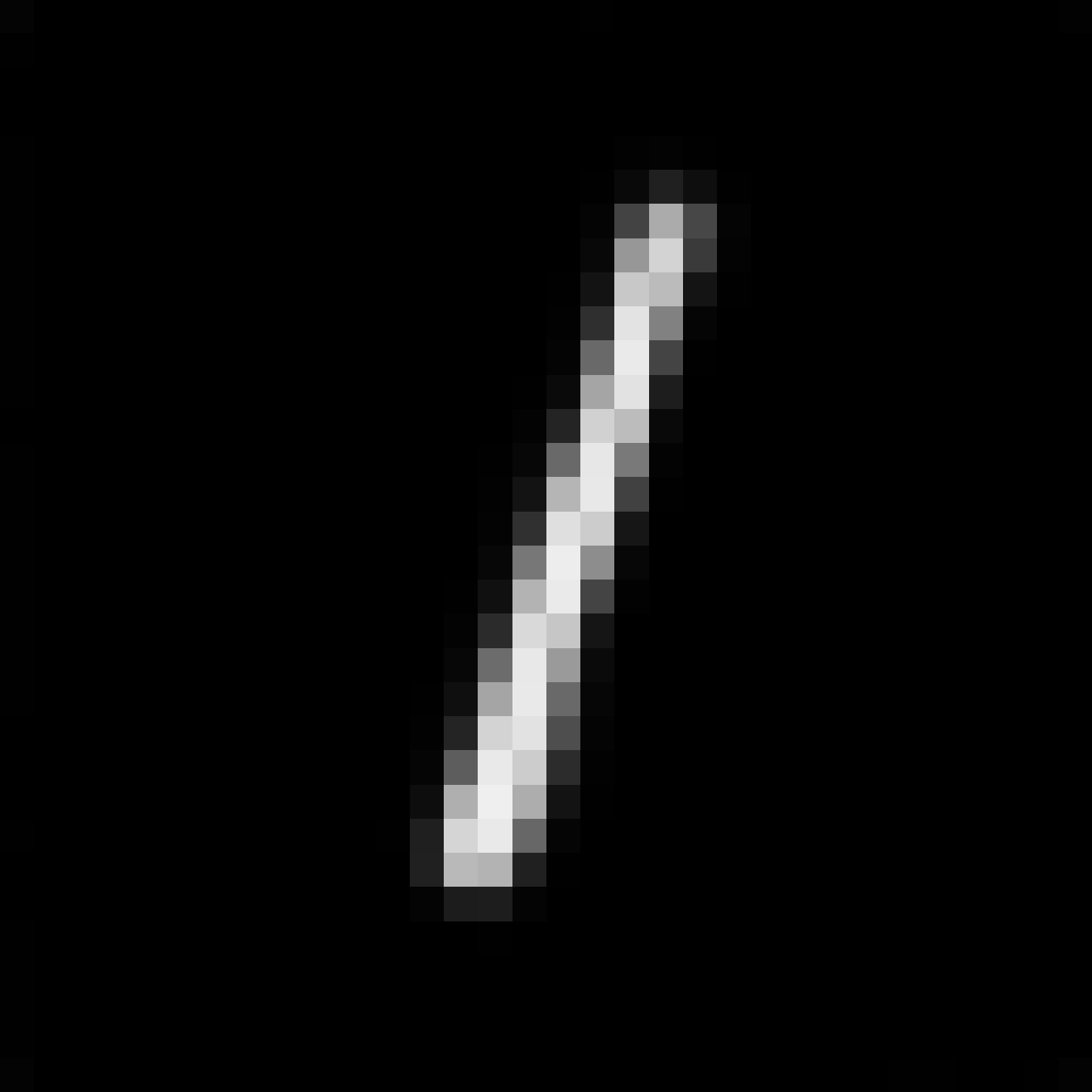} } &
					\subfloat[]{\includegraphics[width=.05\textwidth]{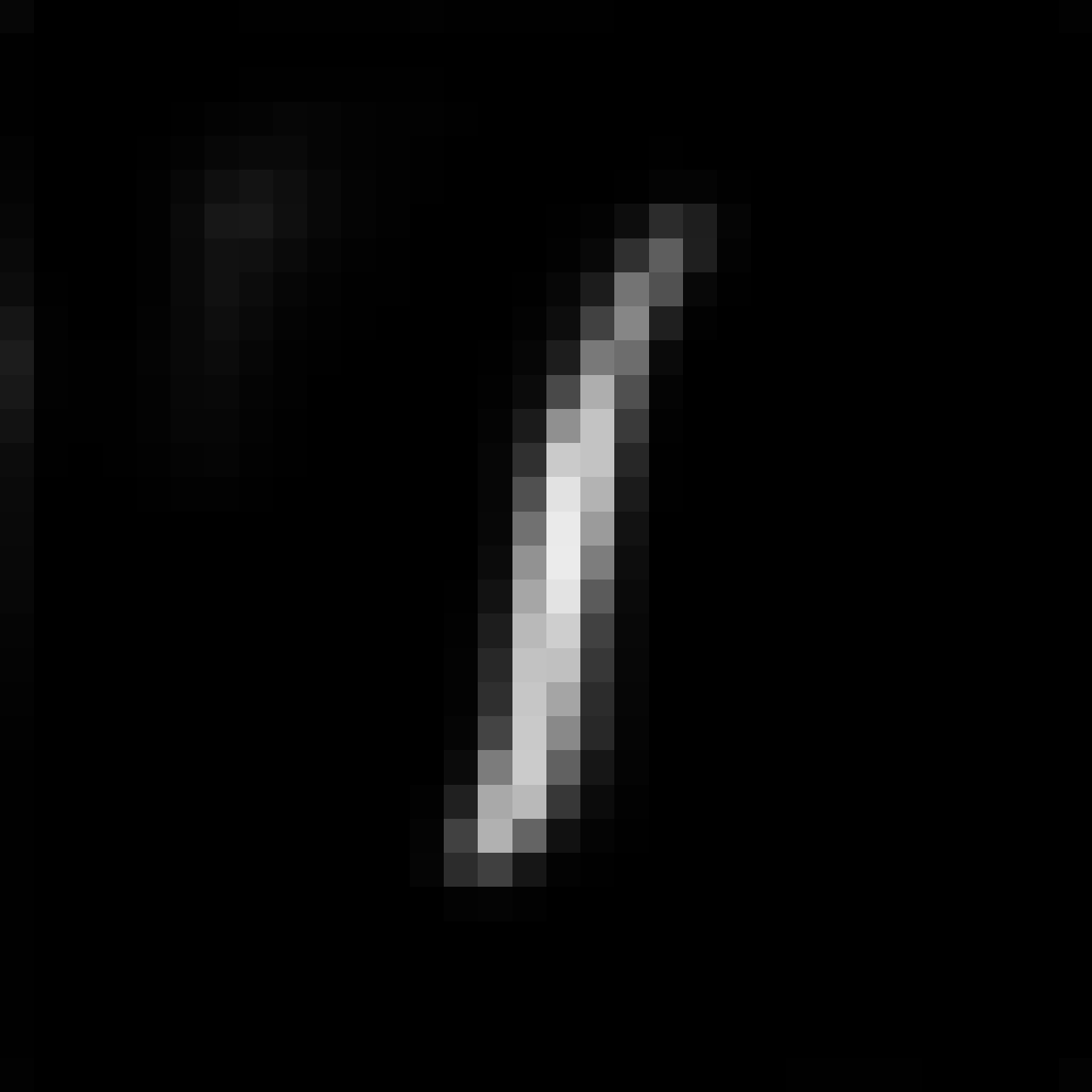} } &
					\subfloat[]{\includegraphics[width=.05\textwidth]{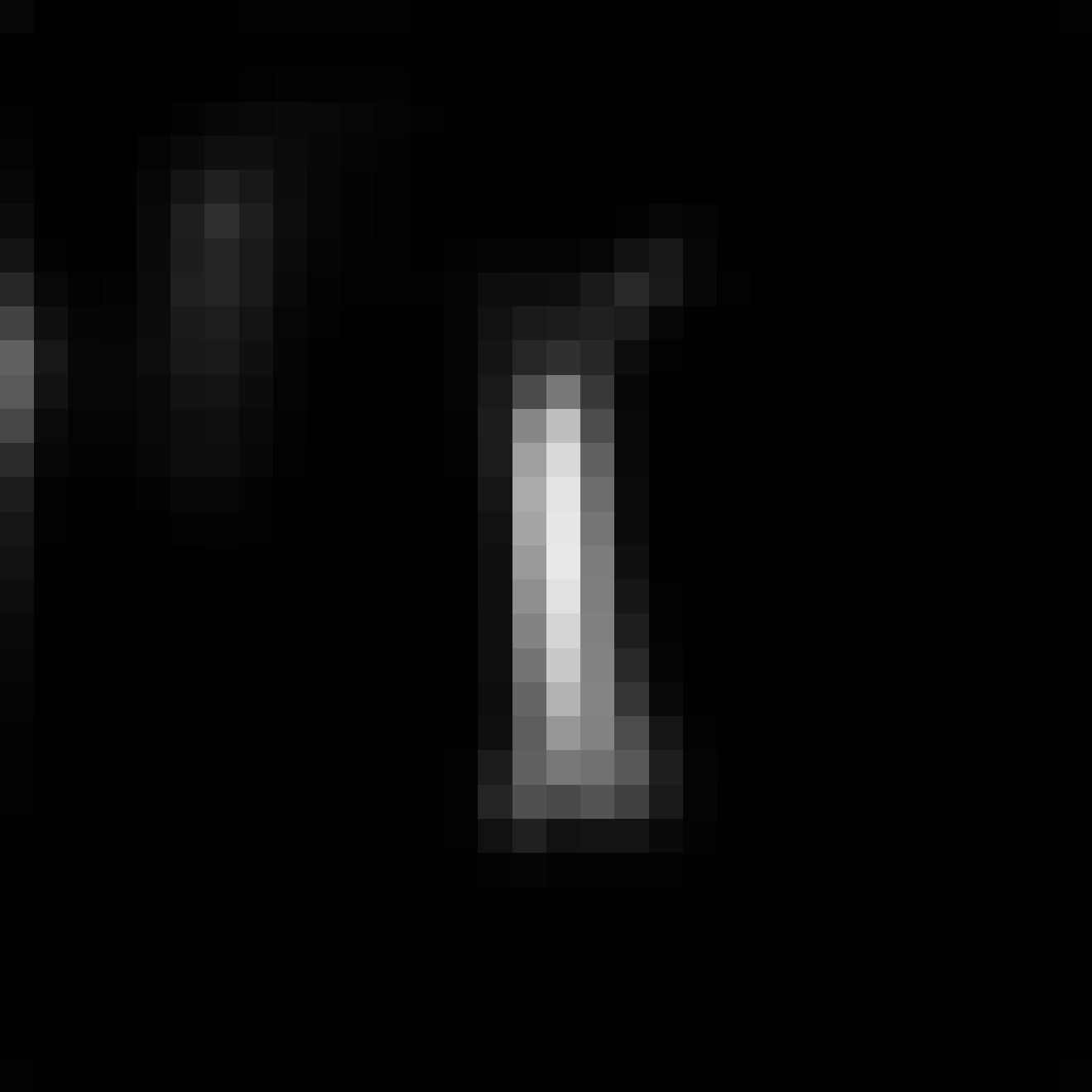} } &
					\subfloat[]{\includegraphics[width=.05\textwidth]{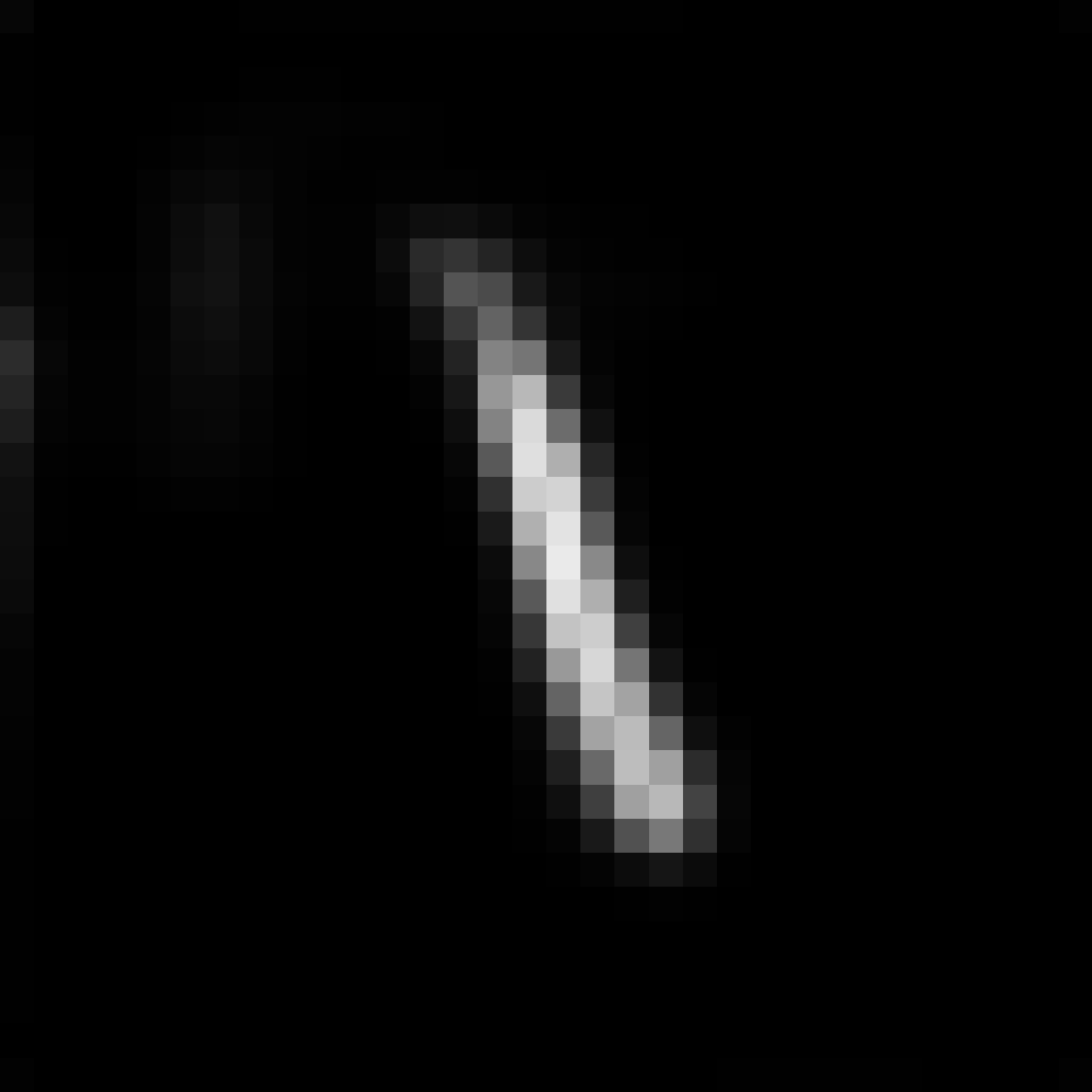} } & 
					\subfloat[]{\includegraphics[width=.05\textwidth]{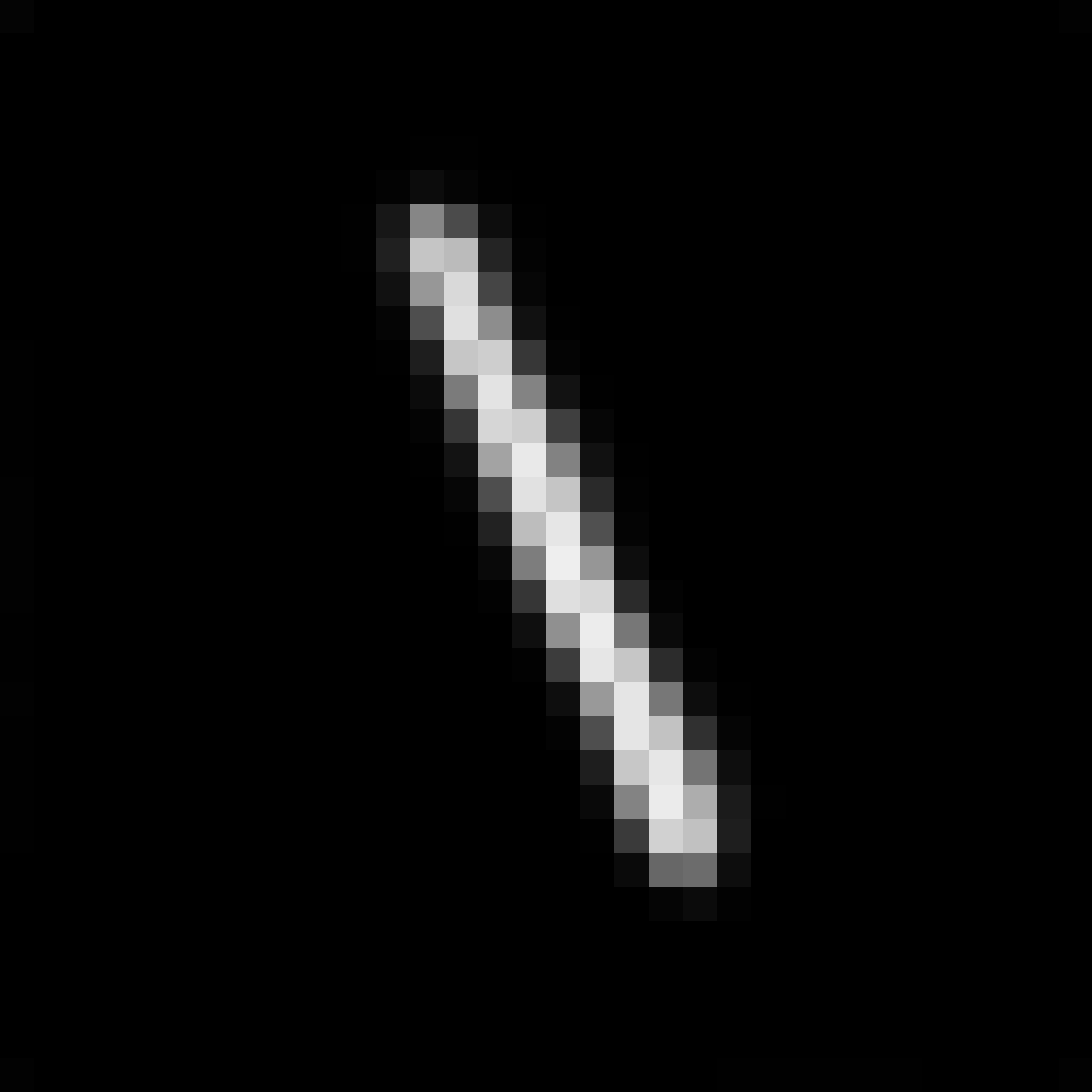} } \\
					
					% row ours	
					\begin{tabular}{@{}c@{}} ours \\  \\   \phantom{x} \end{tabular} &
					\subfloat[]{\includegraphics[width=.05\textwidth]{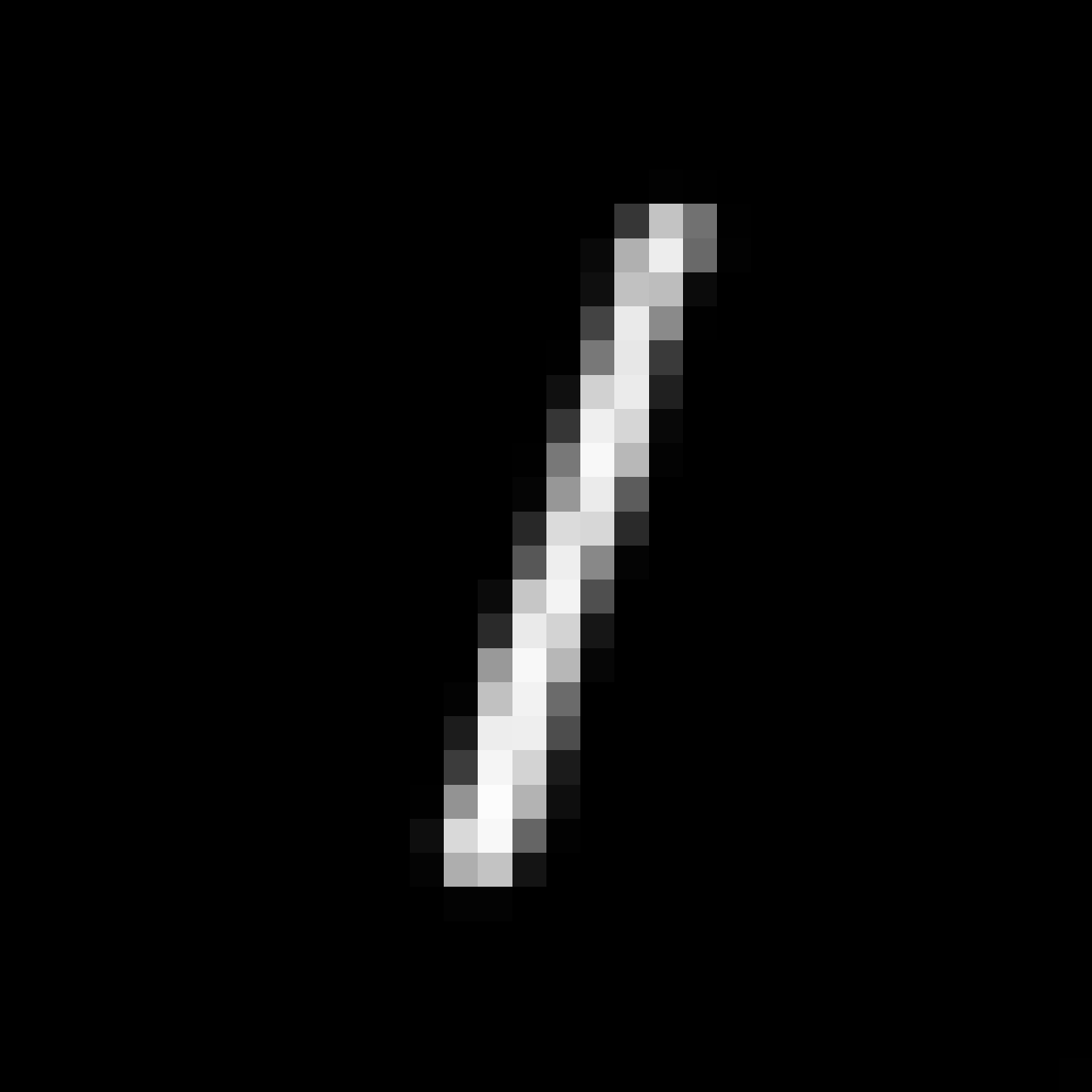} } &
					\subfloat[]{\includegraphics[width=.05\textwidth]{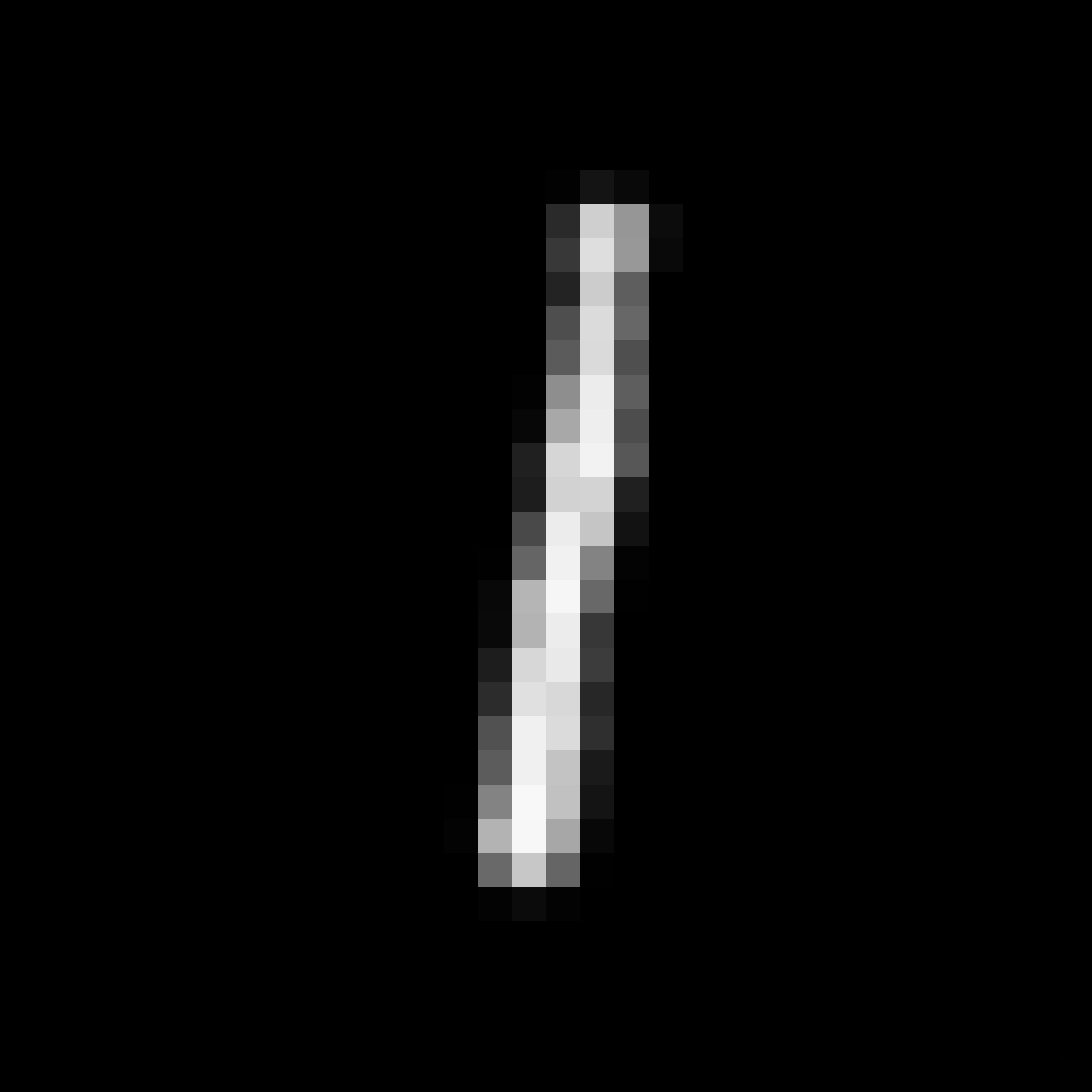} } &
					\subfloat[]{\includegraphics[width=.05\textwidth]{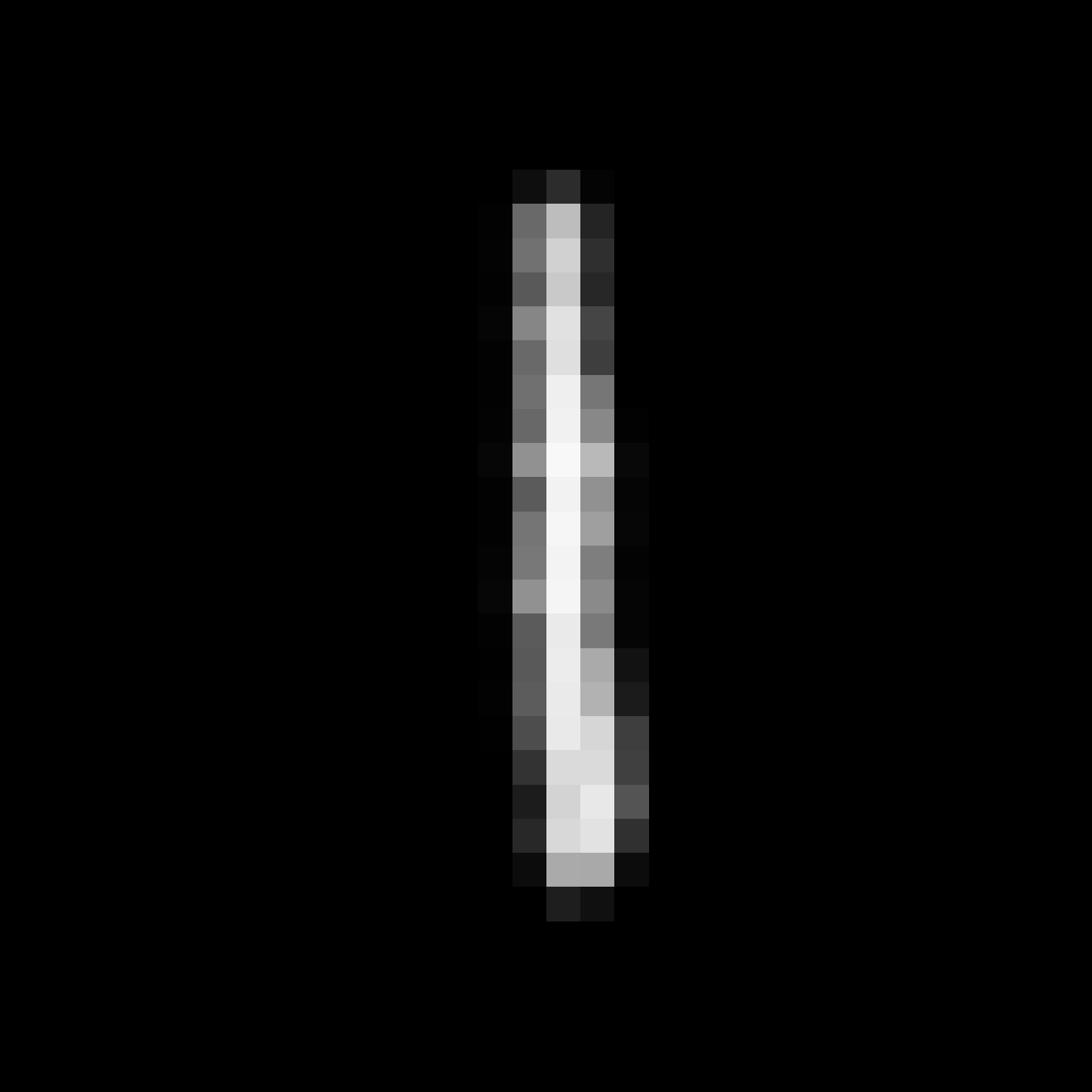} } &
					\subfloat[]{\includegraphics[width=.05\textwidth]{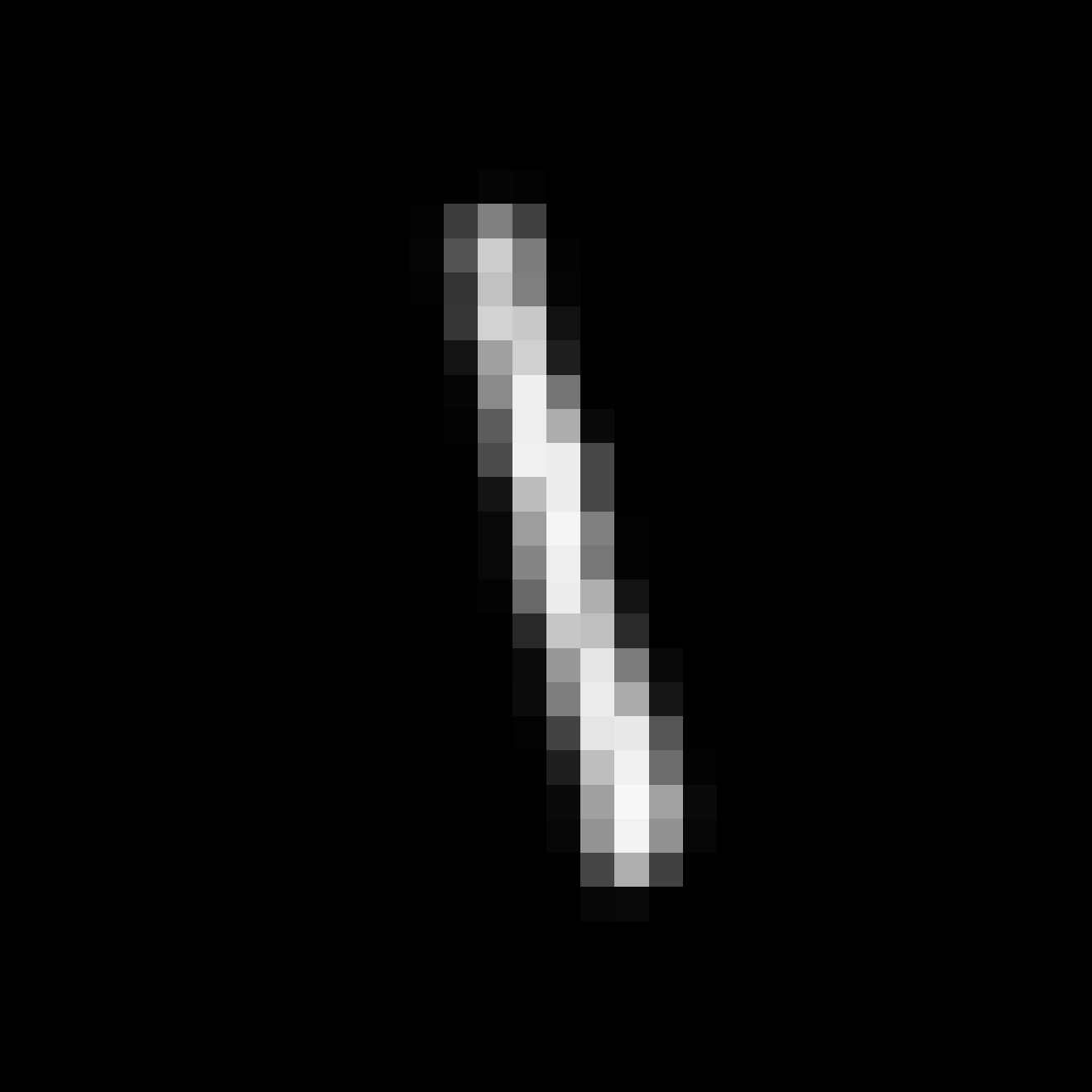} } & 
					\subfloat[]{\includegraphics[width=.05\textwidth]{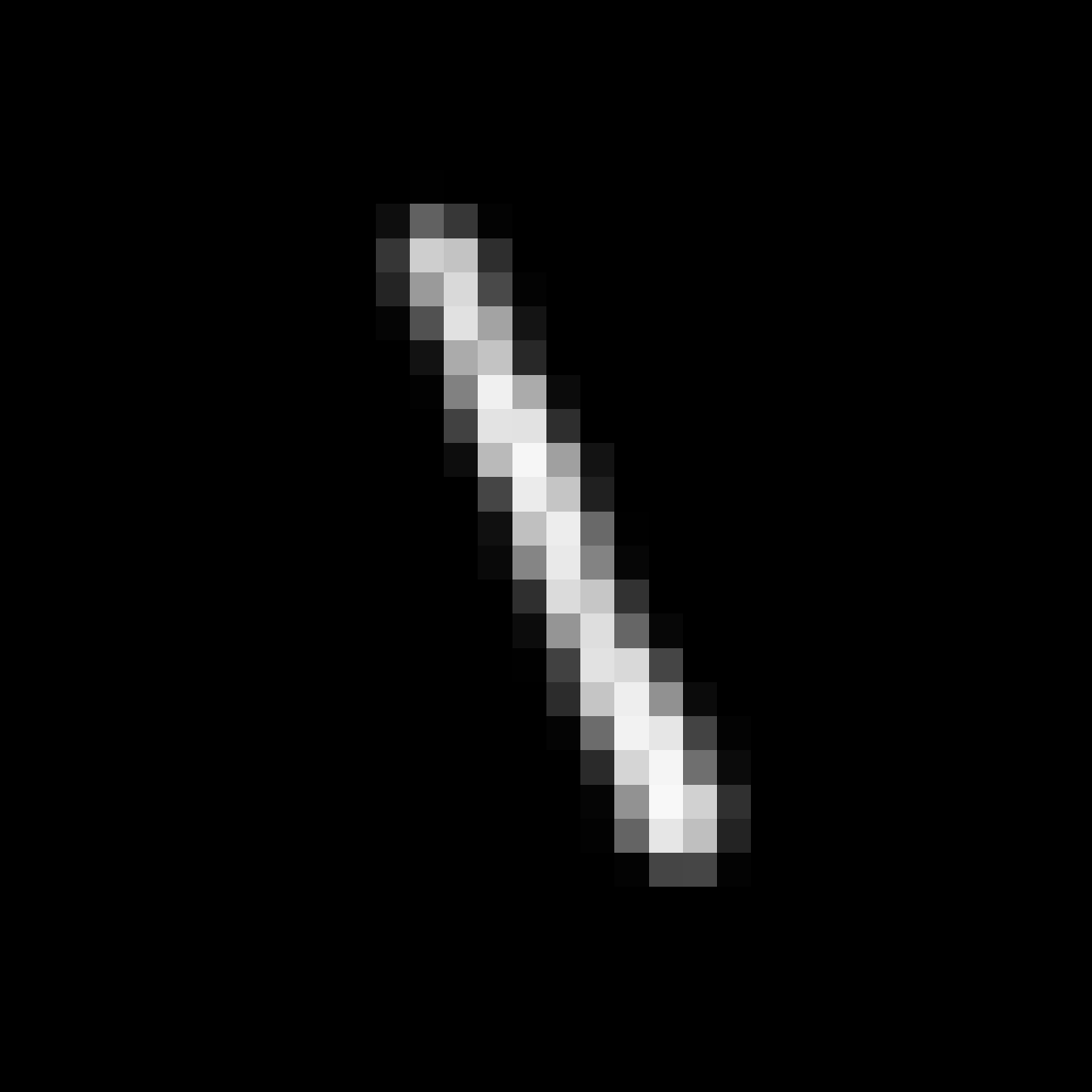} } \\
					
				\end{tabular}
			}  % end subfloat
			% --------------------------------- next MNIST block
			\subfloat[]{
				\begin{tabular}{c c c c c}
					% row Reference
					\num{0}$^{\circ}$ & \num{10}$^{\circ}$ & \num{20}$^{\circ}$ & \num{30}$^{\circ}$ & \num{40}$^{\circ}$  \\
					\subfloat[Reconstructed]{\includegraphics[width=.05\textwidth]{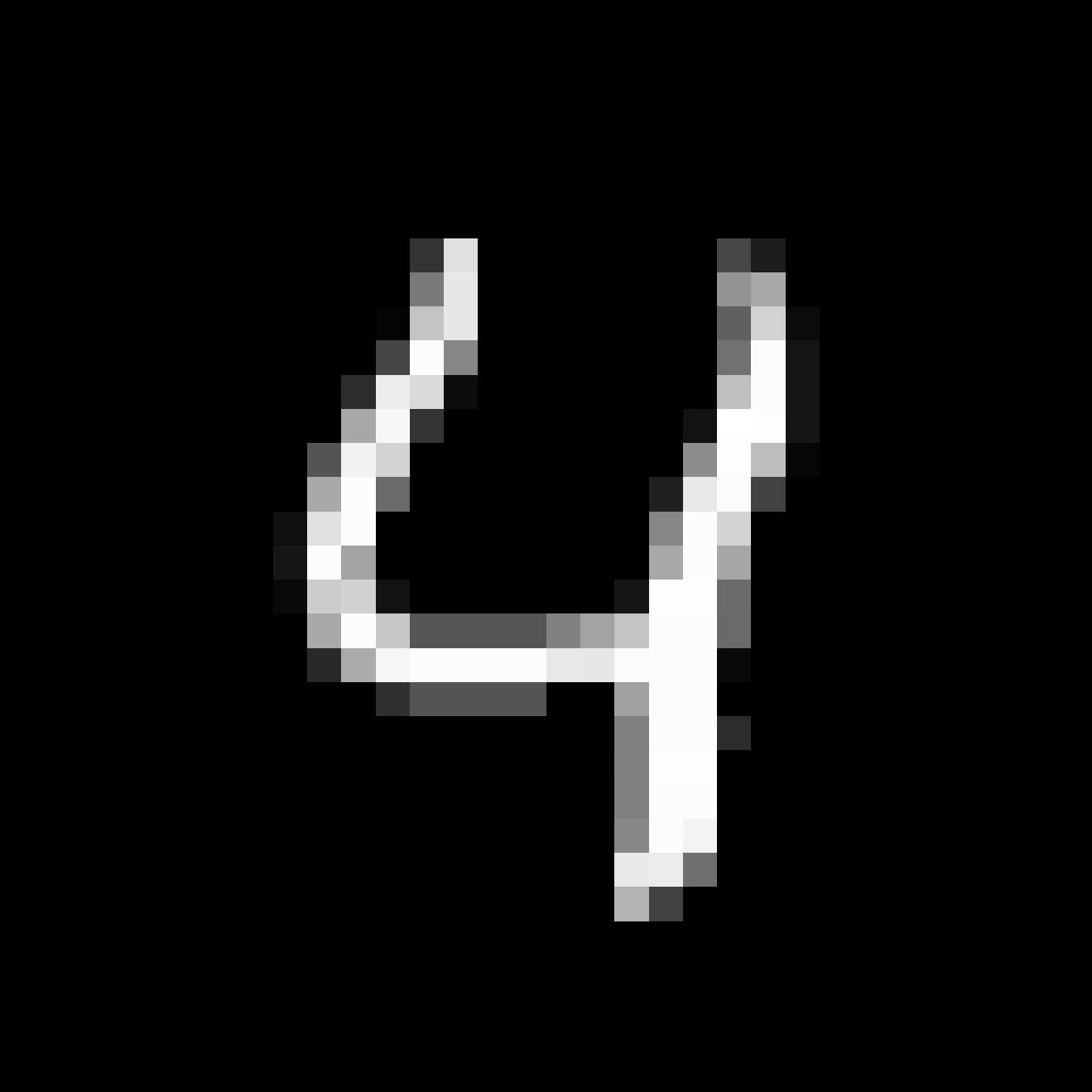} } &
					\subfloat[$\alpha=0.25$]{\includegraphics[width=.05\textwidth]{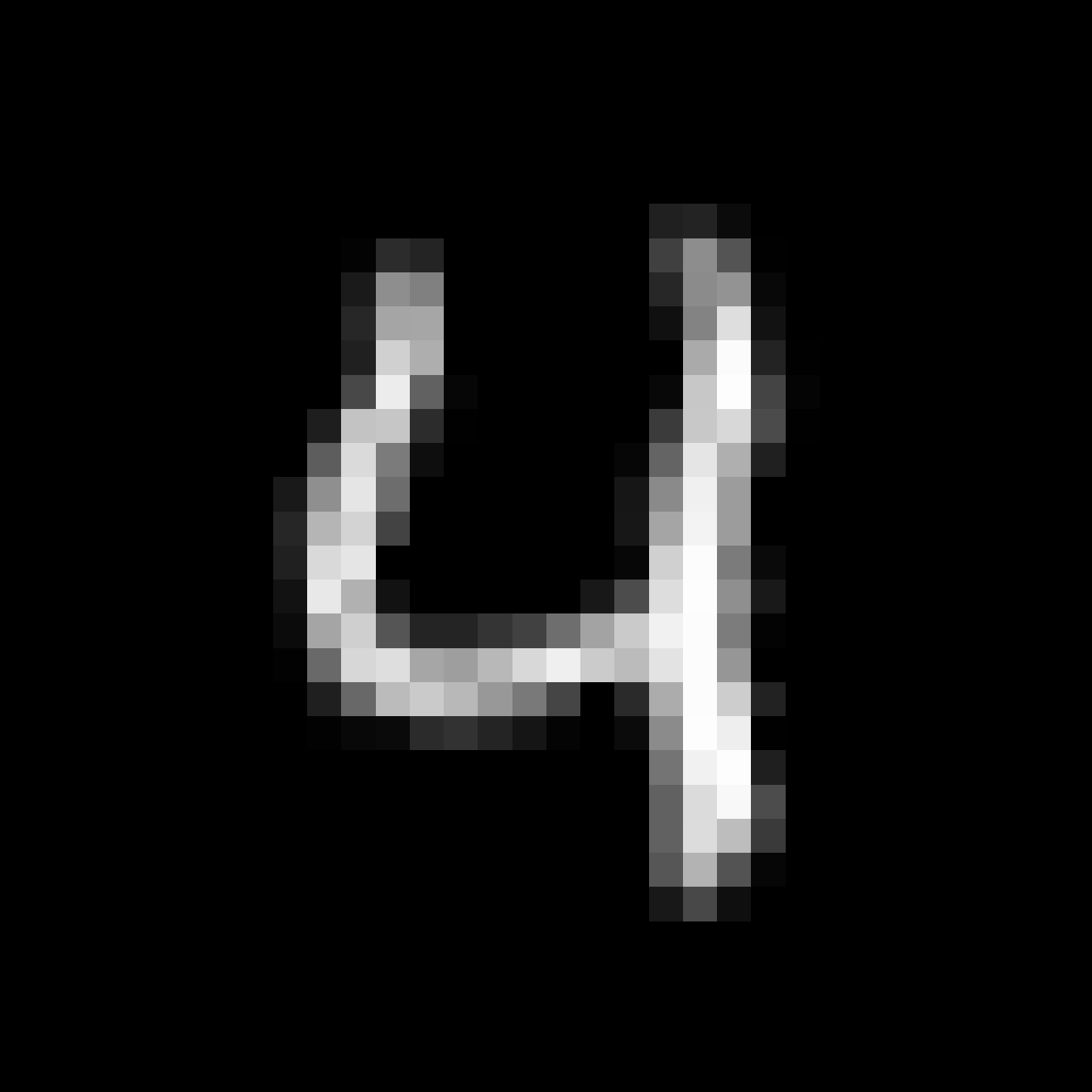} } &
					\subfloat[$\alpha=0.5$]{\includegraphics[width=.05\textwidth]{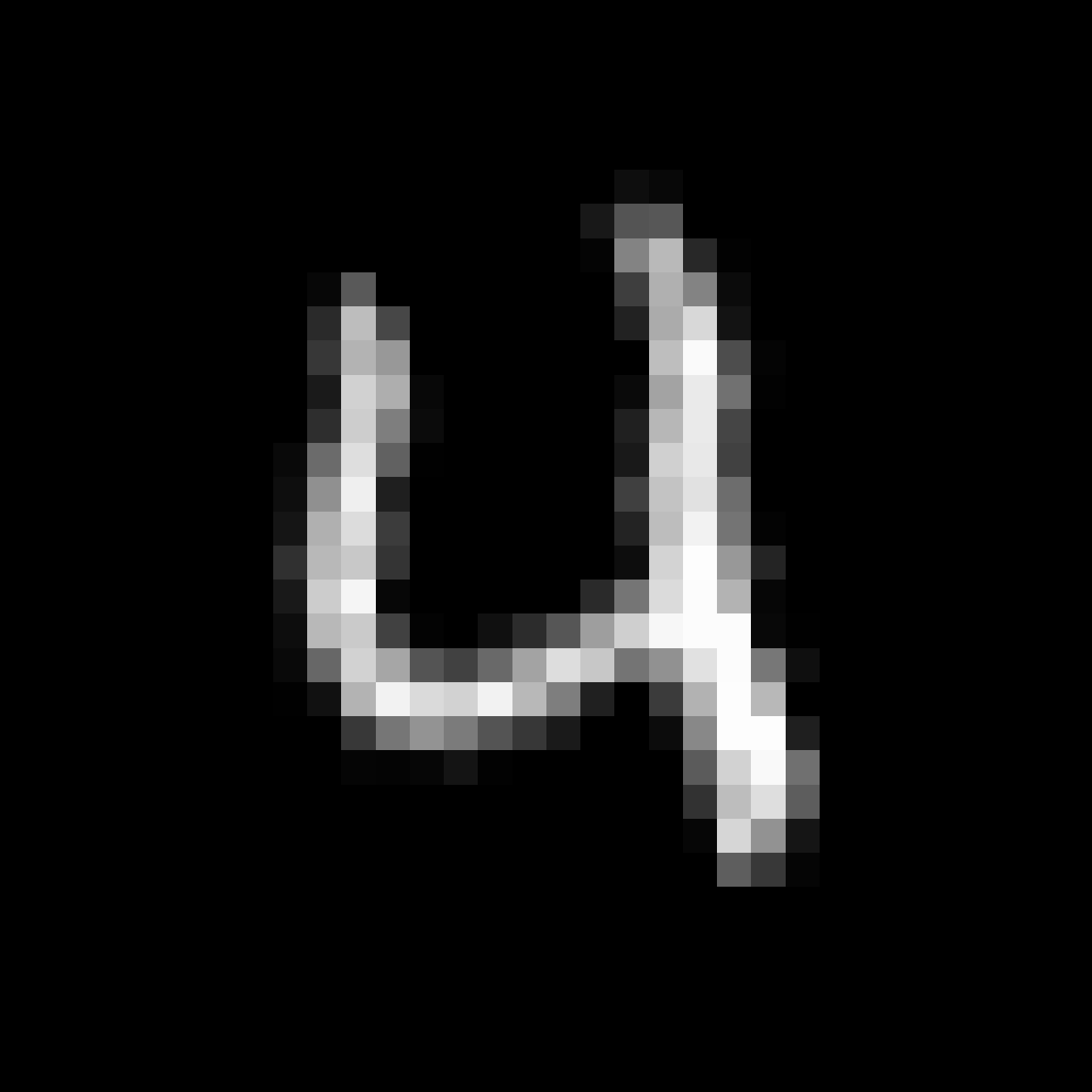} } &
					\subfloat[$\alpha=0.75$]{\includegraphics[width=.05\textwidth]{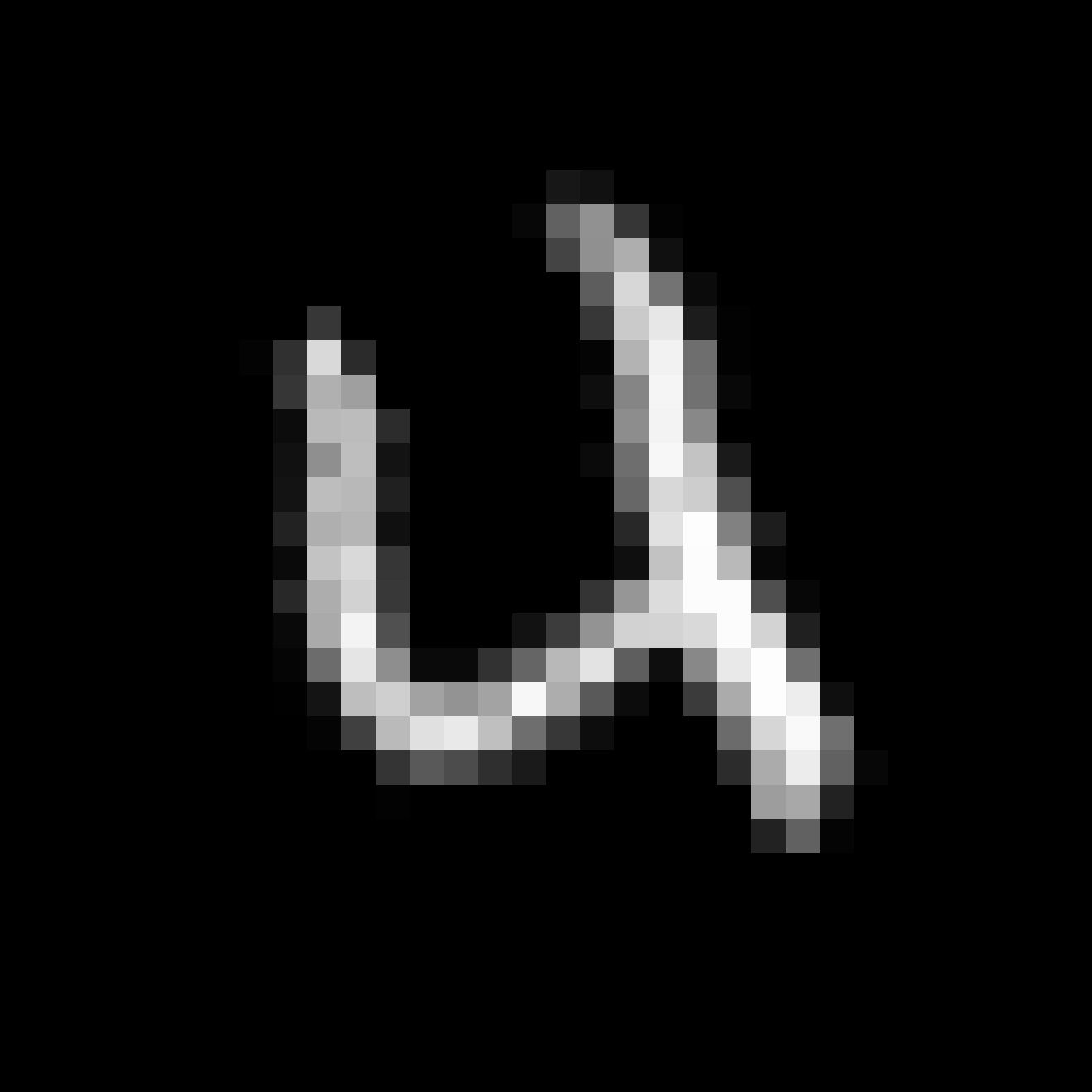} } &
					\subfloat[Reconstructed]{\includegraphics[width=.05\textwidth]{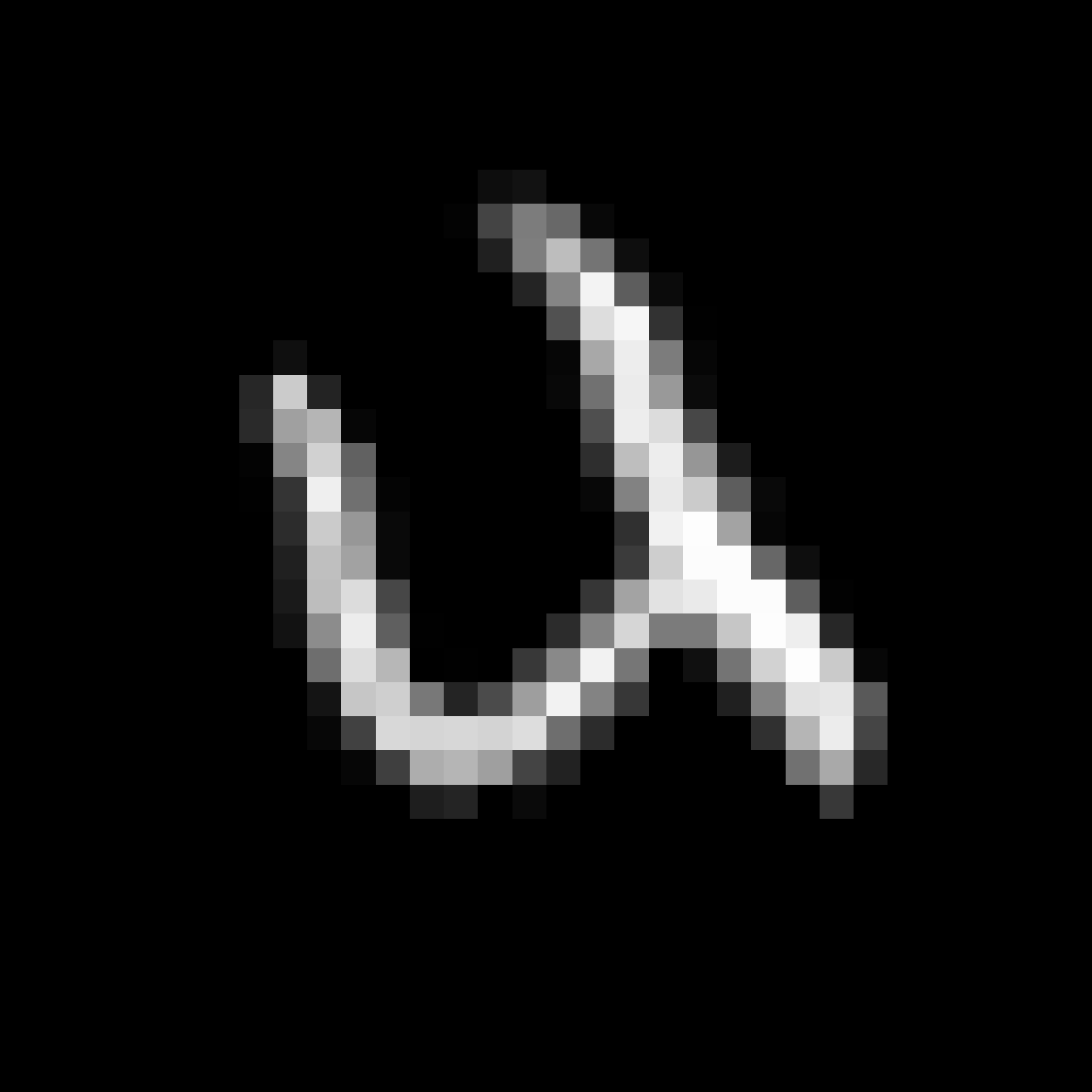} }  \\
					
					% row VAE
					\subfloat[ ]{\includegraphics[width=.05\textwidth]{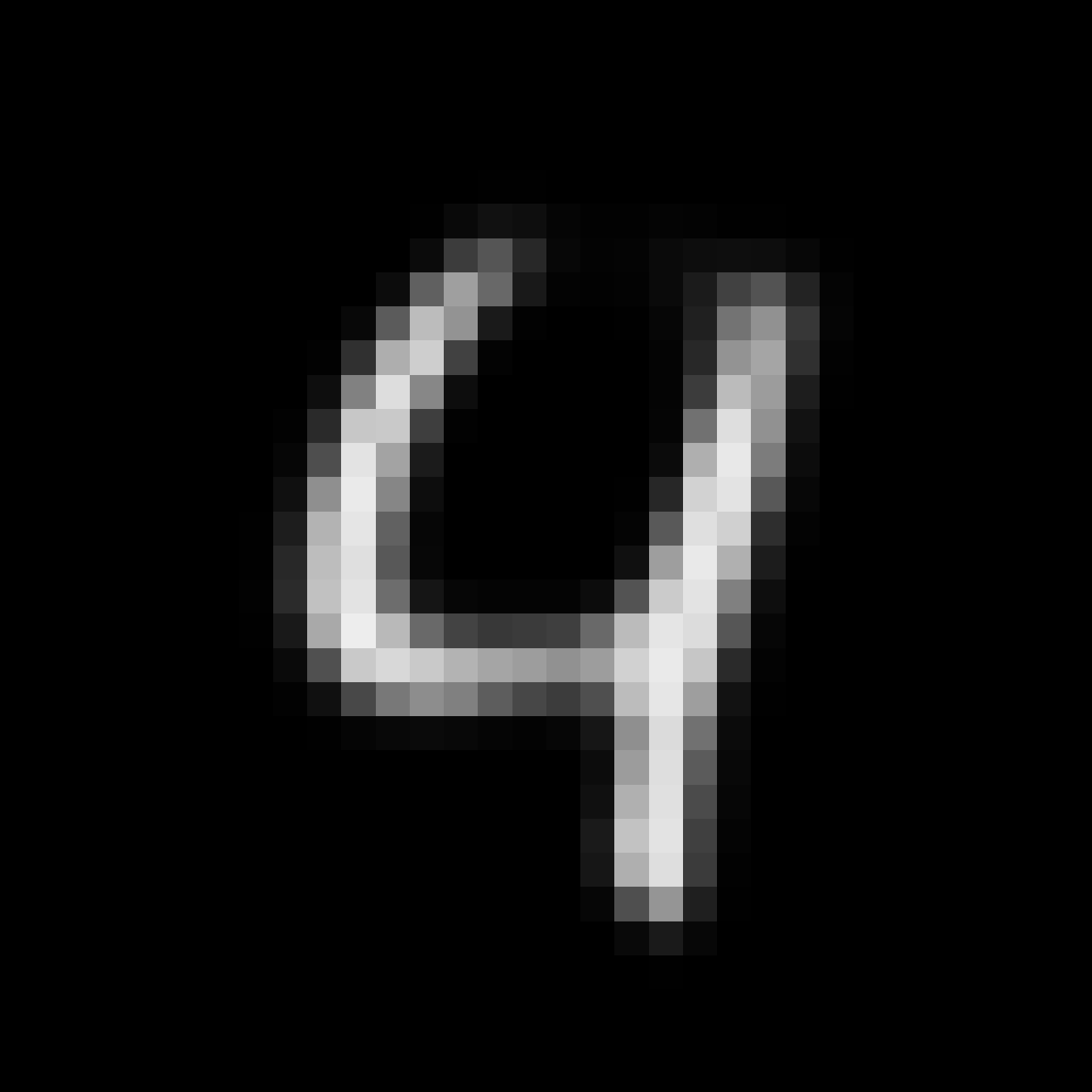} } &
					\subfloat[ ]{\includegraphics[width=.05\textwidth]{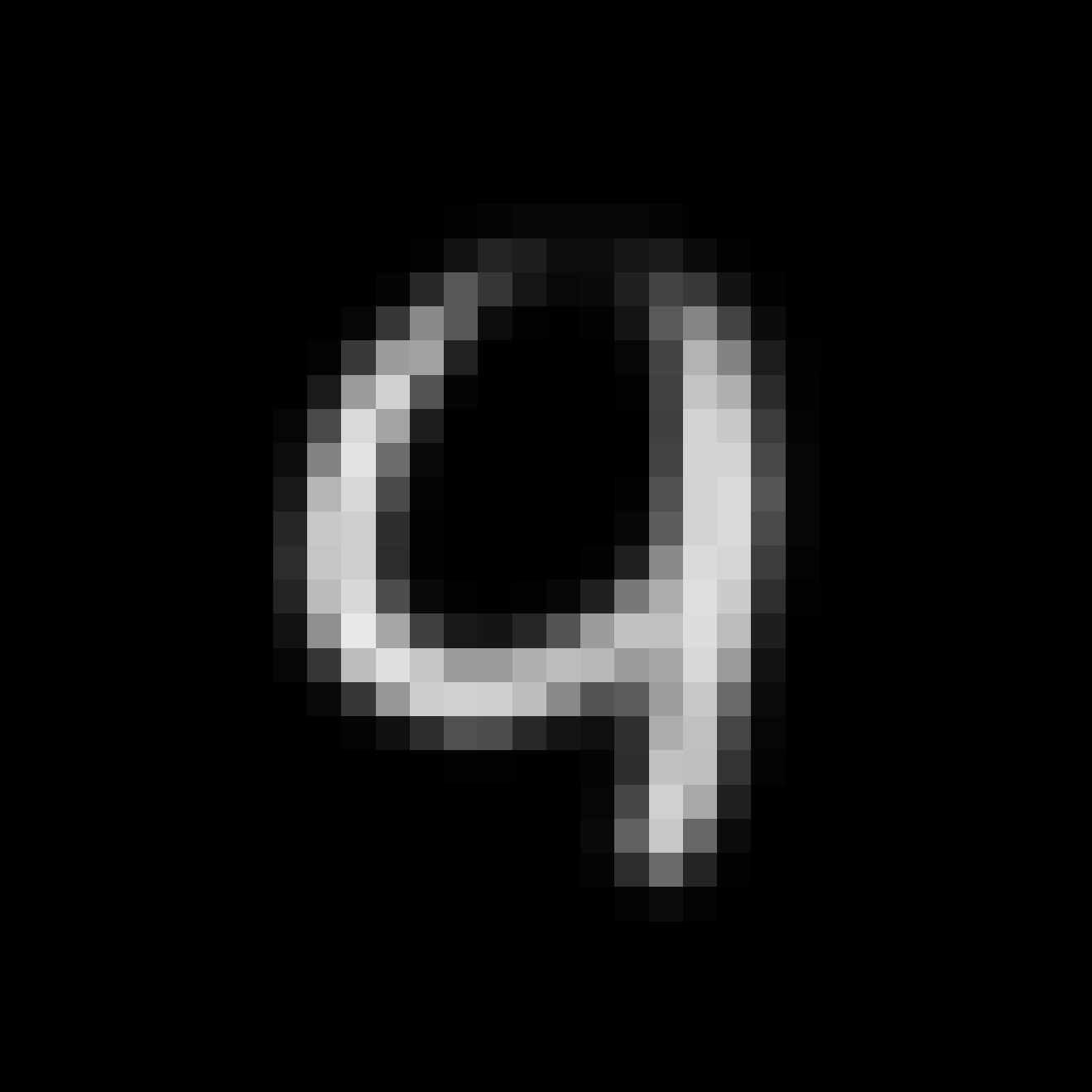} } &
					\subfloat[]{\includegraphics[width=.05\textwidth]{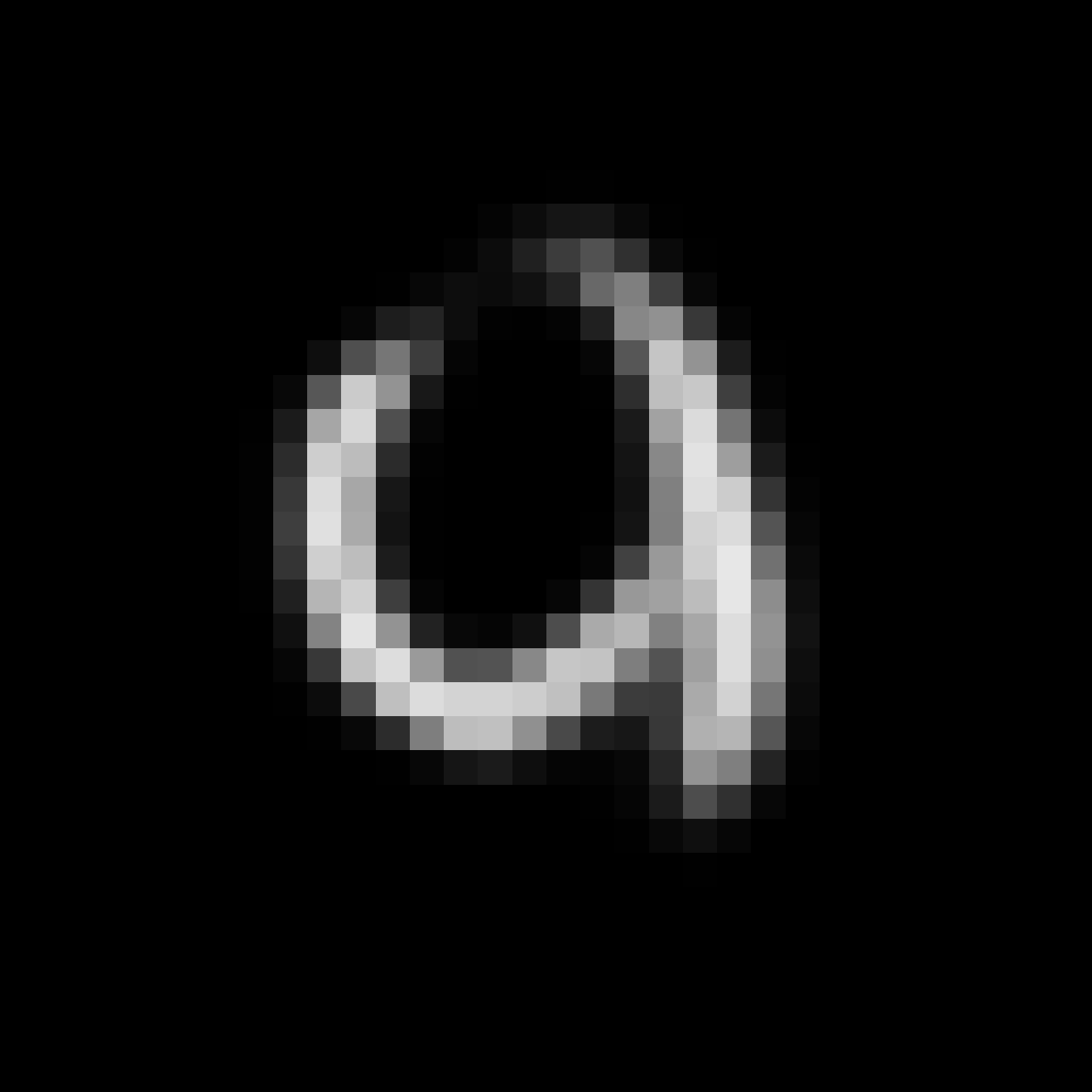} } &
					\subfloat[]{\includegraphics[width=.05\textwidth]{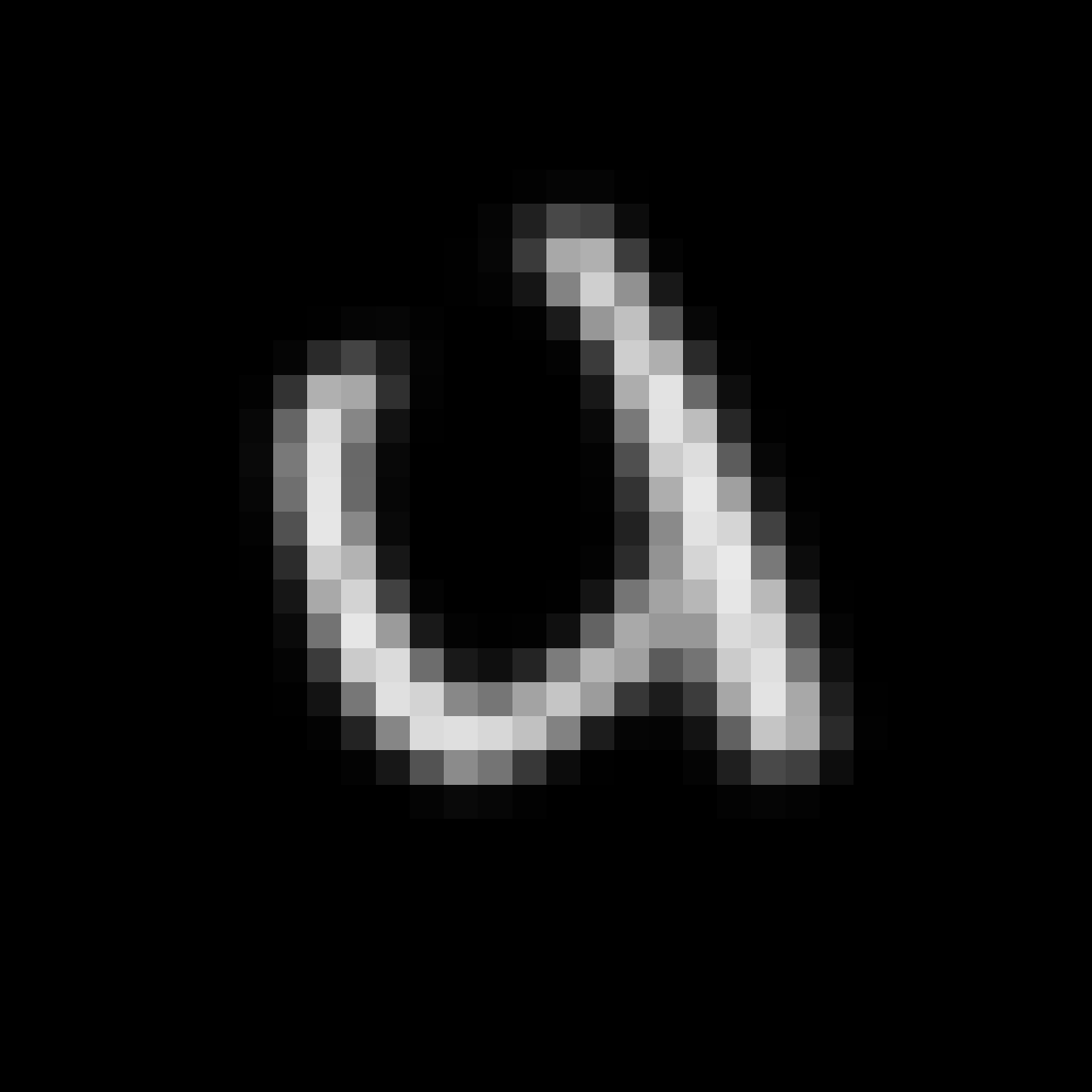} } & 
					\subfloat[]{\includegraphics[width=.05\textwidth]{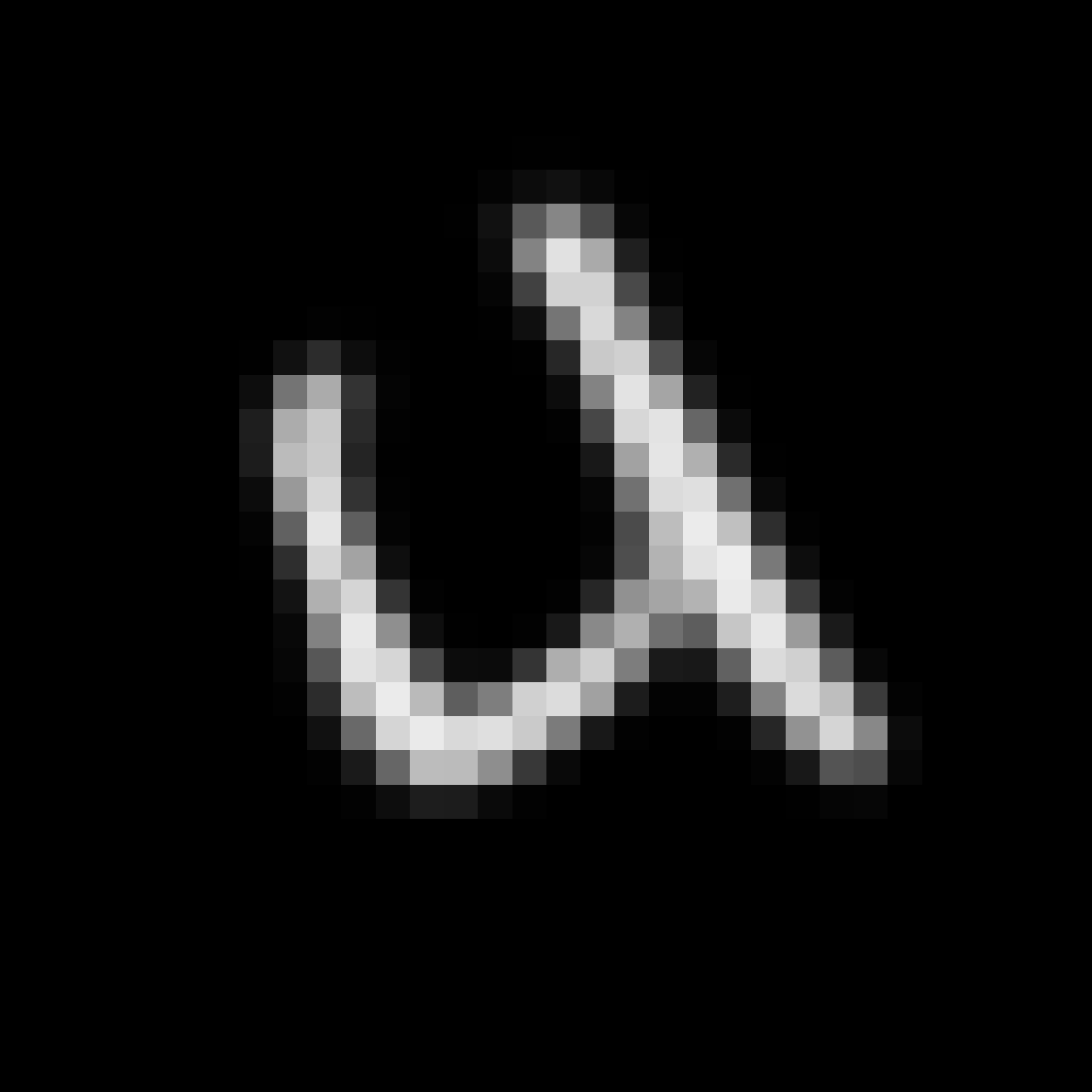} } \\
					
					% row ACAI	
					\subfloat[]{\includegraphics[width=.05\textwidth]{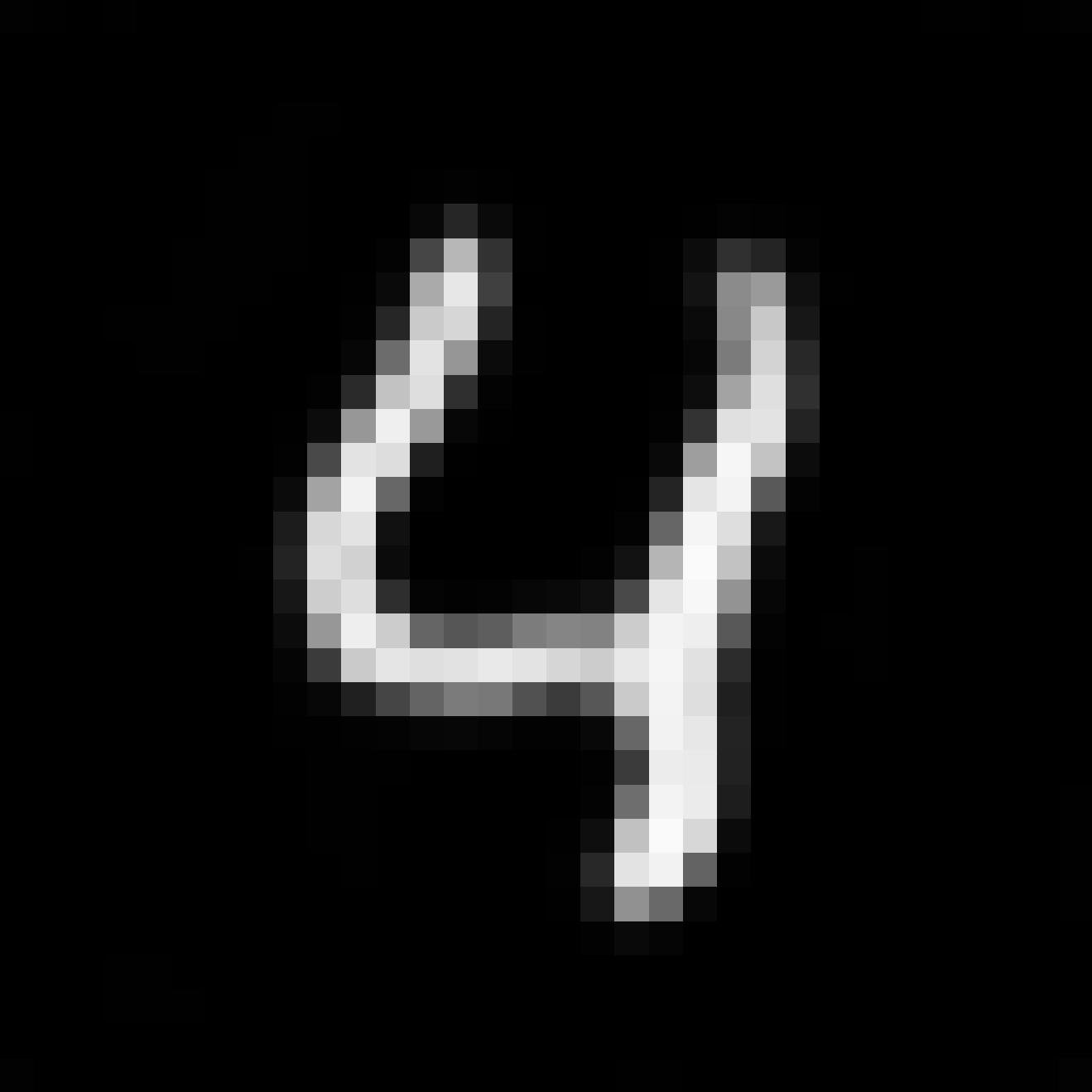} } &
					\subfloat[]{\includegraphics[width=.05\textwidth]{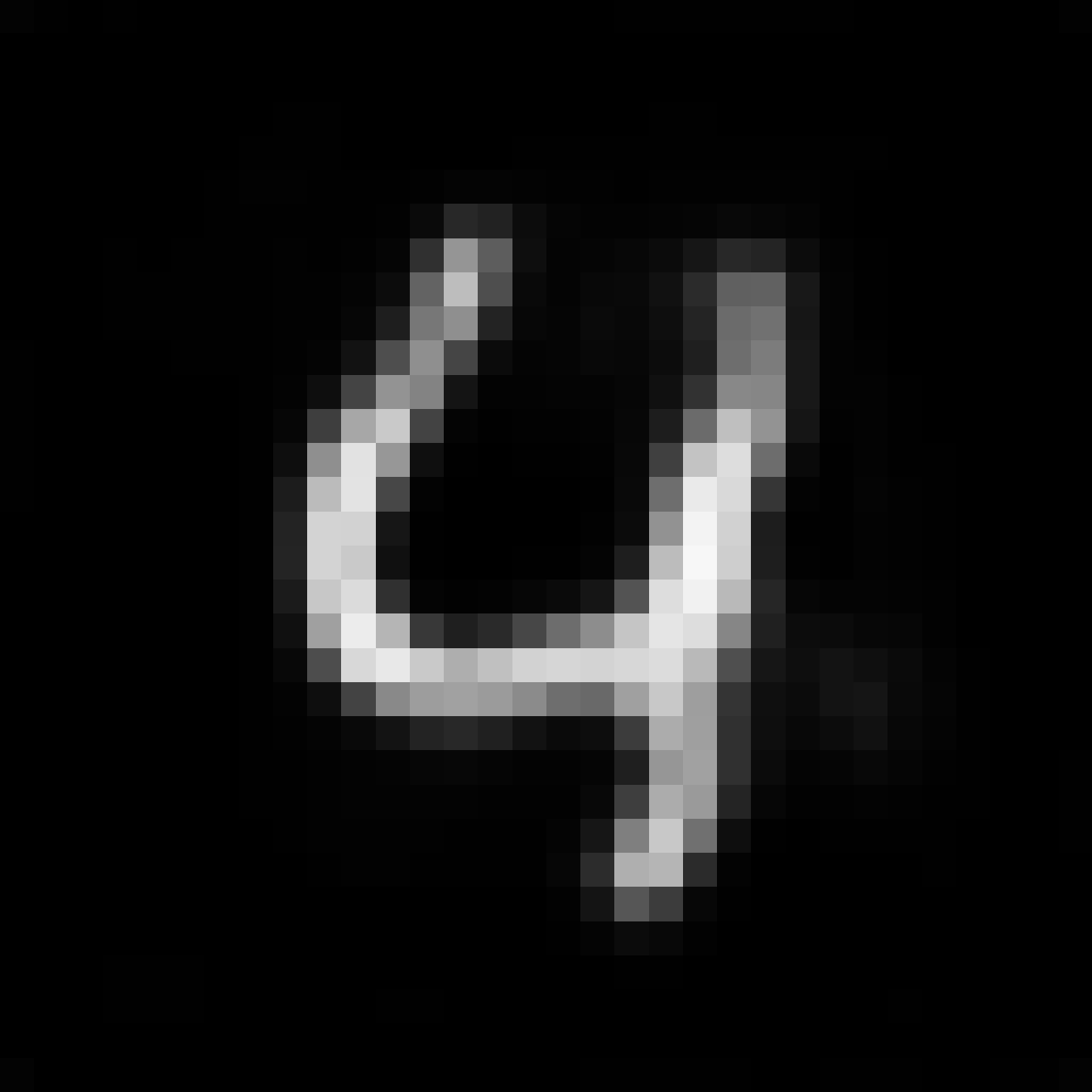} } &
					\subfloat[]{\includegraphics[width=.05\textwidth]{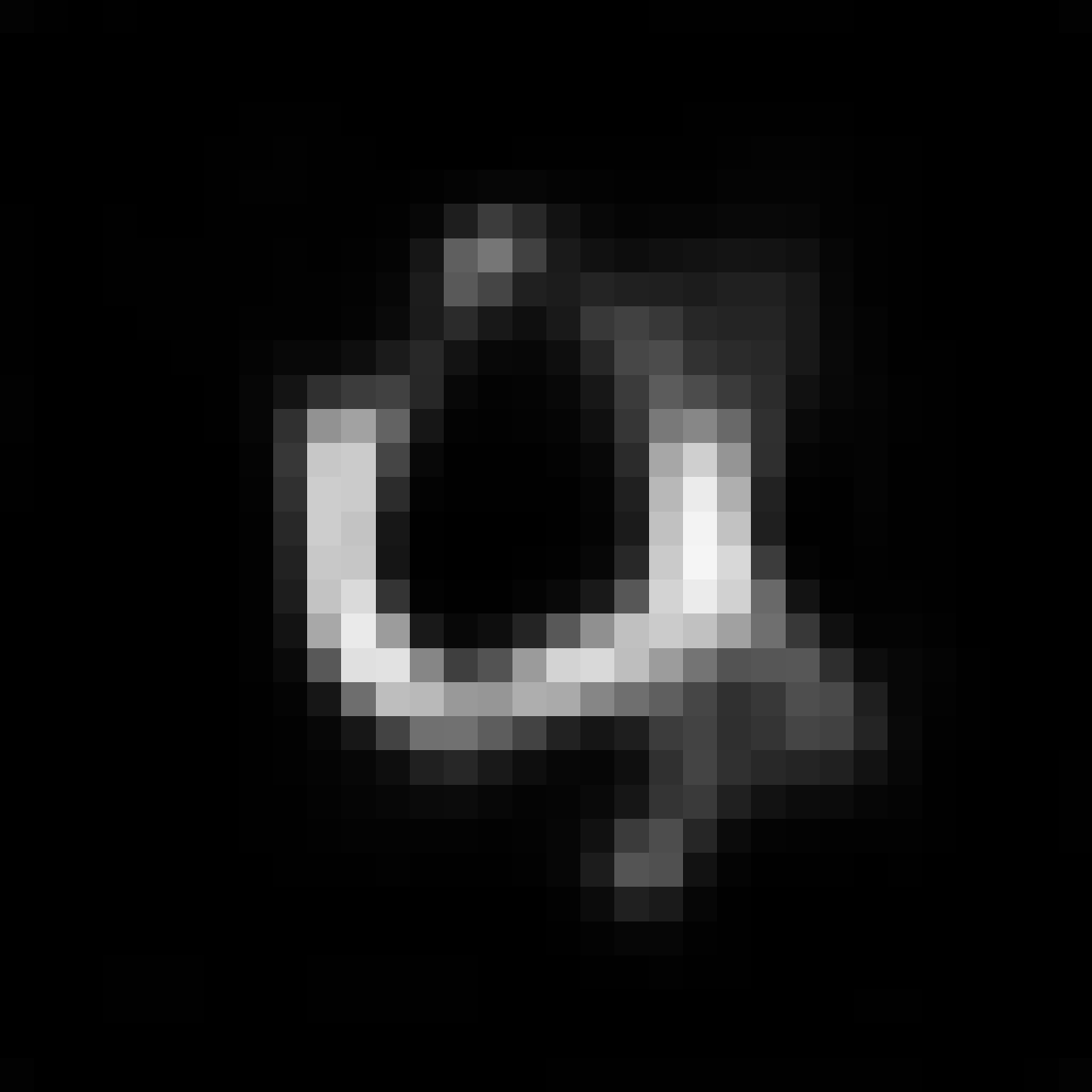} } &
					\subfloat[]{\includegraphics[width=.05\textwidth]{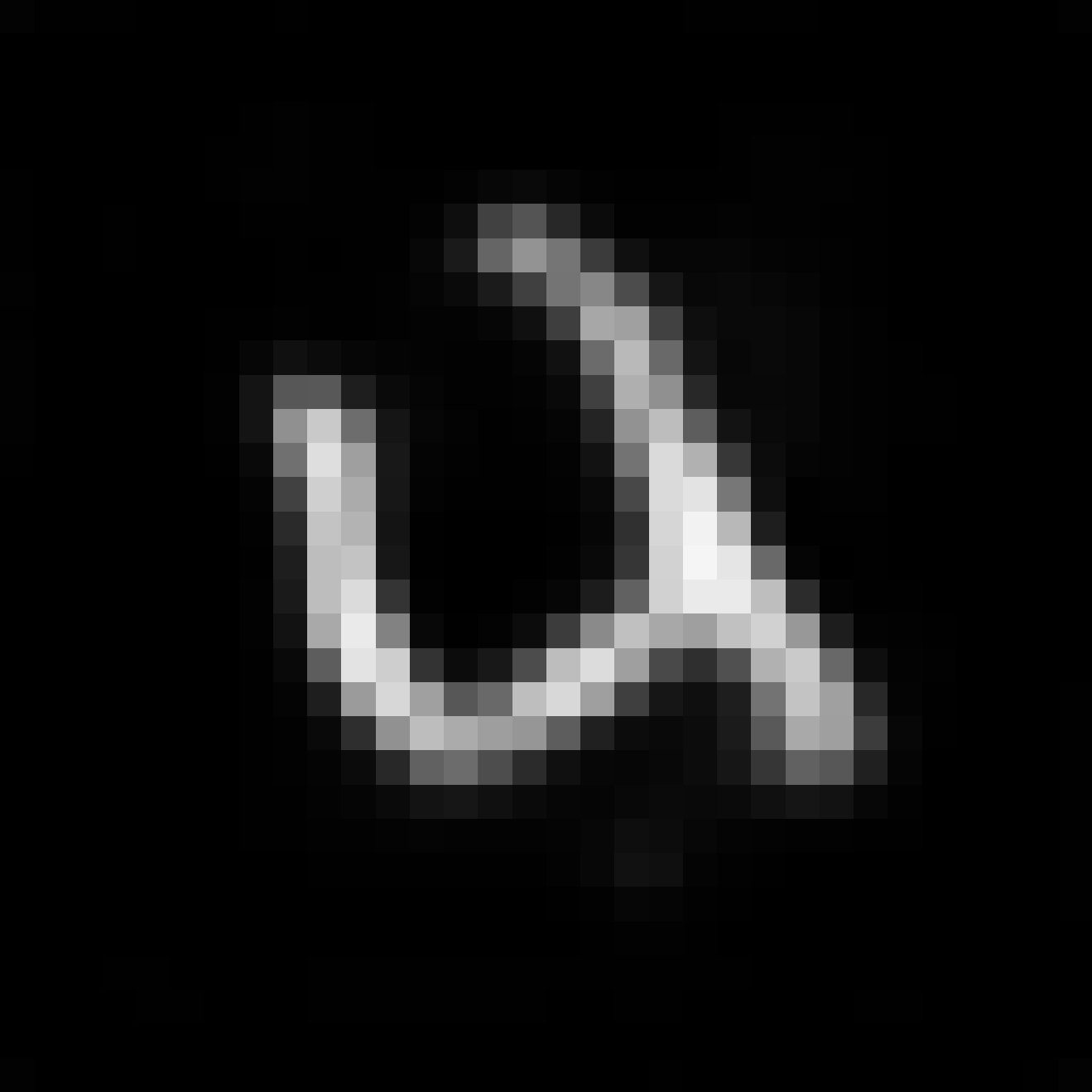} } &
					\subfloat[]{\includegraphics[width=.05\textwidth]{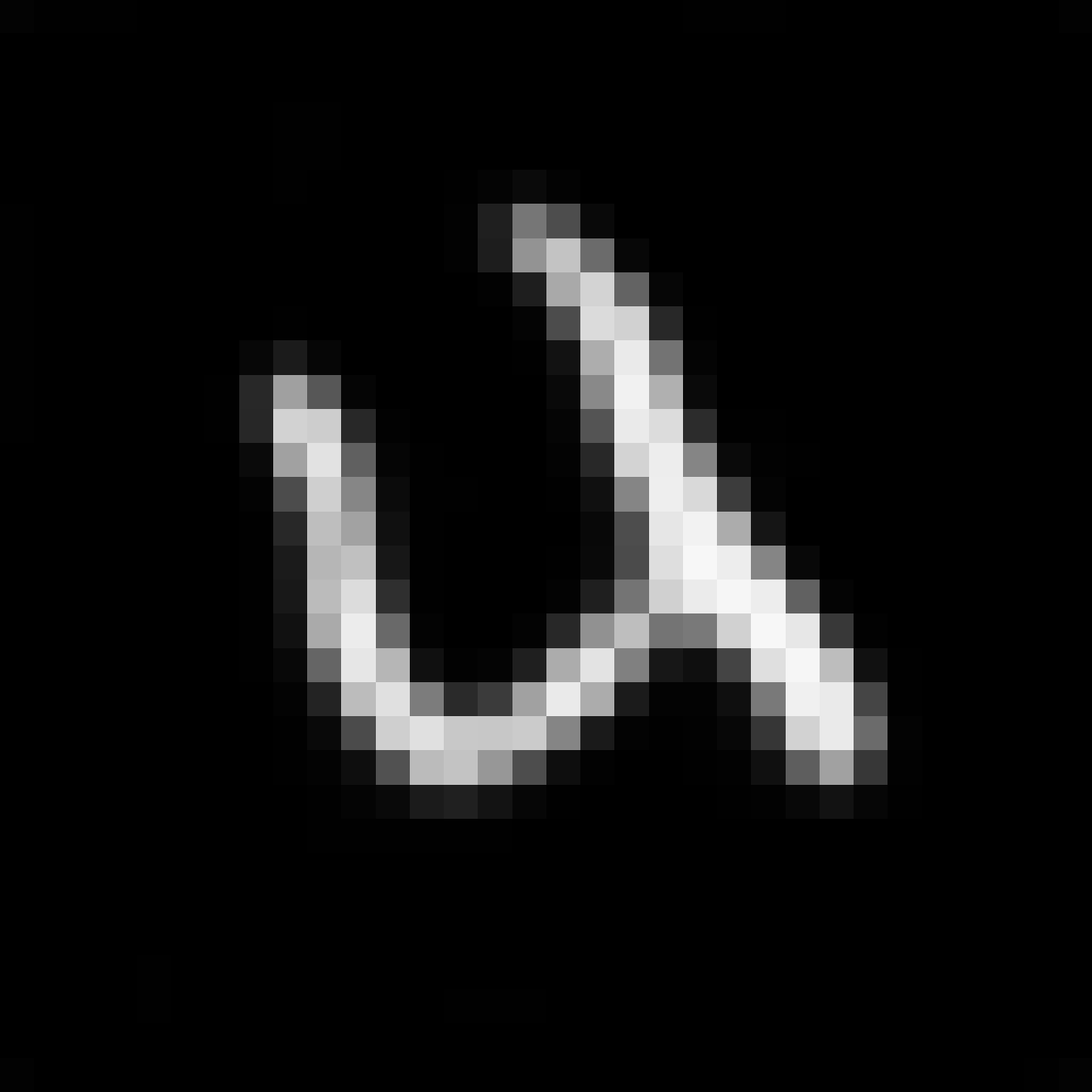} } \\
					
					% row ours	
					\subfloat[]{\includegraphics[width=.05\textwidth]{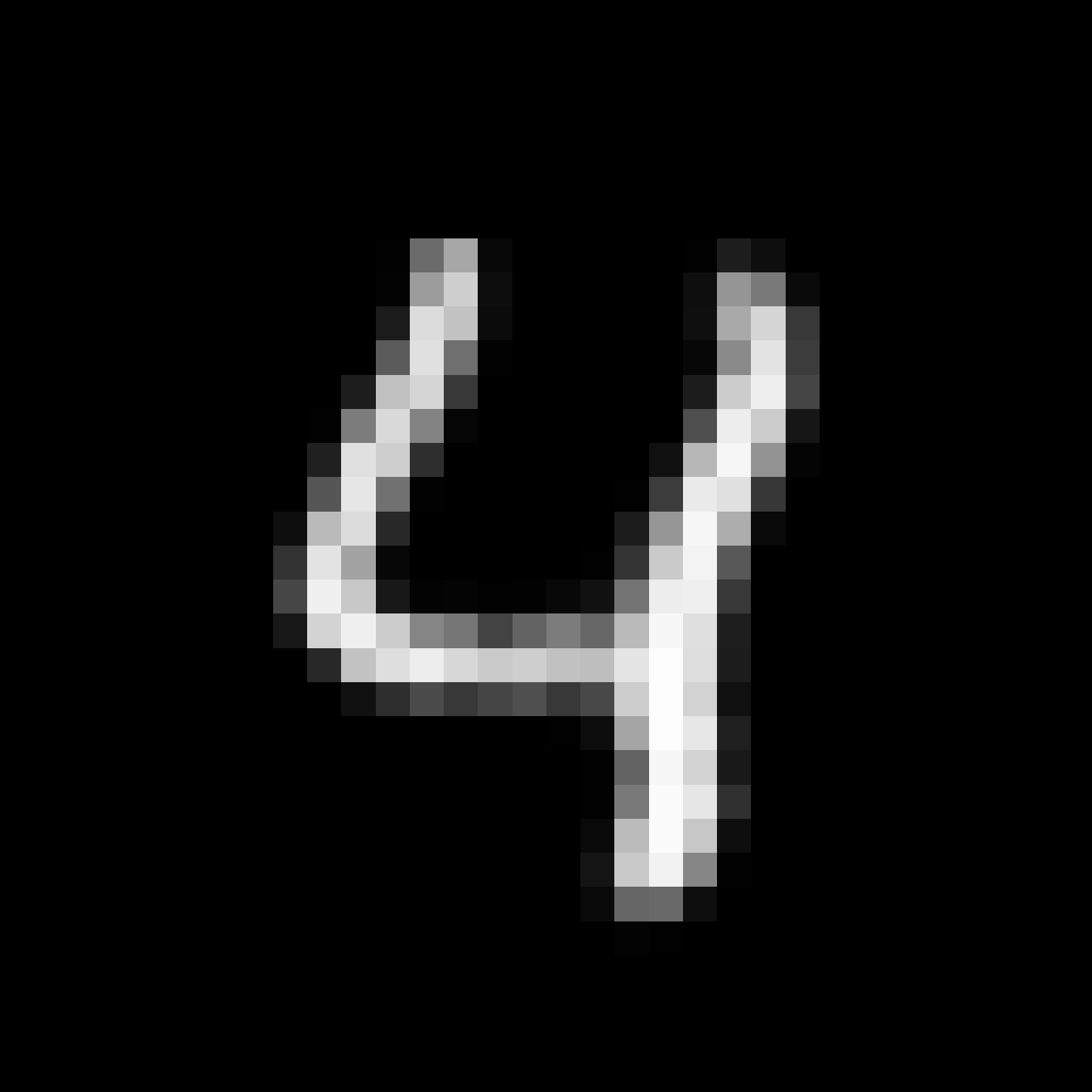} } &
					\subfloat[ ]{\includegraphics[width=.05\textwidth]{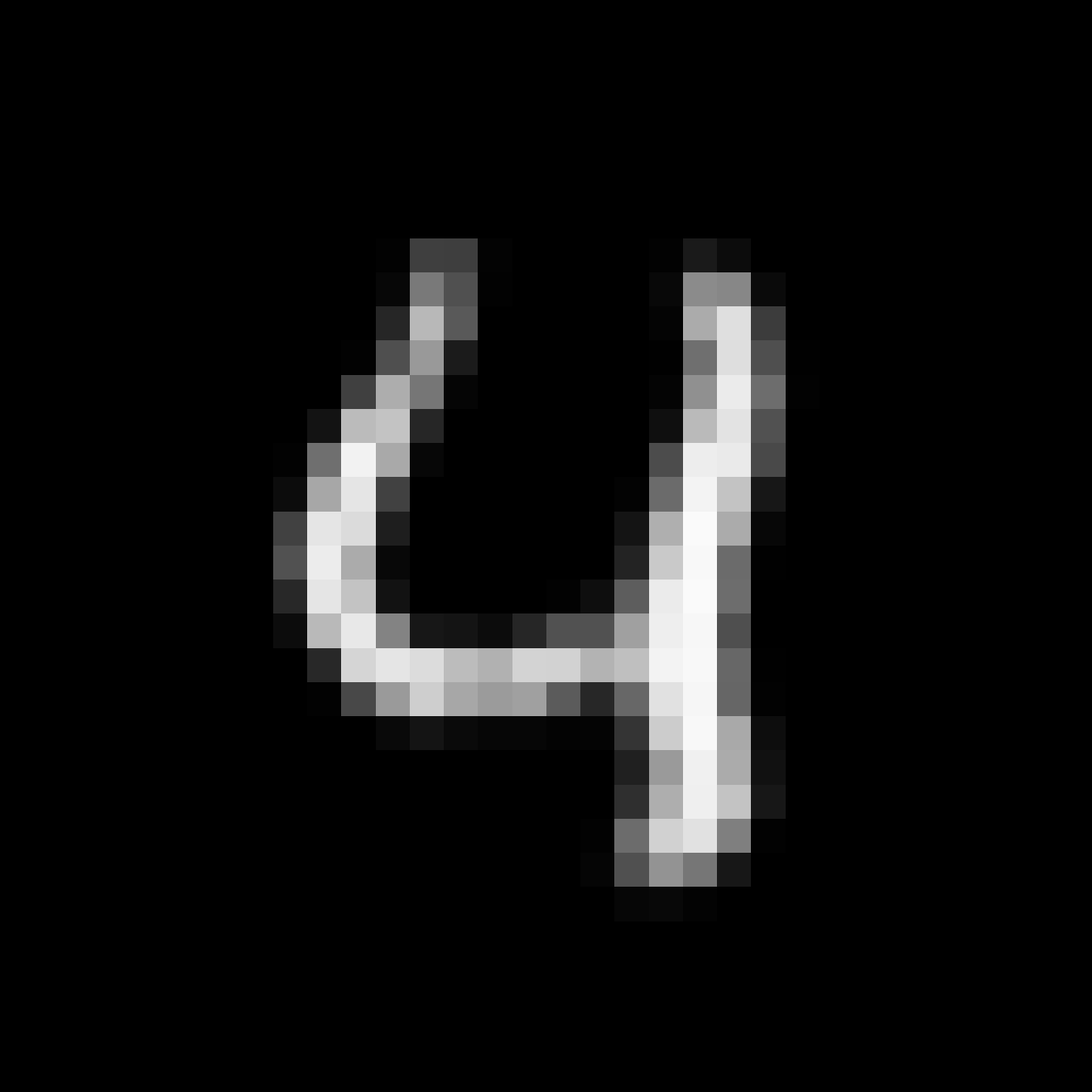} } &
					\subfloat[]{\includegraphics[width=.05\textwidth]{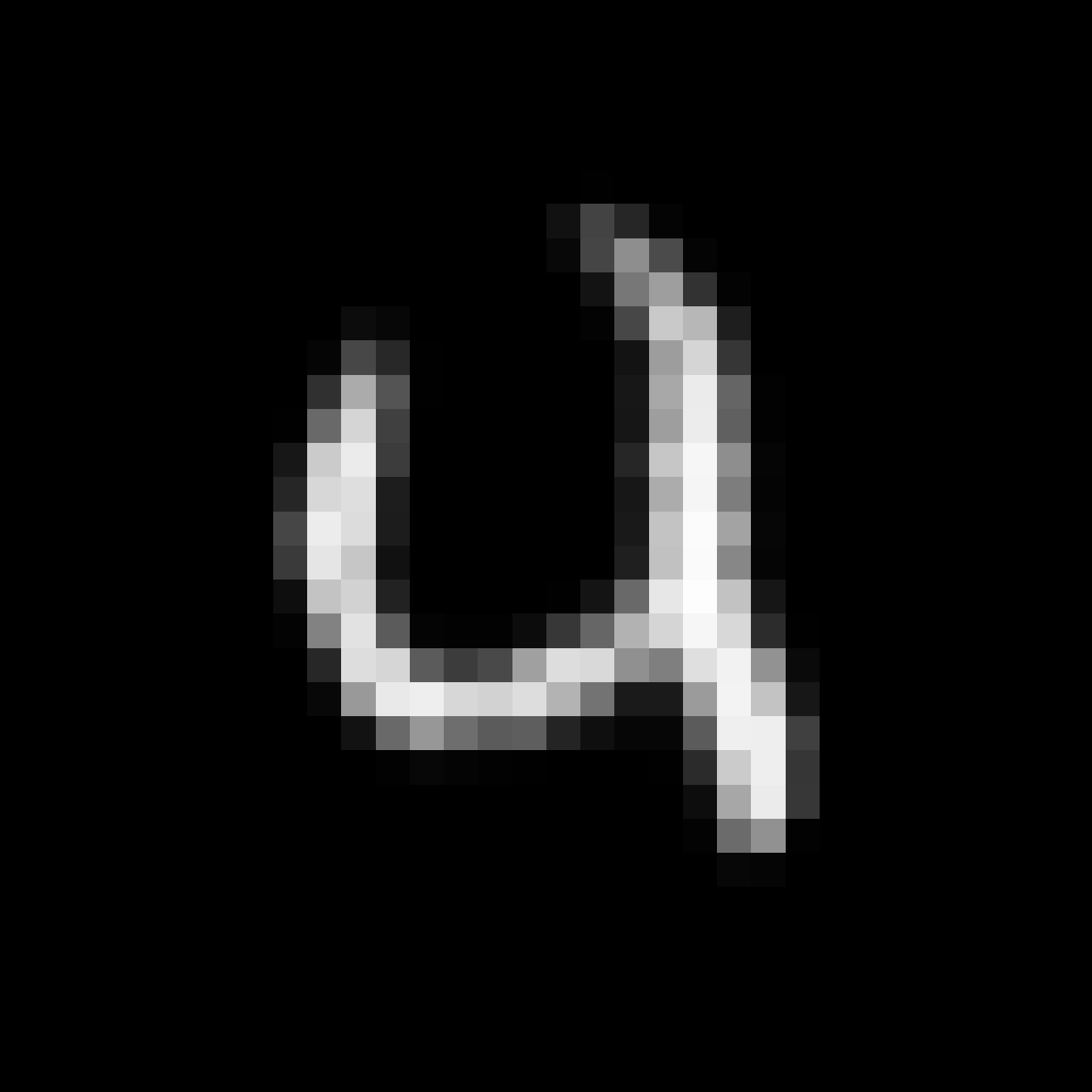} } &
					\subfloat[]{\includegraphics[width=.05\textwidth]{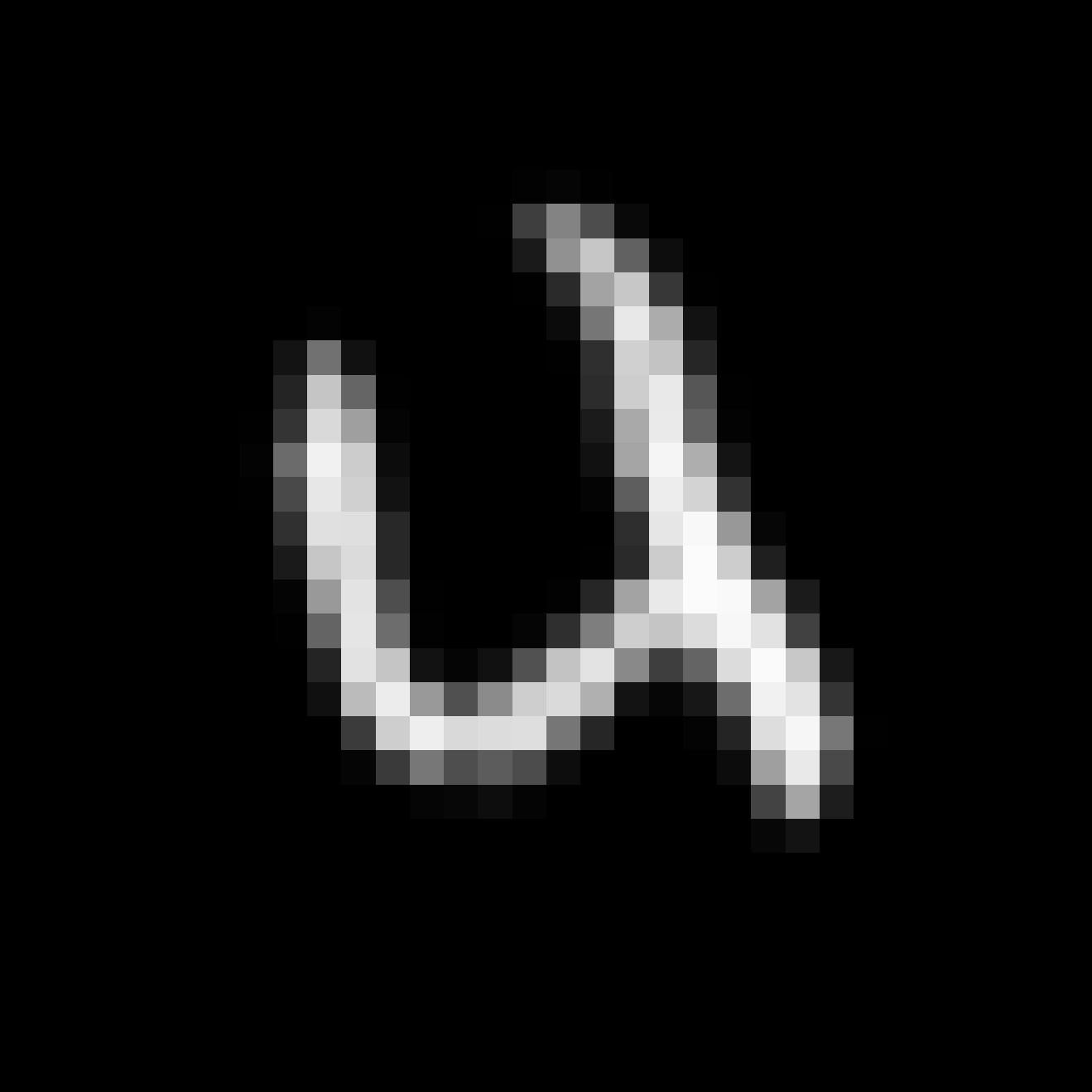} } & 
					\subfloat[]{\includegraphics[width=.05\textwidth]{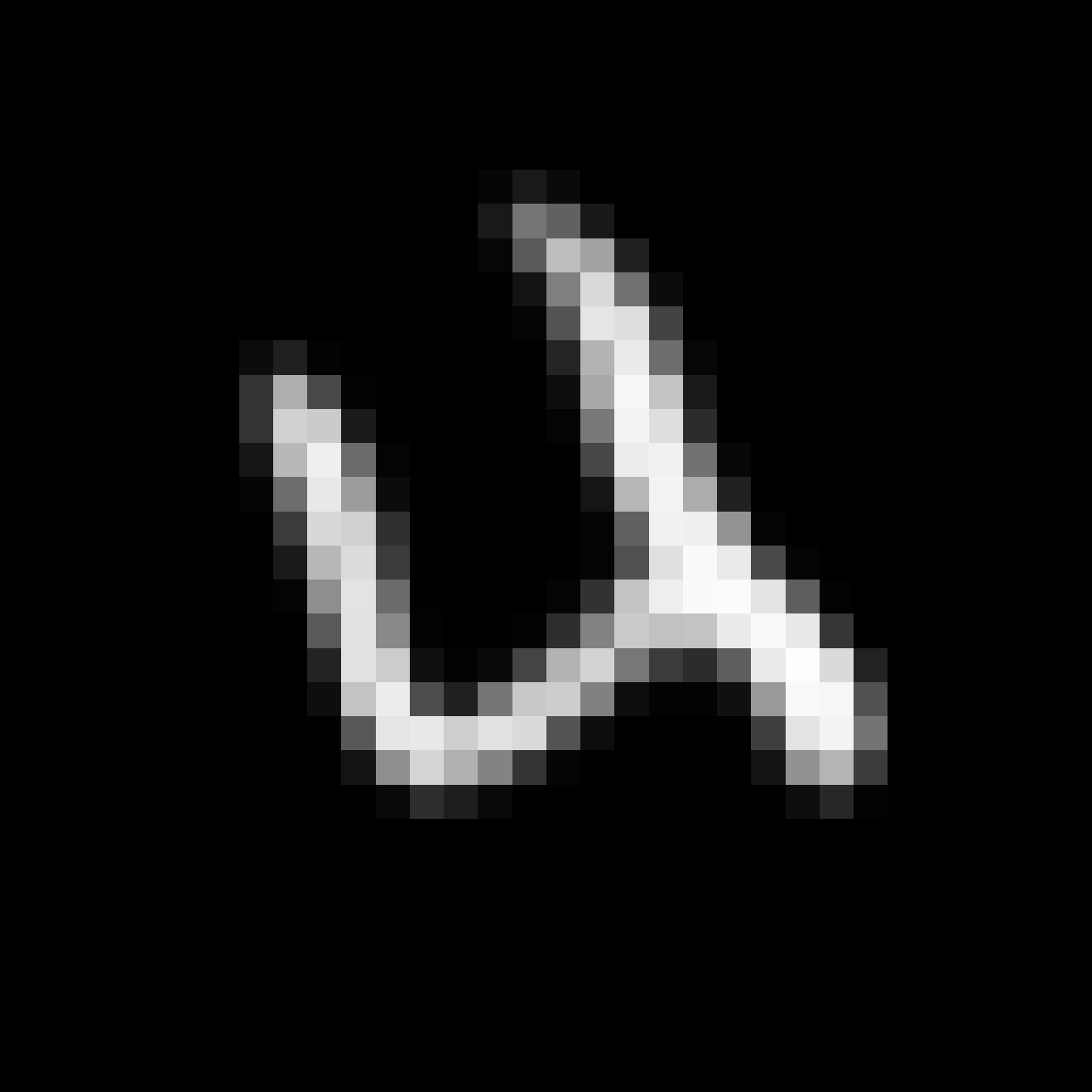} }  \\

				\end{tabular}
			}  % end subfloat
			%%%%%%%%%%%%%%%%%%%%%%%                  3nd exampled  ----------------------
			% --------------------------------- next MNIST block
			\subfloat[]{
				\begin{tabular}{c c c c c}
					% row References
					\num{0}$^{\circ}$ & \num{10}$^{\circ}$ & \num{20}$^{\circ}$ & \num{30}$^{\circ}$ & \num{40}$^{\circ}$  \\
					\subfloat[Reconstructed]{\includegraphics[width=.05\textwidth]{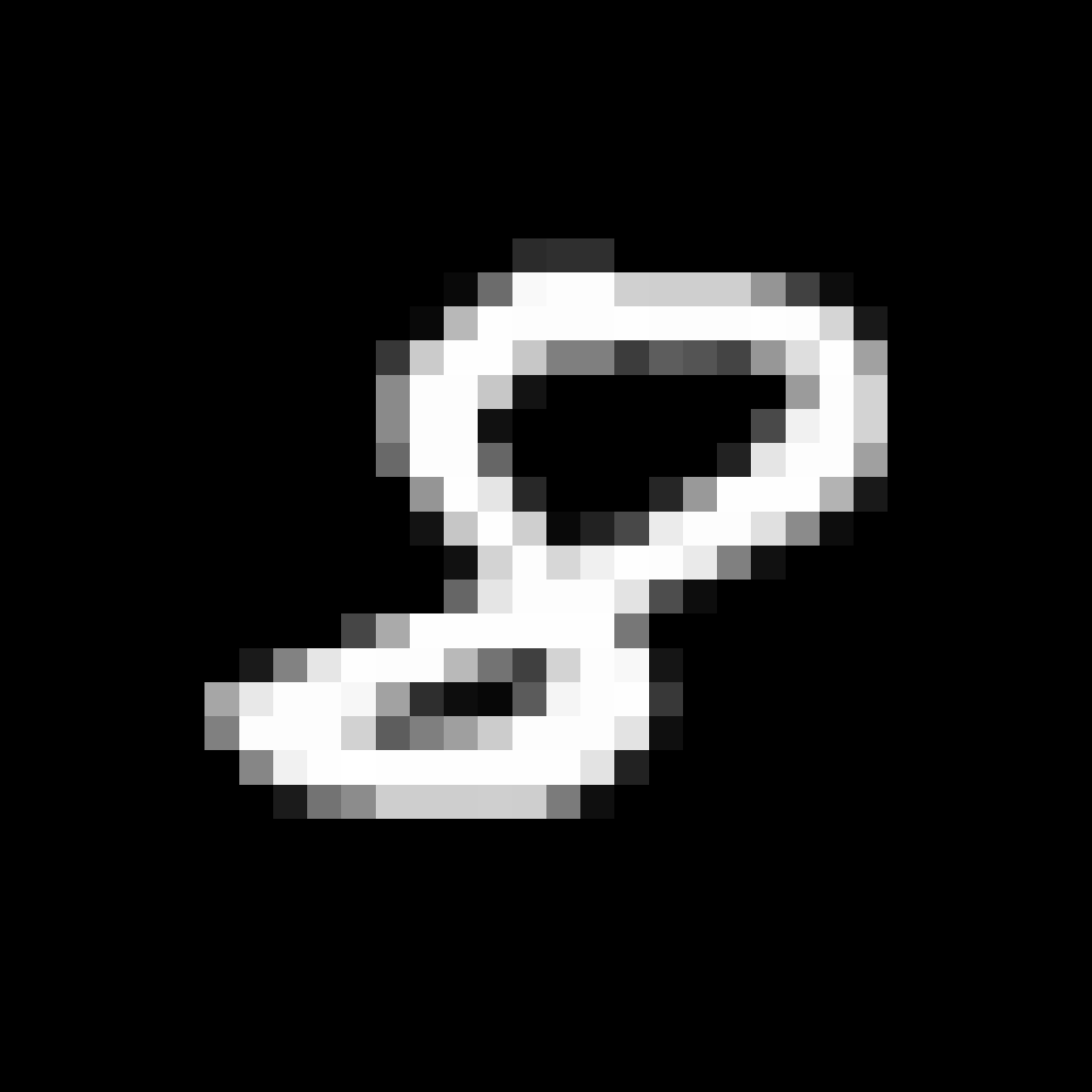} } &
					\subfloat[$\alpha=0.25$]{\includegraphics[width=.05\textwidth]{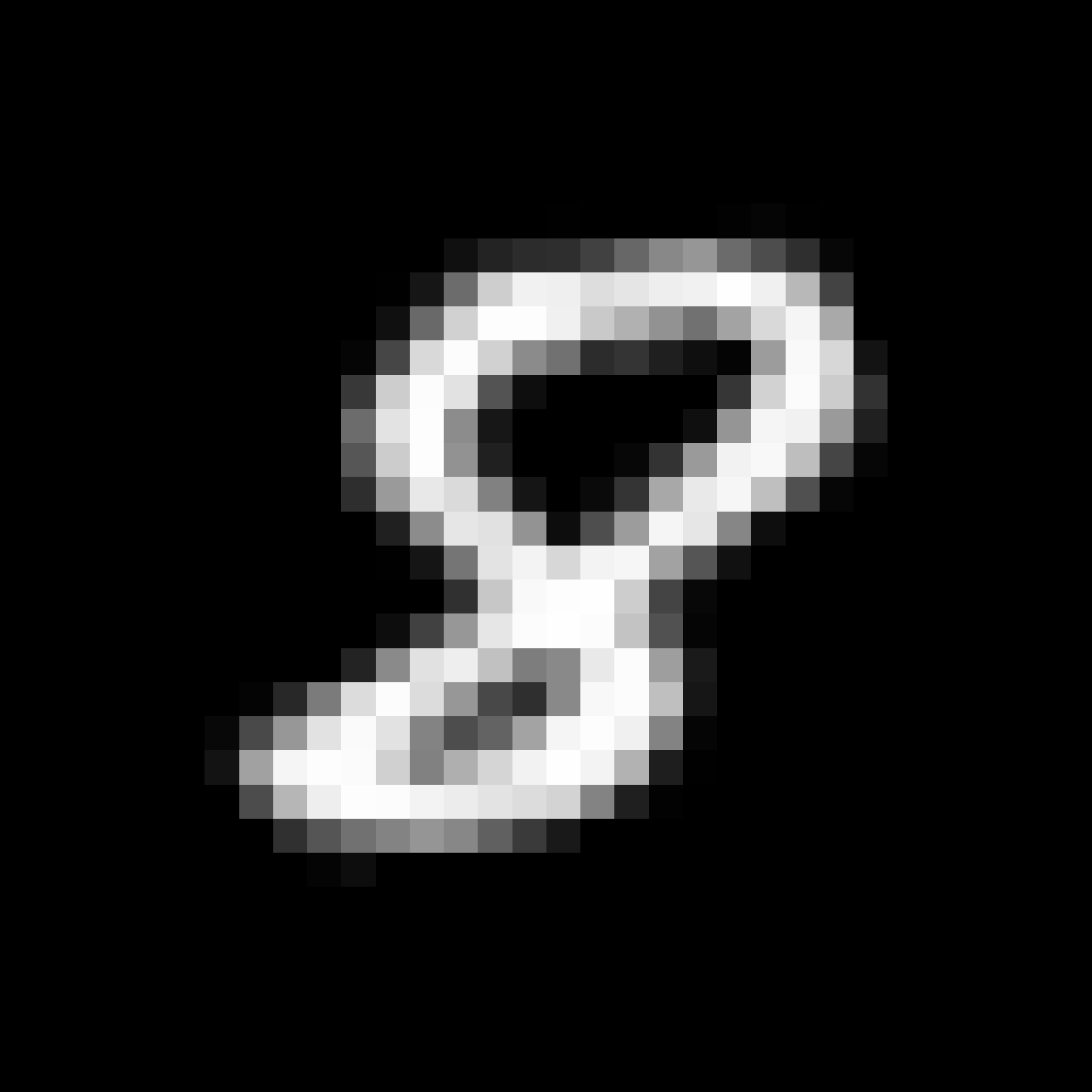} } &
					\subfloat[$\alpha=0.5$]{\includegraphics[width=.05\textwidth]{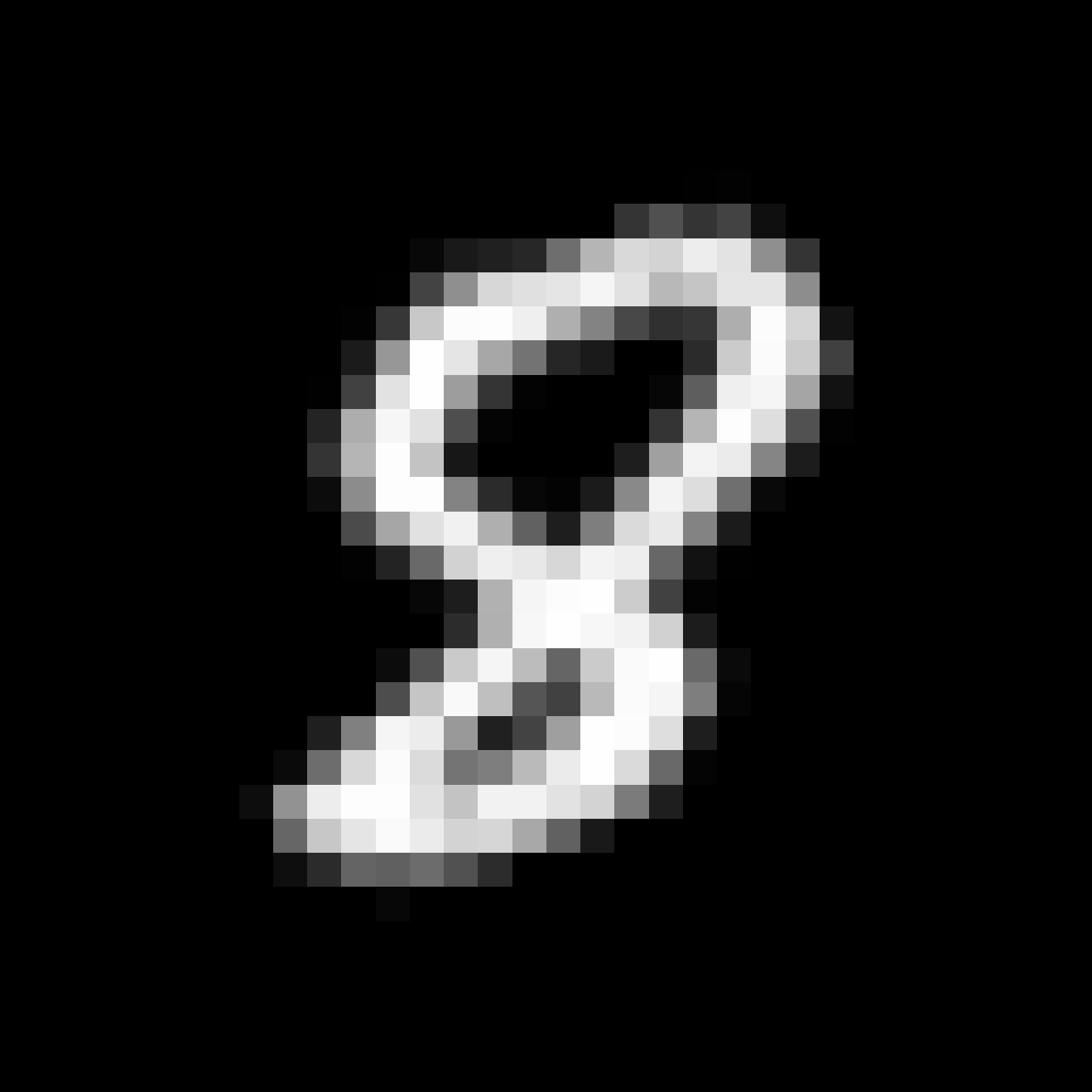} } &
					\subfloat[$\alpha=0.75$]{\includegraphics[width=.05\textwidth]{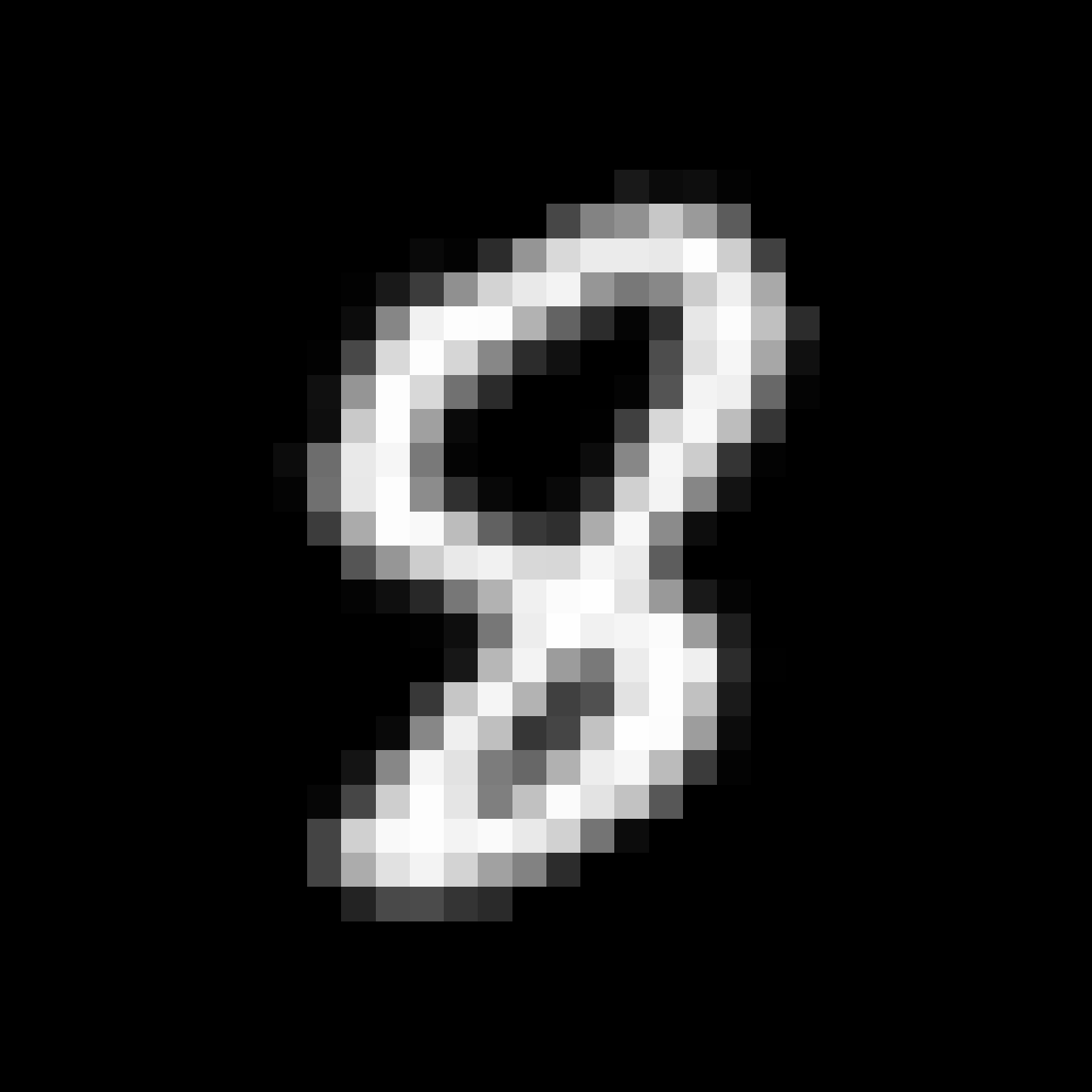} } &
					\subfloat[Reconstructed]{\includegraphics[width=.05\textwidth]{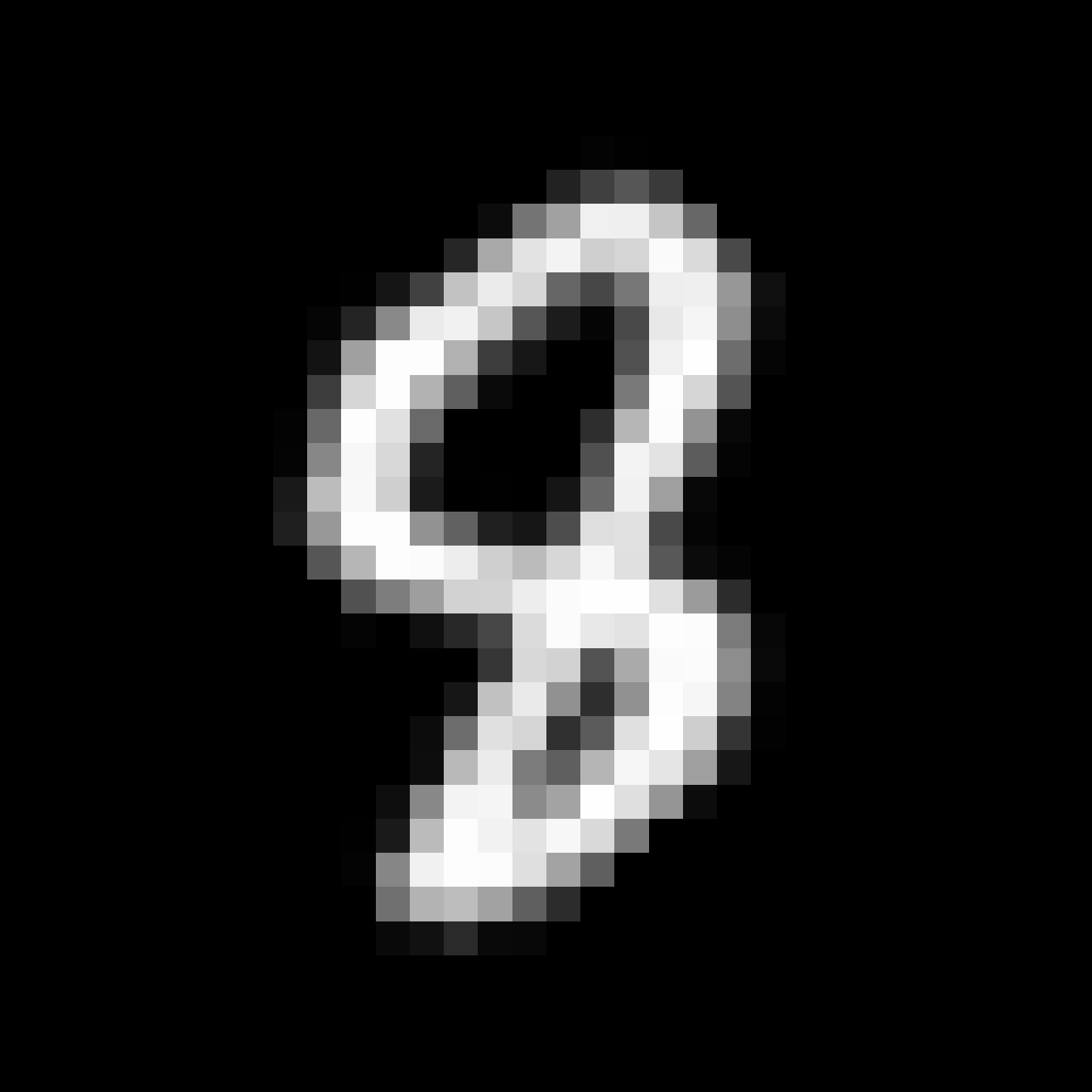} }  \\
					
					% row VAE
					\subfloat[ ]{\includegraphics[width=.05\textwidth]{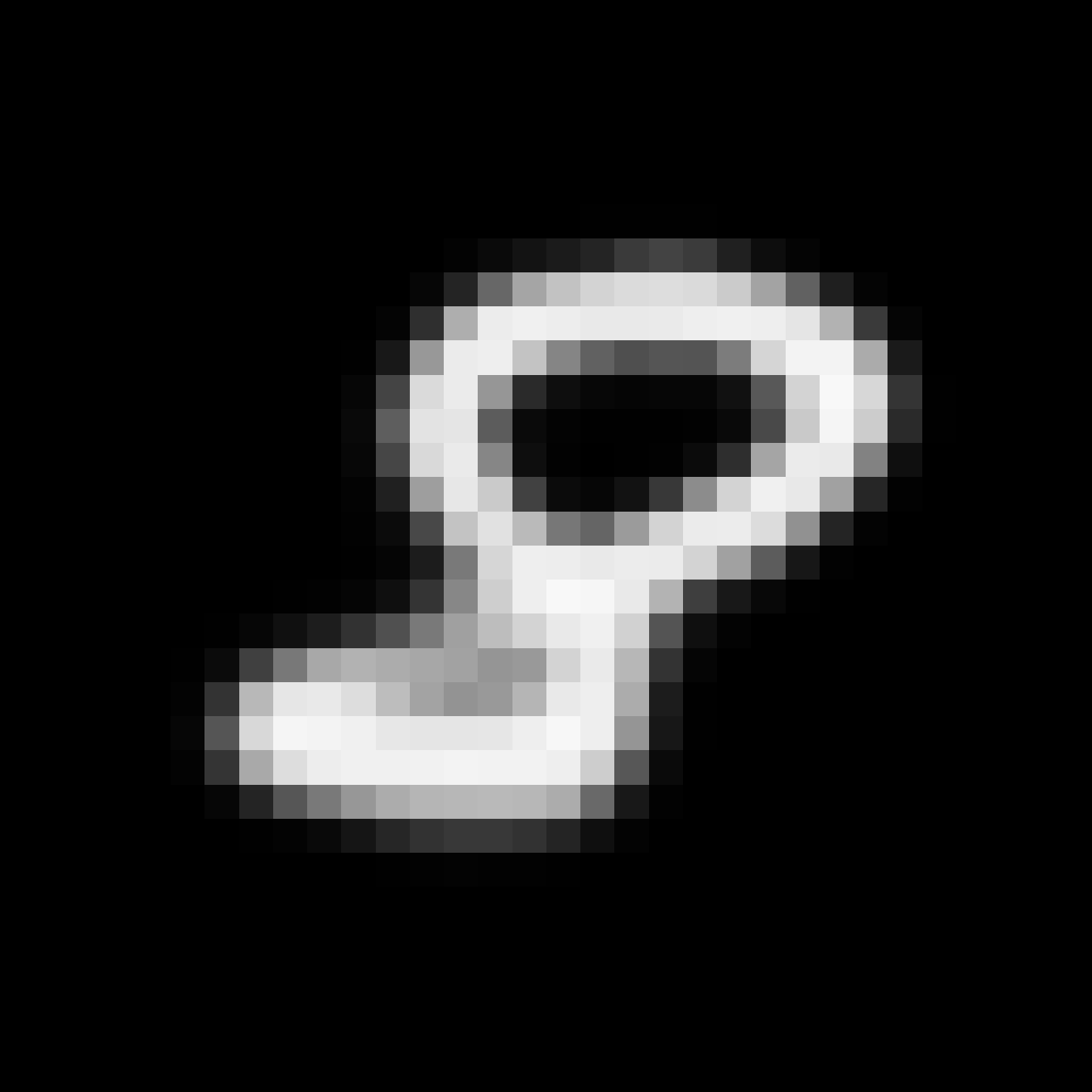} } &
					\subfloat[ ]{\includegraphics[width=.05\textwidth]{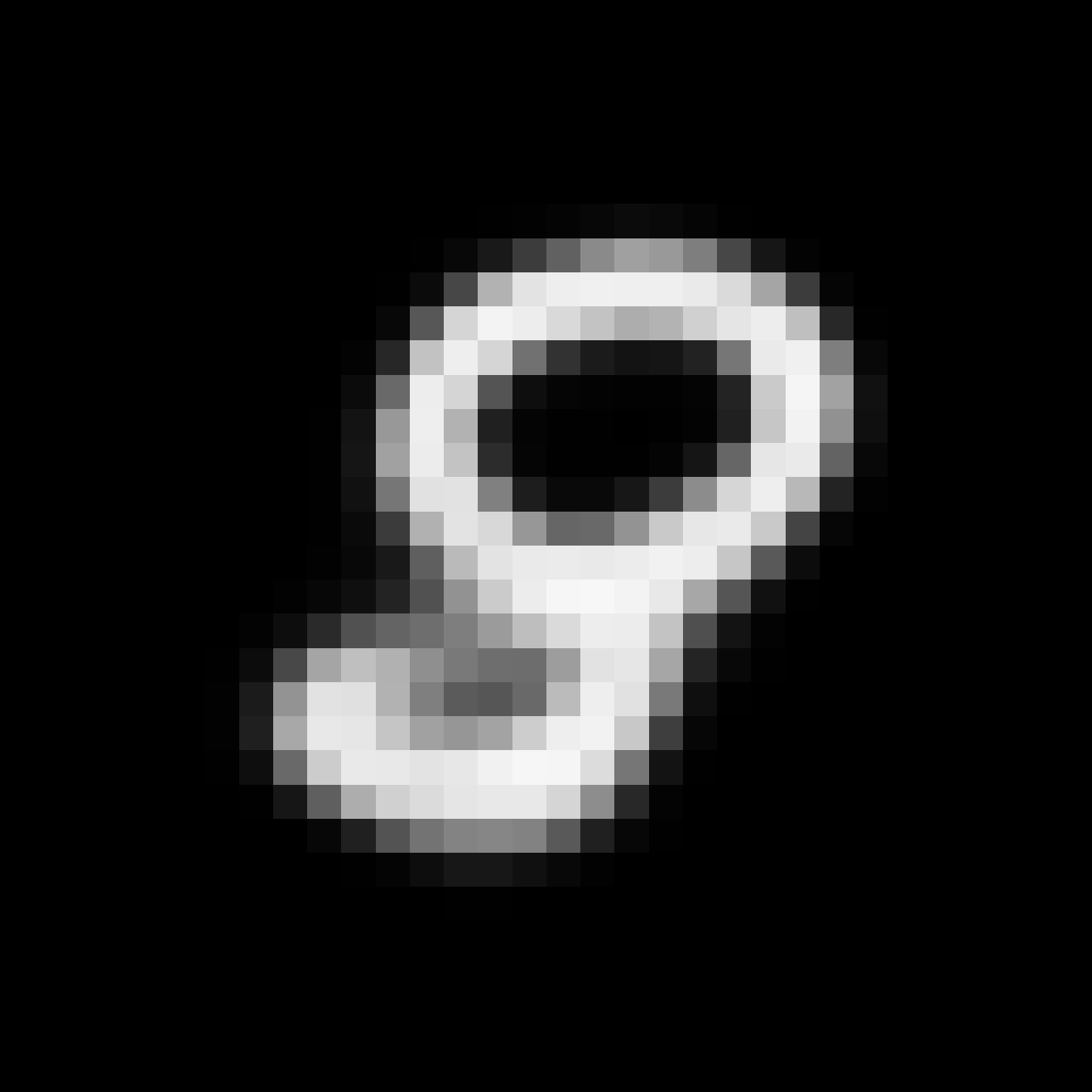} } &
					\subfloat[]{\includegraphics[width=.05\textwidth]{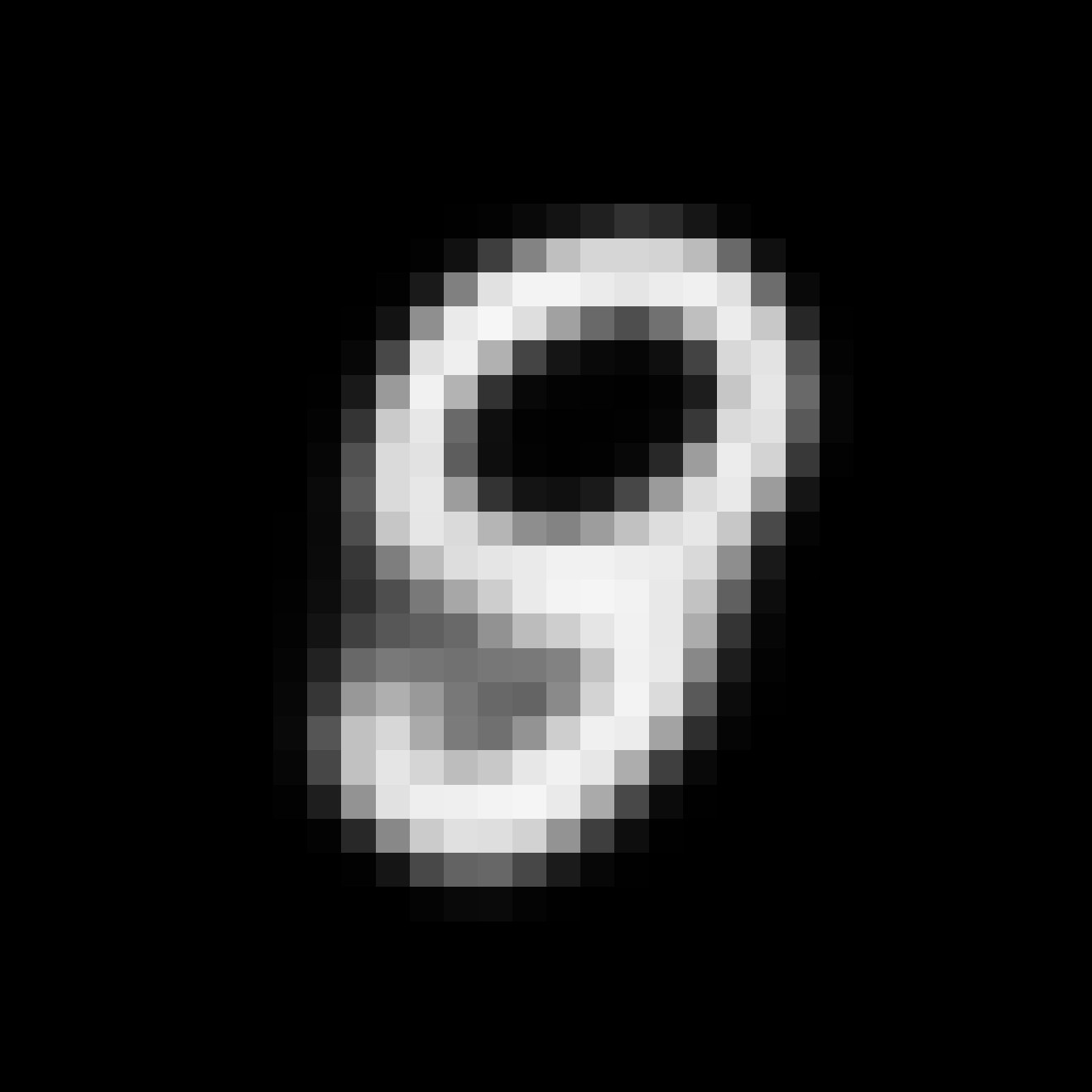} } &
					\subfloat[]{\includegraphics[width=.05\textwidth]{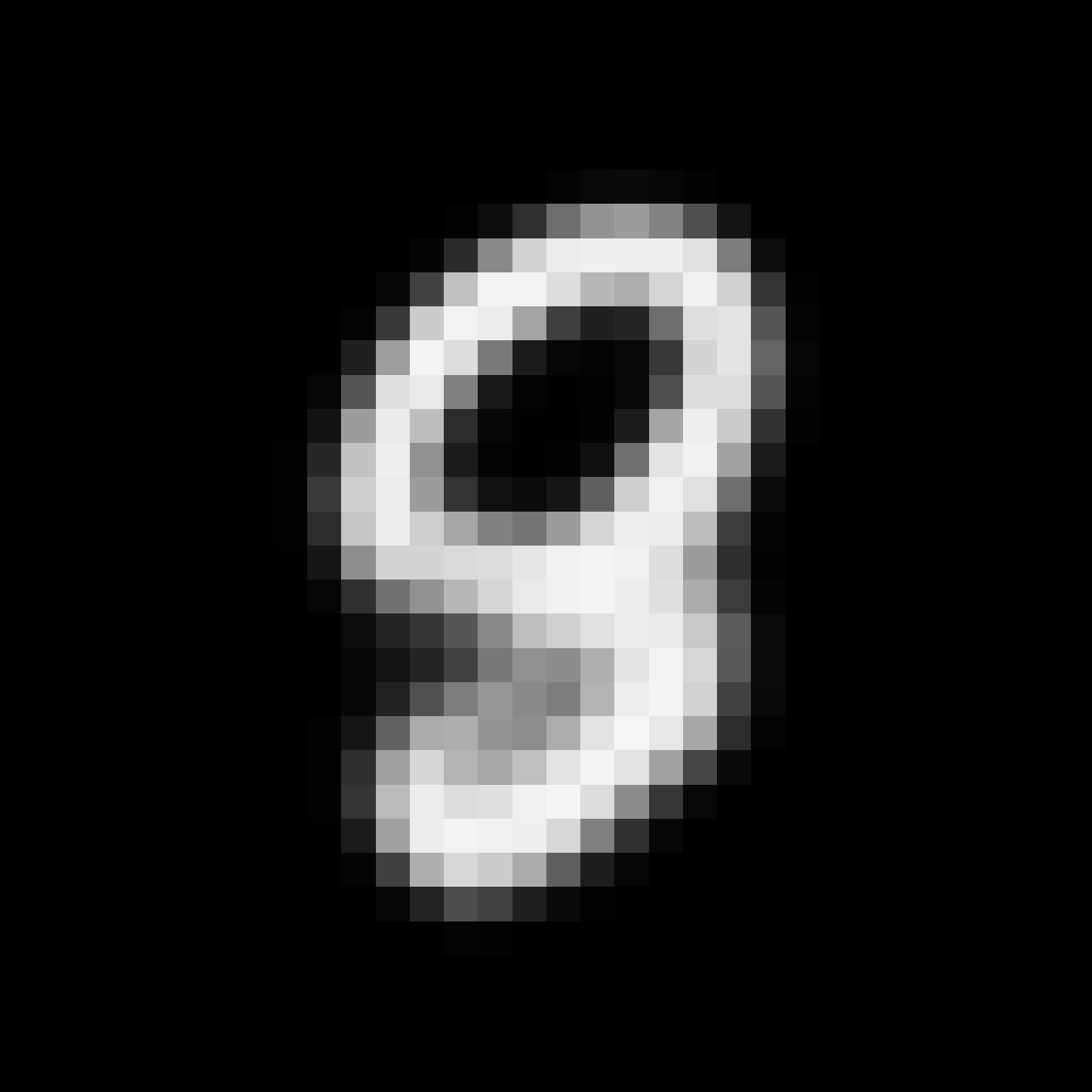} } & 
					\subfloat[]{\includegraphics[width=.05\textwidth]{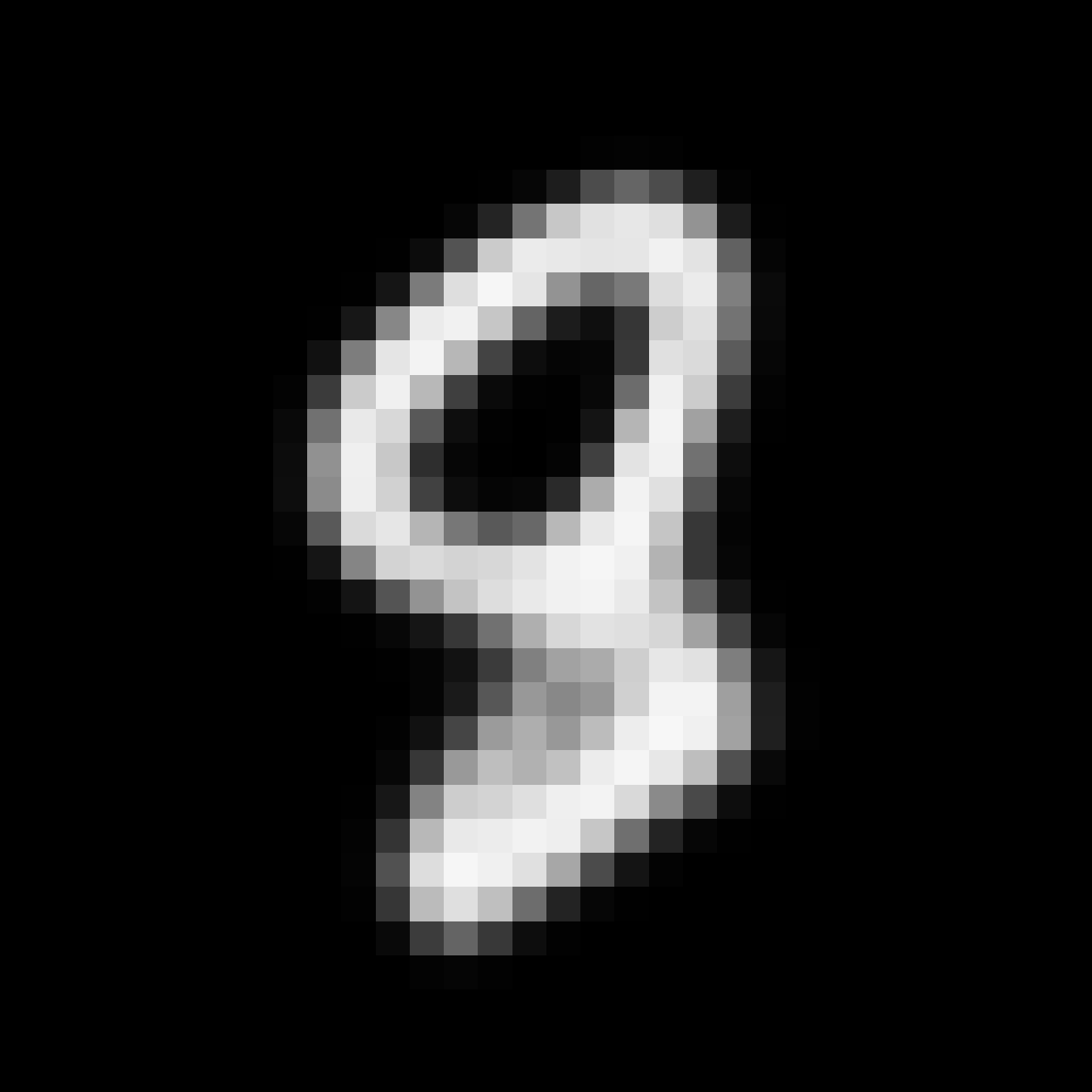} } \\
					
					% row ACAI	
					\subfloat[]{\includegraphics[width=.05\textwidth]{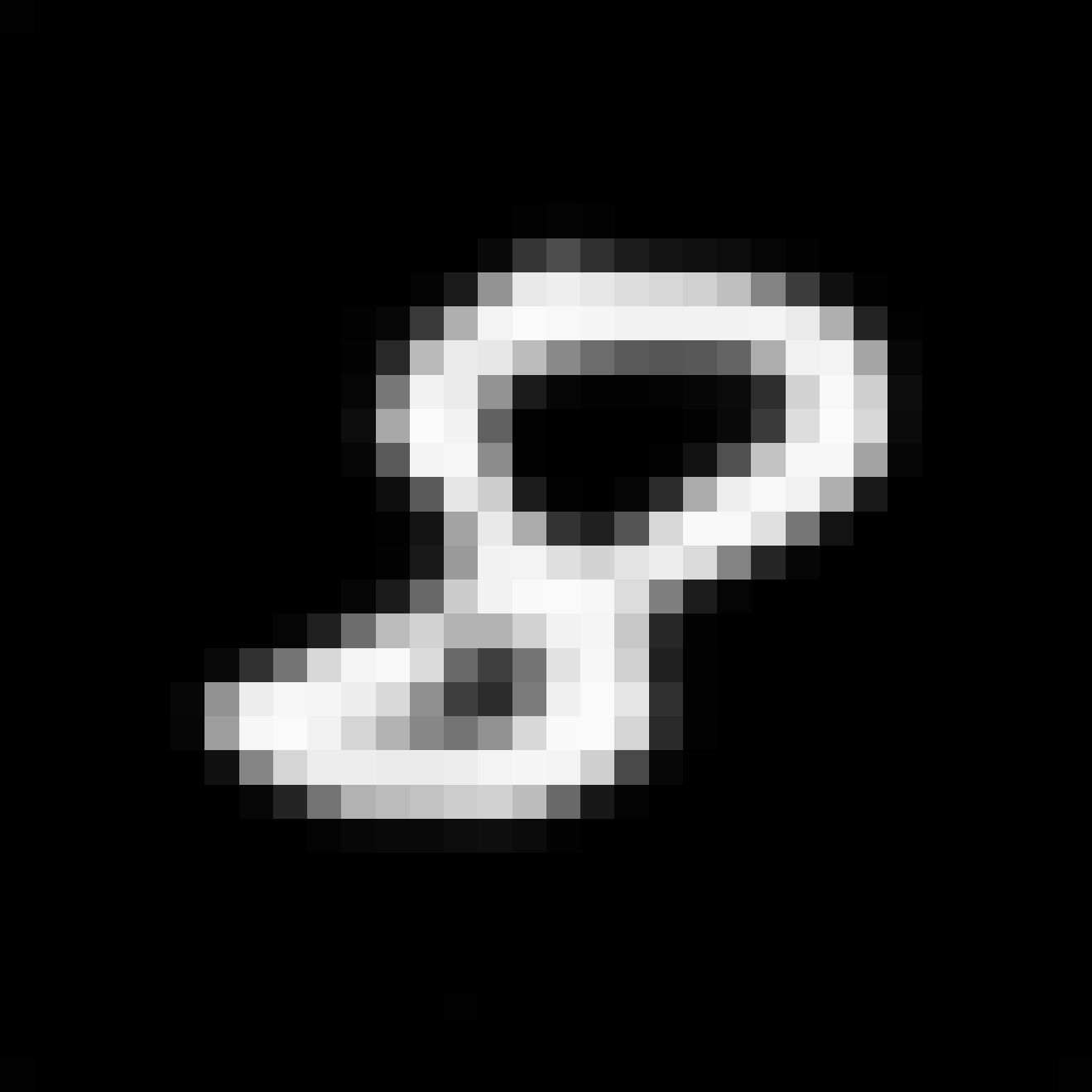} } &
					\subfloat[]{\includegraphics[width=.05\textwidth]{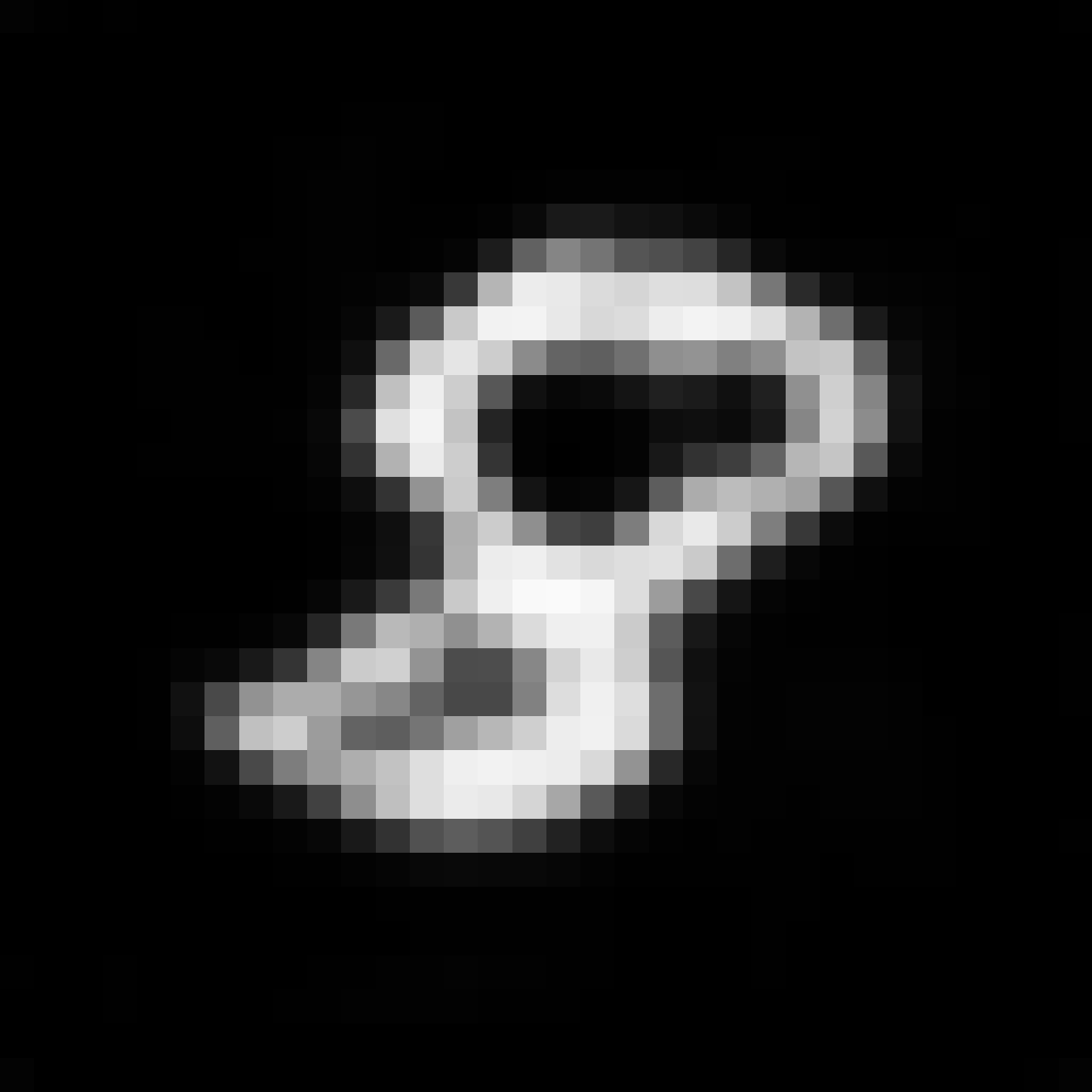} } &
					\subfloat[]{\includegraphics[width=.05\textwidth]{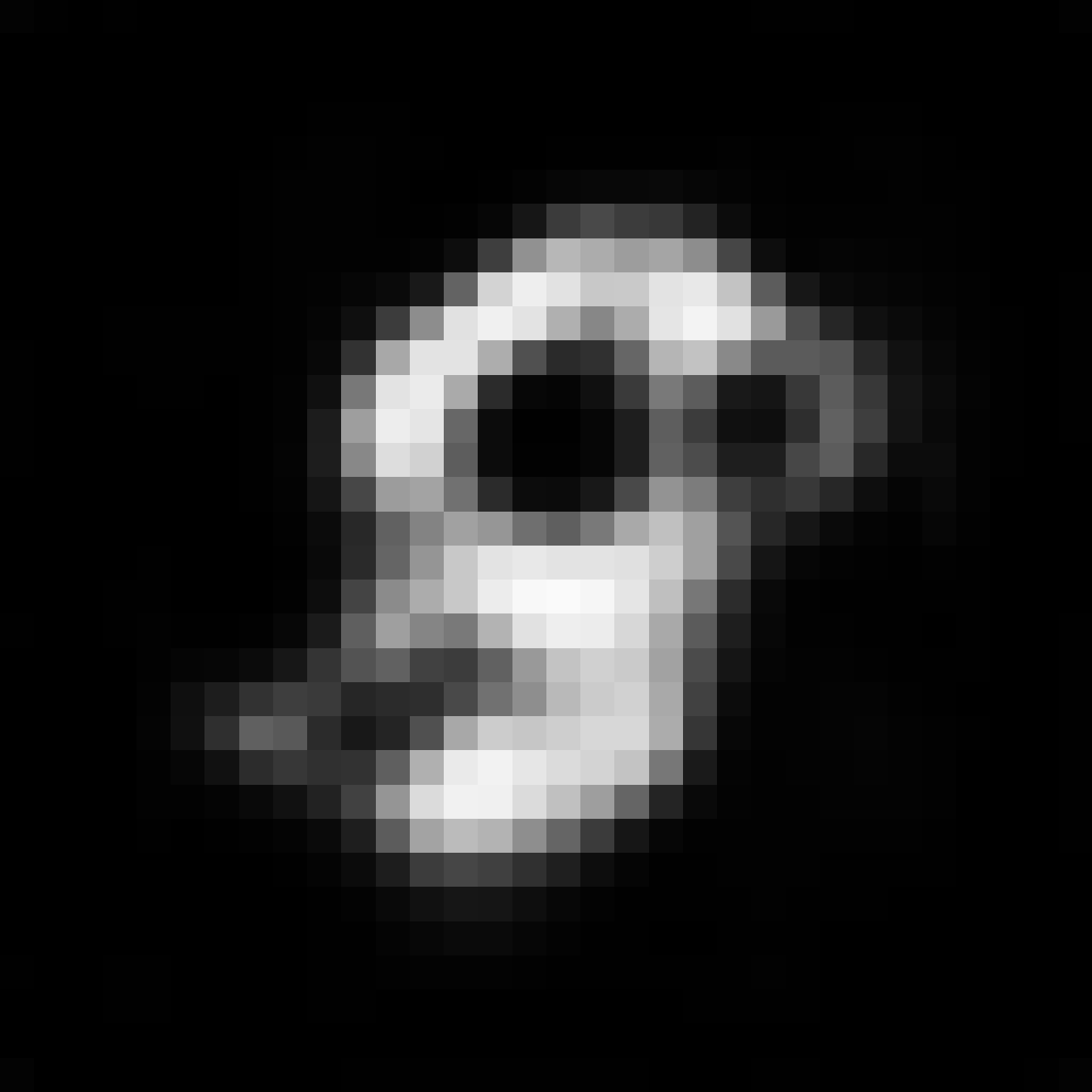} } &
					\subfloat[]{\includegraphics[width=.05\textwidth]{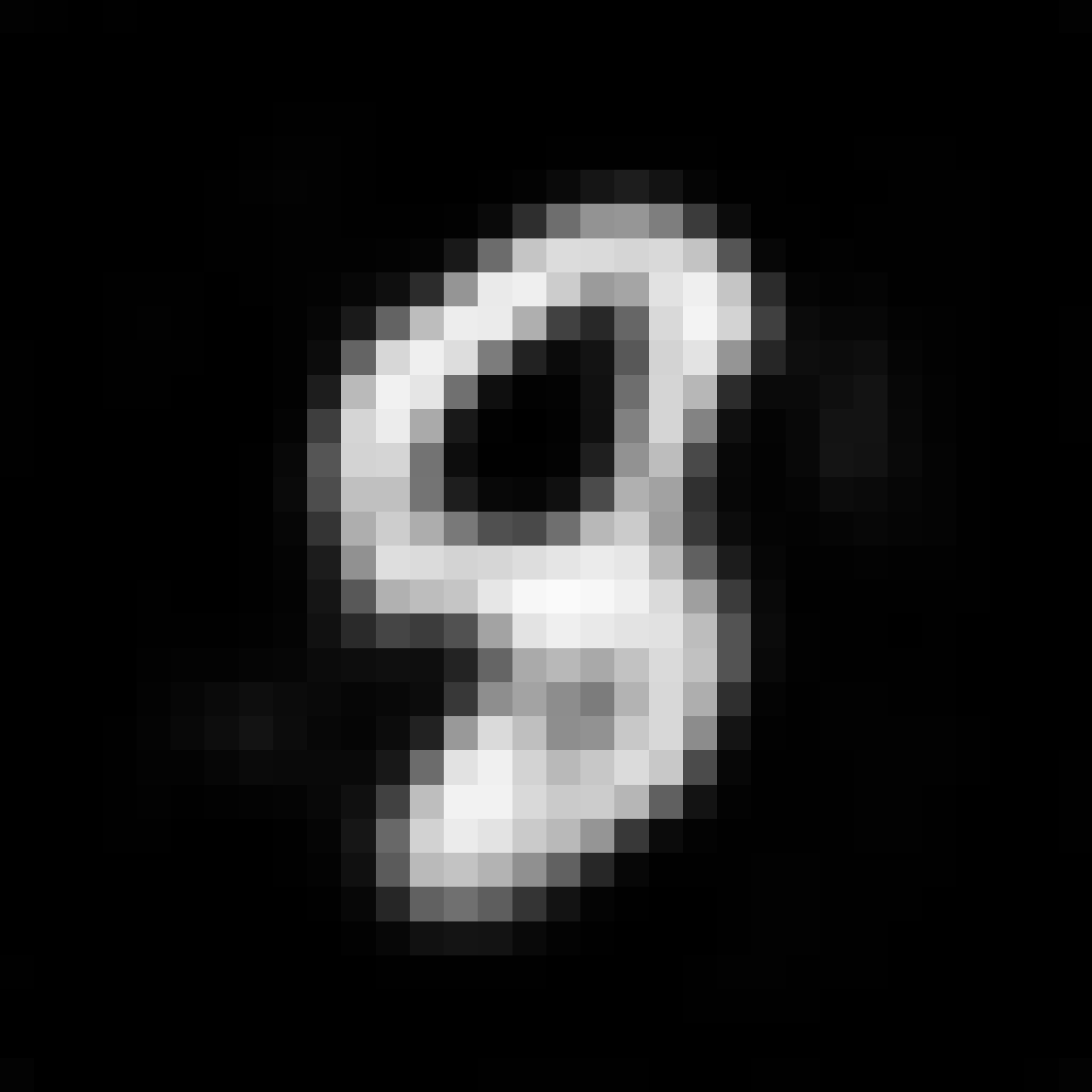} } &
					\subfloat[]{\includegraphics[width=.05\textwidth]{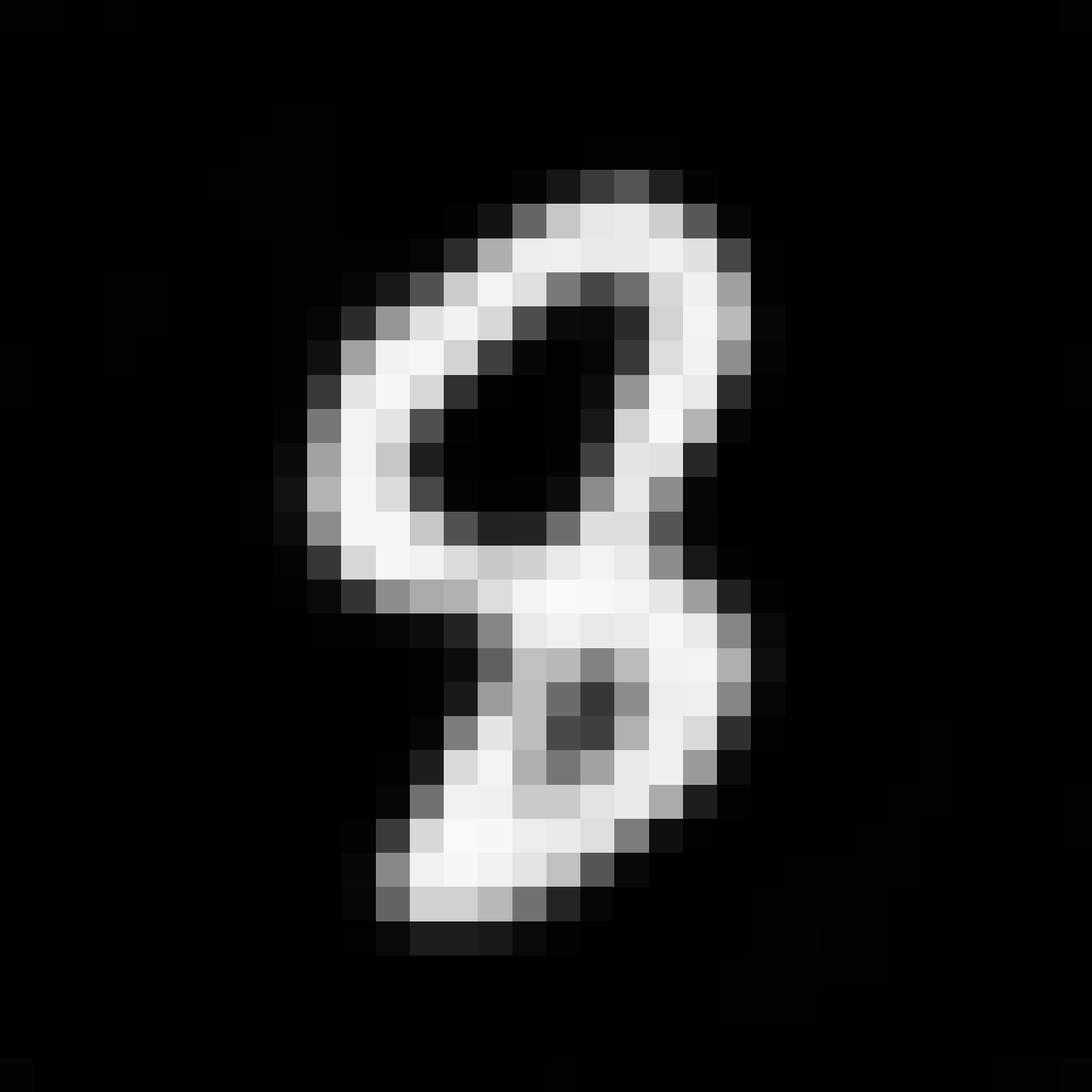} } \\
					
					% row ours
					\subfloat[]{\includegraphics[width=.05\textwidth]{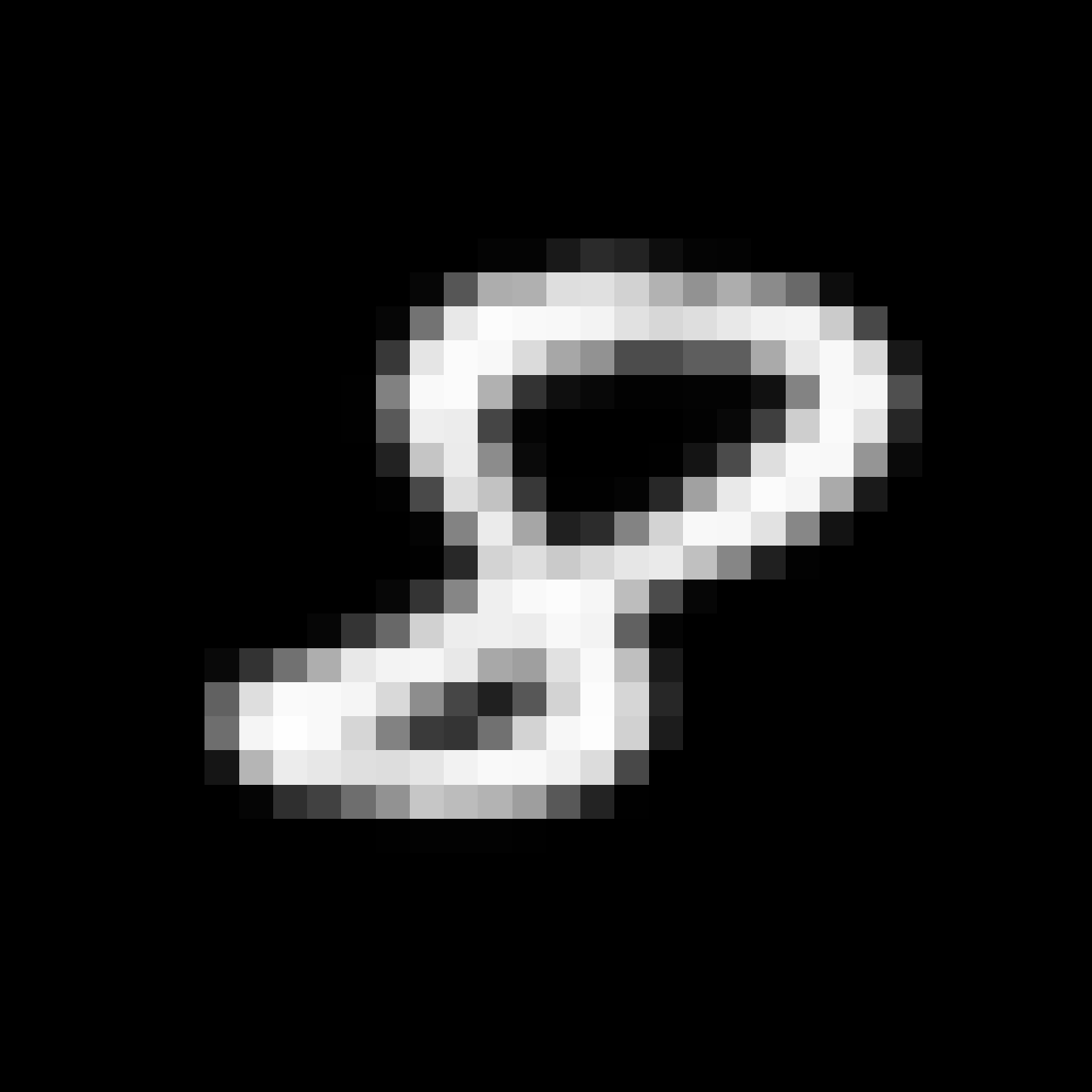} } &
					\subfloat[ ]{\includegraphics[width=.05\textwidth]{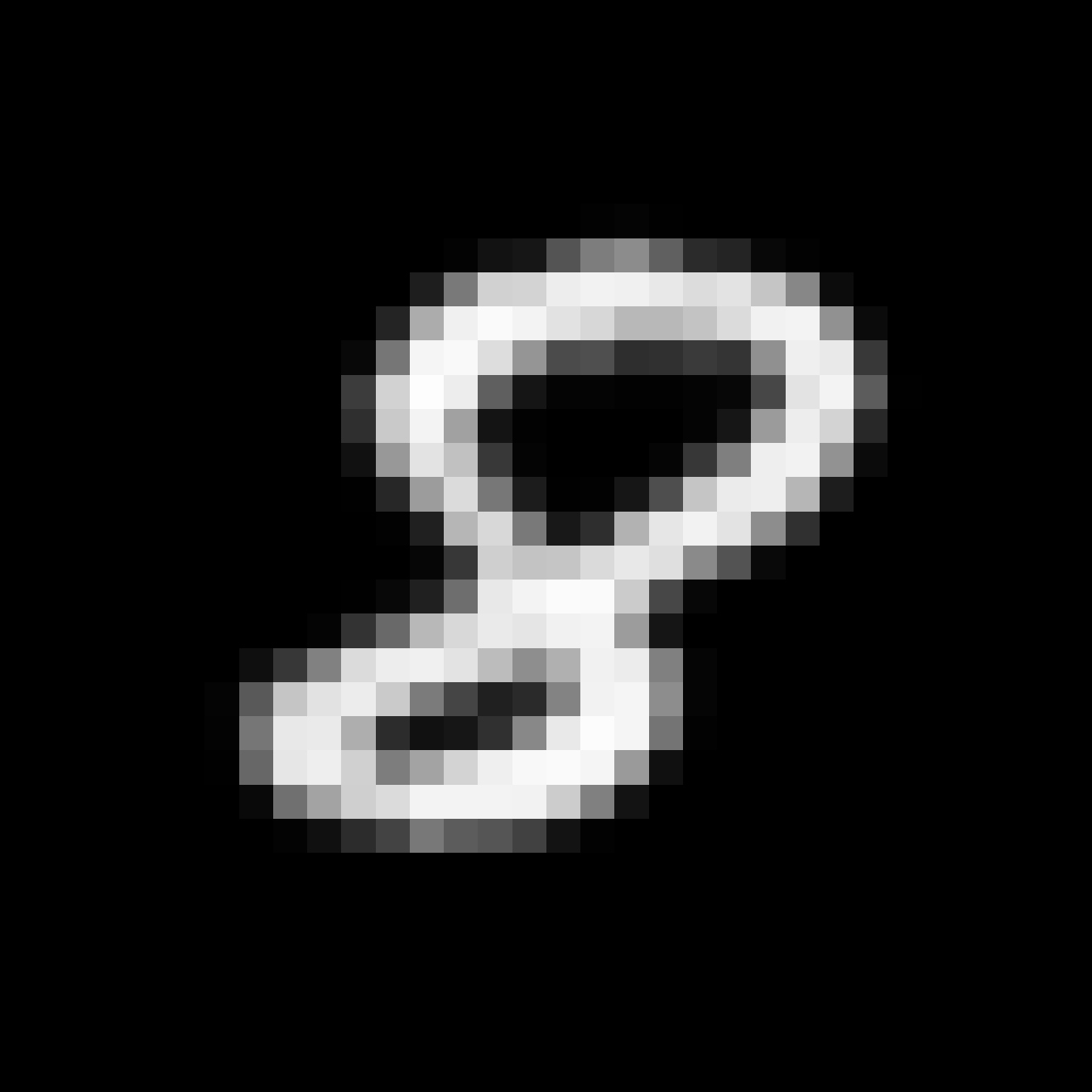} } &
					\subfloat[]{\includegraphics[width=.05\textwidth]{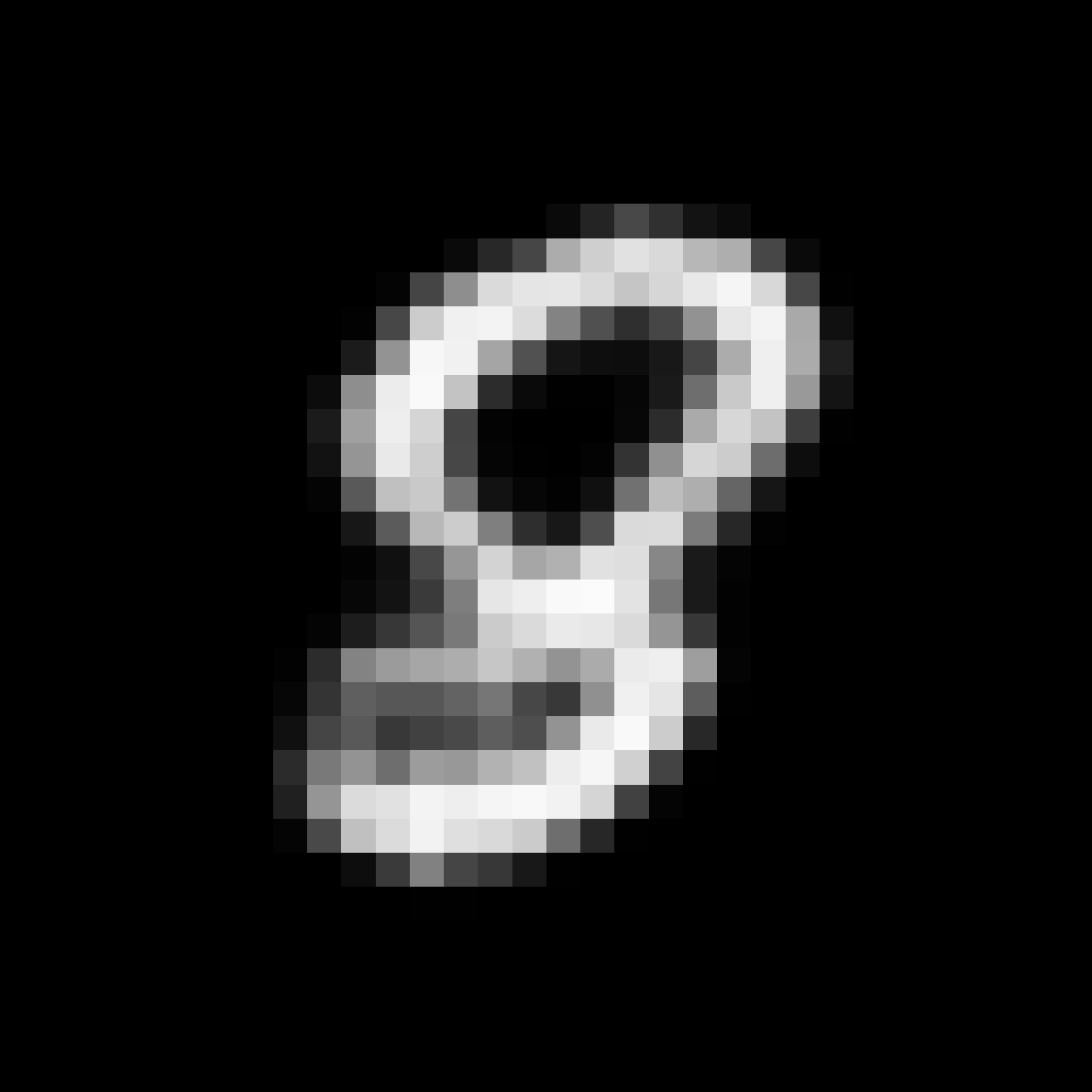} } &
					\subfloat[]{\includegraphics[width=.05\textwidth]{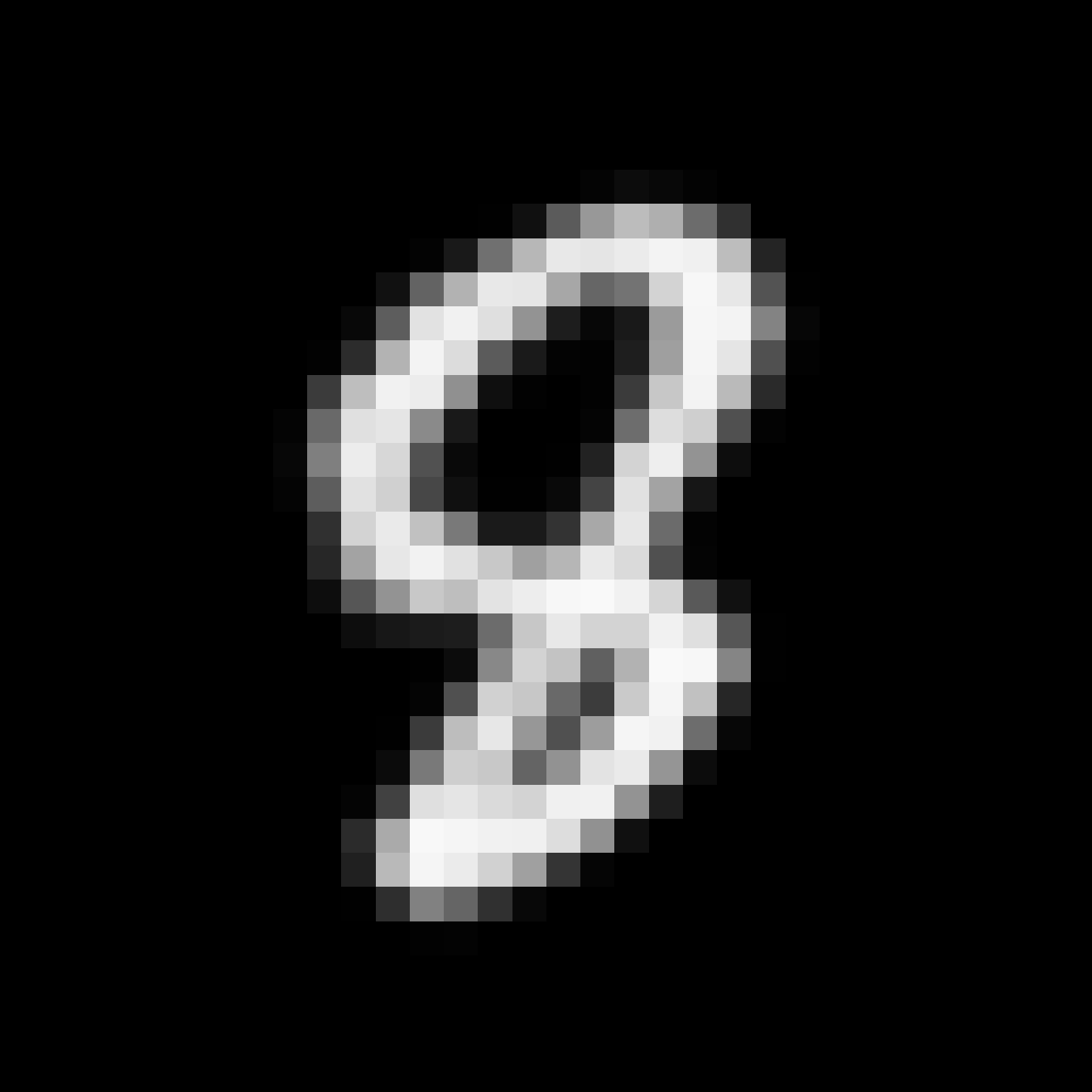} } & 
					\subfloat[]{\includegraphics[width=.05\textwidth]{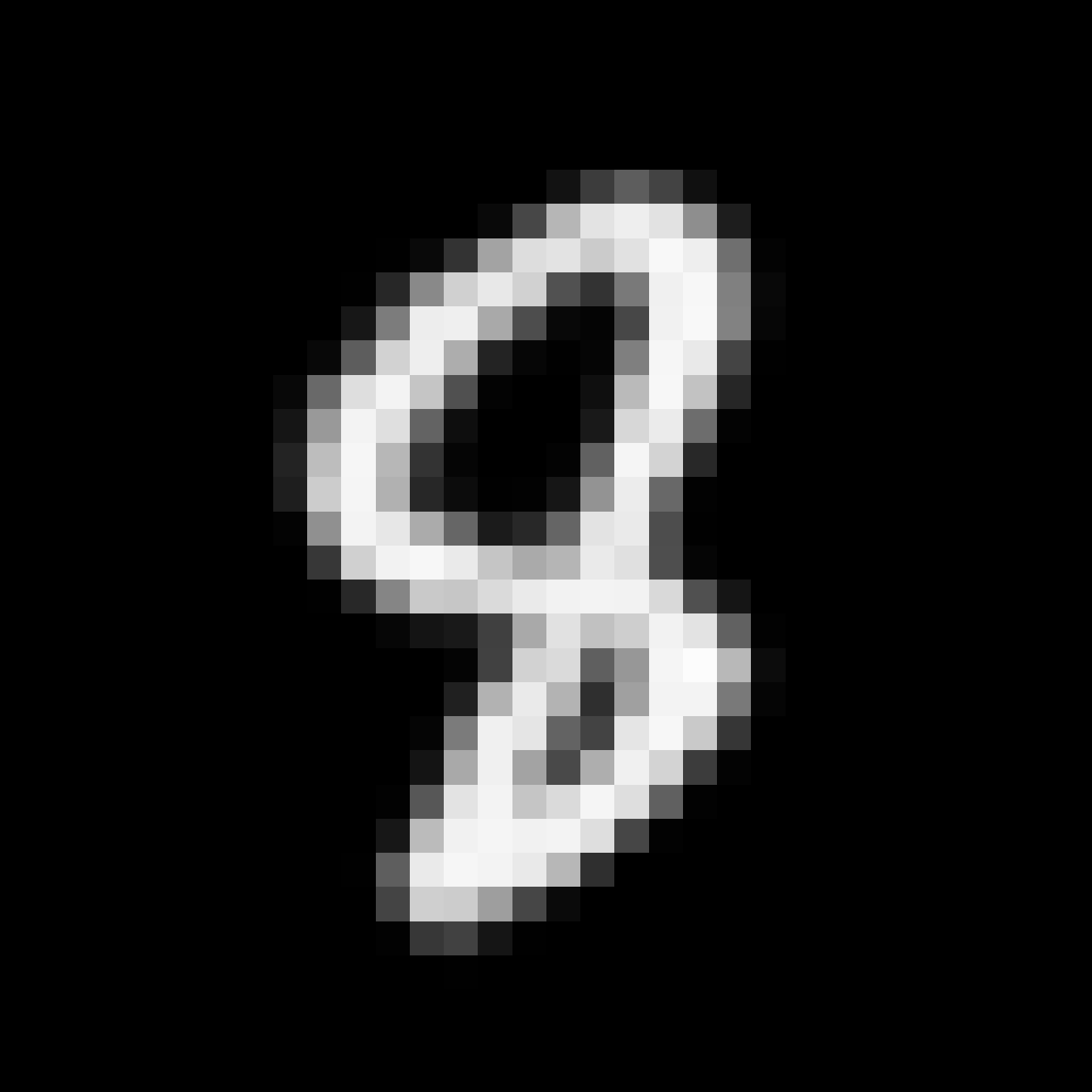} }  \\
					
				\end{tabular}
			}  % end subfloat
		\end{center}
		\caption{Examples of rotation performed by interpolating the latent space representation of MNIST digits. The top row shows the original input images (rotated \num{0}$^{\circ}$ and \num{40}$^{\circ}$) and the intermediate target rotations (\num{10}$^{\circ}$, \num{20}$^{\circ}$, \num{30}$^{\circ}$). The results in the rows below show reconstructions (most left and right) and synthesis (intermediate rotation angles) of intermediate rotations using latent space interpolation. Note that our approach follows the original image rotation more closely than VAE \citep{kingma2014stochastic} and ACAI approach \citep{berthelot2018understanding}. $\alpha$ denotes the mixing coefficient as specified in Equation~\ref{eq_convex_combination}.}
		\label{fig_mnist_roto_qual1}
	\end{figure*}

	\renewcommand{\arraystretch}{1} 
	% Interpolation performance of proposed model (ASI) compared with Variational Autoencoder (VAE) and Adversarially Constrained Autoencoder Interpolation (ACAI) approach on rotated MNIST images. Shown are three examples. Images in first row show original MNIST images (first column) and four rotated versions of the same image using rotation steps of \num{10}$^\circ$). Second to fourth rows show reconstructed and synthesized images using latent space encodings of the two neighboring images (first and fifth column for each of the three examples).
	
	% ---------------------- IMPLEMENTATION DETAILS
	
	\section{Implementation details} \label{section_implementation}
	
	The method was implemented using the PyTorch framework (\cite{paszke2017automatic}) and trained on one Nvidia GTX Titan X GPU with \num{12} GB memory. Model weights were initialized as zero-mean Gaussian random variables with a standard deviation of $1/\sqrt{n_l \; (1+0.2^2)}$ set in accordance with the leaky ReLU slope of \num{0.2}. $n_l$ denotes the number of incoming network connections to layer $l$. 
	
	A model was trained using mini-batch stochastic gradient descent with a learning rate of \num{1e-5}. Image slices were provided once per epoch to the autoencoder in random order. In each experiment the training set was augmented by \num{90} degree rotations of the images and random intensity changes. Network parameters were optimized using the Adam optimizer (\cite{kingmadp}) minimizing the reconstruction and synthesis loss.
	
	% To compute the reconstruction loss, this work used the pixel-wise mean squared error between original $x_{n}$ and reconstructed $\hat{x}_{n}$ image. In addition, to compute the synthesis loss between reference $x_{n}$ and synthesized image $\hat{x}_{n}^{\alpha=0.5}$ the Learned Perceptual Image Patch Similarity (LPIPS) metric (\cite{zhang2018perceptual}) was used. The LPIPS metric is a perceptually-based pairwise image distance that is calculated as a weighted difference between the VGG-16 (\cite{simonyan2014very}) embedding of the reference and synthesized image. LPIPS uses the embeddings of VGG-16 layers \texttt{conv\_1} to \texttt{conv\_5}. The VGG-16 CNN is pretrained on ImageNet and the weights to compute the weighted difference were fit so that the metric agrees with human perceptual similarity judgments. 
	
	To compute the synthesis loss as described in Section~\ref{method_combined_loss}, this study used the LPIPS metric implementation\footnote{\url{https://github.com/richzhang/PerceptualSimilarity}} of \cite{zhang2018perceptual}. 
	Furthermore, in order to compute the synthesis loss mini-batches of $T$ slice pairs were randomly selected from the training set. A slice pair consisted of two slices ($x_{n-1}$ and $x_{n+1}$) originating from the same volume that are spatially separated by one in-between slice $x_n$. To determine the optimal value for $\lambda$ i.e. the contribution of the synthesis loss to the overall loss, a separate line search was performed for each dataset.
	
	Finally, in all experiments model selection was performed on the validation set. The test set was not used during method development in any way.
	
	%where $d_{MSE}$ denotes the pixel-wise mean squared error between original $x_{n}$ and reconstructed $\hat{x}_{n}$ image and $d_{\text{LPIPS}}$ the perceptual loss between original and synthesized $\hat{x}_{n}^{\alpha=0.5}$ image as described in Section~\ref{method_loss_combined}. Moreover, $\lambda$ was set to zero when training a model with the AISR baseline approach. A separate line search was performed for each dataset to determine the optimal value for $\lambda$ when training a model with the CAISR approach.
	
	% ---------------------------- EVALUATION 

	\begin{table}[ht]
		\caption{Quantitative comparison of reconstruction and synthesis performance of cardiac cine MRIs (ACDC dataset) in terms of SSIM, PSNR, and VIF between proposed model trained with \textit{reconstruction} loss only (ASI$_{\lambda=0}$) and model trained with combination of \textit{reconstruction} and \textit{synthesis} loss (ASI$_{\lambda=0.05}$). A higher score indicates better performance. Measures (mean$\pm$standard deviation) are computed on cardiac short-axis slices. \textit{Rec} denotes reconstructed and \textit{Syn} synthesized slices. Synthesis performance was assessed on downsampled test volumes using a factor of \num{2} in through-plane direction. Best performance is indicated in bold.}
		\label{table_acdc_ae_caisr_recon_versus_synth_1}
		\centering
		
		\begin{tabular}{l  C{0.7cm} C{0.7cm} C{0.7cm} C{0.7cm} C{0.7cm} C{0.7cm} }
			\toprule
			\textbf{Method} & \multicolumn{2}{c}{\textbf{SSIM}} & \multicolumn{2}{c}{\textbf{PSNR} } & \multicolumn{2}{c}{\textbf{VIF} } \\
			& Rec & Syn & Rec & Syn  & Rec & Syn  \\
			
			ASI$_{\lambda=0}$ & \begin{tabular}{@{}c@{}}\textbf{0.994} \\ $\pm$0.01 \end{tabular} & \begin{tabular}{@{}c@{}}0.572 \\ $\pm$0.09 \end{tabular} & \begin{tabular}{@{}c@{}}\textbf{41.34} \\ $\pm$1.66 \end{tabular} & \begin{tabular}{@{}c@{}}17.94 \\ $\pm$2.01 \end{tabular} & \begin{tabular}{@{}c@{}}\textbf{0.960} \\ $\pm$0.01 \end{tabular} & \begin{tabular}{@{}c@{}}\textbf{0.815} \\ $\pm$0.01 \end{tabular} \\
			ASI$_{\lambda=0.05}$ & \begin{tabular}{@{}c@{}}0.968 \\ $\pm$0.01 \end{tabular} & \begin{tabular}{@{}c@{}}\textbf{0.650} \\ $\pm$0.07 \end{tabular} & \begin{tabular}{@{}c@{}}32.83 \\ $\pm$1.31 \end{tabular} & \begin{tabular}{@{}c@{}}\textbf{19.01} \\ $\pm$1.89 \end{tabular} & \begin{tabular}{@{}c@{}}0.891 \\ $\pm$0.01 \end{tabular} & \begin{tabular}{@{}c@{}}0.810 \\ $\pm$0.01 \end{tabular} \\
			\bottomrule
		\end{tabular}
	\end{table}
	
	% FIGURE 2
	\begin{figure}[ht]
		\captionsetup[subfigure]{justification=centering}
		\centering
		\subfloat[Original]{\includegraphics[width=.16\textwidth]{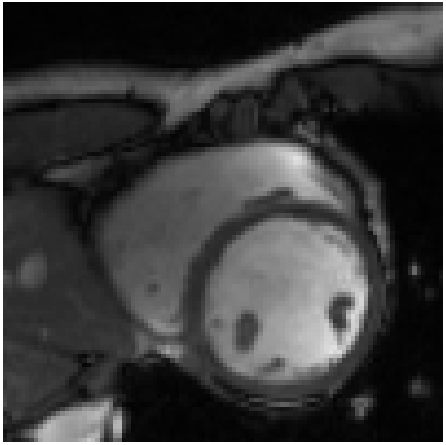} }
		\subfloat[Reconstruction]{\includegraphics[width=.16\textwidth]{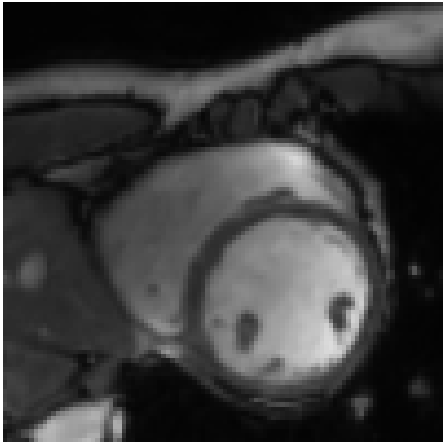} }
		\subfloat[Differences]{\includegraphics[width=.16\textwidth]{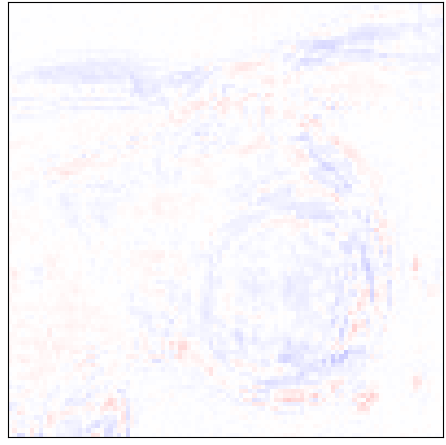} 
			\label{fig_qualitative_reconstruction_acdc_diff}  }
		\caption{Qualitative evaluation of reconstruction performance of our method on cardiac cine MRI (ACDC dataset). (a) Original cardiac MRI scan; (b) Its reconstruction and (c) Differences between original (minuend) and corresponding reconstructed (subtrahend) slice. Note that to reconstruct a slice $x_n$ the mixing coefficient $\alpha$ in Equation~\ref{eq_convex_combination} is set to zero. Blue corresponds to negative and red to positive differences. Image intensities are scaled to a $[0,1]$ range. All difference images use the same color scale $[-1, 1]$.}
		\label{fig_qualitative_reconstruction_acdc}
	\end{figure}
	
	\vspace{1ex}
	
	% Figure 3
	\begin{figure*}
		\captionsetup[subfigure]{justification=centering, labelformat=empty}
		\setlength{\tabcolsep}{1pt}
		\begin{center}
			\begin{tabular}{c c c c c c c c}
				% row 1
				\subfloat[Neighboring slice 1]{\includegraphics[width=.1\textwidth]{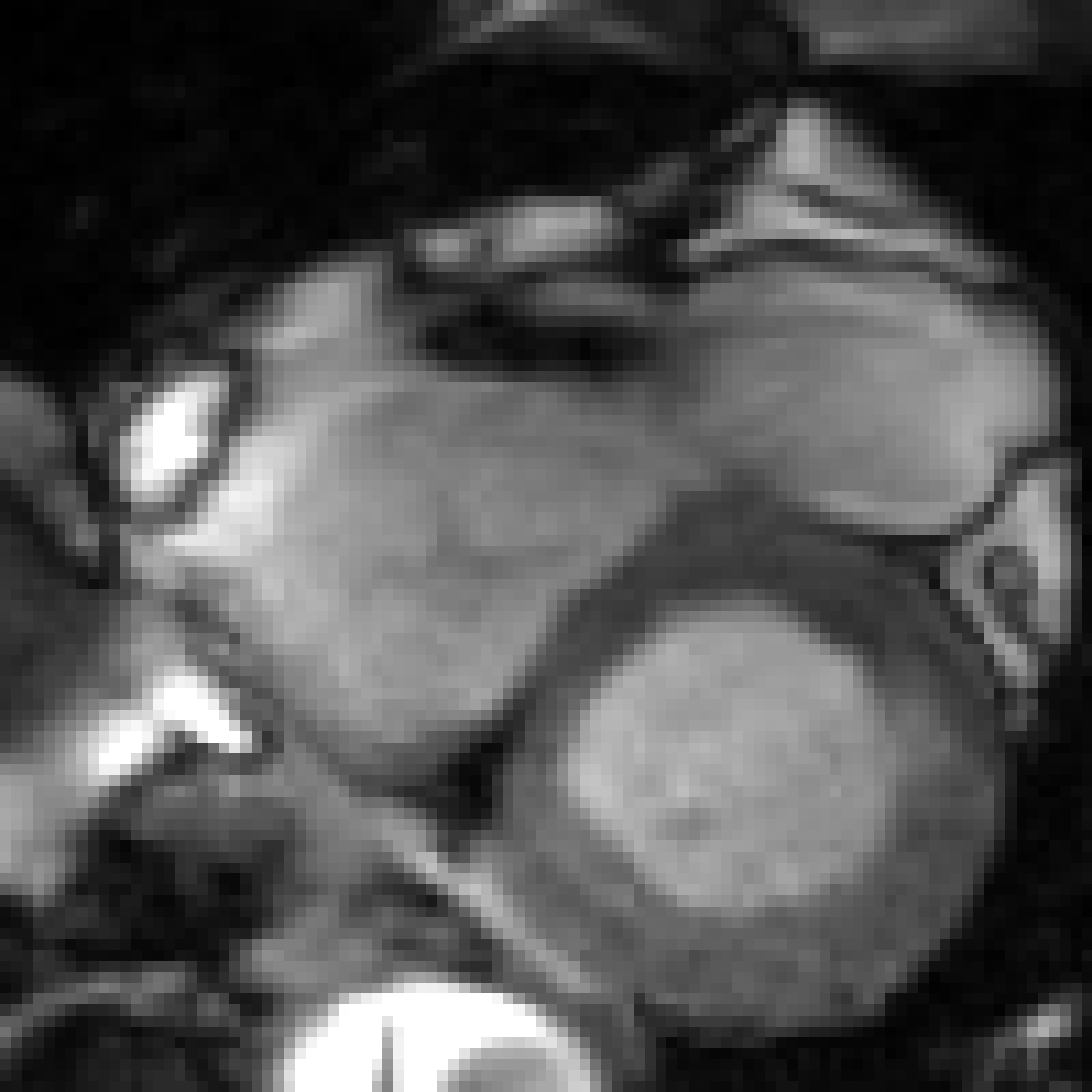} } & 
				\subfloat[$\alpha=$ \nicefrac{1}{7}]{\includegraphics[width=.1\textwidth]{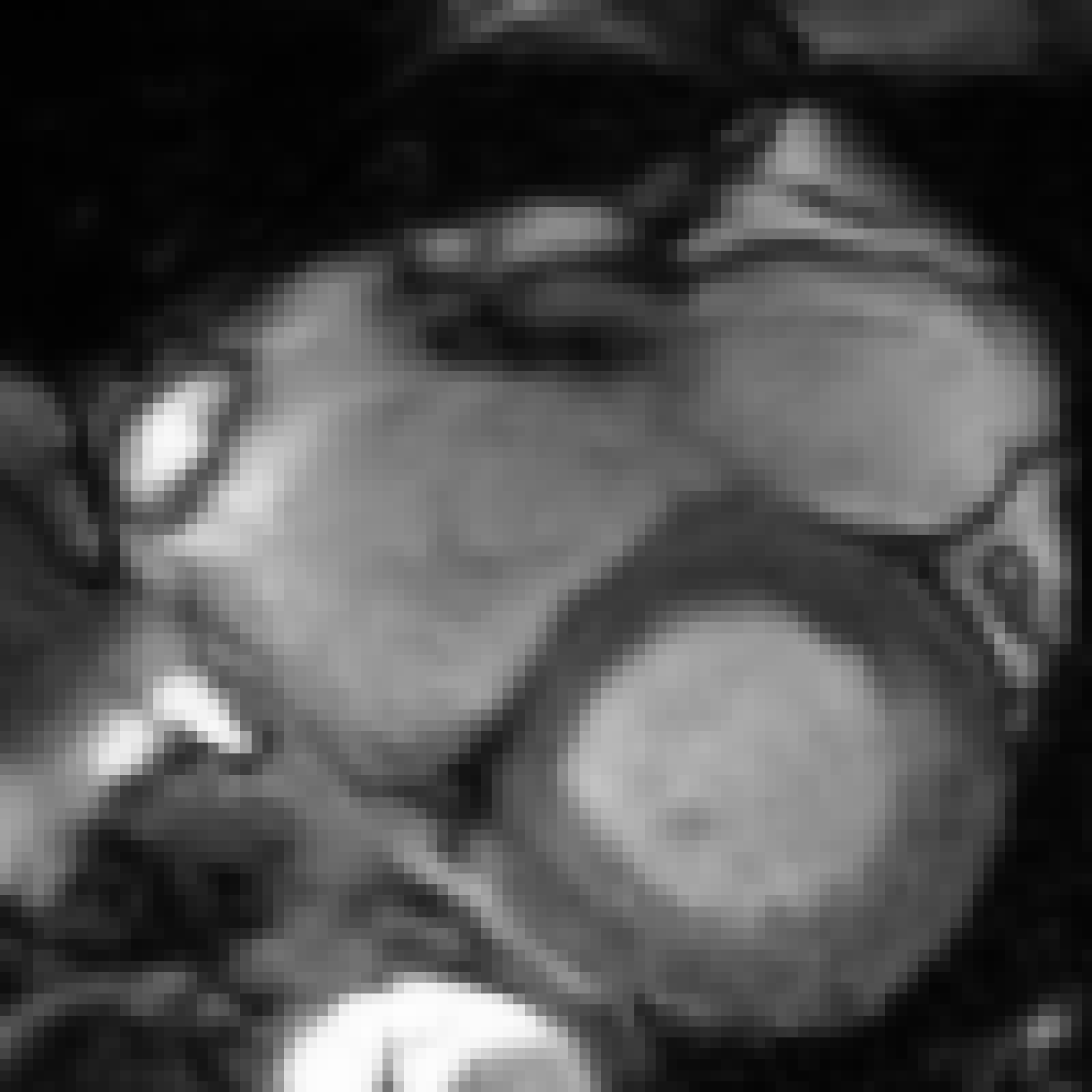} } &
				\subfloat[$\alpha=$ \nicefrac{2}{7}]{\includegraphics[width=.1\textwidth]{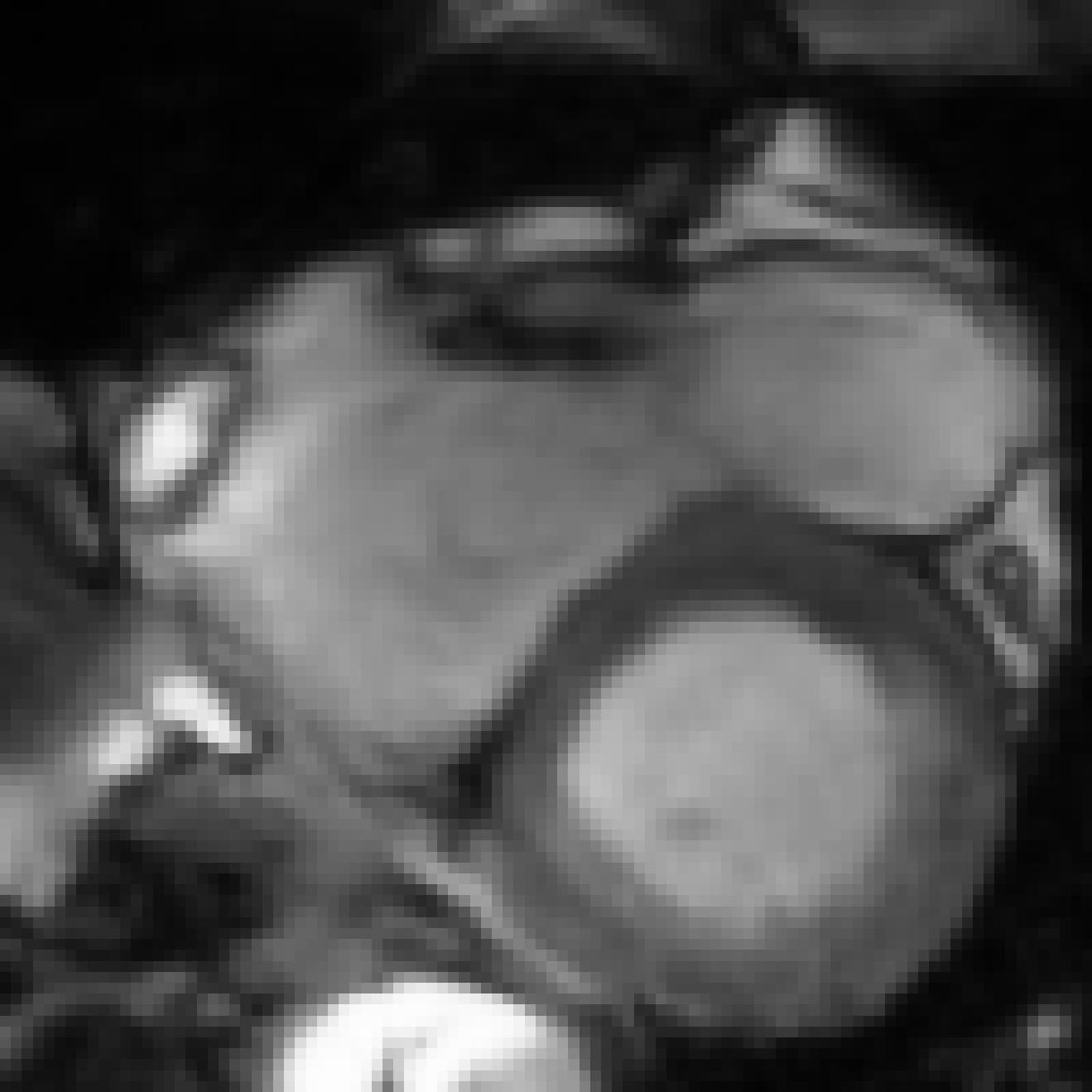} } &
				\subfloat[$\alpha=$ \nicefrac{3}{7}]{\includegraphics[width=.1\textwidth]{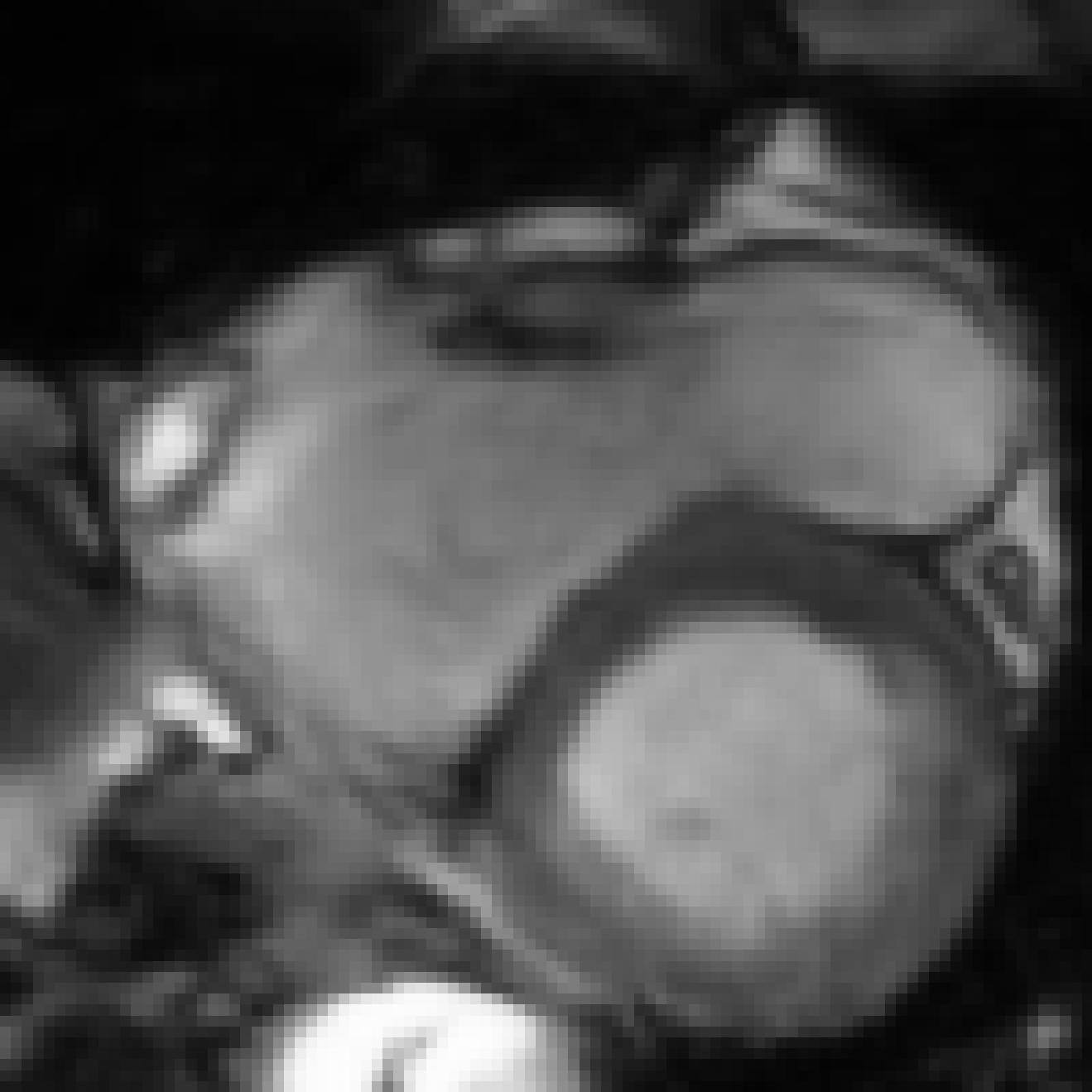} } &
				\subfloat[$\alpha=$ \nicefrac{4}{7}]{\includegraphics[width=.1\textwidth]{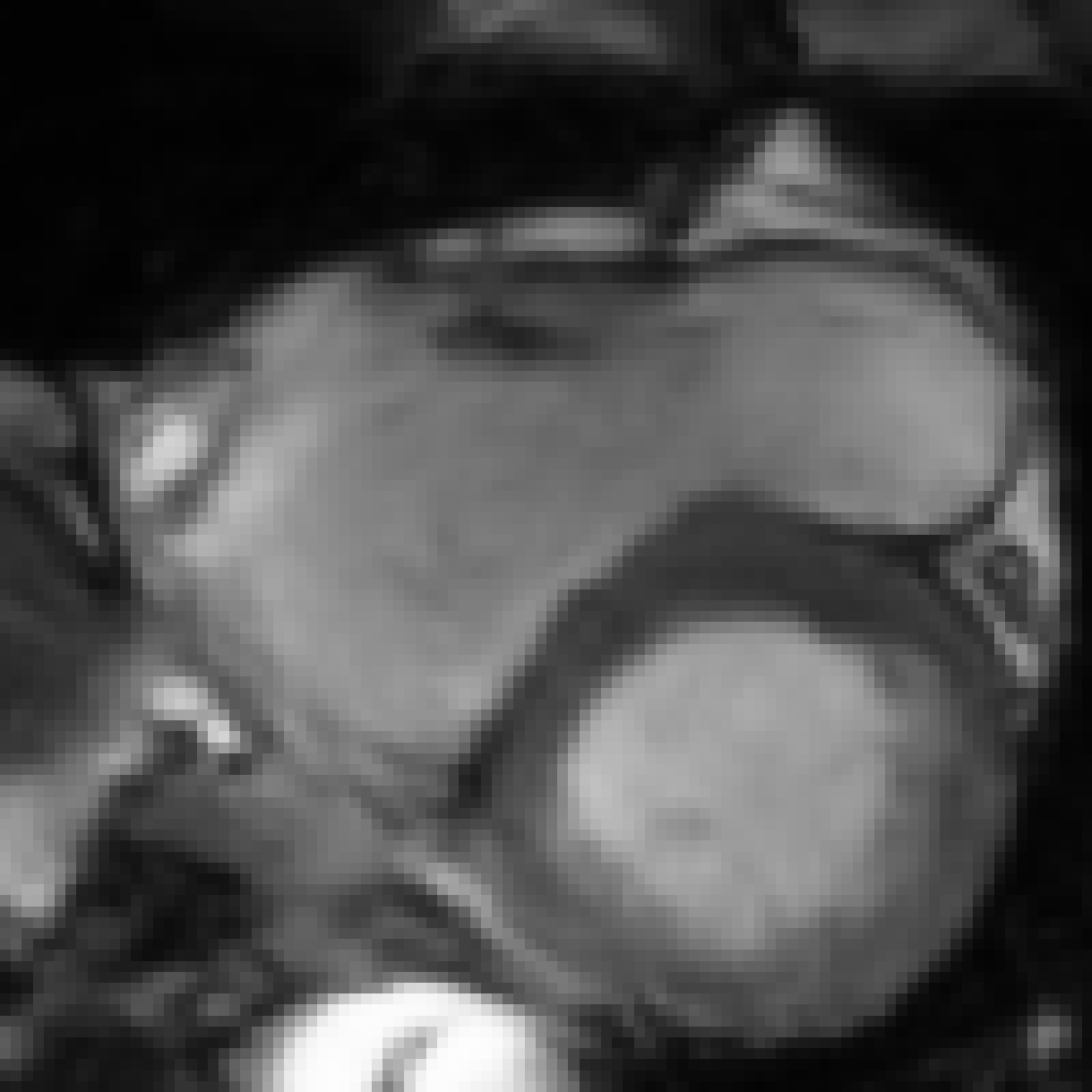} } &
				\subfloat[$\alpha=$ \nicefrac{5}{7}]{\includegraphics[width=.1\textwidth]{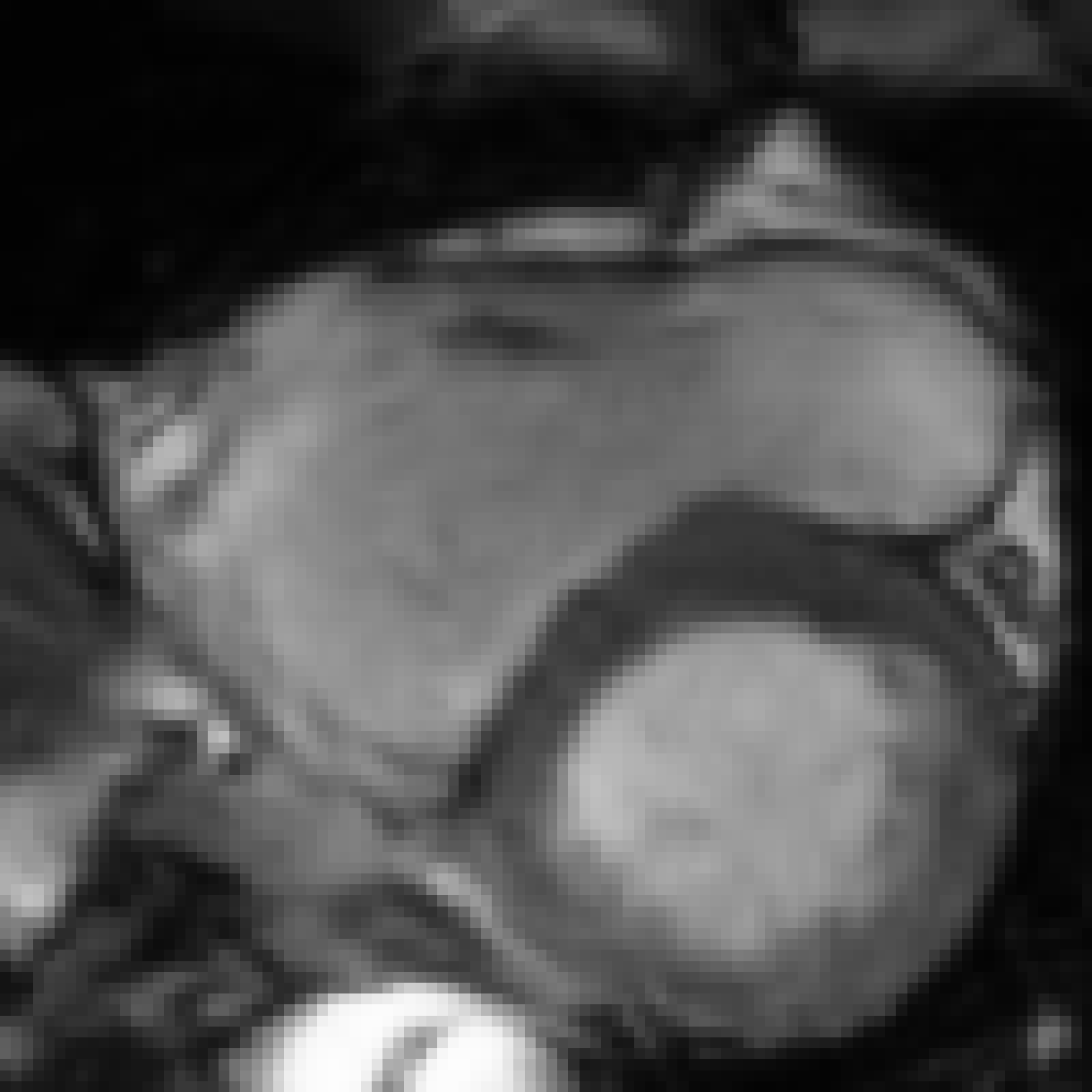} } &
				\subfloat[$\alpha=$ \nicefrac{6}{7}]{\includegraphics[width=.1\textwidth]{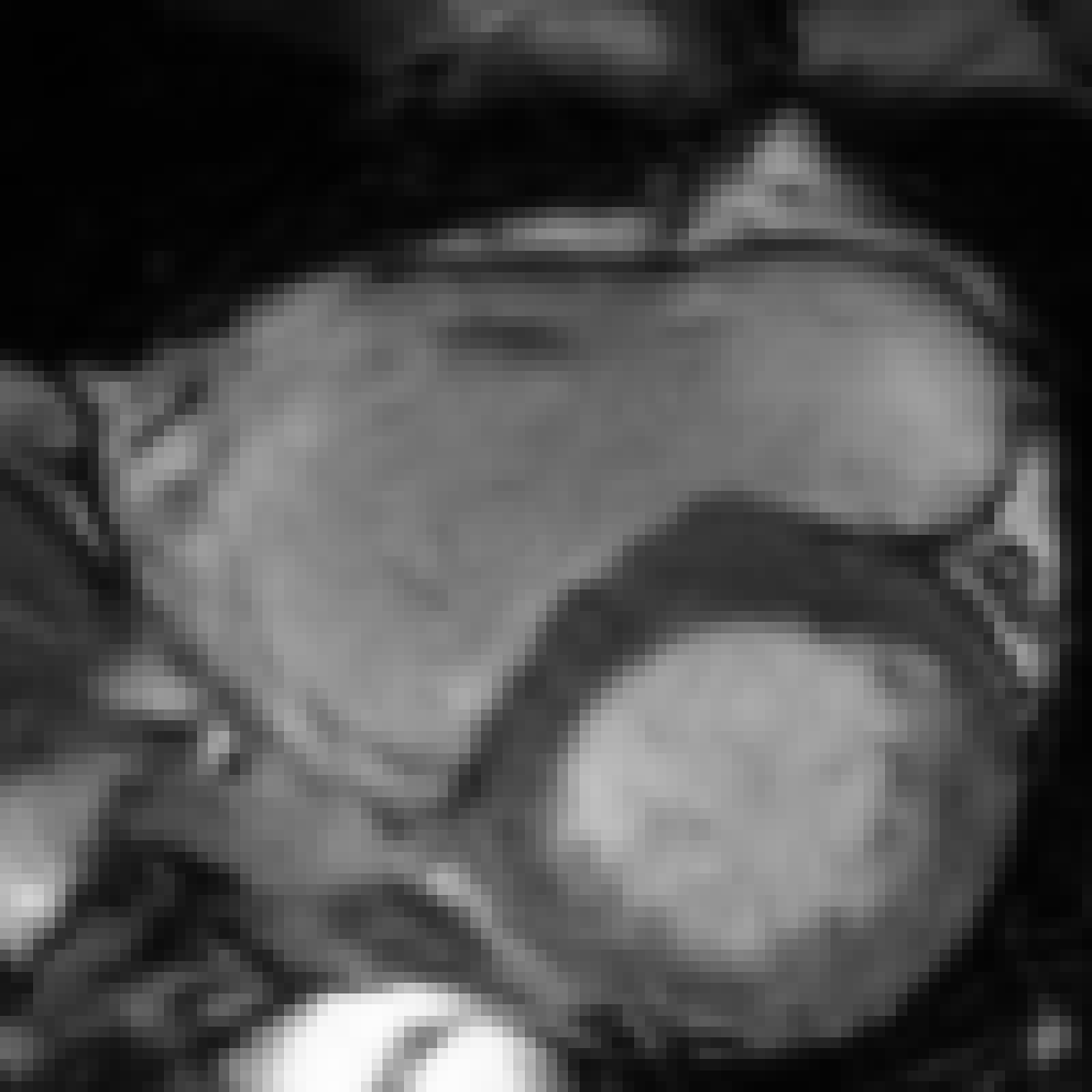} } &
				\subfloat[Neighboring slice 2]{\includegraphics[width=.1\textwidth]{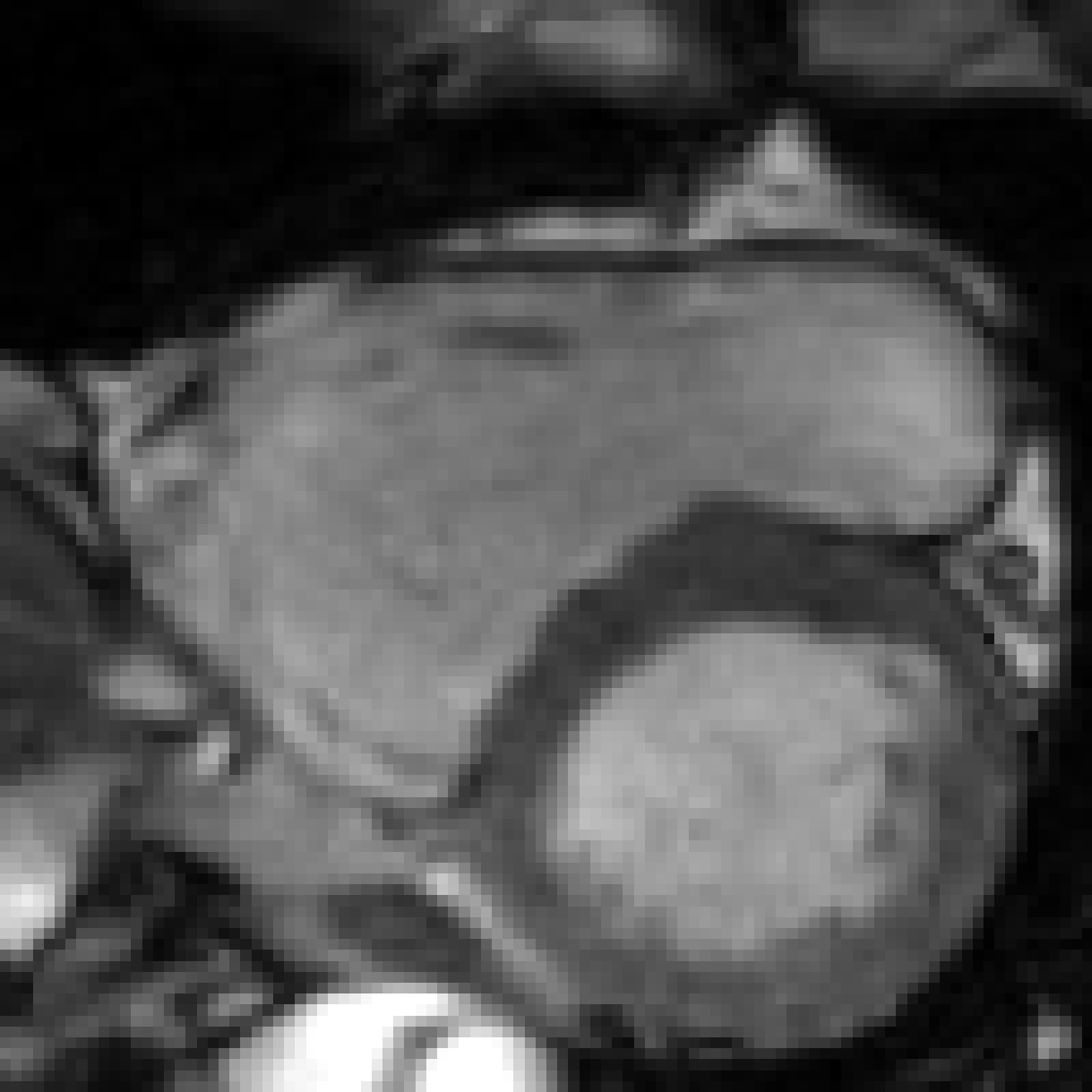} } \\
				% row 2
				\subfloat[Neighboring slice 1]{\includegraphics[width=.1\textwidth]{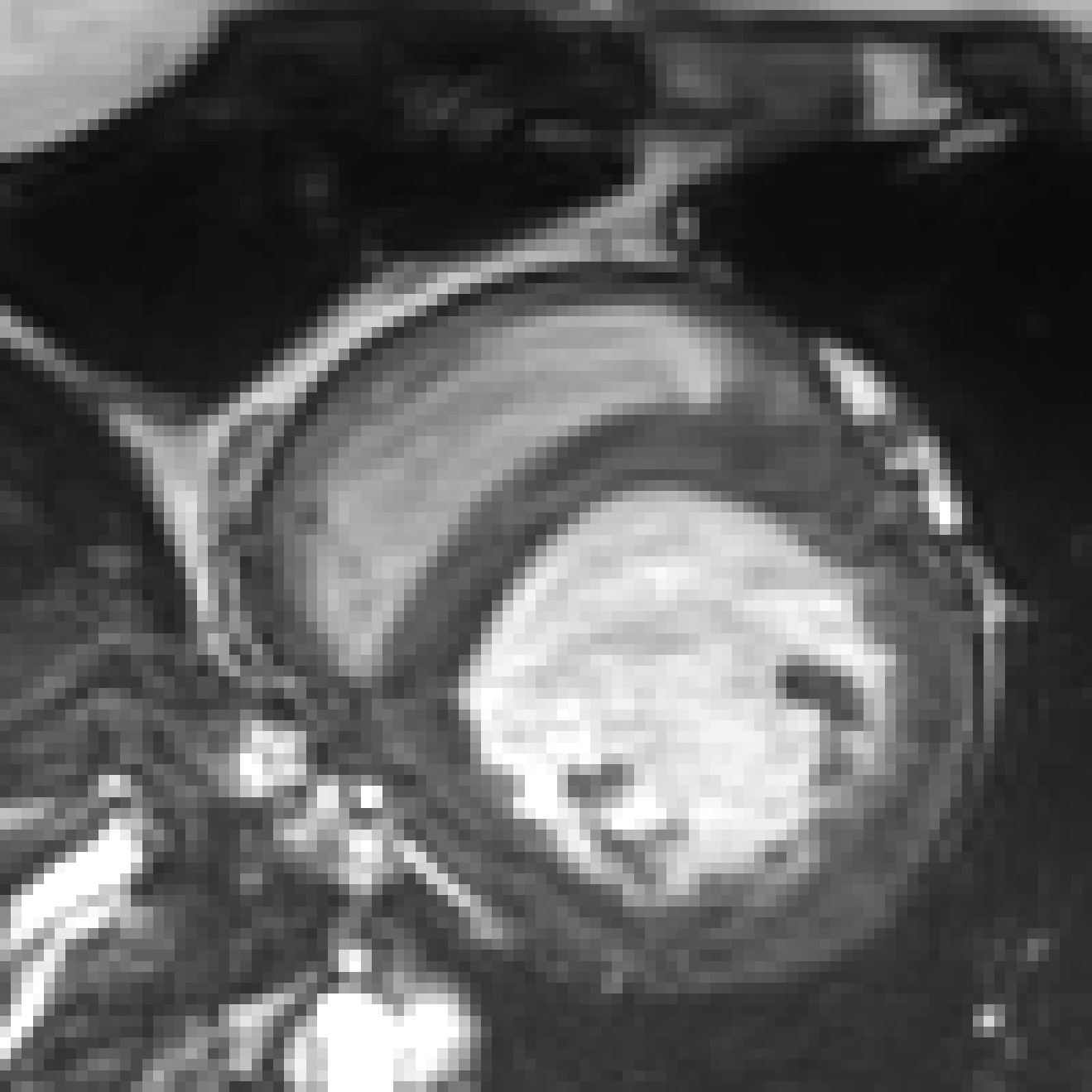} } & 
				\subfloat[$\alpha=$ \nicefrac{1}{7}]{\includegraphics[width=.1\textwidth]{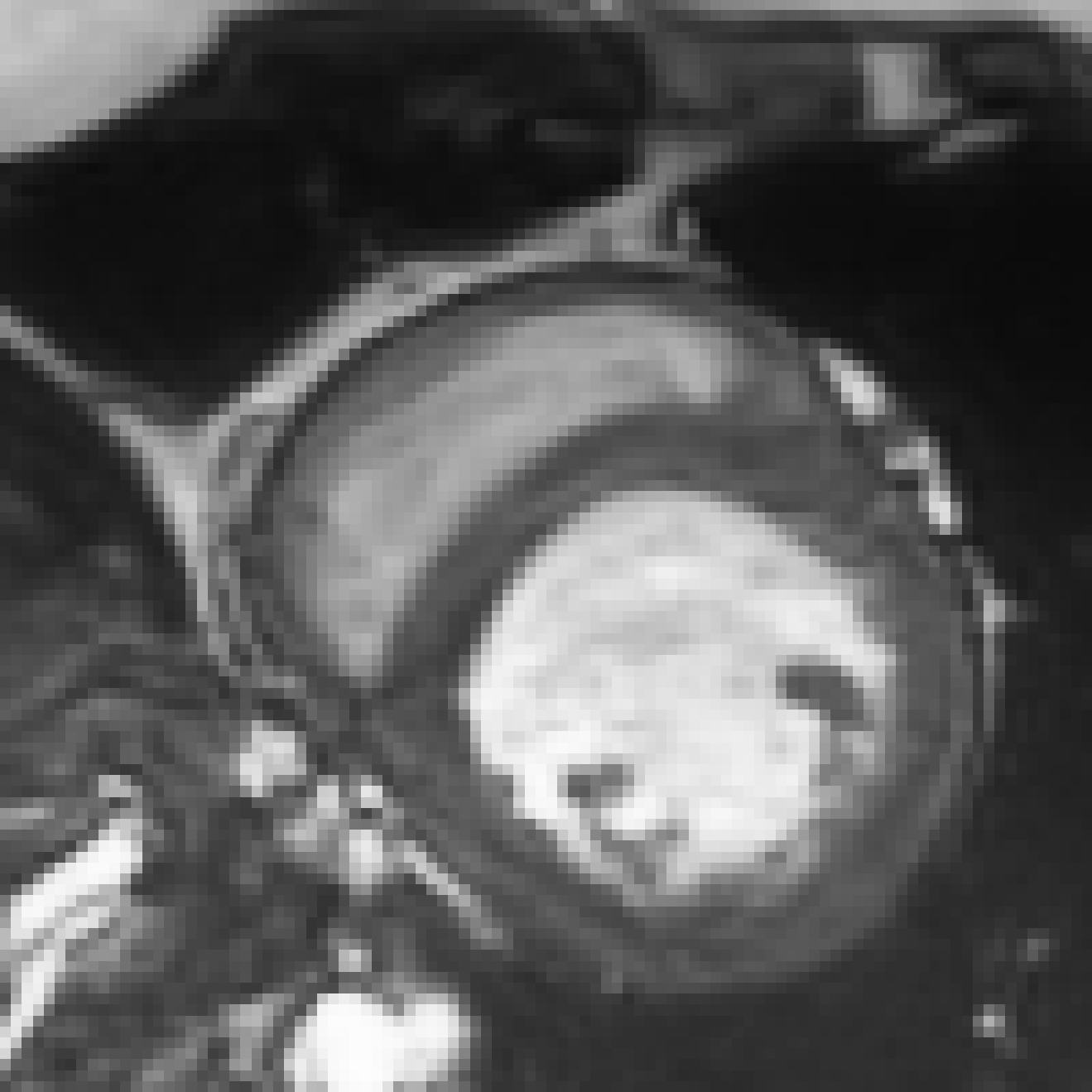} } &
				\subfloat[$\alpha=$ \nicefrac{2}{7}]{\includegraphics[width=.1\textwidth]{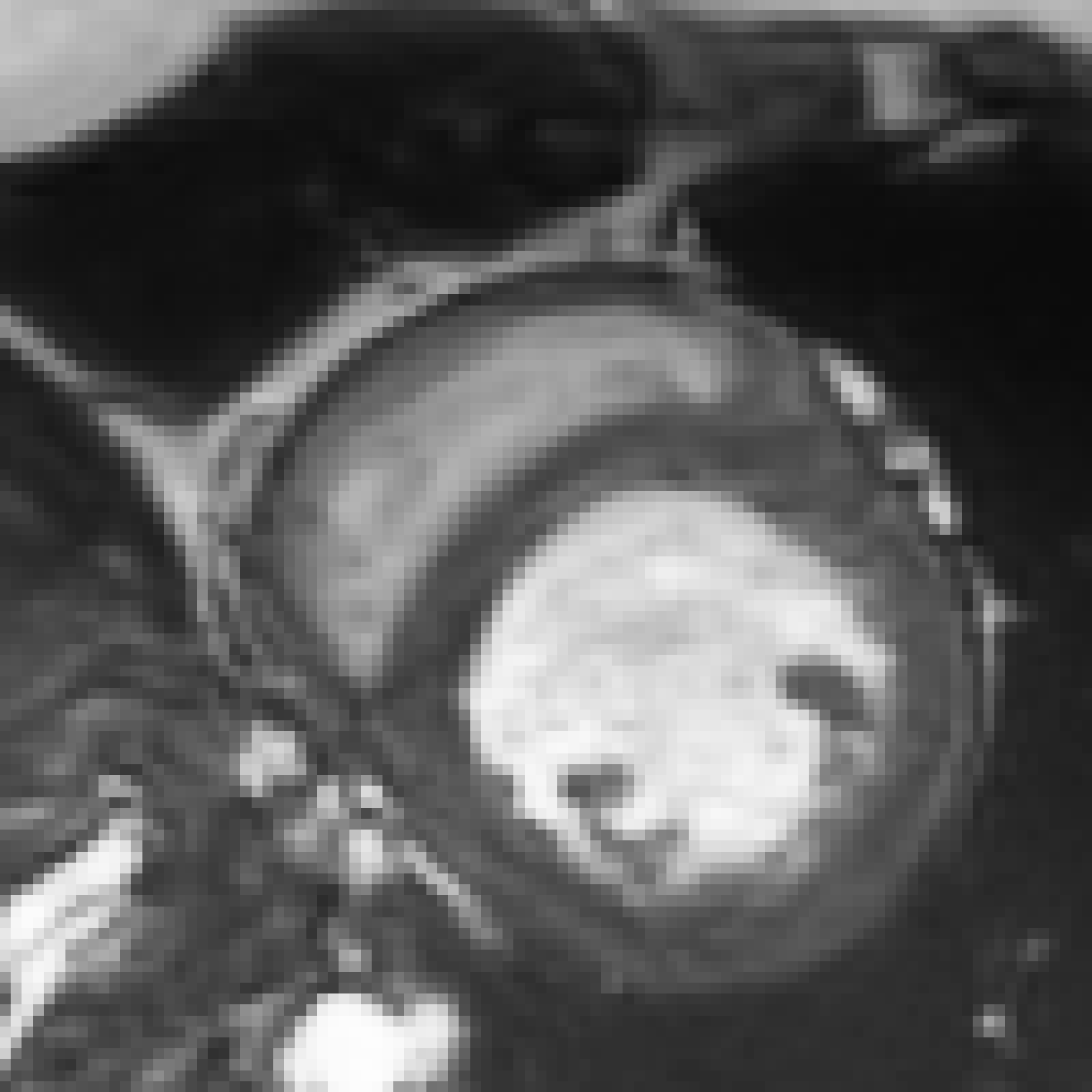} } &
				\subfloat[$\alpha=$ \nicefrac{3}{7}]{\includegraphics[width=.1\textwidth]{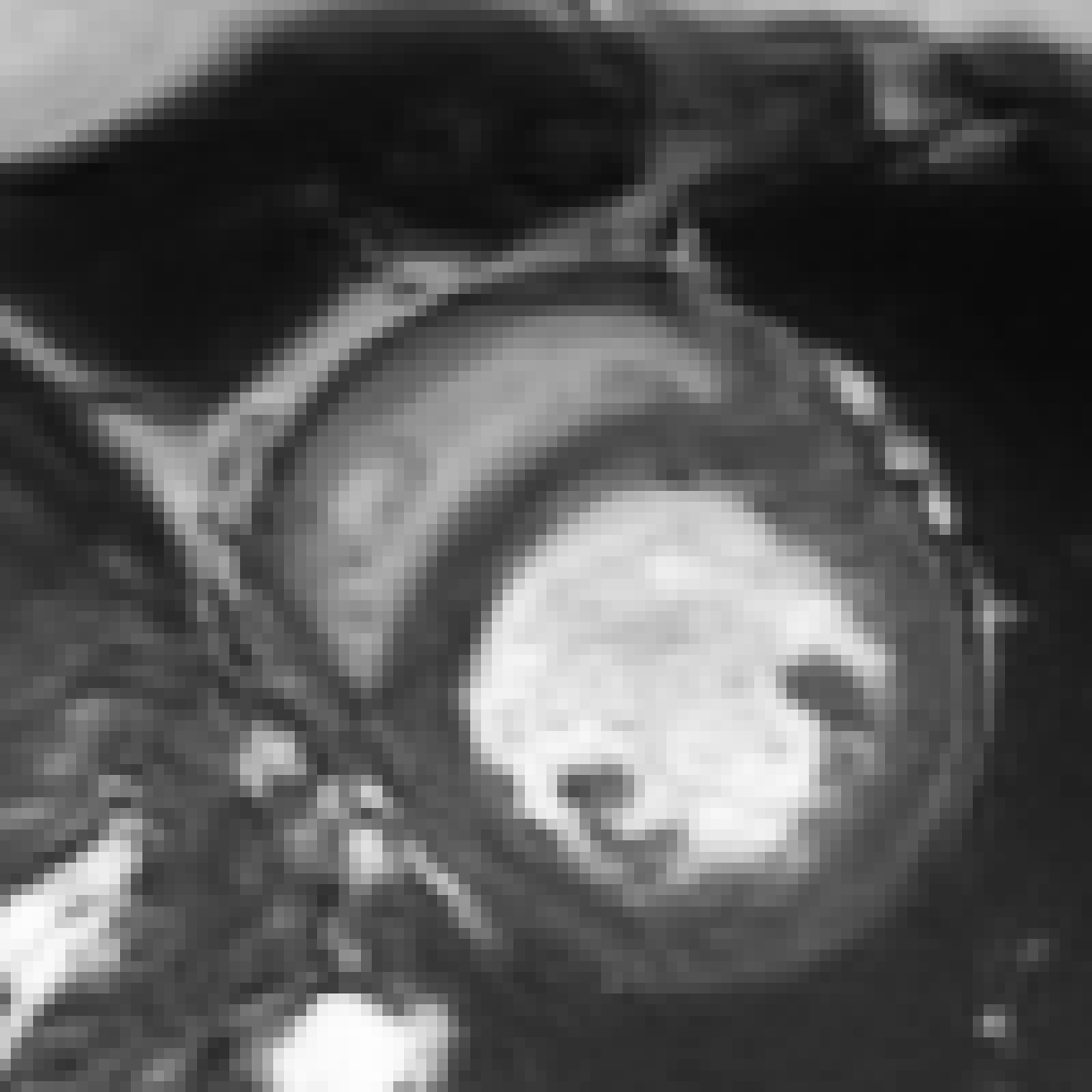} } &
				\subfloat[$\alpha=$ \nicefrac{4}{7}]{\includegraphics[width=.1\textwidth]{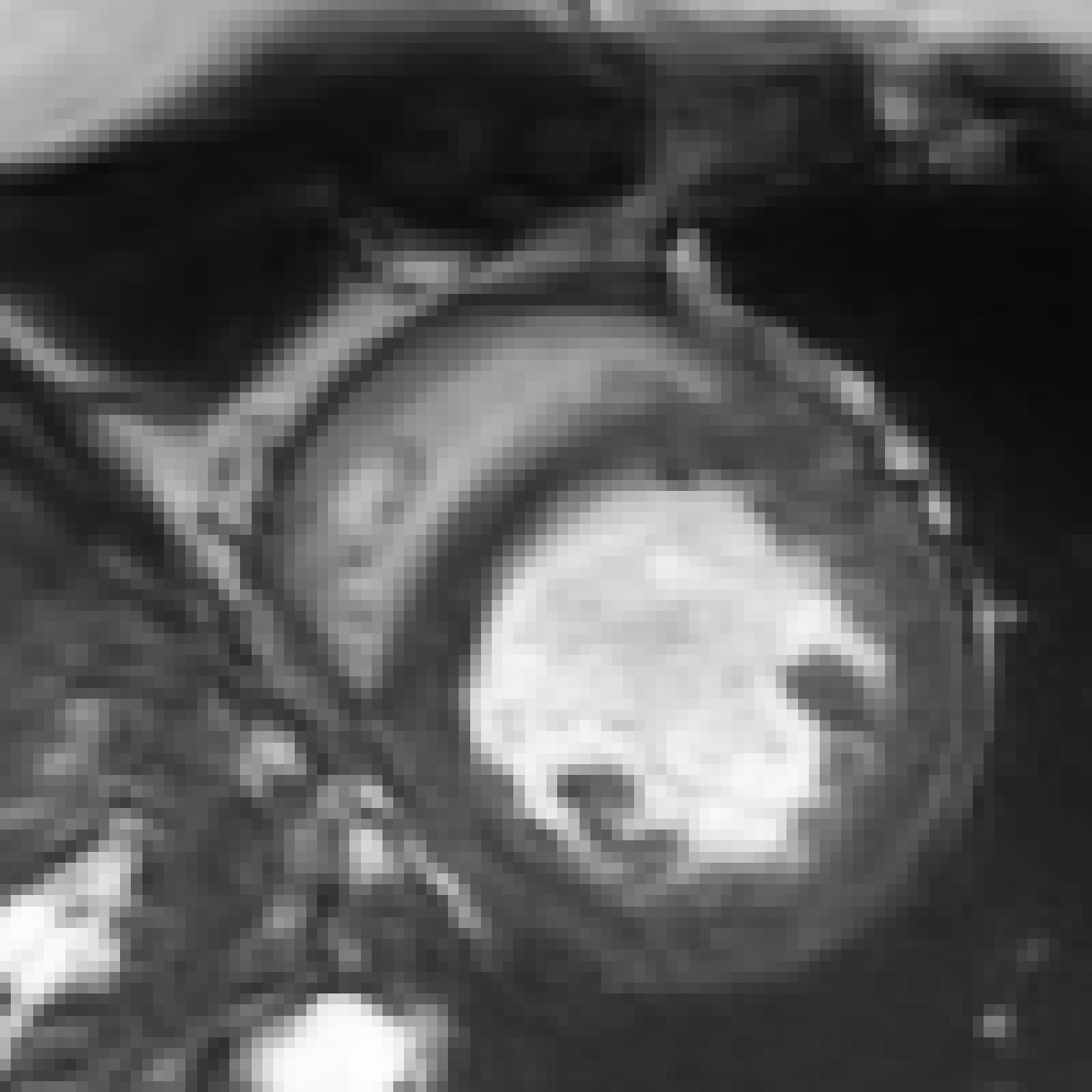} } &
				\subfloat[$\alpha=$ \nicefrac{5}{7}]{\includegraphics[width=.1\textwidth]{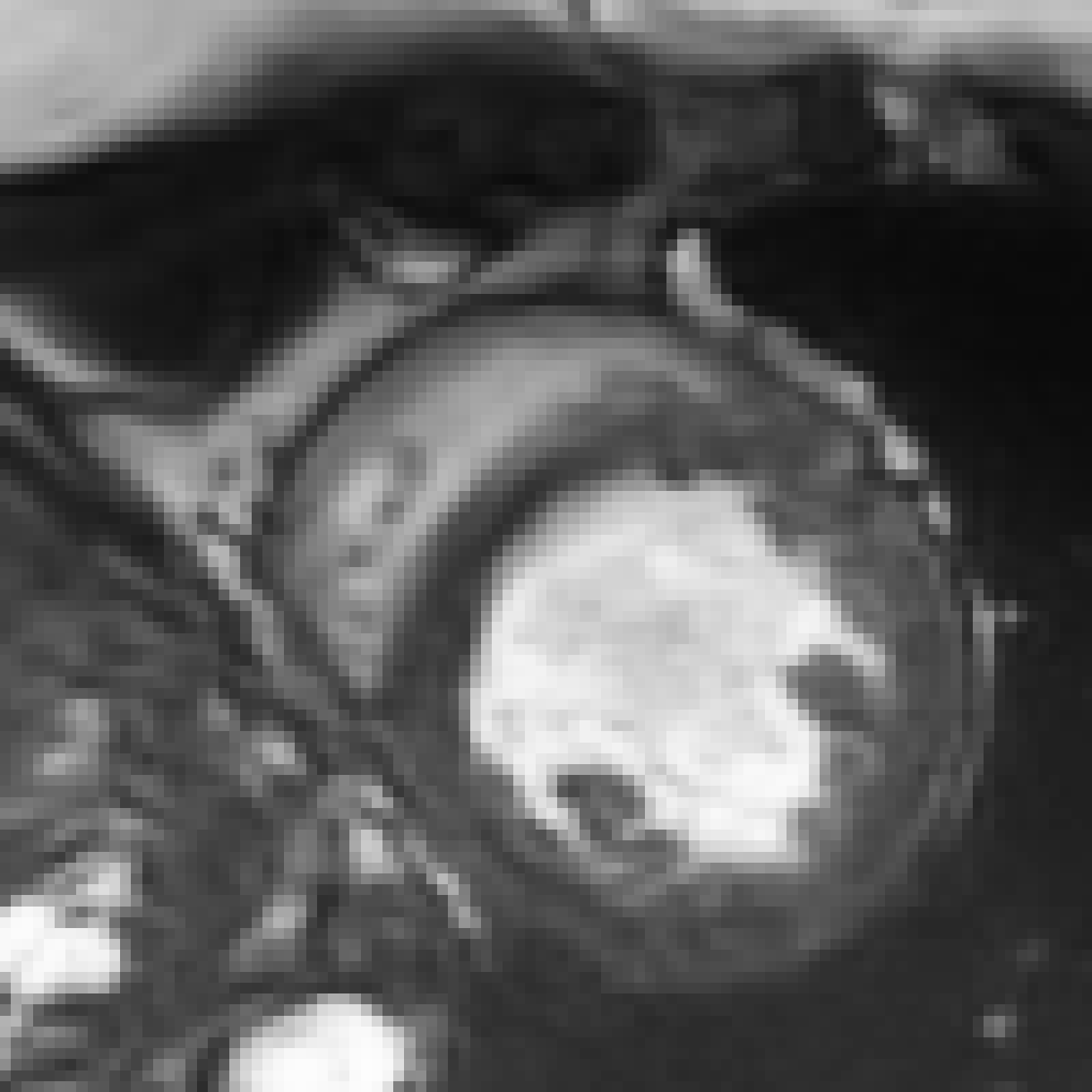} } &
				\subfloat[$\alpha=$ \nicefrac{6}{7}]{\includegraphics[width=.1\textwidth]{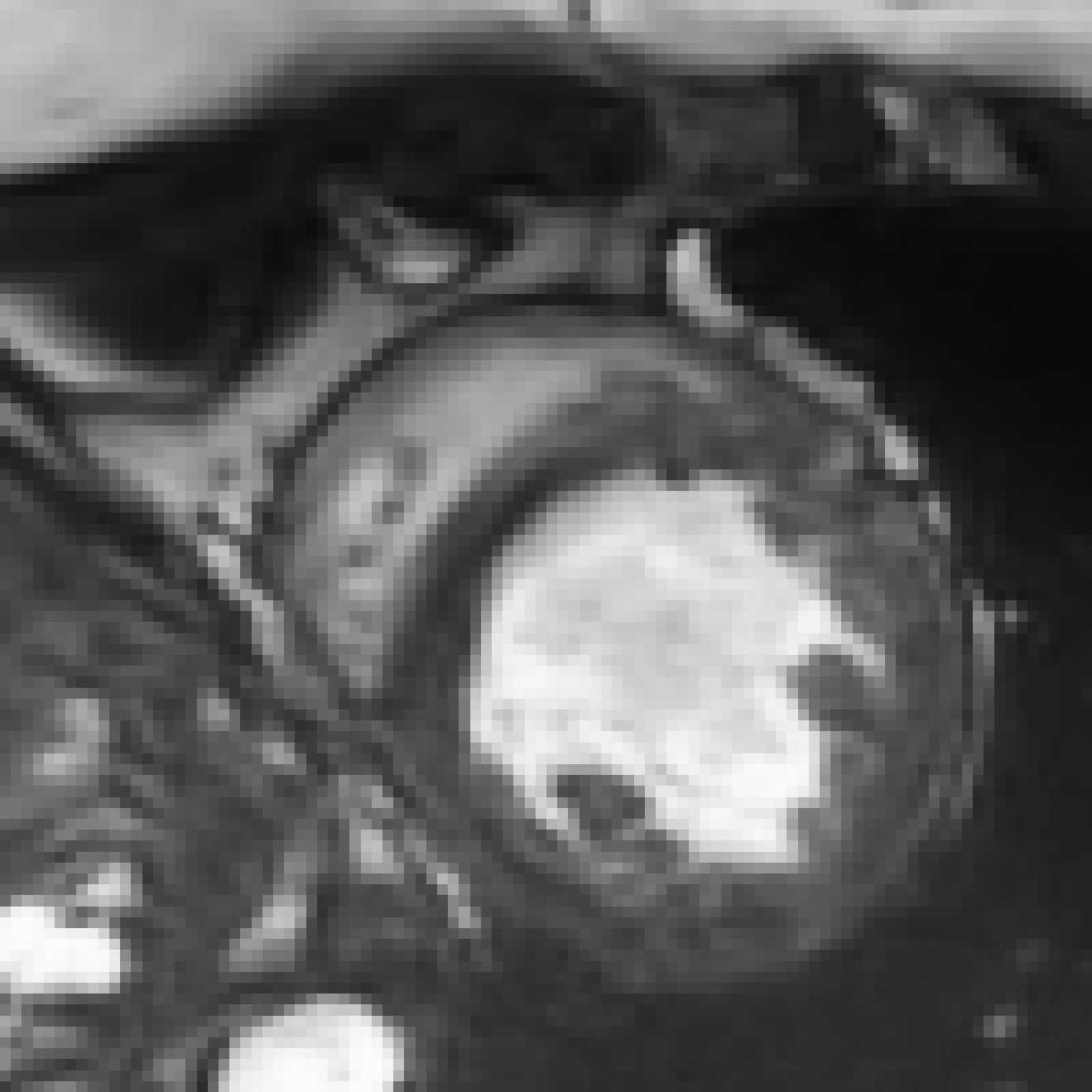} } &
				\subfloat[Neighboring slice 2]{\includegraphics[width=.1\textwidth]{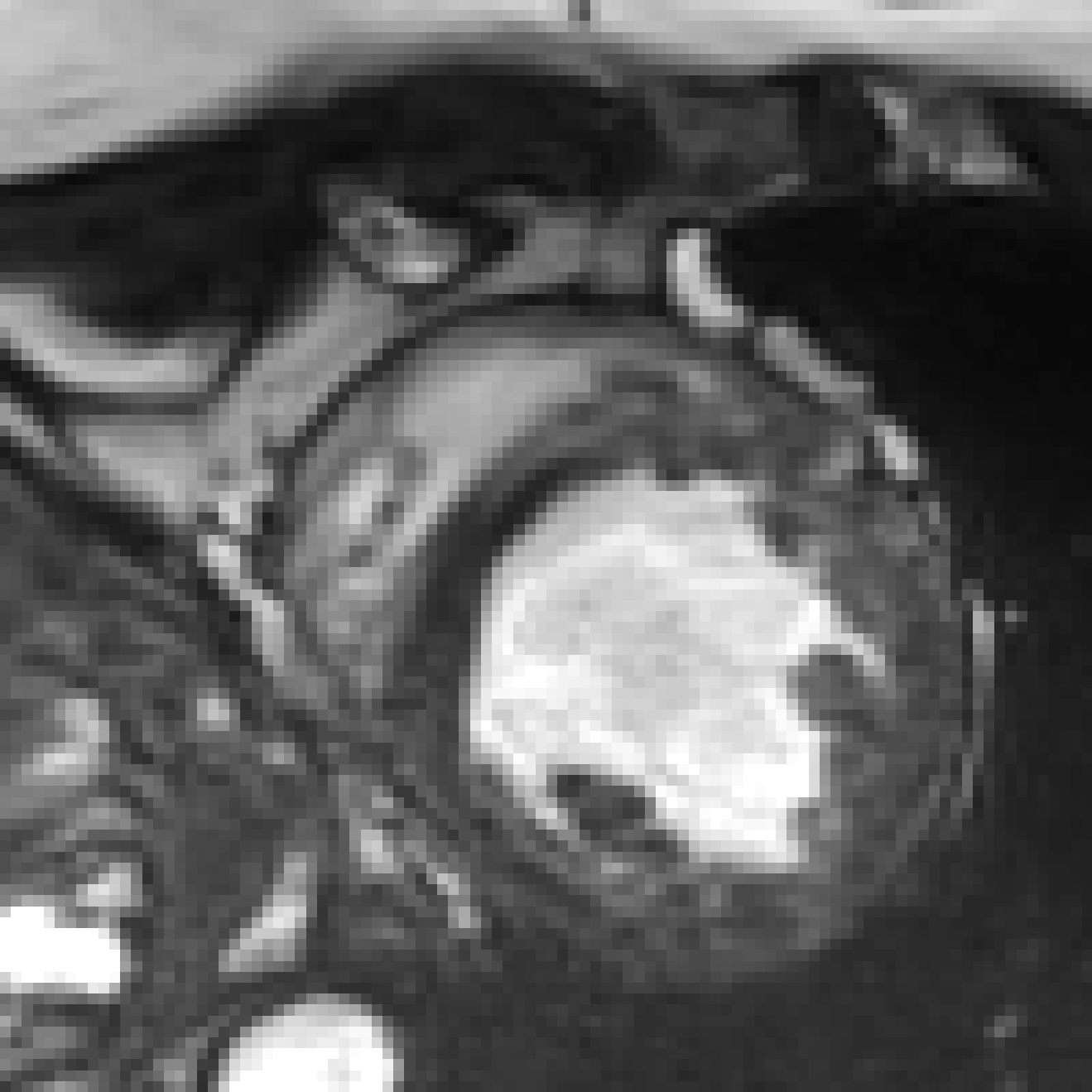} } \\
				% row 3
				\subfloat[Neighboring slice 1]{\includegraphics[width=.1\textwidth]{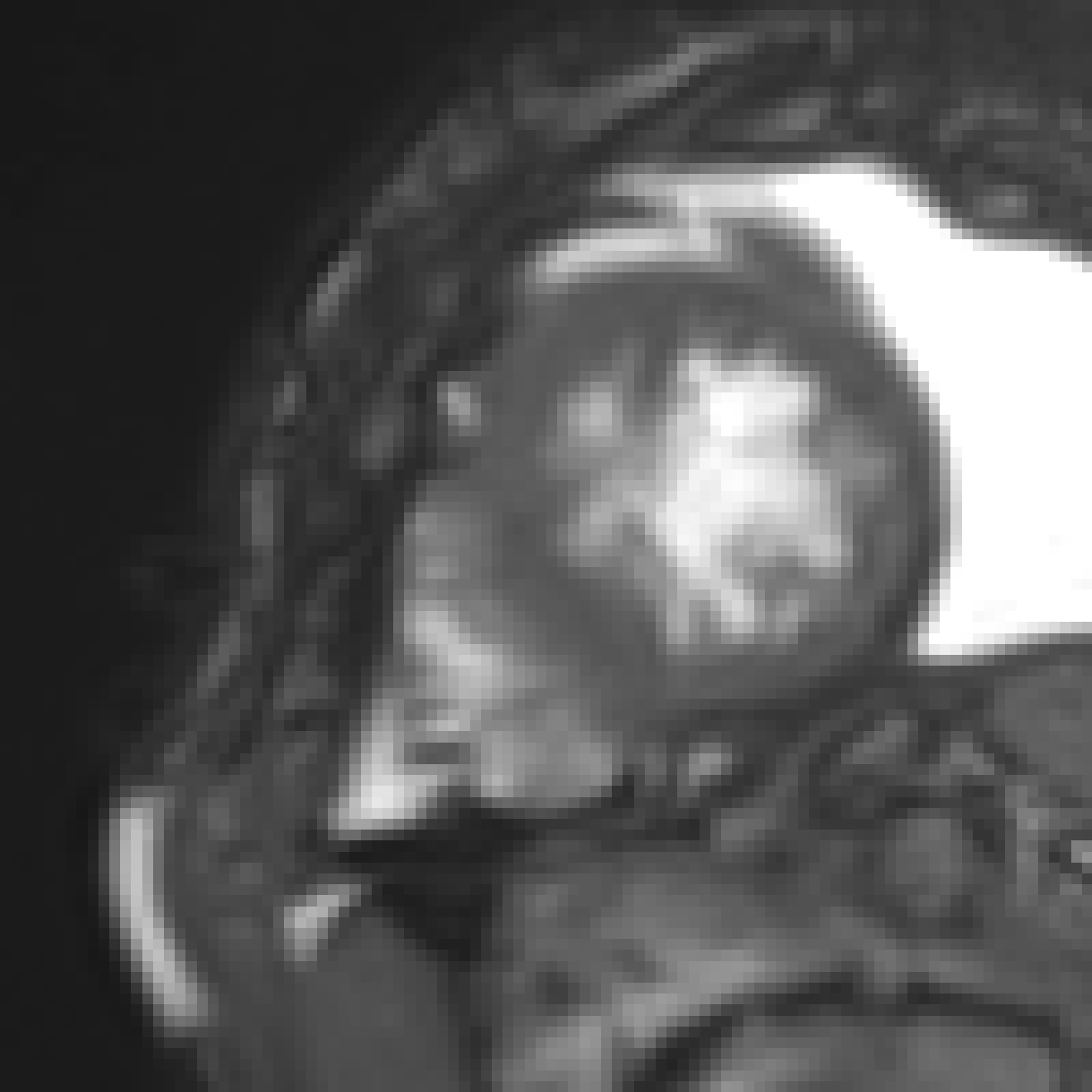} } & 
				\subfloat[$\alpha=$ \nicefrac{1}{7}]{\includegraphics[width=.1\textwidth]{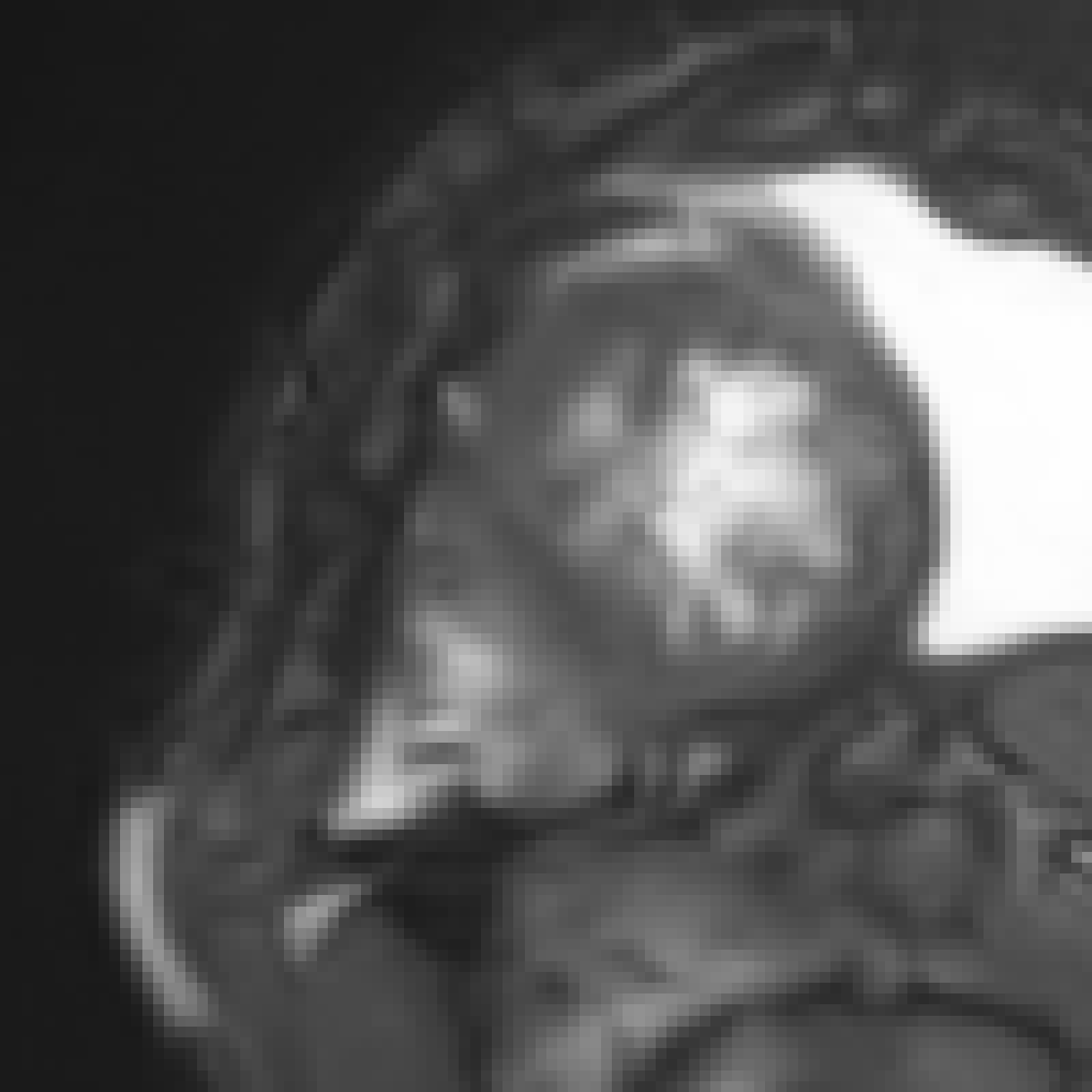} } &
				\subfloat[$\alpha=$ \nicefrac{2}{7}]{\includegraphics[width=.1\textwidth]{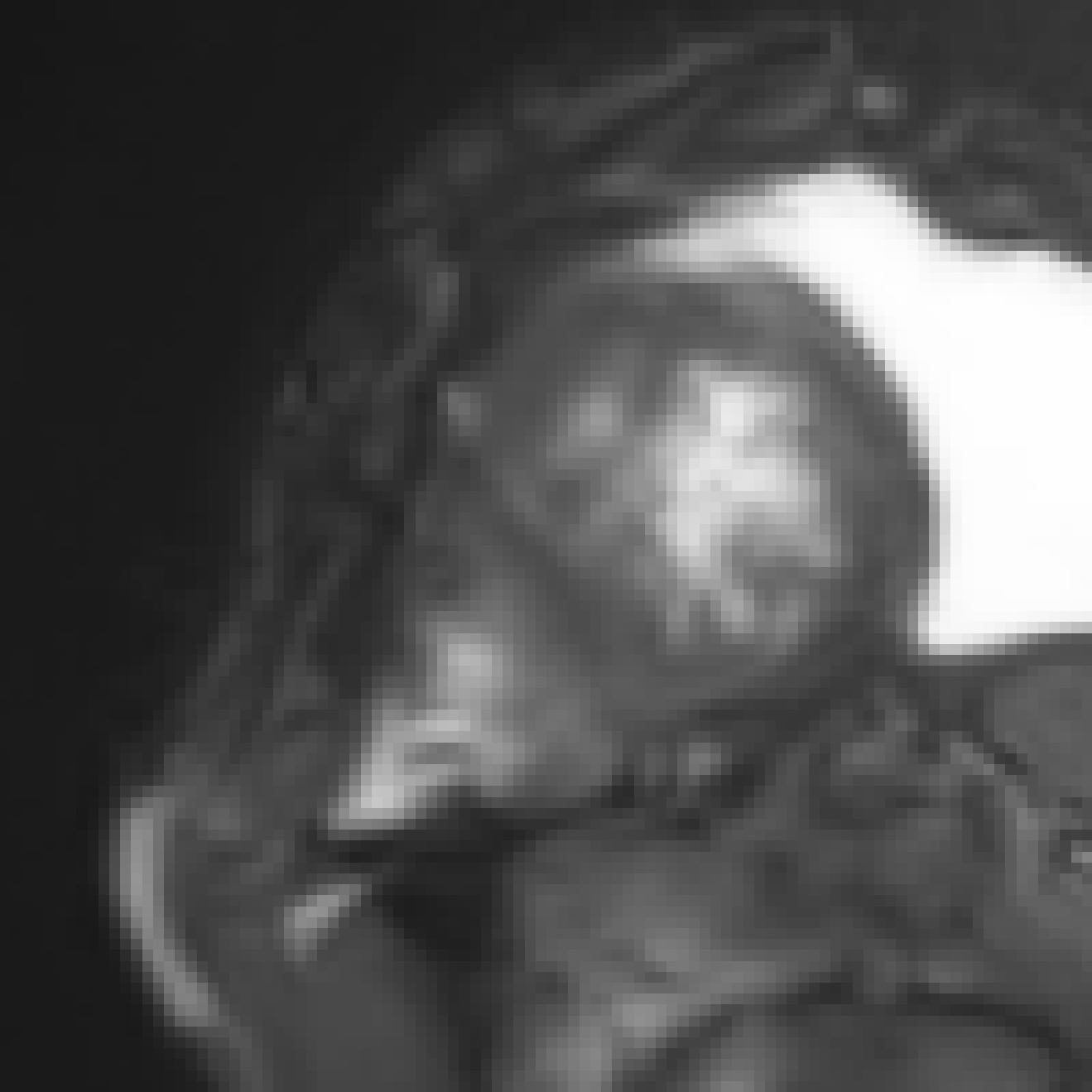} } &
				\subfloat[$\alpha=$ \nicefrac{3}{7}]{\includegraphics[width=.1\textwidth]{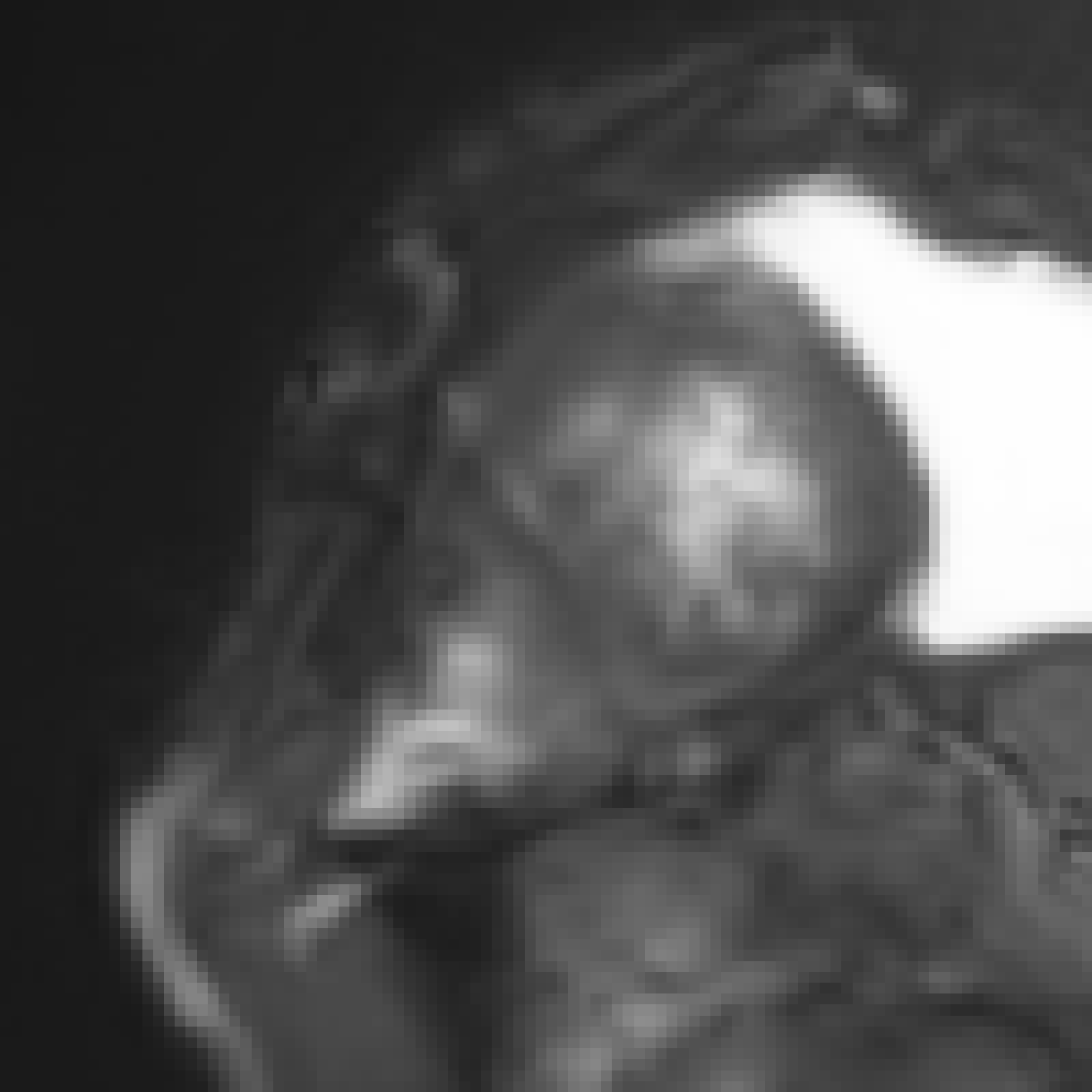} } &
				\subfloat[$\alpha=$ \nicefrac{4}{7}]{\includegraphics[width=.1\textwidth]{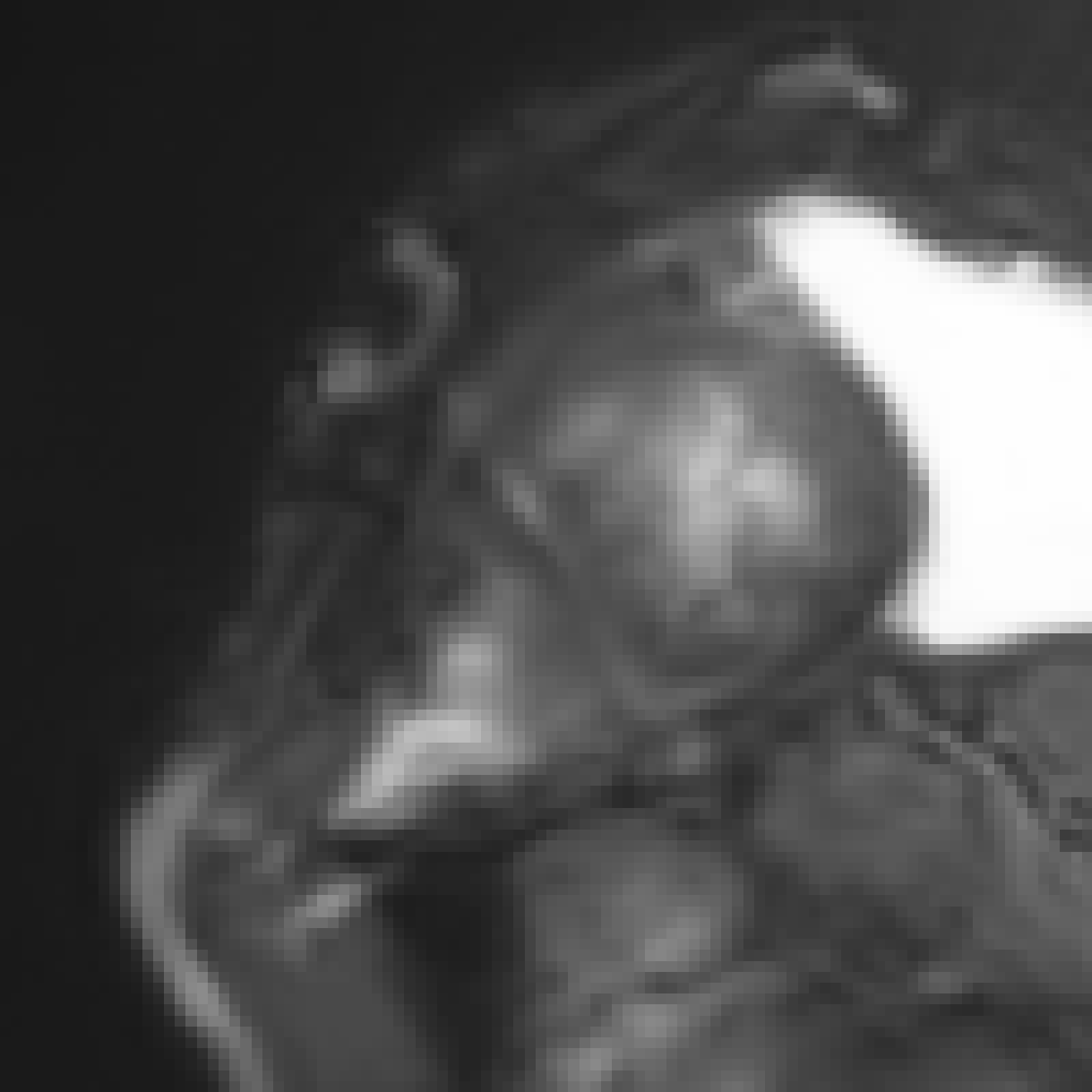} } &
				\subfloat[$\alpha=$ \nicefrac{5}{7}]{\includegraphics[width=.1\textwidth]{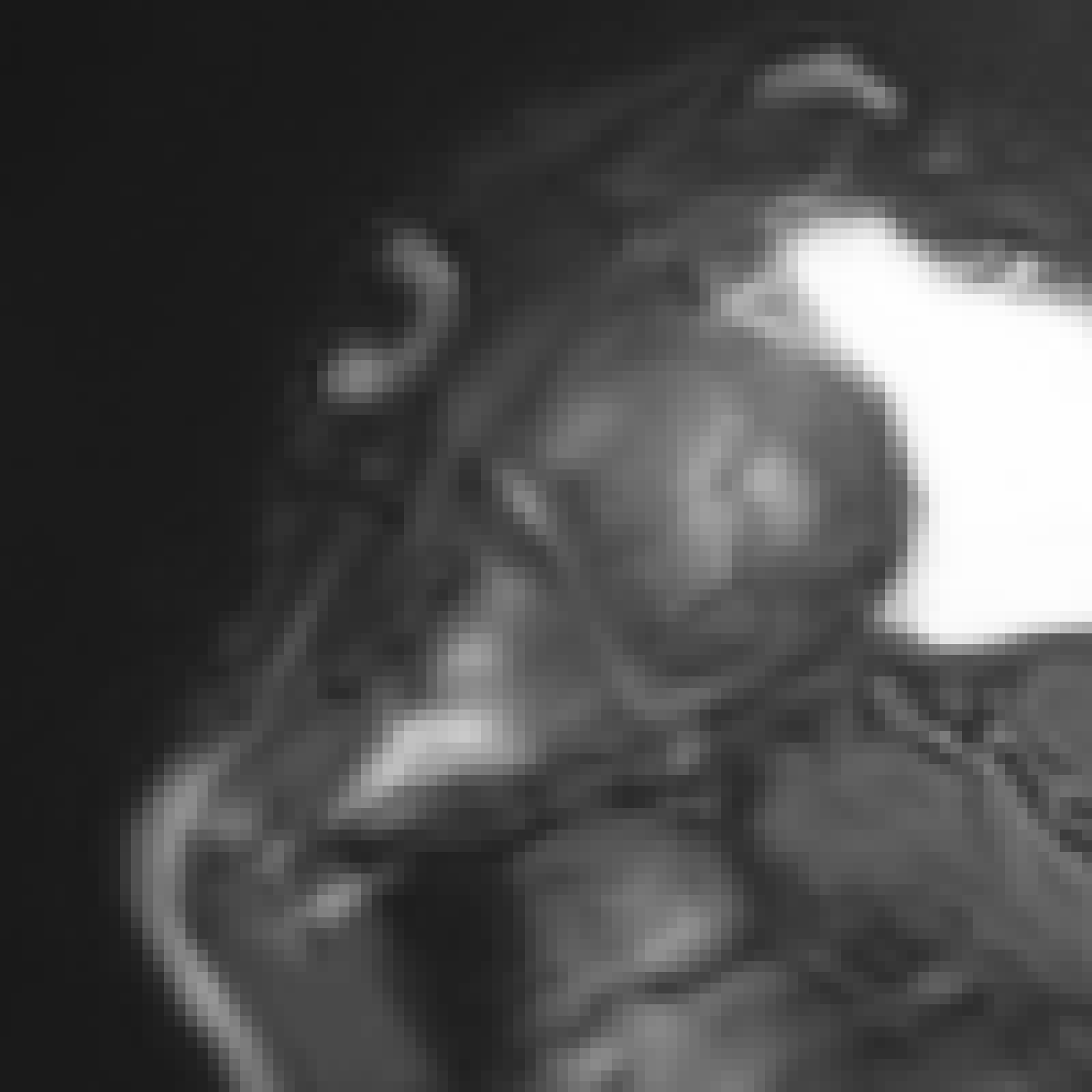} } &
				\subfloat[$\alpha=$ \nicefrac{6}{7}]{\includegraphics[width=.1\textwidth]{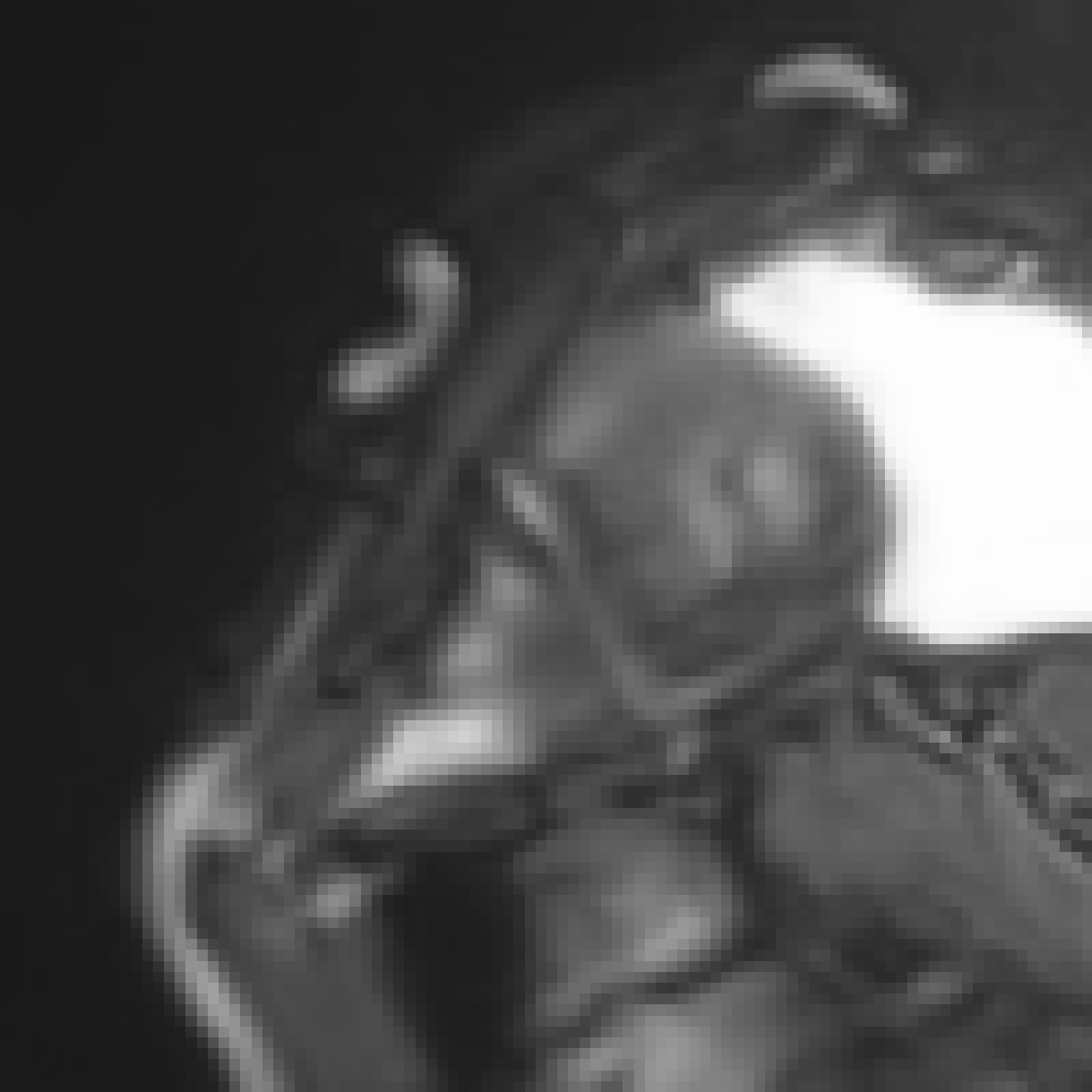} } &
				\subfloat[Neighboring slice 2]{\includegraphics[width=.1\textwidth]{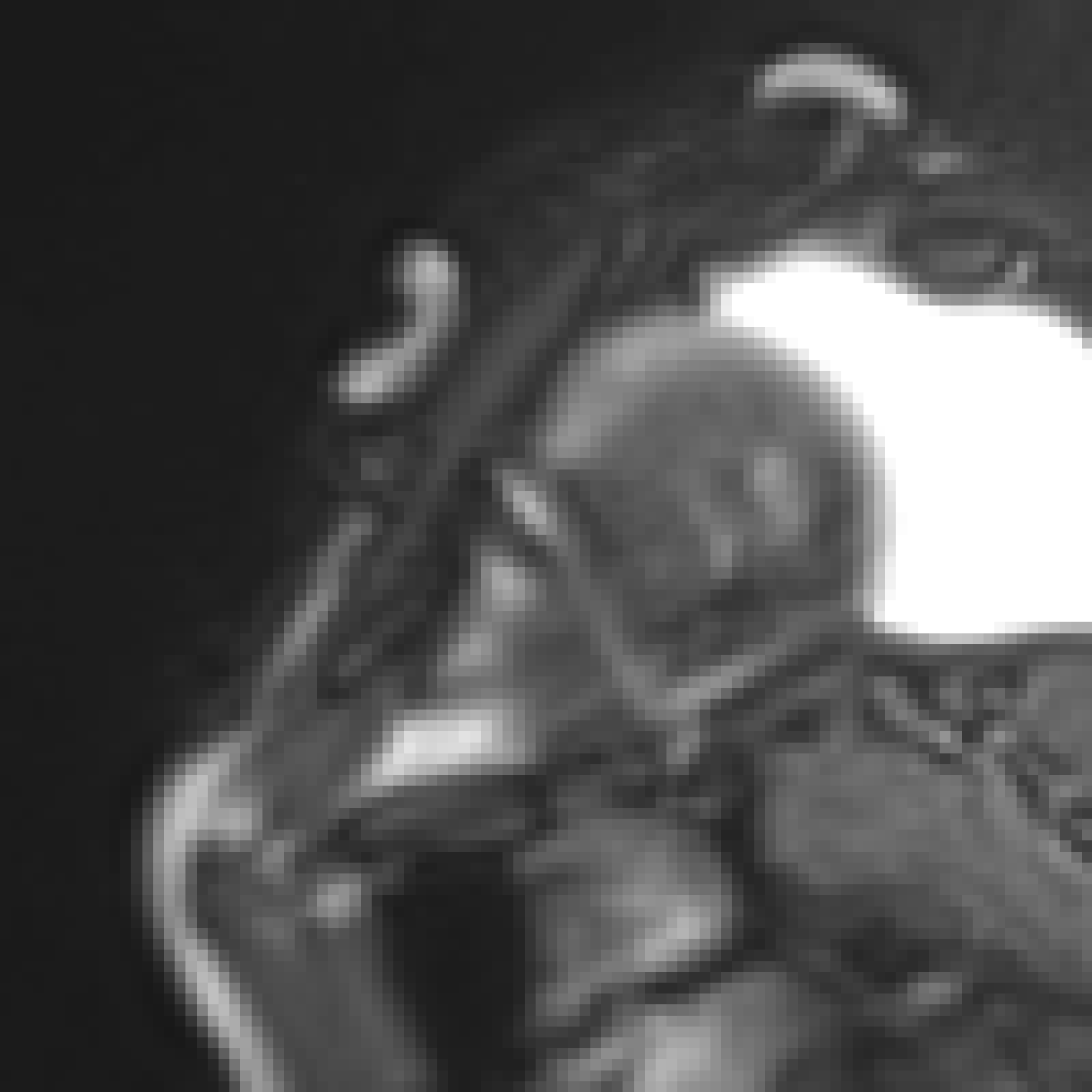} } 
				
			\end{tabular}
			
		\end{center}
		
		\caption{Qualitative evaluation of image synthesis performance of proposed method on cardiac cine MRI (ACDC dataset). Slice spacing was improved from \num{10} to \SI{1.43}{\milli\meter} by synthesizing six intermediate slices (second to penultimate columns) using latent space encodings of the two neighboring slices (first and last column). $\alpha$ denotes the mixing coefficient as specified in Equation~\ref{eq_convex_combination}.}
		
		\label{fig_qualitative_synthesis_acdc}
	\end{figure*}
	
	% FIGURE 4
	% p-value annotation legend:
	% ns: 5.00e-02 < p <= 1.00e+00
	% *: 1.00e-02 < p <= 5.00e-02
	% **: 1.00e-03 < p <= 1.00e-02
	% ***: 1.00e-04 < p <= 1.00e-03
	% ****: p <= 1.00e-04
	\begin{figure}[ht]
		\captionsetup[subfigure]{justification=centering}
		\centering
		\includegraphics[width=.48\textwidth]{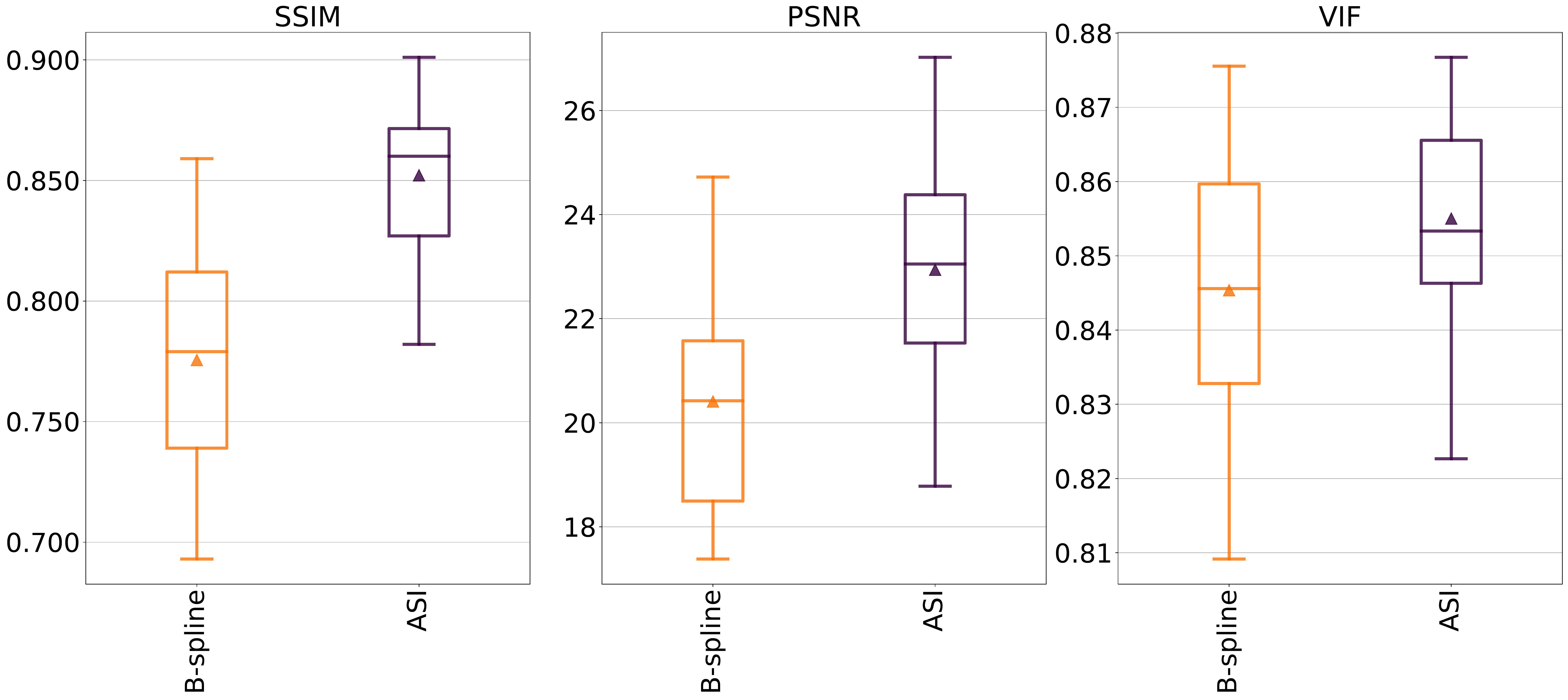} 
		\caption{Boxplots comparing upsampling performance for cubic B-spline compared with proposed method (ASI) in terms of SSIM, PSNR, and VIF. Cardiac cine MRIs of \num{20} patients from the ACDC dataset were upsampled with factor \num{2} in through-plane direction. A higher score indicates better performance. Measures were computed on sagittal slices through short-axis volume. The proposed method achieved higher performance when evaluated by all measures compared with cubic B-spline interpolation. The differences between proposed and conventional method are statistically significant ($p < 0.0001$) using the one-sided Wilcoxon signed-rank test. Triangle indicates mean value.}		
		\label{fig_acdc_quantitative_results} 
	\end{figure}
	
	% FIGURE 5
	\begin{figure}[ht]
		\captionsetup[subfigure]{justification=centering, labelformat=empty}
		\setlength{\tabcolsep}{1pt}
		\begin{center}
			\begin{tabular}{c c c}
				% row 1
				% \setcounter{subfigure}{0}
				\subfloat[Original slice to be synthesized]{\includegraphics[width=.15\textwidth]{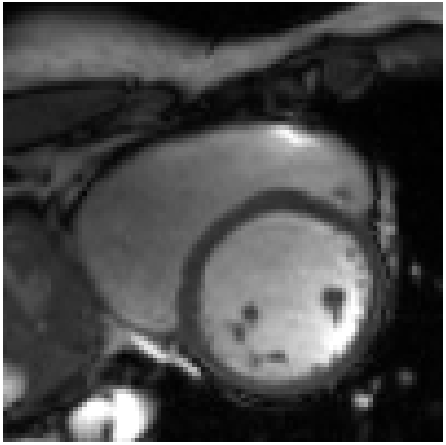} }  &
				\subfloat[B-spline ]{\includegraphics[width=.15\textwidth]{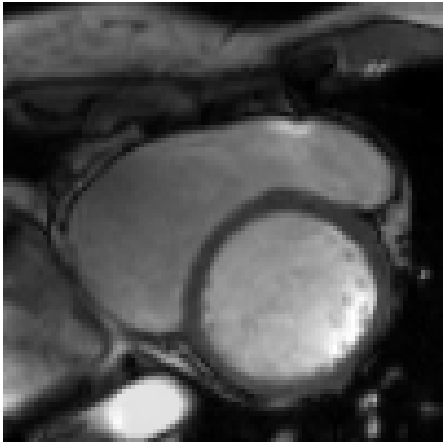} } & 
				\subfloat[ours]{\includegraphics[width=.15\textwidth]{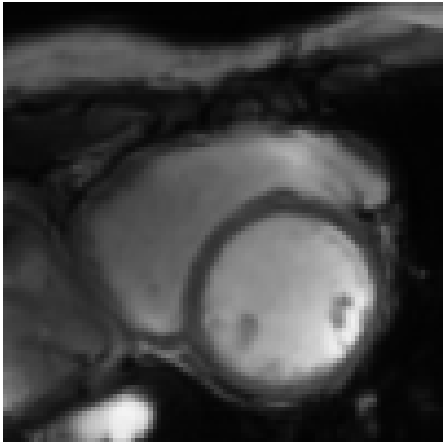} }	\\  
				% row 2
				&
				\subfloat{\includegraphics[width=.15\textwidth]{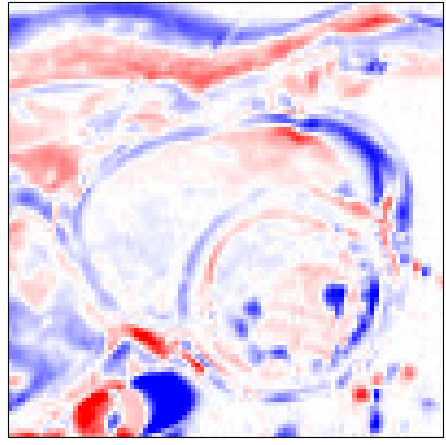}  } &
				\subfloat{\includegraphics[width=.15\textwidth]{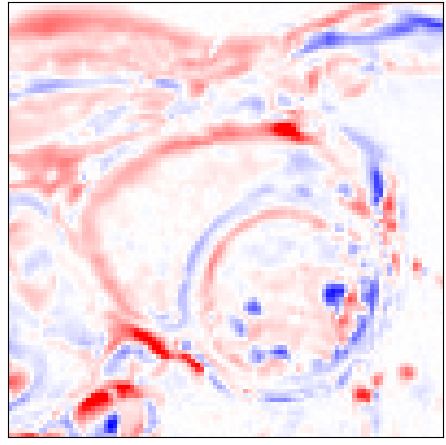}  } 
			\end{tabular}
			
		\end{center}
		
		\caption{Qualitative comparison between cubic B-spline and proposed approach on cardiac MRI (ACDC dataset) for upsampling factor \num{2}. First column shows slice from reference volume. Second column depicts image generated by conventional interpolation method. Last column shows synthesized slice using proposed method. Bottom row: Differences between original (minuend) and synthesized slice (subtrahend). Blue corresponds to negative and red to positive differences. Image intensities are scaled to a $[0,1]$ range. All difference images use the same color scale $[-1, 1]$.}
		\label{fig_acdc_qualitative_comparison_other_methods}
	\end{figure}

	% FIGURE 6
	\begin{figure*}
		\captionsetup[subfigure]{justification=centering}
		\centering
		\subfloat[Original sagittal view with \num{10}mm slice spacing]{\begin{tikzpicture}
			\node[above right, inner sep=0] (image) at (0,0)  {\includegraphics[width=0.33\textwidth]{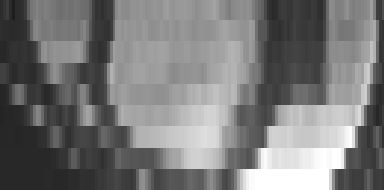}  
				\label{fig_qualitative_acdc_lax1}  }  ;
			% Create scope with normalized axes
			\begin{scope}[
			x={($0.05*(image.south east)$)},
			y={($0.1*(image.north west)$)}]
			% Grid
			% \draw[lightgray,step=1] (image.south west) grid (image.north east);
			\draw[thick,orange] (11.5,0.2) rectangle (15, 5) ;
			\draw[thick,orange] (3.,3) rectangle (5., 6.5) ;
			\end{scope}
			\end{tikzpicture} }
		\subfloat[B-spline]{\begin{tikzpicture}
			\node[above right, inner sep=0] (image) at (0,0)  {\includegraphics[width=0.33\textwidth]{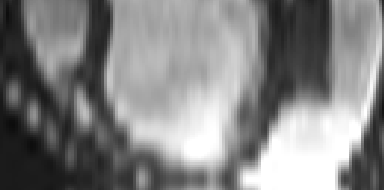}  
				\label{fig_qualitative_acdc_lax3}  }  ; % \vspace{2ex} }  
			% Create scope with normalized axes
			\begin{scope}[
			x={($0.05*(image.south east)$)},
			y={($0.1*(image.north west)$)}]
			% Grid
			% \draw[lightgray,step=1] (image.south west) grid (image.north east);
			\draw[thick,orange] (11.5,0.2) rectangle (15, 5) ;
			\draw[thick,orange] (3.,3) rectangle (5., 6.5) ;
			\end{scope}
			\end{tikzpicture} }
		\subfloat[ours]{\begin{tikzpicture}
			\node[above right, inner sep=0] (image) at (0,0)  {\includegraphics[width=0.33\textwidth]{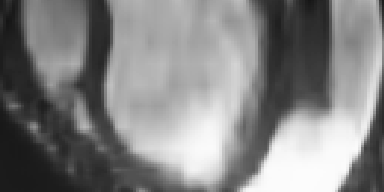}  
				\label{fig_qualitative_acdc_lax6} }  ;
			% Create scope with normalized axes
			\begin{scope}[
			x={($0.05*(image.south east)$)},
			y={($0.1*(image.north west)$)}]
			% Grid
			% \draw[lightgray,step=1] (image.south west) grid (image.north east);
			\draw[thick,orange] (11.5,0.2) rectangle (15, 5) ;
			\draw[thick,orange] (3.,3) rectangle (5., 6.5) ;
			\end{scope}
			\end{tikzpicture} }		
		\caption{Qualitative comparison between cubic B-spline and proposed method for upsampled cardiac short-axis stacks (ACDC dataset). Volumes were upsampled with factor \num{10} from \num{10}mm (original) to \num{1}mm slice spacing. Shown are sagittal views through upsampled short-axis stacks. The proposed method can generate new slices that result in smoother anatomical transitions compared to slices generated by conventional interpolation method. The performance difference is more pronounced for the structures of the left ventricle myocardium. Note that a feature located at the apex of the right ventricle seems to appear more accurate in the upsampled image using cubic B-spline interpolation method compared to the proposed approach.}
		
		\label{fig_qualitative_ae_acdc_long_axis}
	\end{figure*}

	\section{Evaluation} \label{section_evaluation}
	
	Performance of the method was quantitatively evaluated in terms of Structural Similarity Index Measure (SSIM), Peak Signal-to-Noise Ratio (PSNR) and Visual Information Fidelity (VIF) (\cite{sheikh2005information}). Recent work of \cite{mason2019comparison} conveyed that Visual Information Fidelity demonstrates a high correlation with radiologists’ opinions of MRI quality. The proposed method was compared against performance of cubic B-spline interpolation which is known to outperform methods like Nearest-Neighbor or Linear interpolation \citep{meijering2001quantitative, lehmann2001addendum}. Statistical significance of performance differences between evaluated methods was tested using the one-sided Wilcoxon signed-rank test.
	% frequently used state-of-the-art conventional interpolation method in the medical imaging domain.
	
	In addition, upsampling performance was qualitatively evaluated by visually inspecting the reconstructed and synthesized slices. Visual inspection mainly focused on \textit{anatomical plausibility} and \textit{semantic similarity} of synthesized slices compared to corresponding reference slices. Furthermore, generated images were visually examined for smoothness of interpolation. % We found this evaluation important because the quantitative measures do not capture any notion of semantic similarity and anatomical plausibility.

	%---------------------------------------------------------------------
	%   SECTION EXPERIMENTS
	%---------------------------------------------------------------------
	
	\section{Experiments and Results} \label{section_expers_results}
	
	\subsection{Comparison of autoencoding approaches}  \label{exp_mnist}
	
	Before applying the proposed approach to cardiac and brain MRI scans, several autoencoder approaches were investigated for latent space interpolation using MNIST data \citep{lecun1998mnist}. Given any digit and its \num{40} degree rotated variant, referred to as input images, intermediate rotations were synthesized by mixing the latent space encodings of the two input images. Results were compared with digits that were rotated in the image space. Three different approaches were compared: Variational Autoencoder (VAE) \citep{kingma2014stochastic, rezende2014stochastic}, Adversarially Constrained Autoencoder Interpolation (ACAI) \cite{berthelot2018understanding} and the proposed approach (ASI). Interpolation performance was qualitatively evaluated using images of the MNIST dataset.
	
	% To compare the ability of different autoencoder approaches to interpolate in latent space, the proposed method was compared with two autoencoder structures; 
	
	\subsubsection{Experimental details}
	
	The dataset was randomly split into training (\num{60000} images), validation (\num{1000} images), and test sets (\num{9000} images). To train the models, patches of \num{32}$\times$\num{32} were randomly chosen from the training set in mini batches of \num{32} images. The training set was augmented by random rotations $\gamma \in [0, 2\pi]$ of the images. Models were trained for \num{100} epochs. The proposed model was implemented as described in section~\ref{subsection_ae_architecture} except that \num{16} kernels were used for the latent space representation. Furthermore, $\lambda$ as specified in Equation~\ref{eq_combined_loss} was set to \num{10} after performing a line search ($\lambda \in \{0.05, 0.5, 1, 10, 100, 1000  \}$).
	
	To compute the synthesis loss as specified in Equation~\ref{eq_combined_loss} each training image $x_{n}$ was augmented with two \textit{neighboring} images $\{ x_{n-1}, x_{n+1} \}$. $x_{n-1}$ denotes a \num{15} degree counterclockwise rotation of image $x_{n}$ and $x_{n+1}$ a \num{15} degree clockwise rotation of image $x_{n}$. This enabled synthesizing image $\hat{x}_n^{\alpha=0.5}$ in Equation~\ref{eq_combined_loss} using a convex combination of the neighboring image encodings $\{ z_{n-1}, z_{n+1} \}$ where $\alpha$ (the mixing coefficient) in Equation~\ref{eq_convex_combination} was set to \num{0.5}.
	
	During evaluation, for each test image $x$ three new images were synthesized by interpolating between the image and a \num{40} degree counterclockwise rotation of the same image $x_{\text{rot}40^{\circ}}$. For this, the set of mixing coefficients $\mathcal{A}$ as specified in Equation~\ref{eq_alpha_set} was set to $\{$0.25, 0.5, 0.75$\}$. As a result, the three synthesized in-between images should be rotated versions of the original image $x$ in steps of \num{10} degree $\{x_{\text{rot}10^{\circ}}, x_{\text{rot}20^{\circ}}, x_{\text{rot}30^{\circ}} \}$. Three examples are shown in Figure~\ref{fig_mnist_roto_qual1}.
	
	%For this, an experiment was designed using the MNIST dataset \citep{deng2012mnist} in which a correct interpolation between two datapoints is unambiguously defined.
	% $x_{\text{rot}40^{\circ}}$
	
	\vspace{1ex}
	\textbf{Variational Autoencoder (VAE) \phantom{x} }To improve smoothness of the latent space of an autoencoder \cite{kingma2014stochastic, rezende2014stochastic} proposed to model the latent representations as a random variable distributed according to a prior distribution. The latent distribution constraint is enforced by an additional loss term which measures the Kullback-Leibler divergence between approximate posterior, modelled by the encoder, and prior distribution. In this work the prior was equal to a Gaussian distribution with diagonal covariance matrix. Implementation of the autoencoder was identical to the proposed approach except for the encoder that was extended with two linear layers to model the mean and covariance matrix of the posterior Gaussian distribution.

	\textbf{Adversarially Constrained Autoencoder Interpolation (ACAI) \phantom{x} } To improve the ability of a convolutional autoencoder to interpolate in latent space \cite{berthelot2018understanding} proposed to regularize the autoencoder by means of an adversarial training objective. Using a discriminator the autoencoder is encouraged to generate interpolated images that appear to be indistinguishable from reconstructions of real images. In this work, the approach was implemented following implementation details as described in \cite{berthelot2018understanding}.

	\subsubsection{Results}
	
	Qualitative comparison of autoencoding approaches shown in Figure~\ref{fig_mnist_roto_qual1} demonstrates that our proposed model achieved the best interpolation performance. Performance differences become most apparent for interpolated images using a mixing coefficient equal to \num{0.5}. Additionally, one can observe that linear steps taken in latent space using the set of mixing coefficients can approximate \textit{rotation steps} in image space.

	%% !!!!!!!!!!!!!!!!!!!!!!!!!!!!! NEONATAL BRAIN MRI FIGURES !!!!!!!!!!!!!!!!!!!!!!!!!!!!!!!!!!!!!!!!!!!
	% FIGURE 7
	\begin{figure*}
		\captionsetup[subfigure]{justification=centering, labelformat=empty}
		\setlength{\tabcolsep}{1pt}
		\begin{center}
			\begin{tabular}{c c c c c c c c}
				% row 1
				\subfloat[Neighboring slice 1]{ \begin{tikzpicture}
					\node[above right, inner sep=0] (image) at (0,0)  { \includegraphics[width=.1\textwidth]{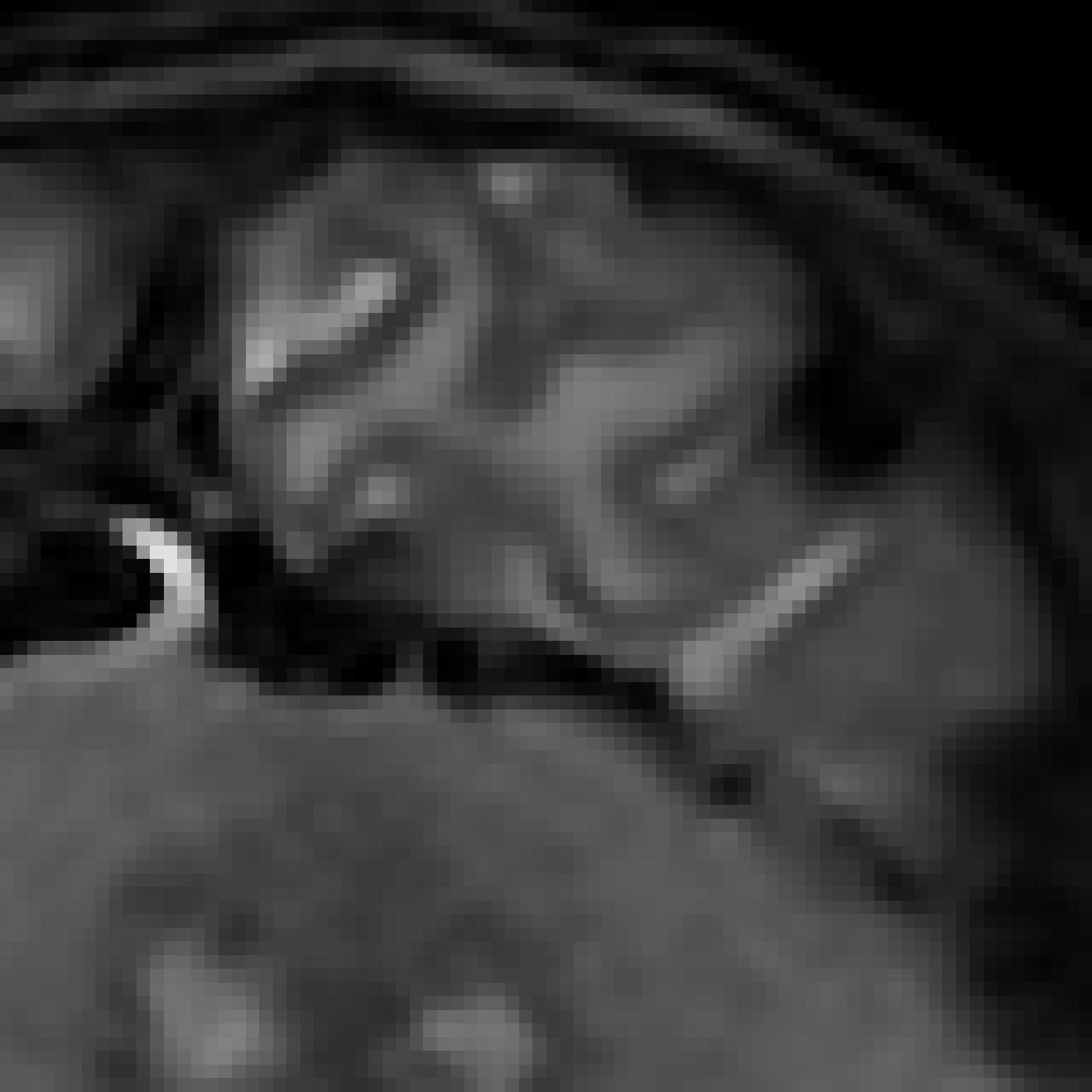} }; 
					% Create scope with normalized axes
					\begin{scope}[
					x={($0.1*(image.south east)$)},
					y={($0.1*(image.north west)$)}]
					% Grid
					% \draw[lightgray,step=1] (image.south west) grid (image.north east);
					\draw[thick,orange] (6,1) rectangle (9.5, 7) ;
					\end{scope}
					\end{tikzpicture} }
				& 
				\subfloat[$\alpha=$ \nicefrac{1}{7}]{ \begin{tikzpicture}
					\node[above right, inner sep=0] (image) at (0,0)  { \includegraphics[width=.1\textwidth]{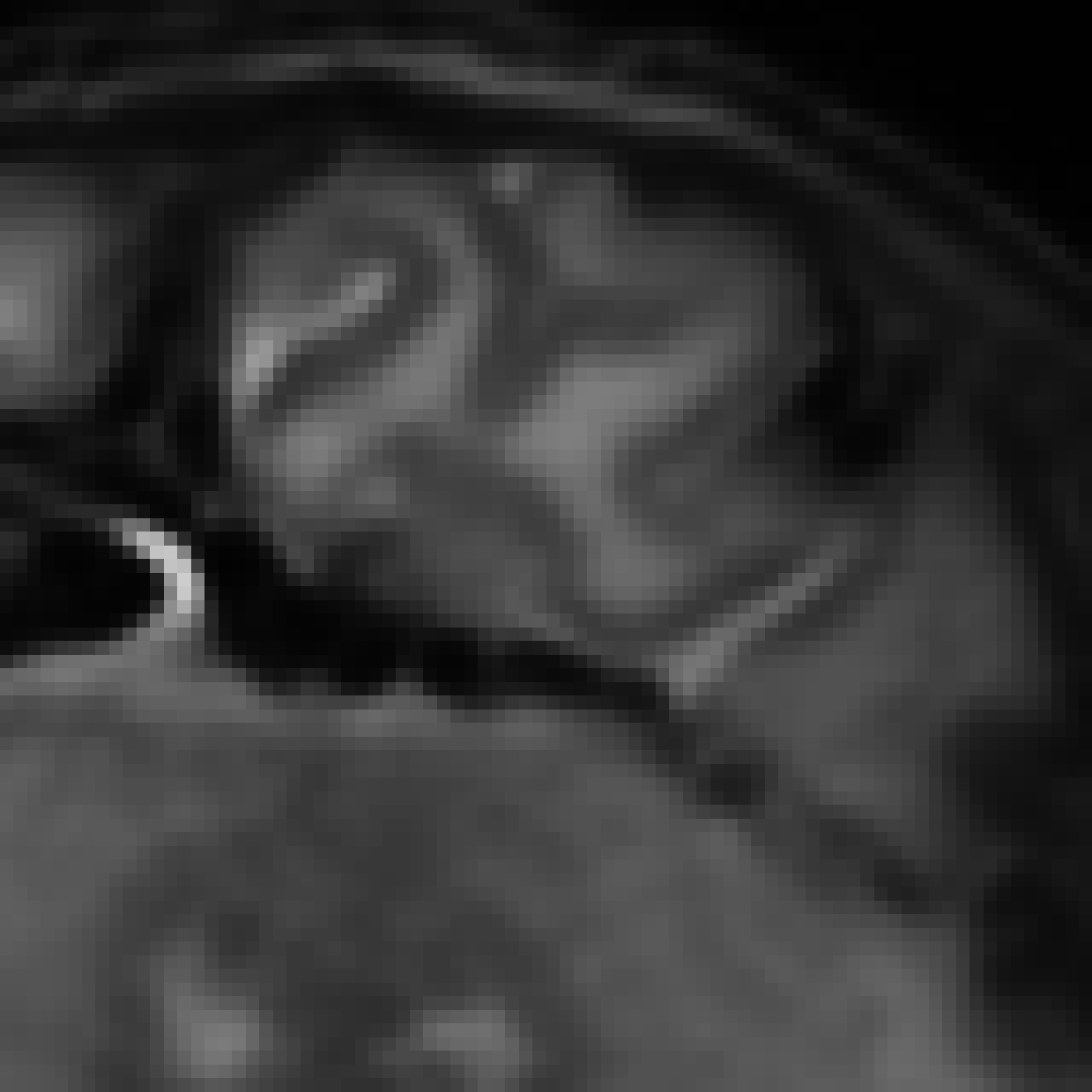} }; 
					% Create scope with normalized axes
					\begin{scope}[
					x={($0.1*(image.south east)$)},
					y={($0.1*(image.north west)$)}]
					% Grid
					% \draw[lightgray,step=1] (image.south west) grid (image.north east);
					\draw[thick,orange] (6,1) rectangle (9.5, 7) ;
					\end{scope}
					\end{tikzpicture} } &
				\subfloat[$\alpha=$ \nicefrac{2}{7}]{ \begin{tikzpicture}
					\node[above right, inner sep=0] (image) at (0,0)  { \includegraphics[width=.1\textwidth]{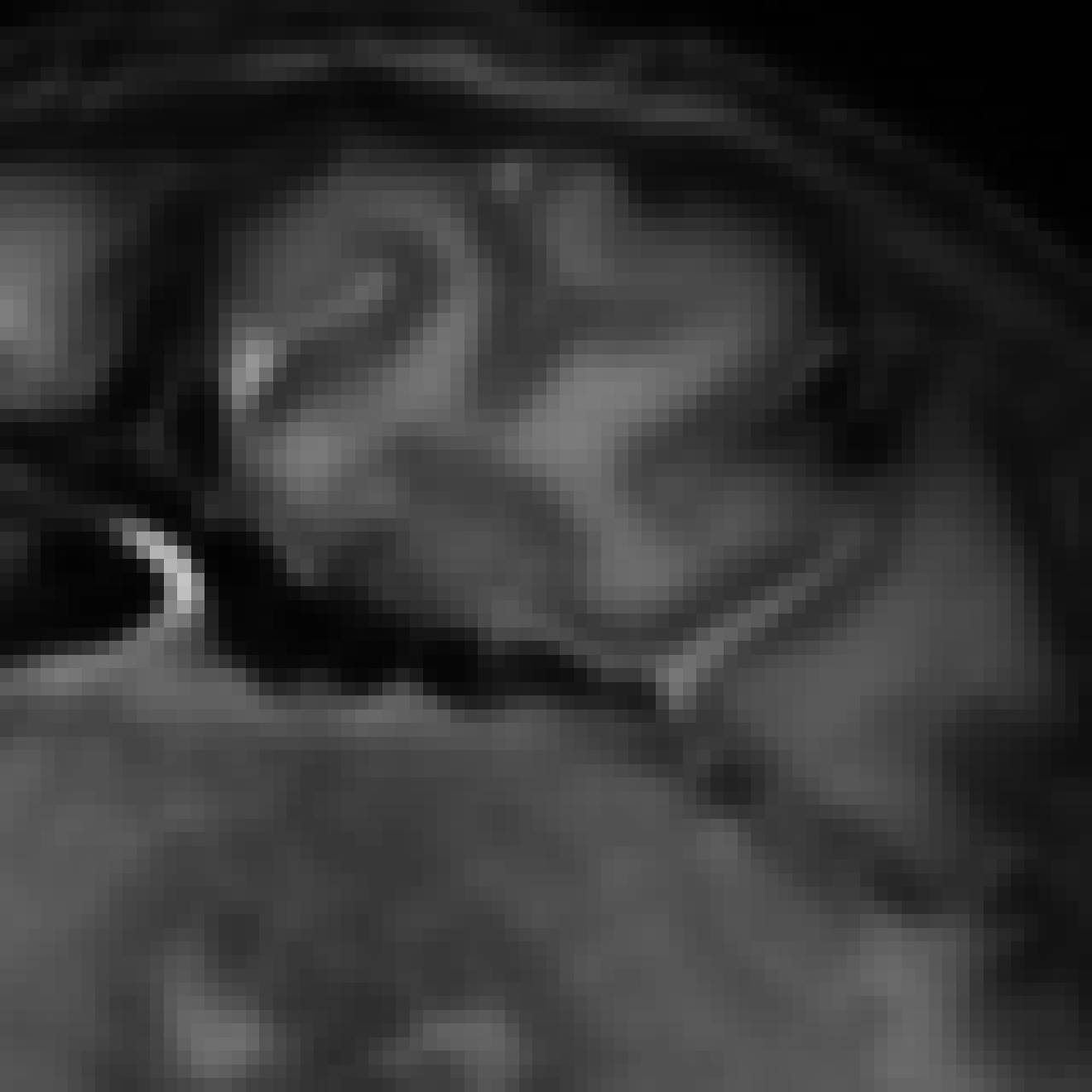} }; 
					% Create scope with normalized axes
					\begin{scope}[
					x={($0.1*(image.south east)$)},
					y={($0.1*(image.north west)$)}]
					% Grid
					% \draw[lightgray,step=1] (image.south west) grid (image.north east);
					\draw[thick,orange] (6,1) rectangle (9.5, 7) ;
					\end{scope}
					\end{tikzpicture} } &
				\subfloat[$\alpha=$ \nicefrac{3}{7}]{ \begin{tikzpicture}
					\node[above right, inner sep=0] (image) at (0,0)  { \includegraphics[width=.1\textwidth]{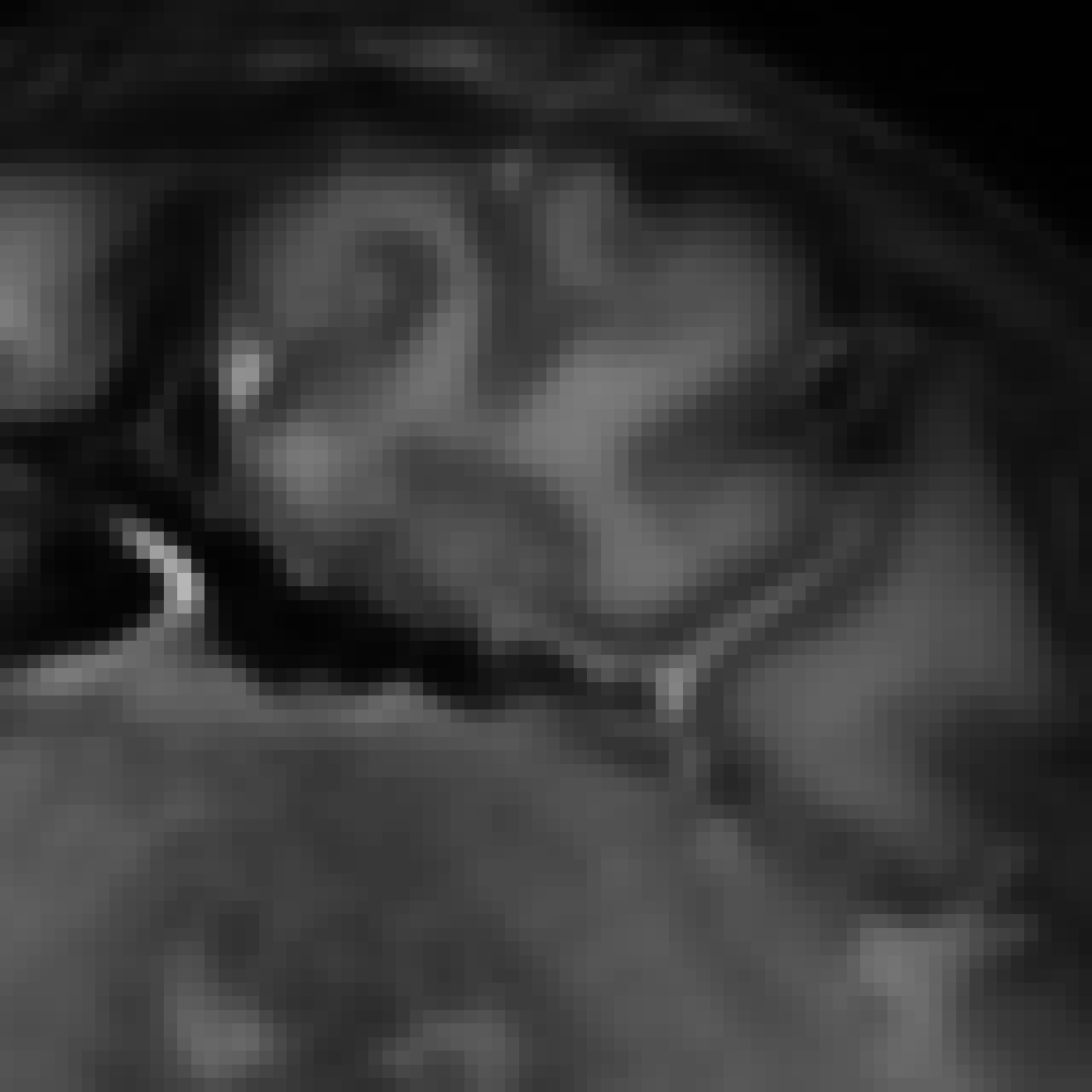} }; 
					% Create scope with normalized axes
					\begin{scope}[
					x={($0.1*(image.south east)$)},
					y={($0.1*(image.north west)$)}]
					% Grid
					% \draw[lightgray,step=1] (image.south west) grid (image.north east);
					\draw[thick,orange] (6,1) rectangle (9.5, 7) ;
					\end{scope}
					\end{tikzpicture} } &
				\subfloat[$\alpha=$ \nicefrac{4}{7}]{ \begin{tikzpicture}
					\node[above right, inner sep=0] (image) at (0,0)  { \includegraphics[width=.1\textwidth]{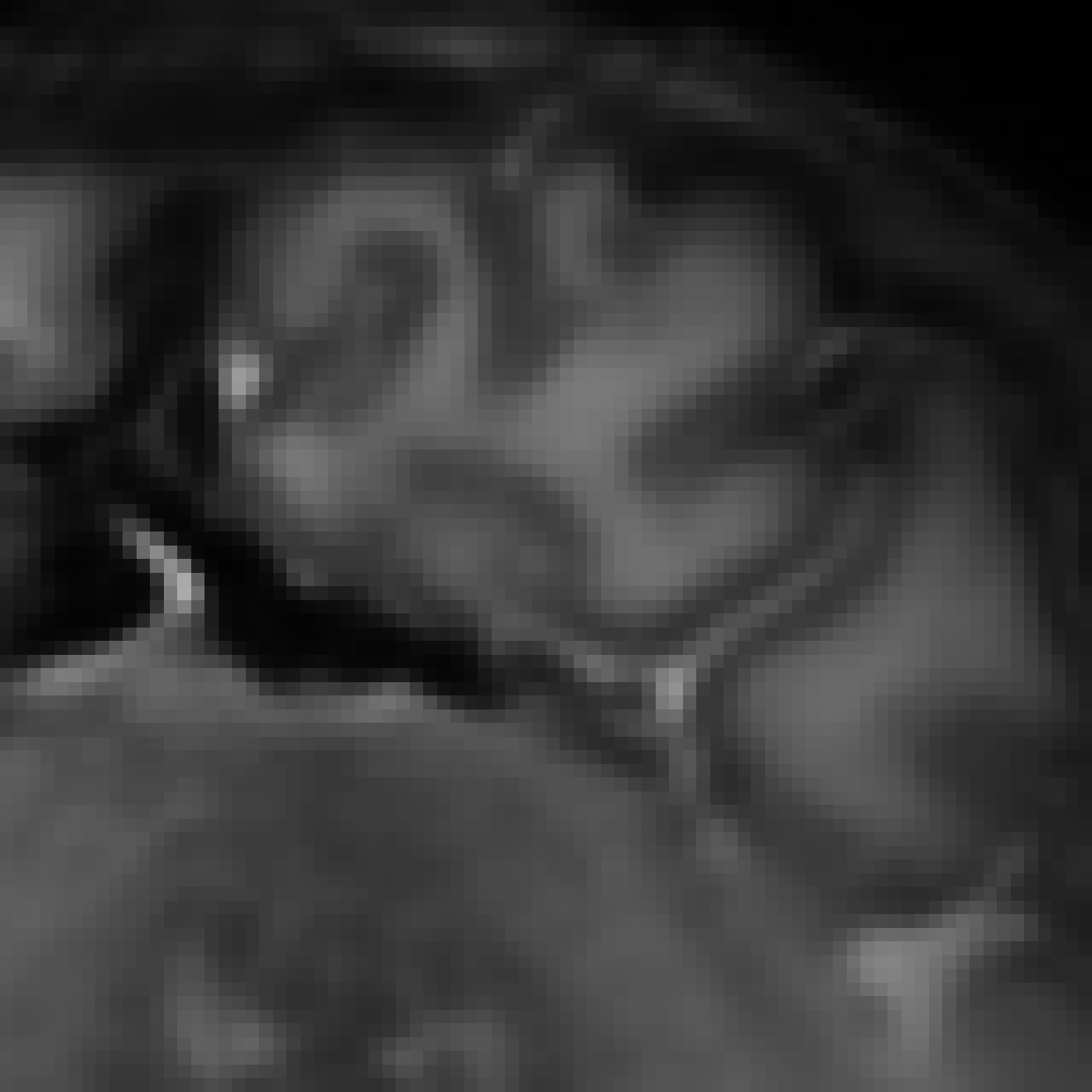} }; 
					% Create scope with normalized axes
					\begin{scope}[
					x={($0.1*(image.south east)$)},
					y={($0.1*(image.north west)$)}]
					% Grid
					% \draw[lightgray,step=1] (image.south west) grid (image.north east);
					\draw[thick,orange] (6,1) rectangle (9.5, 7) ;
					\end{scope}
					\end{tikzpicture} } &
				\subfloat[$\alpha=$ \nicefrac{5}{7}]{ \begin{tikzpicture}
					\node[above right, inner sep=0] (image) at (0,0)  { \includegraphics[width=.1\textwidth]{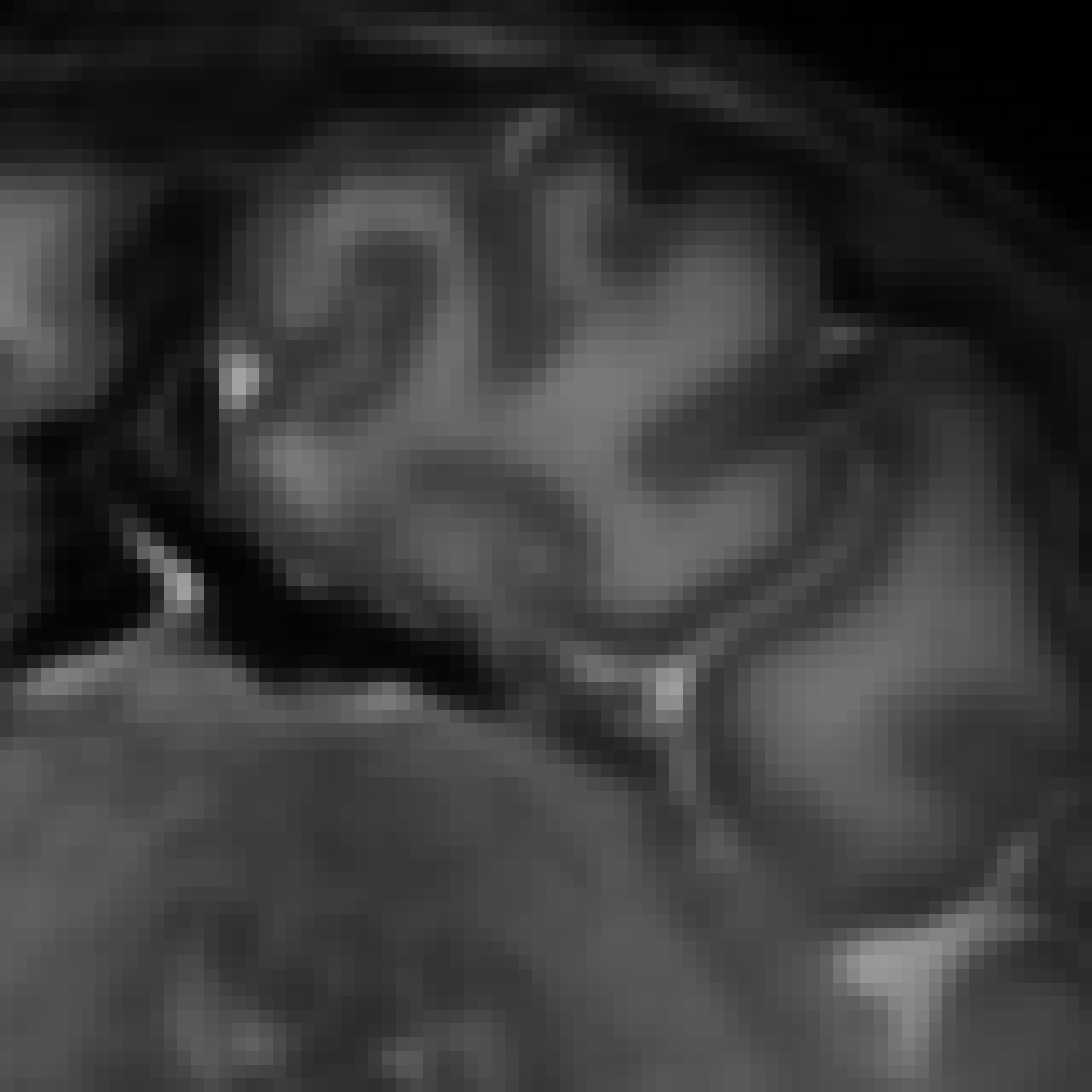} }; 
					% Create scope with normalized axes
					\begin{scope}[
					x={($0.1*(image.south east)$)},
					y={($0.1*(image.north west)$)}]
					% Grid
					% \draw[lightgray,step=1] (image.south west) grid (image.north east);
					\draw[thick,orange] (6,1) rectangle (9.5, 7) ;
					\end{scope}
					\end{tikzpicture} } &
				\subfloat[$\alpha=$ \nicefrac{6}{7}]{ \begin{tikzpicture}
					\node[above right, inner sep=0] (image) at (0,0)  { \includegraphics[width=.1\textwidth]{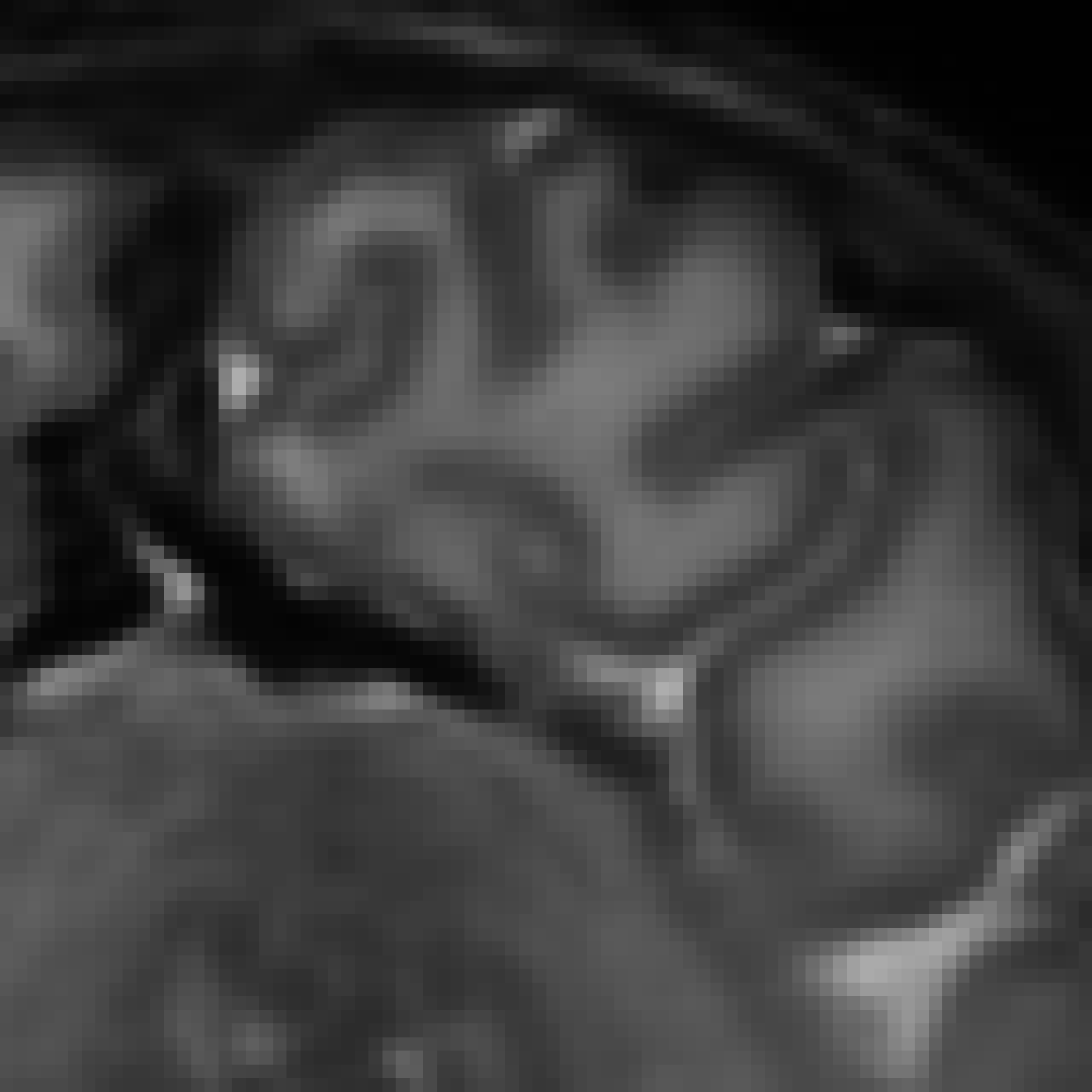} }; 
					% Create scope with normalized axes
					\begin{scope}[
					x={($0.1*(image.south east)$)},
					y={($0.1*(image.north west)$)}]
					% Grid
					% \draw[lightgray,step=1] (image.south west) grid (image.north east);
					\draw[thick,orange] (6,1) rectangle (9.5, 7) ;
					\end{scope}
					\end{tikzpicture} } &
				\subfloat[Neighboring slice 2] { \begin{tikzpicture}
					\node[above right, inner sep=0] (image) at (0,0)  { \includegraphics[width=.1\textwidth]{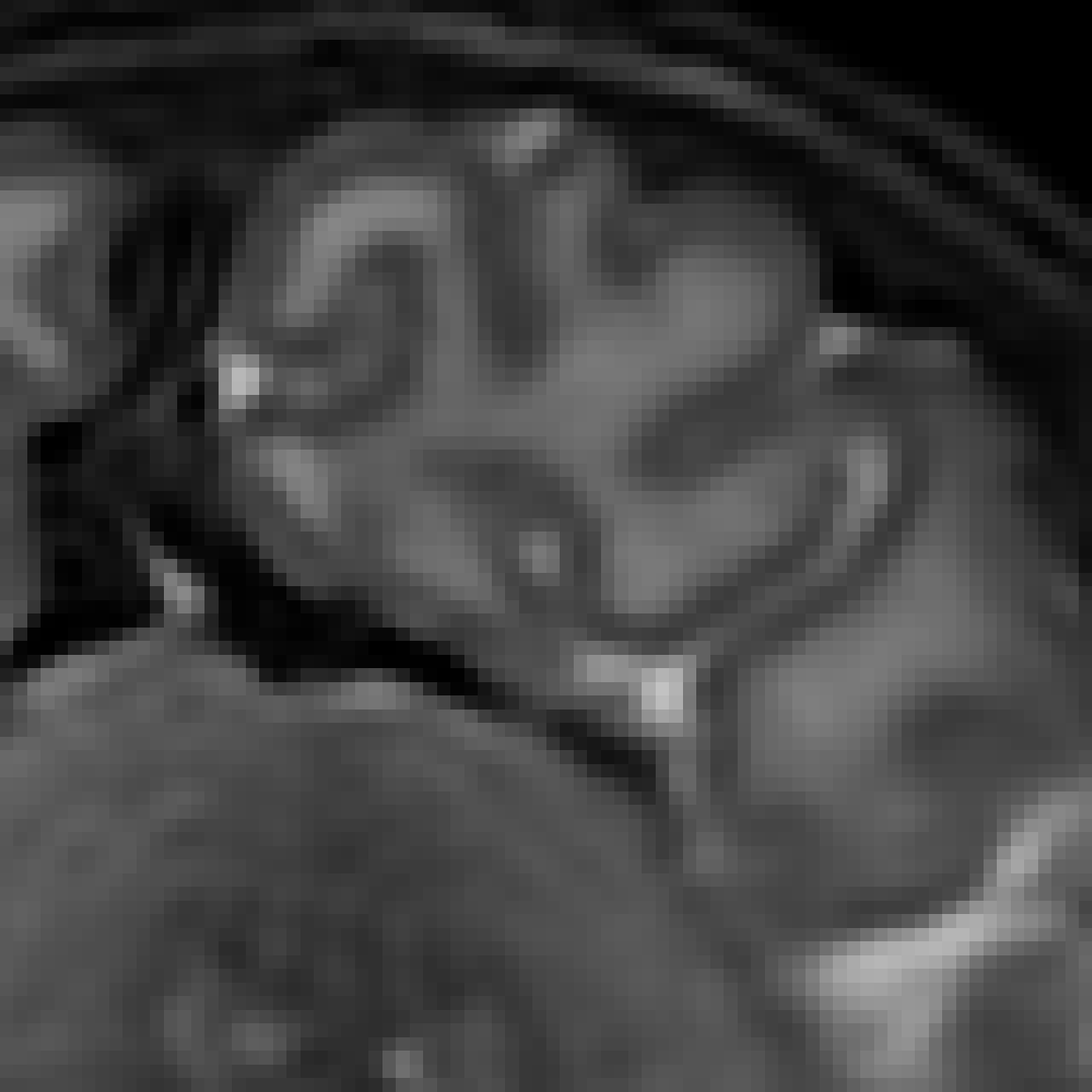} }; 
					% Create scope with normalized axes
					\begin{scope}[
					x={($0.1*(image.south east)$)},
					y={($0.1*(image.north west)$)}]
					% Grid
					% \draw[lightgray,step=1] (image.south west) grid (image.north east);
					\draw[thick,orange] (6,1) rectangle (9.5, 7) ;
					\end{scope}
					\end{tikzpicture} }  \\
				% row 2
				\subfloat[Neighboring slice 1]
				{ \begin{tikzpicture}
					\node[above right, inner sep=0] (image) at (0,0)  { \includegraphics[width=.1\textwidth]{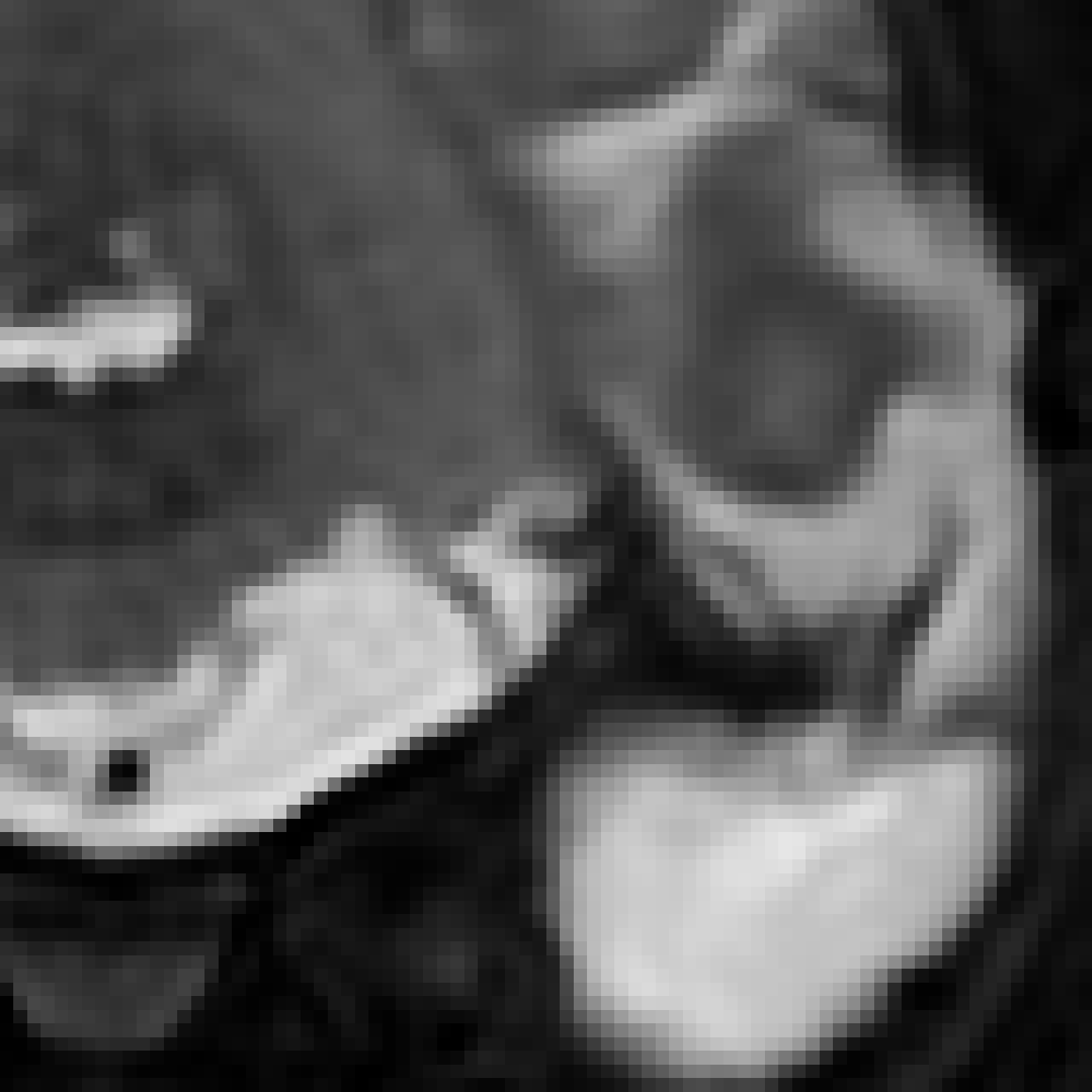} }; 
					% Create scope with normalized axes
					\begin{scope}[
					x={($0.1*(image.south east)$)},
					y={($0.1*(image.north west)$)}]
					% Grid
					% \draw[lightgray,step=1] (image.south west) grid (image.north east);
					\draw[thick,orange] (4.5,5) rectangle (9.5, 9.5) ;
					\end{scope}
					\end{tikzpicture} }  & 
				\subfloat[$\alpha=$ \nicefrac{1}{7}]{\begin{tikzpicture}
					\node[above right, inner sep=0] (image) at (0,0)  {\includegraphics[width=.1\textwidth]{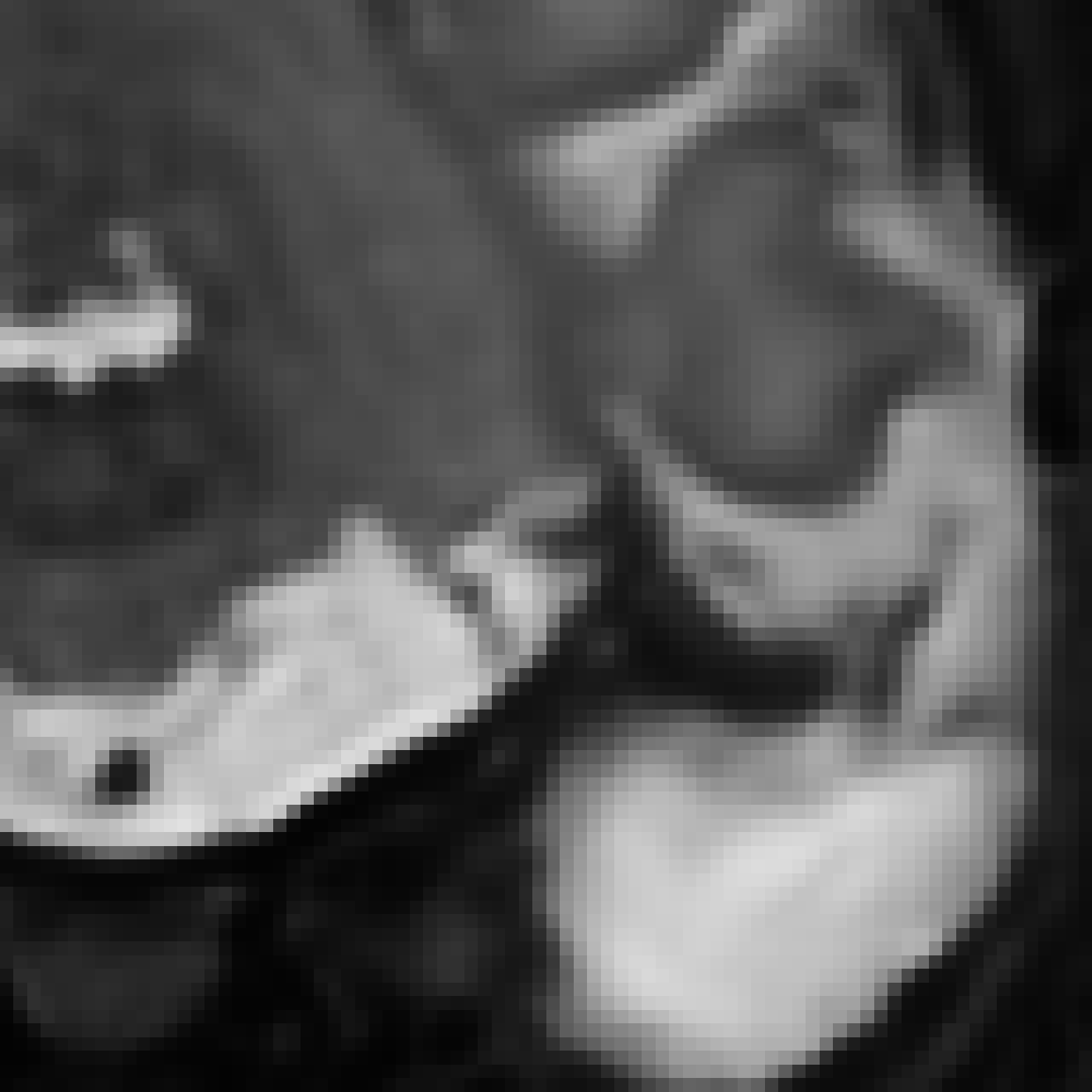} }; 
					% Create scope with normalized axes
					\begin{scope}[
					x={($0.1*(image.south east)$)},
					y={($0.1*(image.north west)$)}]
					% Grid
					% \draw[lightgray,step=1] (image.south west) grid (image.north east);
					\draw[thick,orange] (4.5,5) rectangle (9.5, 9.5) ;
					\end{scope}
					\end{tikzpicture} } &
				\subfloat[$\alpha=$ \nicefrac{2}{7}]{\begin{tikzpicture}
					\node[above right, inner sep=0] (image) at (0,0)  {\includegraphics[width=.1\textwidth]{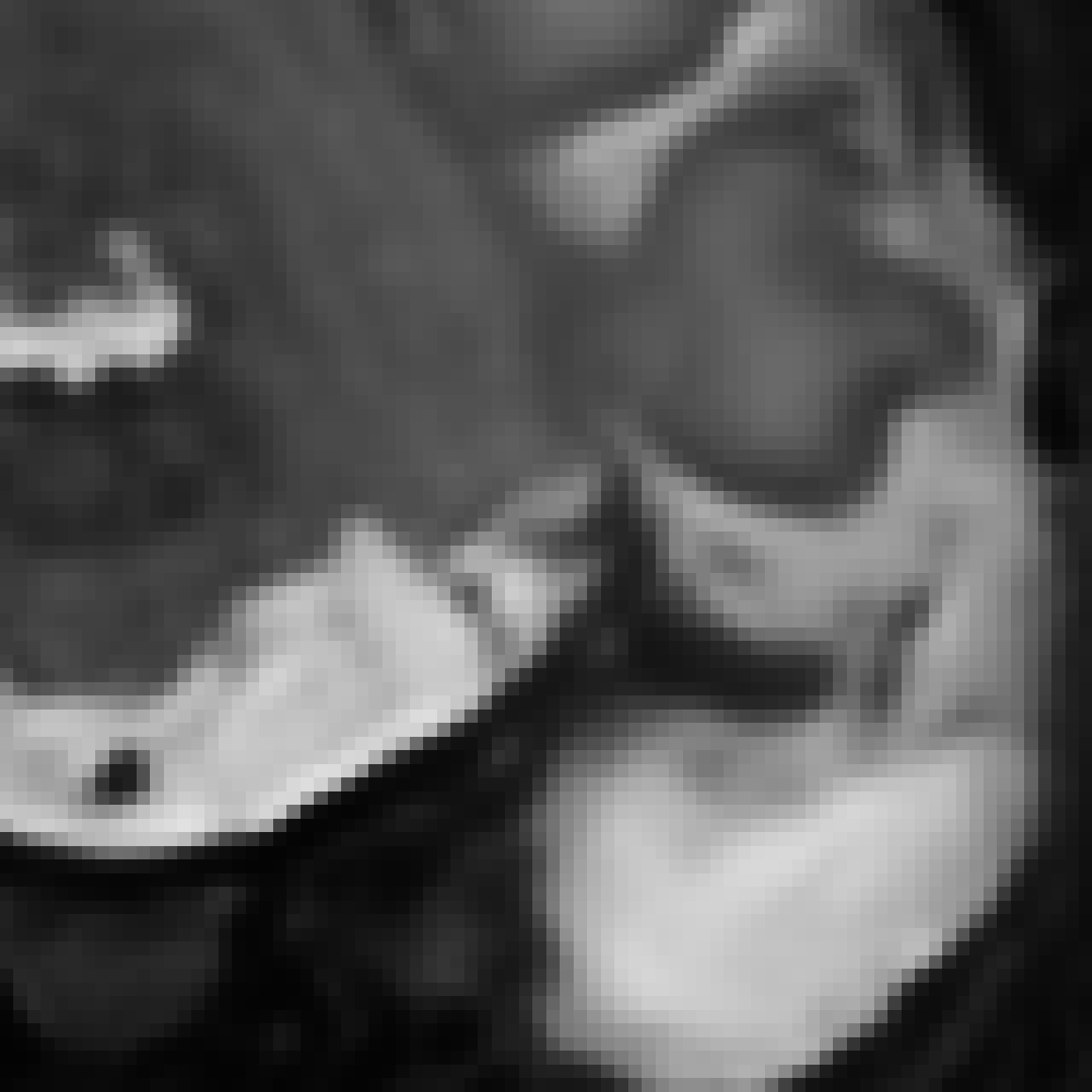} }; 
					% Create scope with normalized axes
					\begin{scope}[
					x={($0.1*(image.south east)$)},
					y={($0.1*(image.north west)$)}]
					% Grid
					% \draw[lightgray,step=1] (image.south west) grid (image.north east);
					\draw[thick,orange] (4.5,5) rectangle (9.5, 9.5) ;
					\end{scope}
					\end{tikzpicture} } &
				\subfloat[$\alpha=$ \nicefrac{3}{7}]{\begin{tikzpicture}
					\node[above right, inner sep=0] (image) at (0,0)  {\includegraphics[width=.1\textwidth]{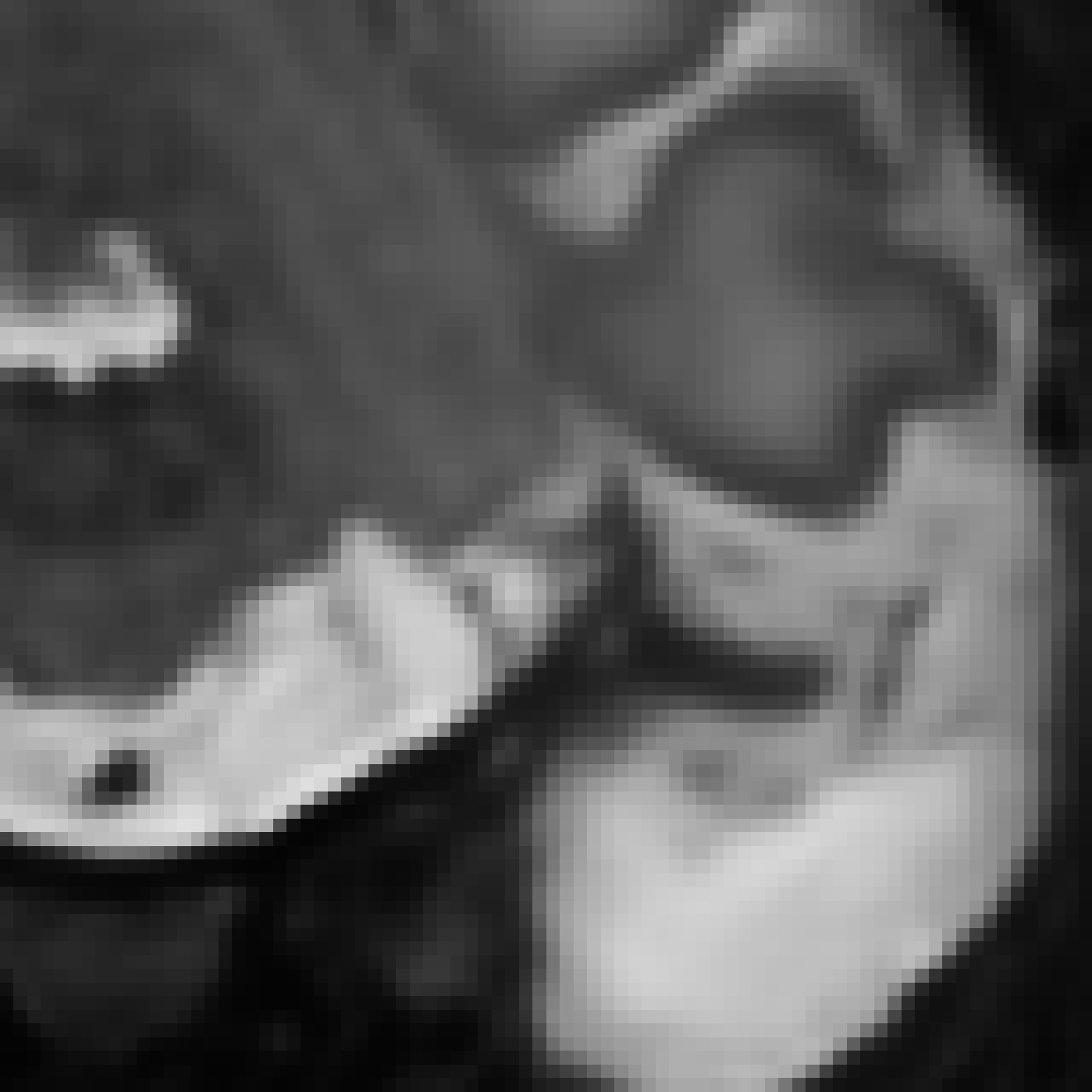} }; 
					% Create scope with normalized axes
					\begin{scope}[
					x={($0.1*(image.south east)$)},
					y={($0.1*(image.north west)$)}]
					% Grid
					% \draw[lightgray,step=1] (image.south west) grid (image.north east);
					\draw[thick,orange] (4.5,5) rectangle (9.5, 9.5) ;
					\end{scope}
					\end{tikzpicture} } &
				\subfloat[$\alpha=$ \nicefrac{4}{7}]{\begin{tikzpicture}
					\node[above right, inner sep=0] (image) at (0,0)  {\includegraphics[width=.1\textwidth]{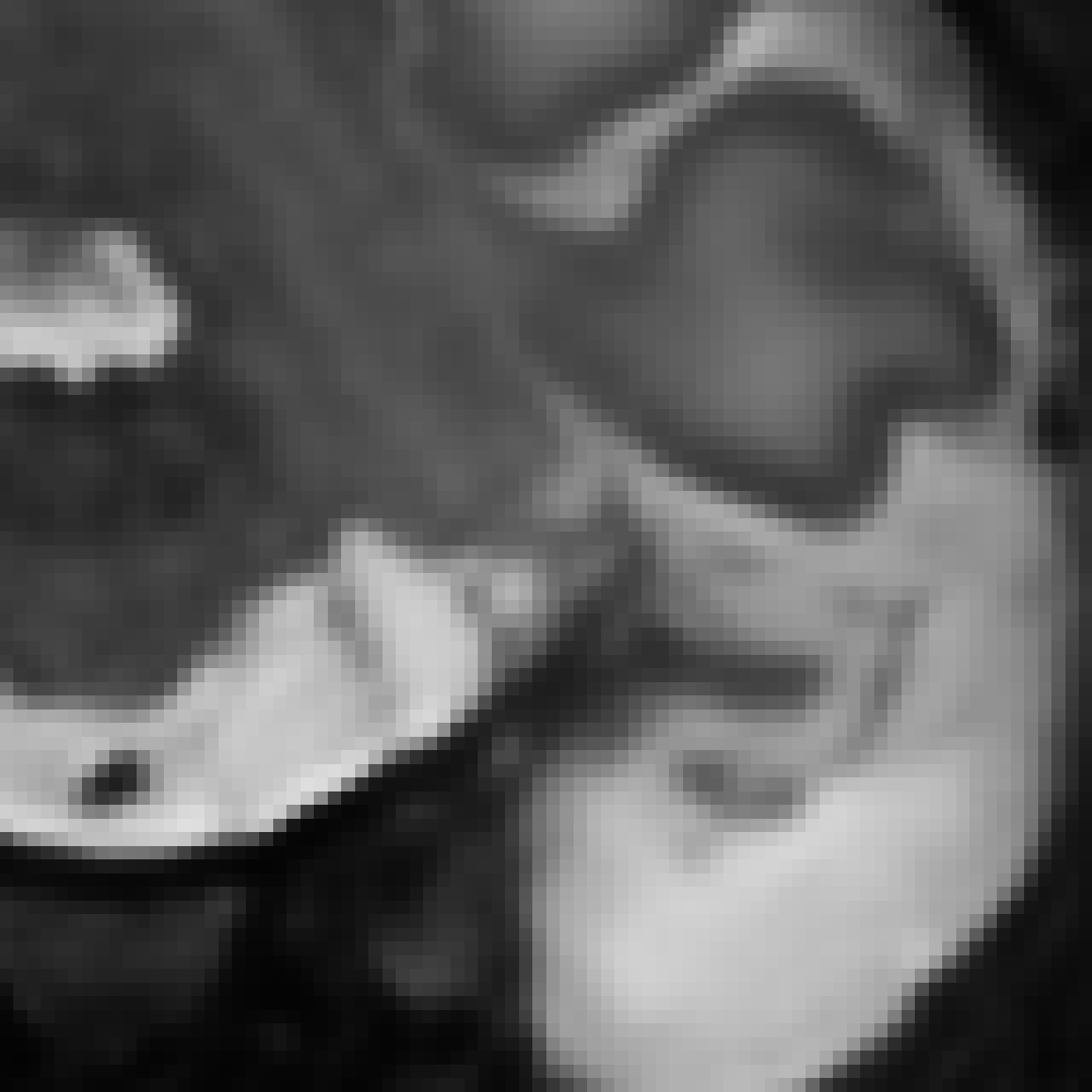} }; 
					% Create scope with normalized axes
					\begin{scope}[
					x={($0.1*(image.south east)$)},
					y={($0.1*(image.north west)$)}]
					% Grid
					% \draw[lightgray,step=1] (image.south west) grid (image.north east);
					\draw[thick,orange] (4.5,5) rectangle (9.5, 9.5) ;
					\end{scope}
					\end{tikzpicture} } &
				\subfloat[$\alpha=$ \nicefrac{5}{7}]{\begin{tikzpicture}
					\node[above right, inner sep=0] (image) at (0,0)  {\includegraphics[width=.1\textwidth]{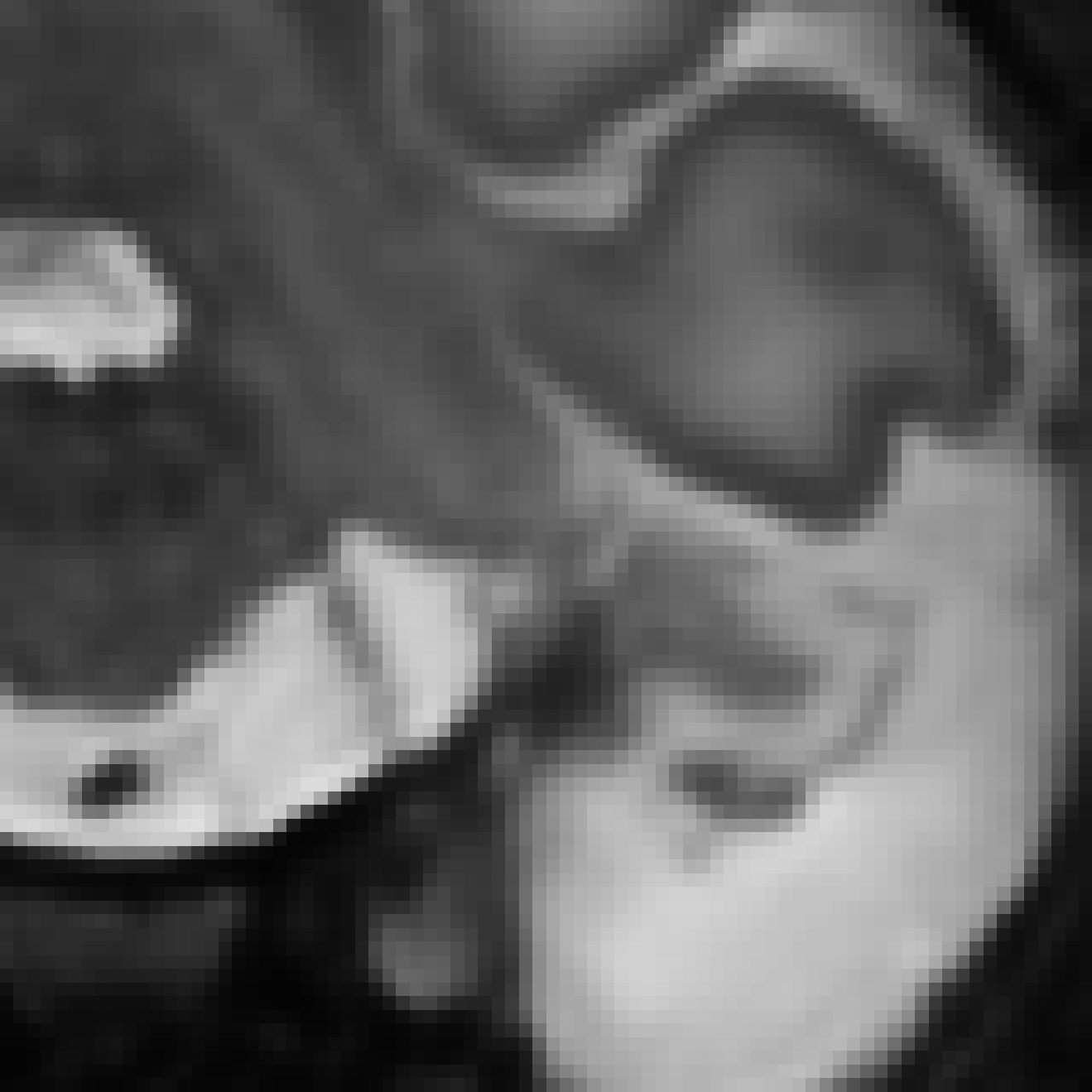} }; 
					% Create scope with normalized axes
					\begin{scope}[
					x={($0.1*(image.south east)$)},
					y={($0.1*(image.north west)$)}]
					% Grid
					% \draw[lightgray,step=1] (image.south west) grid (image.north east);
					\draw[thick,orange] (4.5,5) rectangle (9.5, 9.5) ;
					\end{scope}
					\end{tikzpicture}} &
				\subfloat[$\alpha=$ \nicefrac{6}{7}]{\begin{tikzpicture}
					\node[above right, inner sep=0] (image) at (0,0)  {\includegraphics[width=.1\textwidth]{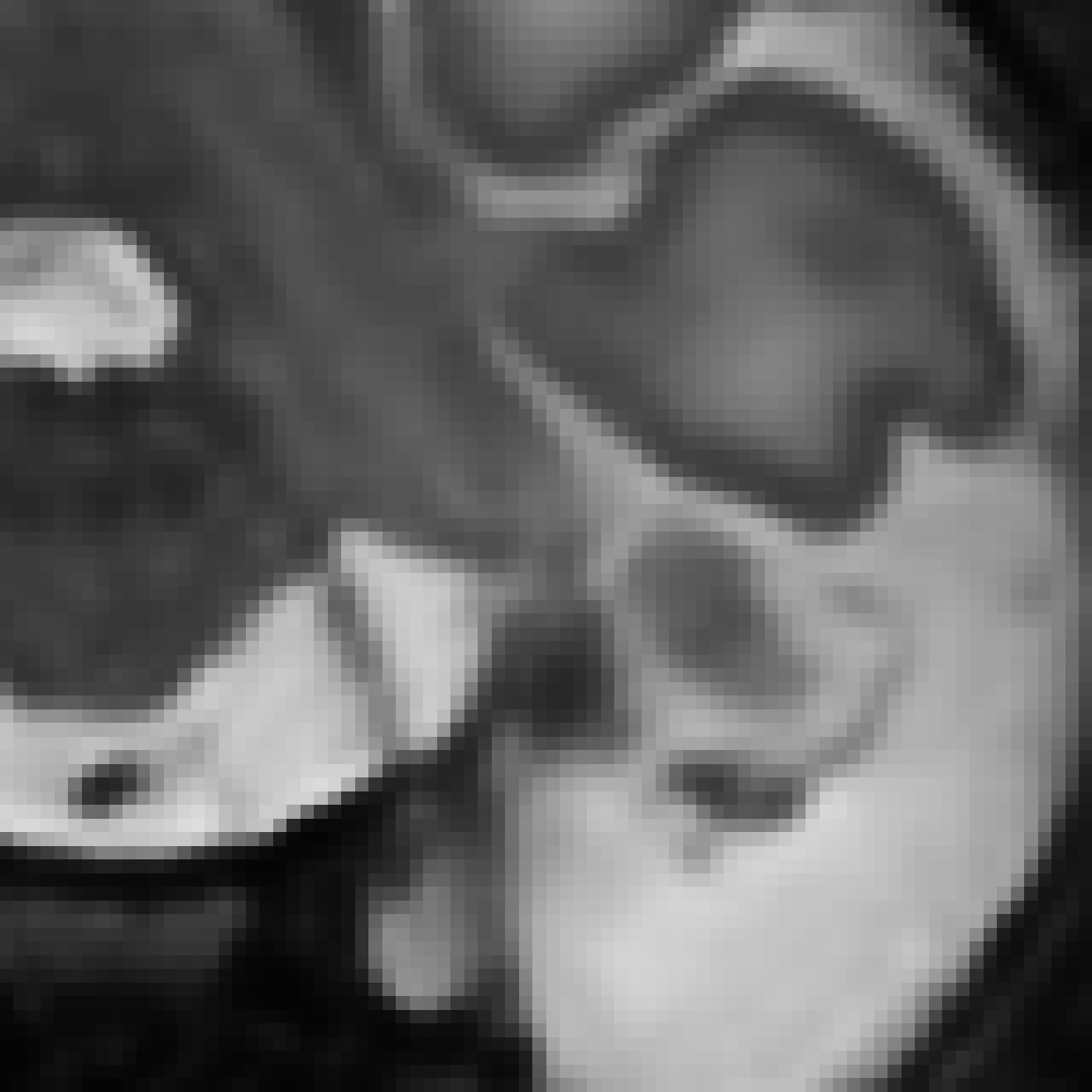} }; 
					% Create scope with normalized axes
					\begin{scope}[
					x={($0.1*(image.south east)$)},
					y={($0.1*(image.north west)$)}]
					% Grid
					% \draw[lightgray,step=1] (image.south west) grid (image.north east);
					\draw[thick,orange] (4.5,5) rectangle (9.5, 9.5) ;
					\end{scope}
					\end{tikzpicture}} &
				\subfloat[Neighboring slice 2]
				{\begin{tikzpicture}
					\node[above right, inner sep=0] (image) at (0,0)  { \includegraphics[width=.1\textwidth]{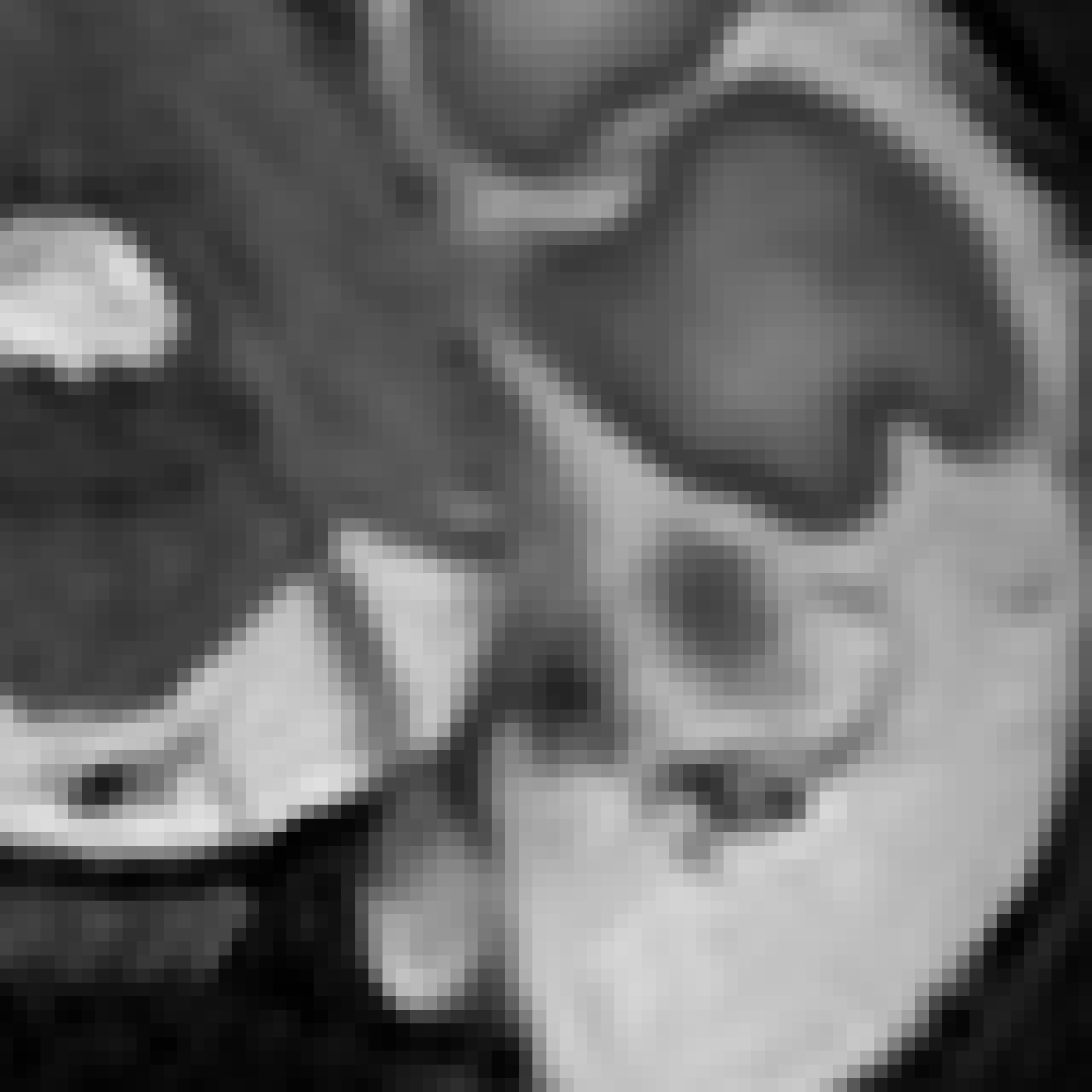} }; 
					% Create scope with normalized axes
					\begin{scope}[
					x={($0.1*(image.south east)$)},
					y={($0.1*(image.north west)$)}]
					% Grid
					% \draw[lightgray,step=1] (image.south west) grid (image.north east);
					\draw[thick,orange] (4.5,5) rectangle (9.5, 9.5) ;
					\end{scope}
					\end{tikzpicture}
				}   
			\end{tabular}
			
		\end{center}
		
		\caption{Zoomed-in images of synthesis results in scans of two different patients (one example per row) using neonatal brain MRIs from the dHCP dataset. Slice spacing was improved from \num{2} to \SI{0.29}{\milli\meter} by synthesizing six intermediate slices (second to penultimate columns). For this, latent space encodings of the two neighboring slices (first and last column) were combined using their convex combination. $\alpha$ denotes the mixing coefficient as specified in Equation~\ref{eq_convex_combination}. Bounding boxes focus on anatomical variations between images in a row. Note that a small \textit{cross-fade} artifact appears in the bright area of the fourth ($\alpha=\nicefrac{3}{7}$) and fifth ($\alpha=\nicefrac{4}{7}$) image in the second row.}
		\label{fig_qualitative_synthesis_dhcp4x}
		
	\end{figure*}
	
	\subsection{Semantic Interpolation of Cardiac Cine MRI} \label{exp_cardiac_acdc}
	
	Short-axis cardiac cine MRIs are acquired to primarily investigate cardiac function. These images have a high temporal and in-plane resolution at the cost of lower through-plane resolution. 
	The functional parameters extracted from these images, such as ejection fraction, may show high variability, which can be explained by the high anisotropic resolution, that may heavily influence volume measurements. These measurements may improve when extracted from upsampled images with smooth cardiac structures in through-plane direction. Therefore, the proposed approach was evaluated on highly anisotropic cardiac cine MRI using the ACDC dataset.
	
	\subsubsection{Experimental details} \label{section_acdc_exp_details}
	The dataset was randomly split into training (\num{70} patients), validation (\num{10} patients), and test sets (\num{20} patients). Table~\ref{table_acdc_train_val_test_split} specifies slice spacing of volumes over the three sets. The test set was only used for the final quantitative as well as qualitative evaluation. 
	
	\begin{table}
		\centering
		\caption{Distribution of patients in ACDC dataset over training, validation and test sets. First column specifies the slice spacing ($\SI{}{\milli\meter}$) of cardiac MRI volumes. The ACDC dataset specifies slice spacing for each image volume while slice thickness is only specified as a range for all volumes.}
		\label{table_acdc_train_val_test_split}
		\begin{tabular}{L{1.5cm}  C{1.2cm} C{1.2cm} C{1.2cm} }
			\toprule
			\multicolumn{1}{c}{\textbf{Slice spacing}} & \multicolumn{1}{c}{\textbf{Training}} & \multicolumn{1}{c}{\textbf{Validation}} & \multicolumn{1}{c}{\textbf{Test}} \\
			\multicolumn{1}{c}{\textbf{($\SI{}{\milli\meter}$})} &  &  & \\
			\hline
			\multicolumn{1}{c}{\phantom{x}\num{5}} & \num{6}  & - & \num{6} \\
			\multicolumn{1}{c}{\phantom{x}\num{6.5}} & -  & \num{1} & - \\
			\multicolumn{1}{c}{\phantom{x}\num{7}} & \num{1}  & - & - \\
			\multicolumn{1}{c}{\num{10}} & \num{63}  & \num{9} & \num{14} \\
			\bottomrule
		\end{tabular}
	\end{table}
	
	To train a model, patches of \num{128}$\times$\num{128} pixels were randomly chosen from the training set in mini-batches of \num{12} slice pairs i.e. \num{24} image slices. A model was trained for \num{900} epochs. Furthermore, $\lambda$ in Equation~\ref{eq_combined_loss} was set to \num{0.05}. %after performing a line search ($\lambda \in \{10, 1, 0.5, 0.1, 0.05, 0.01 \}$). 
	
	Test images were center-cropped to \num{140}$\times$\num{140} pixels covering all cardiac structures of interest. To quantitatively evaluate our method, lower through-plane resolution was mimicked by excluding every other slice in the test images i.e. by increasing slice spacing. These excluded slices were subsequently recovered by synthesizing them using the proposed approach. For this, the upsampling factor $K$ was set to \num{2} and $\mathcal{A}$, the set of mixing coefficients was equal to \num{0.5}. Downsampled test volumes had a slice spacing of $\SI{10}{\milli\meter}$ and $\SI{20}{\milli\meter}$ while slice thickness remained unchanged.

	%%%%    NEO-NATAL brain 
	\vspace{1ex}
	\textbf{Comparison With Conventional Interpolation Method: }\label{exp_dHCP_quant_qual_evaluation} Upsampling performance of the proposed unsupervised approach was quantitatively and qualitatively evaluated and compared with cubic B-spline interpolation.
	
	\begin{figure}
		\captionsetup[subfigure]{justification=centering, labelformat=empty}
		\setlength{\tabcolsep}{1pt}
		% AXIAL SLICES
		\begin{center}
			\begin{tabular}{c c c}
				% row 1 
				% \setcounter{subfigure}{0}
				\subfloat[Original axial slice to be synthesized]{\includegraphics[width=.15\textwidth]{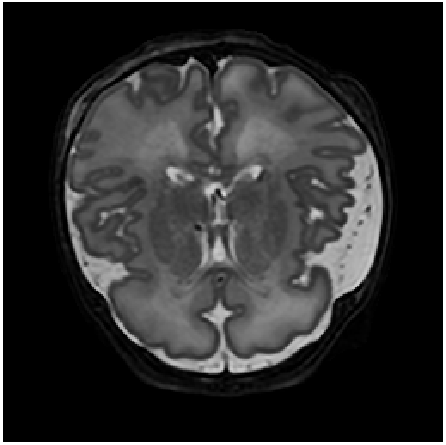} }  &
				\subfloat[B-spline ]{\includegraphics[width=.15\textwidth]{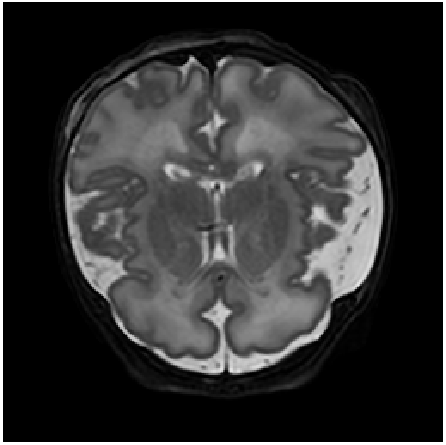} } &
				\subfloat[ours]{\includegraphics[width=.15\textwidth]{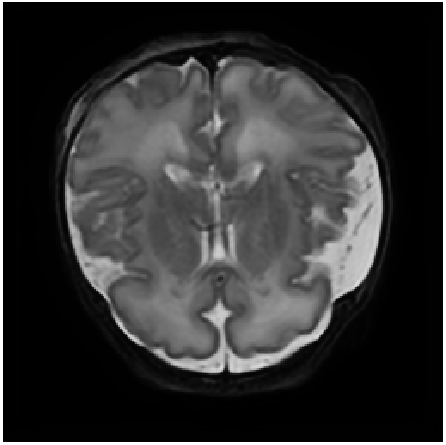} }	\\ 
				% row 2
				&
				\subfloat{\includegraphics[width=.15\textwidth]{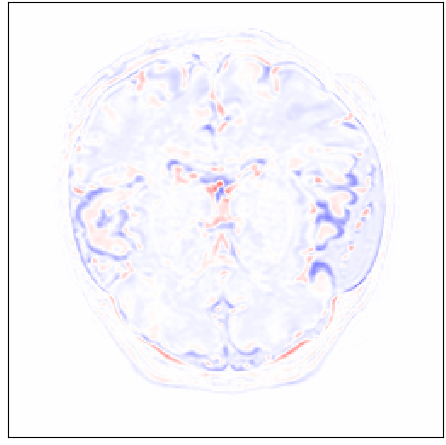}  }  &
				\subfloat{\includegraphics[width=.15\textwidth]{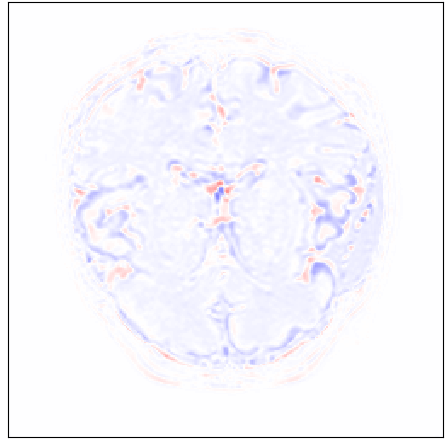}  }  \\
			\end{tabular}
		\end{center}
		
		% CORONAL SLICES
		\begin{center}
			\begin{tabular}{c c c}
				% row 1
				% \setcounter{subfigure}{0}
				\subfloat[Original coronal slice ]{\includegraphics[width=.15\textwidth]{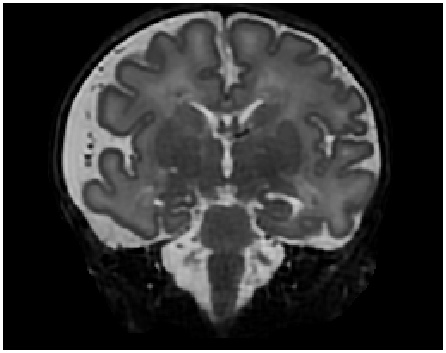} }  &
				\subfloat[B-spline ]{\includegraphics[width=.15\textwidth]{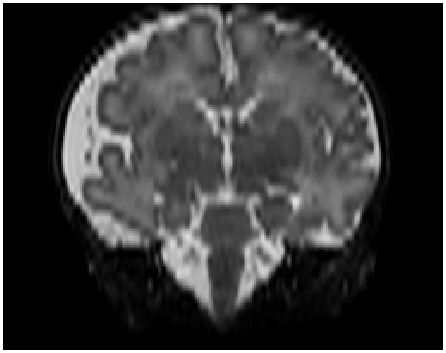} } & 
				\subfloat[ours]{\includegraphics[width=.15\textwidth]{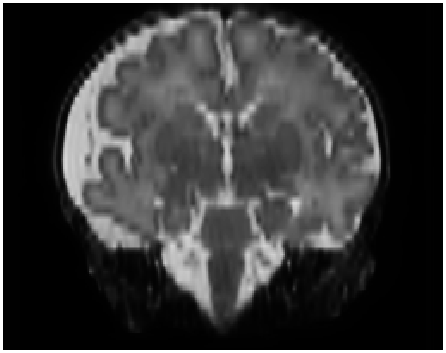} }	\\  
				% row 2 
				&
				\subfloat{\includegraphics[width=.15\textwidth]{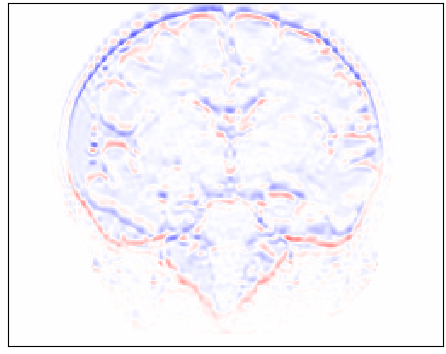}  } &
				\subfloat{\includegraphics[width=.15\textwidth]{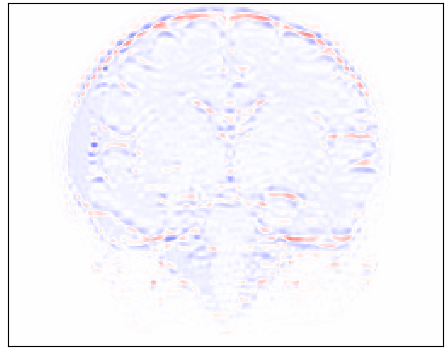}  }  \\
			\end{tabular}
			
		\end{center}
		% SAGITTAL SLICES
		\begin{center}
			\begin{tabular}{c c c}
				% row 1
				% \setcounter{subfigure}{0}
				\subfloat[Original sagittal slice ]{\includegraphics[width=.15\textwidth]{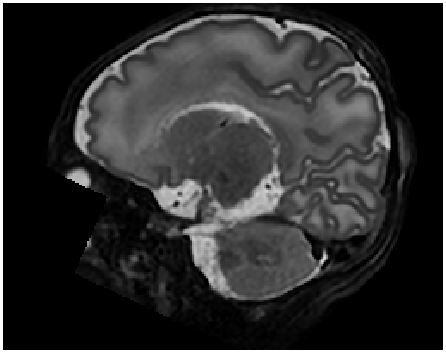} }  &
				\subfloat[B-spline ]{\includegraphics[width=.15\textwidth]{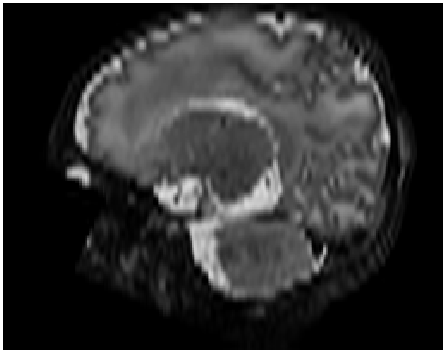} } &
				\subfloat[ours]{\includegraphics[width=.15\textwidth]{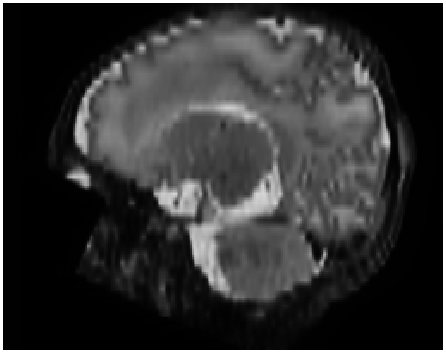} }	\\  
				% row 2 
				&
				\subfloat{\includegraphics[width=.15\textwidth]{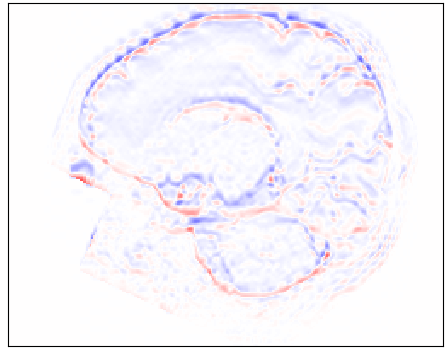}  } &
				\subfloat{\includegraphics[width=.15\textwidth]{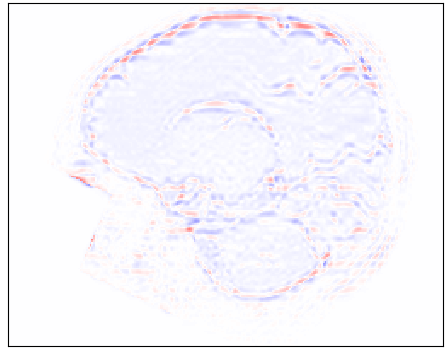}  }  \\
			\end{tabular}
			
		\end{center}
		\caption{Qualitative comparison of image synthesis performance on neonatal brain MRI (dHCP dataset) between conventional interpolation methods and proposed approach. Original volumes with slice thickness and spacing of \SI{0.5}{\milli\meter} were downsampled to \SI{2.5}{\milli\meter} by applying a Gaussian blur before including every fifth slice in the test volume. Differences between reference (minuend) and synthesized slice (subtrahend). Blue corresponds to negative and red to positive differences. Image intensities are scaled to a $[0,1]$ range. All difference images use the same color scale $[-1, 1]$.}
		\label{fig_qualitative_synthesis_compare_dhcp4x}
	\end{figure}
	
	%  Figure rebuttal: comparing SR performance for different upsampling factors dHCP dataset
	\begin{figure*}
		\captionsetup[subfigure]{justification=centering}
		\centering
		\includegraphics[width=.96\textwidth]{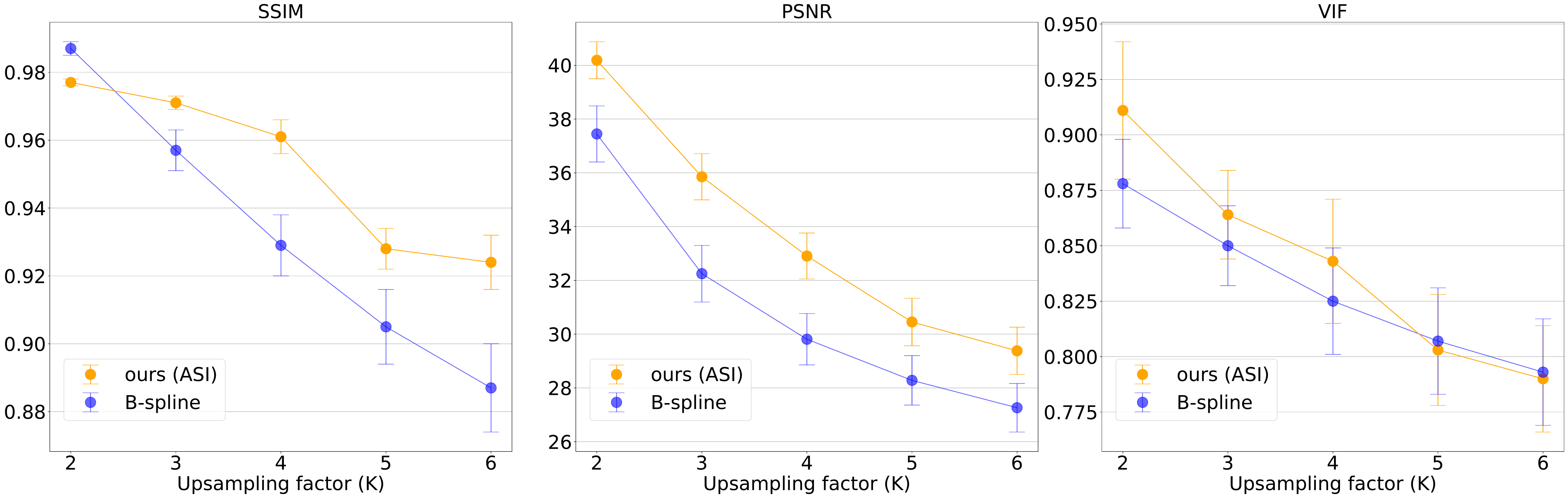} 
		\caption{Comparison of upsampling performance for cubic B-spline interpolation compared with proposed method (ASI) in terms of SSIM, PSNR, and VIF. Neonatal brain MRIs of \num{20} subjects from the dHCP dataset were upsampled with factor $K \in \{ 2, 3, 4, 5, 6 \}$ in through-plane direction. A higher score indicates better performance. The proposed method achieved higher performance in terms of SSIM and PSNR compared with conventional interpolation method. The differences between proposed and cubic B-spline interpolation approach in terms of SSIM and PSNR are statistically significant ($p < 0.001$) using the one-sided Wilcoxon signed-rank test.}		
		\label{fig_dhcp_compare_upsampling_factor} 
	\end{figure*}
	
	\subsubsection{Results}
	
	The primary goal of the proposed method is to \textit{synthesize} new slices located in-between two spatially adjacent slices. However, the method's capacity to synthesize new slices depends on the ability of the autoencoder to \textit{reconstruct} existing slices. Therefore, we report reconstruction and synthesis performance of the trained autoencoder separately.
	
	\vspace{1ex}
	\textbf{Slice Reconstruction: } Results for reconstructed and synthesized slices listed in Table~\ref{table_acdc_ae_caisr_recon_versus_synth_1} convey that the proposed approach achieved high reconstruction performance especially in terms of SSIM and PSNR. 
	Figure~\ref{fig_qualitative_reconstruction_acdc} depicts qualitative results of reconstruction performance for the proposed method on cardiac MRI. The results show that the trained autoencoder can reconstruct high-quality images i.e. input slices. Nevertheless, difference image shown in Figure~\ref{fig_qualitative_reconstruction_acdc_diff} depicts that some high spatial frequency details of the input slice are lacking in the reconstructed slice.

	\textbf{Slice Synthesis:} Qualitative evaluation of synthesis performance of the proposed method conveys that synthesized slices, i.e. those that are generated using a convex combination of the neighboring slice encodings, show an anatomically and semantically meaningful transition between the two neighboring slices. Moreover, despite large anatomical variations between the neighboring slices for the right ventricle, left ventricle and trabecular structures of the left ventricle, the proposed method can generate slices that depict an anatomically smooth transition between the neighboring slices. Figure~\ref{fig_qualitative_synthesis_acdc} illustrates three example evaluations for basal, mid-ventricular and apical MRI slices, respectively, where upsampling factor $K$ was set to \num{7} and $\mathcal{A}$, the set of mixing coefficients was equal to $\{\nicefrac{1}{7}, \nicefrac{2}{7}, \nicefrac{3}{7}, \nicefrac{4}{7}, \nicefrac{5}{7}, \nicefrac{6}{7}\}$.
	Furthermore, quantitative evaluation of synthesis performance assessed on downsampled cardiac cine MRI scans listed in Table~\ref{table_acdc_ae_caisr_recon_versus_synth_1} reveals that synthesis performance is lower than reconstruction performance.

	\vspace{1ex}
	\textbf{Comparison With Conventional Interpolation Method: } Upsampling performance was compared with cubic B-spline in terms of SSIM, PSNR and VIF. For this, the methods synthesized cardiac slices that were excluded from the test volumes (see Section~\ref{section_acdc_exp_details}). Results for upsampling factor \num{2} are shown in Figure~\ref{fig_acdc_quantitative_results}. We observe that the proposed method achieved better performance when evaluated by all measures compared with the conventional interpolation method. These differences are statistically significant ($p < 0.0001$) using the one-sided Wilcoxon signed-rank test.
	
	Moreover, qualitative comparison of the methods reveals that synthesized images using the proposed method contain fewer errors than images generated by conventional interpolation method. Results shown in Figure~\ref{fig_acdc_qualitative_comparison_other_methods} depict that performance differences are especially pronounced for the left ventricle papillary muscles and right ventricle myocardium.
	
	Finally, methods were qualitatively compared for a through-plane upsampling factor of ten using the original cardiac volumes. Visual inspection of the results discloses that proposed method can generate volumes with a higher image quality than cubic B-spline interpolation. Qualitative comparison shown in Figure~\ref{fig_qualitative_ae_acdc_long_axis} reveals that performance differences are most pronounced for the myocardial structures of the left ventricle. These structures show smoother transitions between adjacent axial slices when generated by proposed method compared to cubic B-spline interpolation. The latter method generates volumes that suffer from aliasing artifacts while the proposed method can mostly suppress such artifacts.
	
	%%% NEO-NATAL brain 
	\begin{table}
		\caption{Comparison of upsampling performance in terms of SSIM and PSNR of proposed \textit{unsupervised} method compared to \textit{supervised} super-resolution approaches on neonatal brain MRI (dHCP dataset) as reported by \cite{pham2017brain, pham2019simultaneous} for upsampling factor \num{3}. Approaches of \cite{pham2017brain, pham2019simultaneous} were evaluated on a subset of \num{20} scans taken from the dHCP dataset. Best performance is indicated in bold.}
		
		\label{table_dHCP_other_methods}
		\centering	
		\begin{tabular}{L{8.5cm}  c c }
			\toprule
			\multicolumn{1}{l}{\textbf{Method}} & \multicolumn{1}{c}{\textbf{SSIM}} & \multicolumn{1}{c}{\textbf{PSNR} }  \\
			\hline
			% \vspace{0.2ex} \\
			\multicolumn{1}{l}{\begin{tabular}{@{}c@{}} \cite{pham2019simultaneous} (supervised) \end{tabular}} & 0.962 & 31.75  \\
			\multicolumn{1}{l}{\begin{tabular}{@{}c@{}} \cite{pham2017brain} (supervised)\end{tabular} } & \textbf{0.977} & 35.84   \\
			% \multicolumn{1}{l}{proposed AE} & 0.969 & 35.42  \\
			\multicolumn{1}{l}{ours (ASI)} & 0.971 & \textbf{35.85}  \\			
			\bottomrule
		\end{tabular}
		
	\end{table}

	% !!!!!!!!!!!!!!!!!!!!!!! OASIS FIGURES !!!!!!!!!!!!!!!!!!!!!!
	% FIGURE 10
	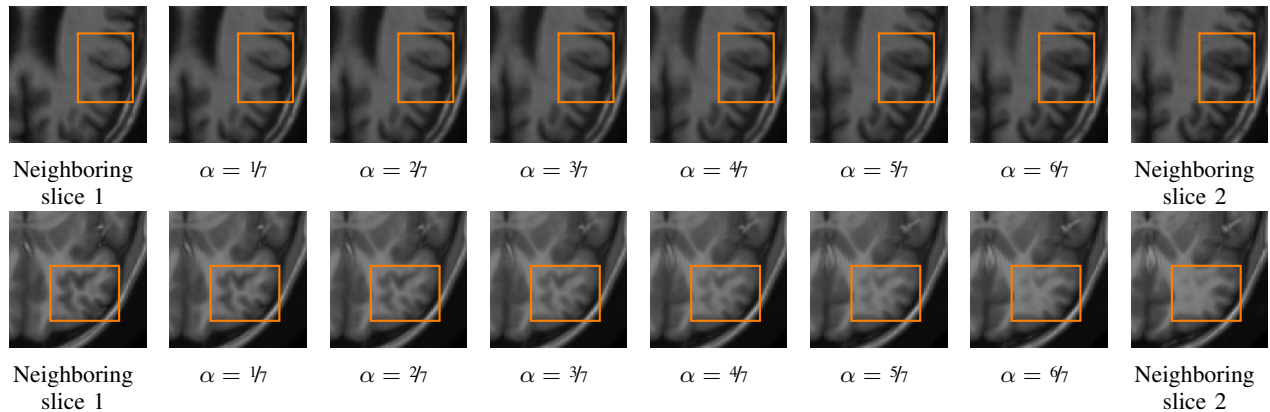
\begin{figure*}
		\captionsetup[subfigure]{justification=centering, labelformat=empty}
		\setlength{\tabcolsep}{1pt}
		\begin{center}
			\begin{tabular}{c c c c c c c c}
				% row 1
				\subfloat[Neighboring slice 1]{ \begin{tikzpicture}
					\node[above right, inner sep=0] (image) at (0,0)  { \includegraphics[width=.1\textwidth]{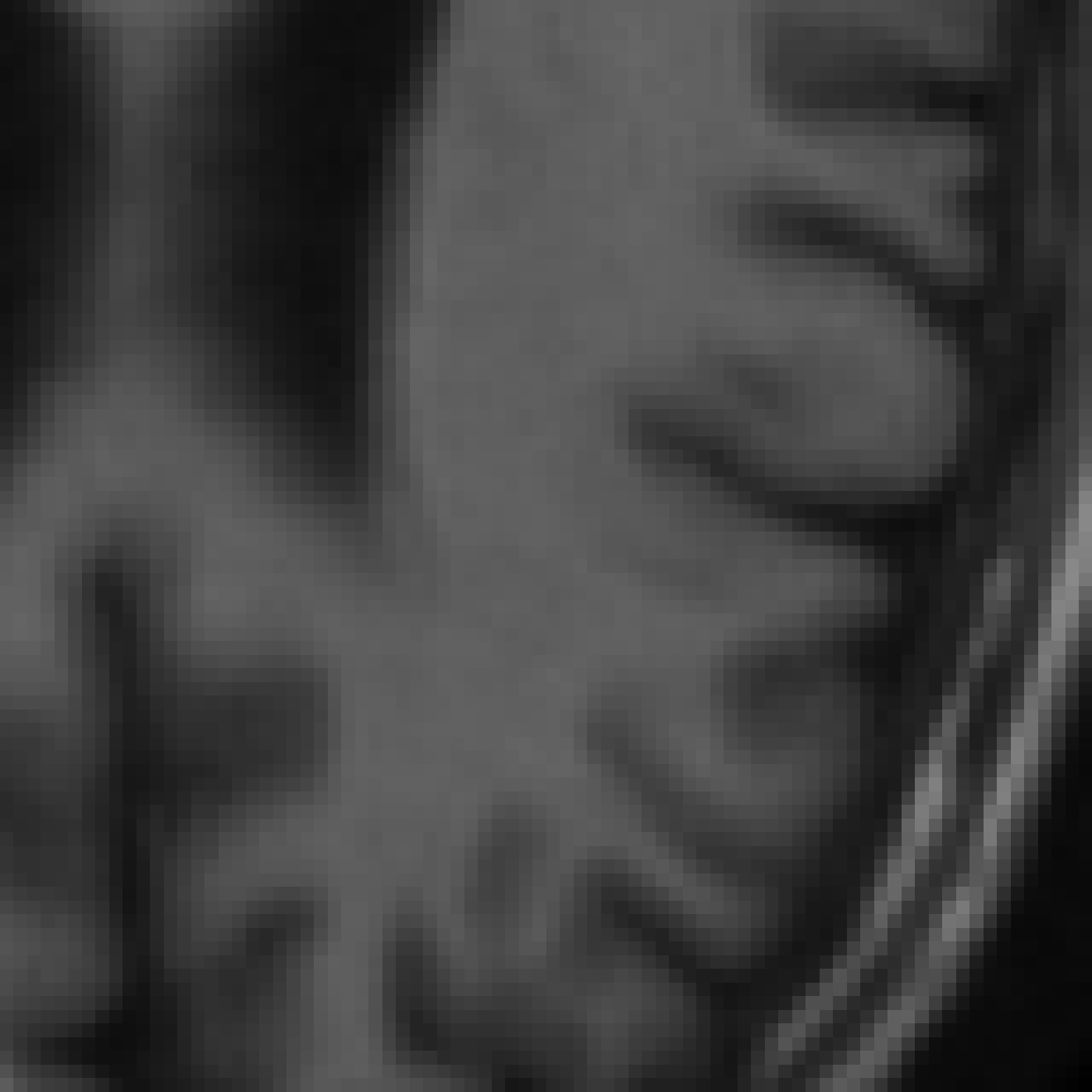} }; 
					% Create scope with normalized axes
					\begin{scope}[
					x={($0.1*(image.south east)$)},
					y={($0.1*(image.north west)$)}]
					% Grid
					% \draw[lightgray,step=1] (image.south west) grid (image.north east);
					\draw[thick,orange] (5,3) rectangle (9, 8) ;
					\end{scope}
					\end{tikzpicture} }
				& 
				\subfloat[$\alpha=$ \nicefrac{1}{7}]{ \begin{tikzpicture}
					\node[above right, inner sep=0] (image) at (0,0)  { \includegraphics[width=.1\textwidth]{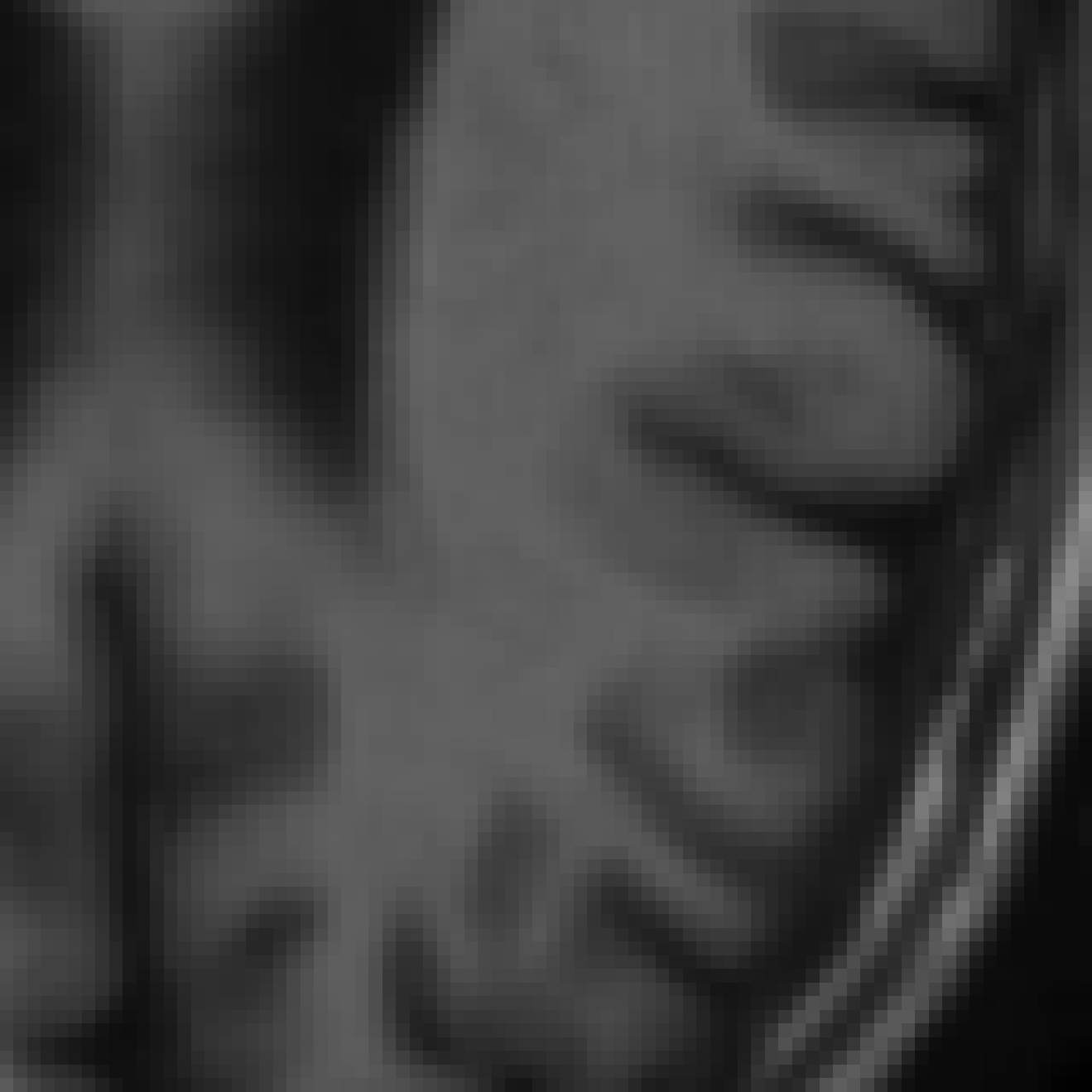} }; 
					% Create scope with normalized axes
					\begin{scope}[
					x={($0.1*(image.south east)$)},
					y={($0.1*(image.north west)$)}]
					% Grid
					% \draw[lightgray,step=1] (image.south west) grid (image.north east);
					\draw[thick,orange] (5,3) rectangle (9, 8) ;
					\end{scope}
					\end{tikzpicture} } &
				\subfloat[$\alpha=$ \nicefrac{2}{7}]{ \begin{tikzpicture}
					\node[above right, inner sep=0] (image) at (0,0)  { \includegraphics[width=.1\textwidth]{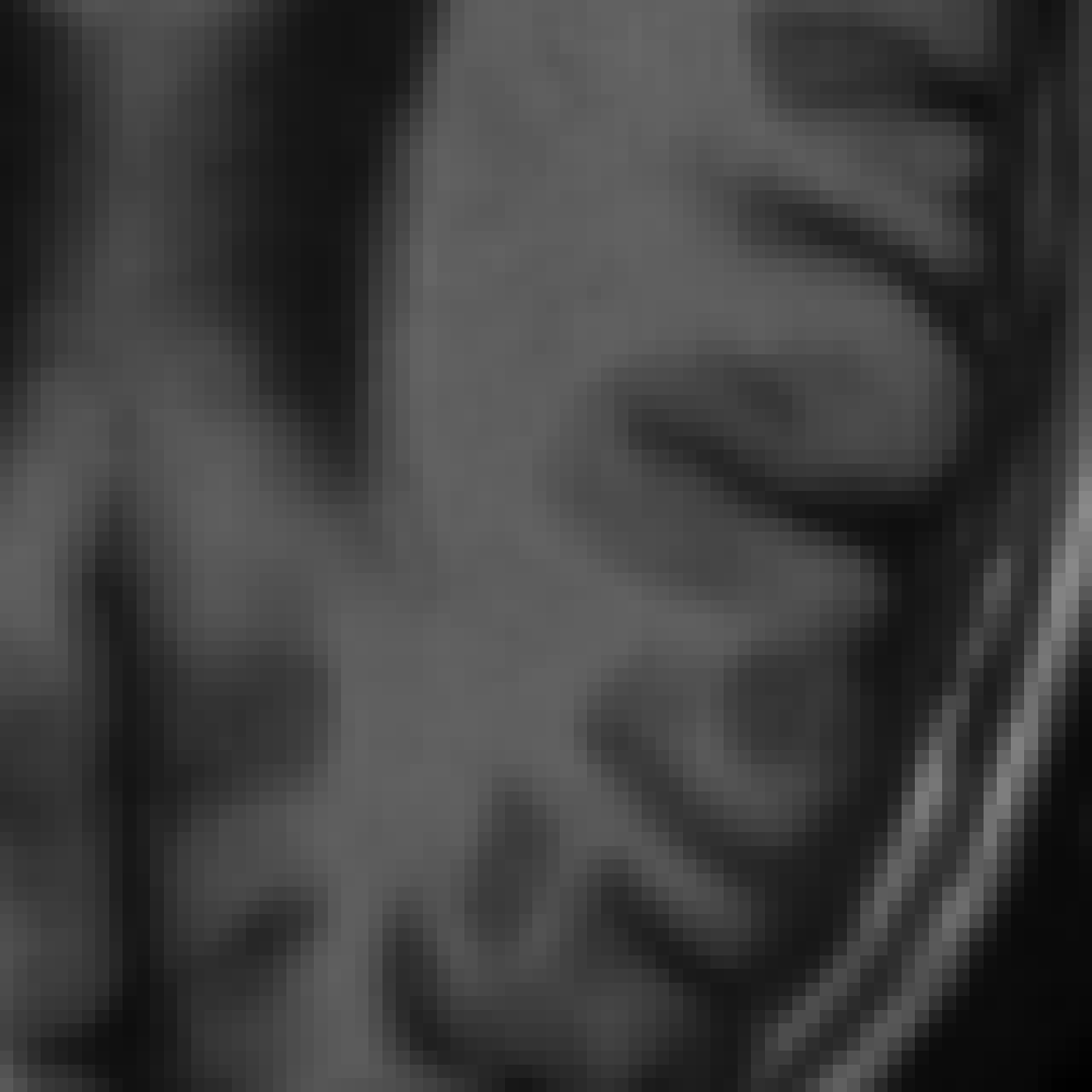} }; 
					% Create scope with normalized axes
					\begin{scope}[
					x={($0.1*(image.south east)$)},
					y={($0.1*(image.north west)$)}]
					% Grid
					% \draw[lightgray,step=1] (image.south west) grid (image.north east);
					\draw[thick,orange] (5,3) rectangle (9, 8) ;
					\end{scope}
					\end{tikzpicture} } &
				\subfloat[$\alpha=$ \nicefrac{3}{7}]{ \begin{tikzpicture}
					\node[above right, inner sep=0] (image) at (0,0)  { \includegraphics[width=.1\textwidth]{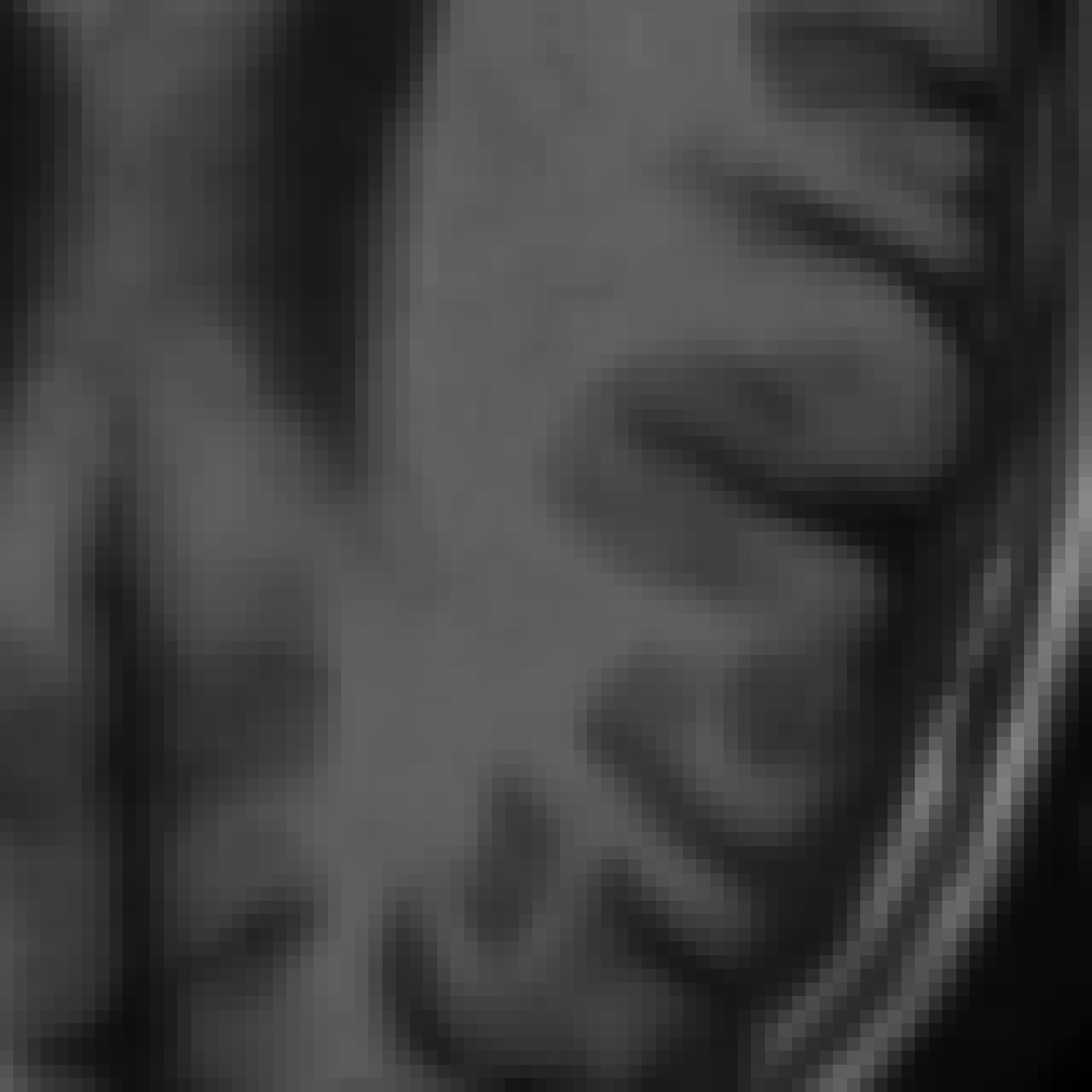} }; 
					% Create scope with normalized axes
					\begin{scope}[
					x={($0.1*(image.south east)$)},
					y={($0.1*(image.north west)$)}]
					% Grid
					% \draw[lightgray,step=1] (image.south west) grid (image.north east);
					\draw[thick,orange] (5,3) rectangle (9, 8) ;
					\end{scope}
					\end{tikzpicture} } &
				\subfloat[$\alpha=$ \nicefrac{4}{7}]{ \begin{tikzpicture}
					\node[above right, inner sep=0] (image) at (0,0)  { \includegraphics[width=.1\textwidth]{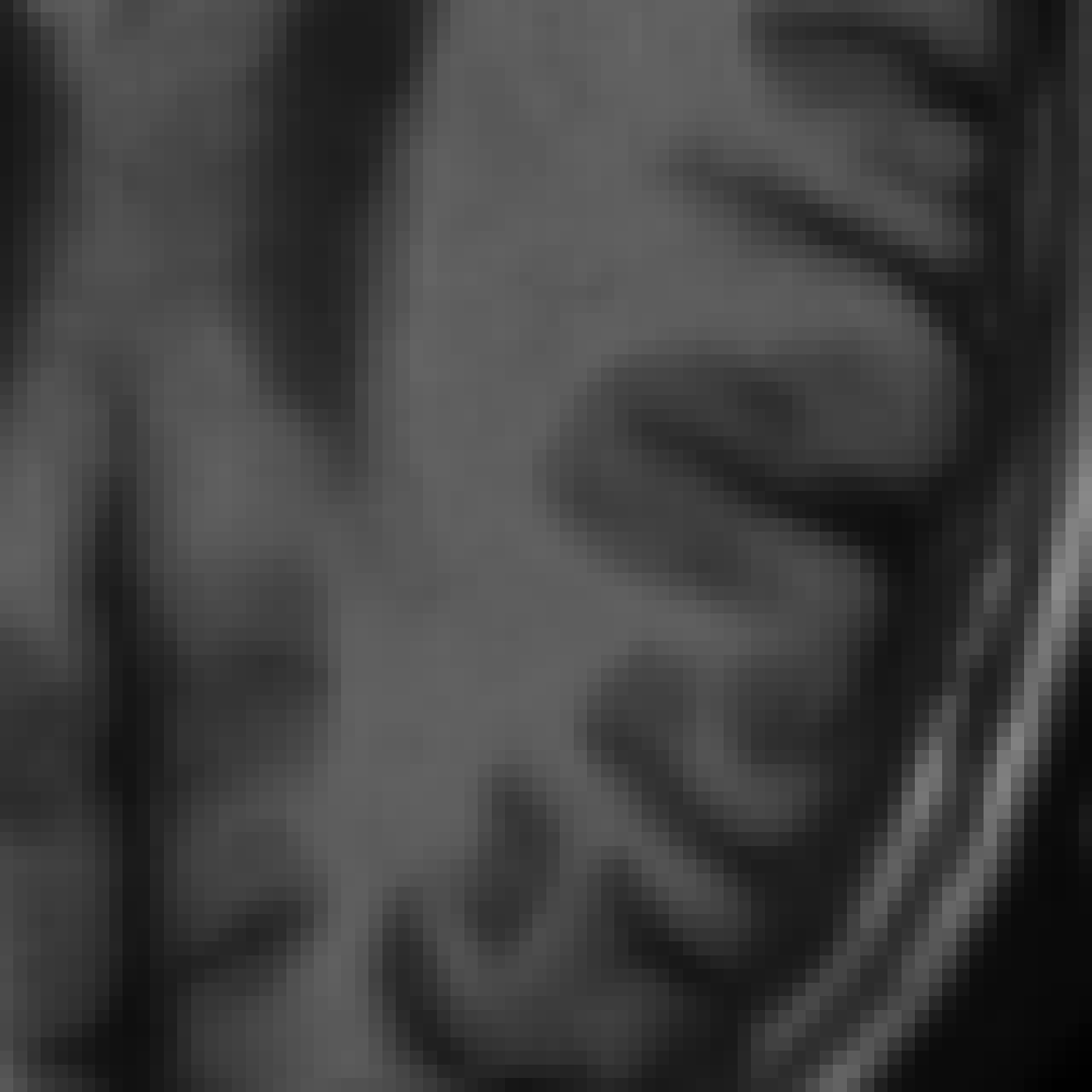} }; 
					% Create scope with normalized axes
					\begin{scope}[
					x={($0.1*(image.south east)$)},
					y={($0.1*(image.north west)$)}]
					% Grid
					% \draw[lightgray,step=1] (image.south west) grid (image.north east);
					\draw[thick,orange] (5,3) rectangle (9, 8) ;
					\end{scope}
					\end{tikzpicture} } &
				\subfloat[$\alpha=$ \nicefrac{5}{7}]{ \begin{tikzpicture}
					\node[above right, inner sep=0] (image) at (0,0)  { \includegraphics[width=.1\textwidth]{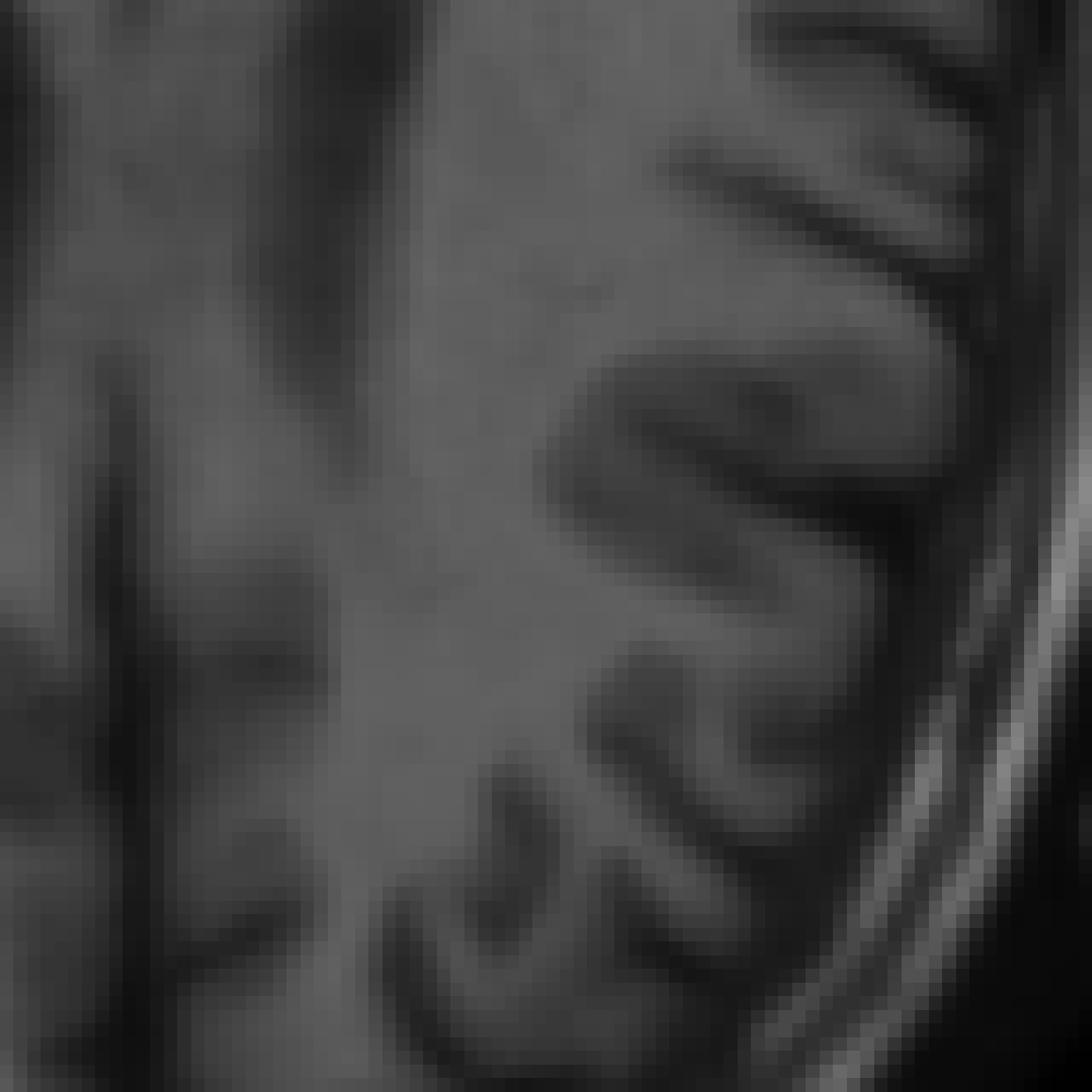} }; 
					% Create scope with normalized axes
					\begin{scope}[
					x={($0.1*(image.south east)$)},
					y={($0.1*(image.north west)$)}]
					% Grid
					% \draw[lightgray,step=1] (image.south west) grid (image.north east);
					\draw[thick,orange] (5,3) rectangle (9, 8) ;
					\end{scope}
					\end{tikzpicture} } &
				\subfloat[$\alpha=$ \nicefrac{6}{7}]{ \begin{tikzpicture}
					\node[above right, inner sep=0] (image) at (0,0)  { \includegraphics[width=.1\textwidth]{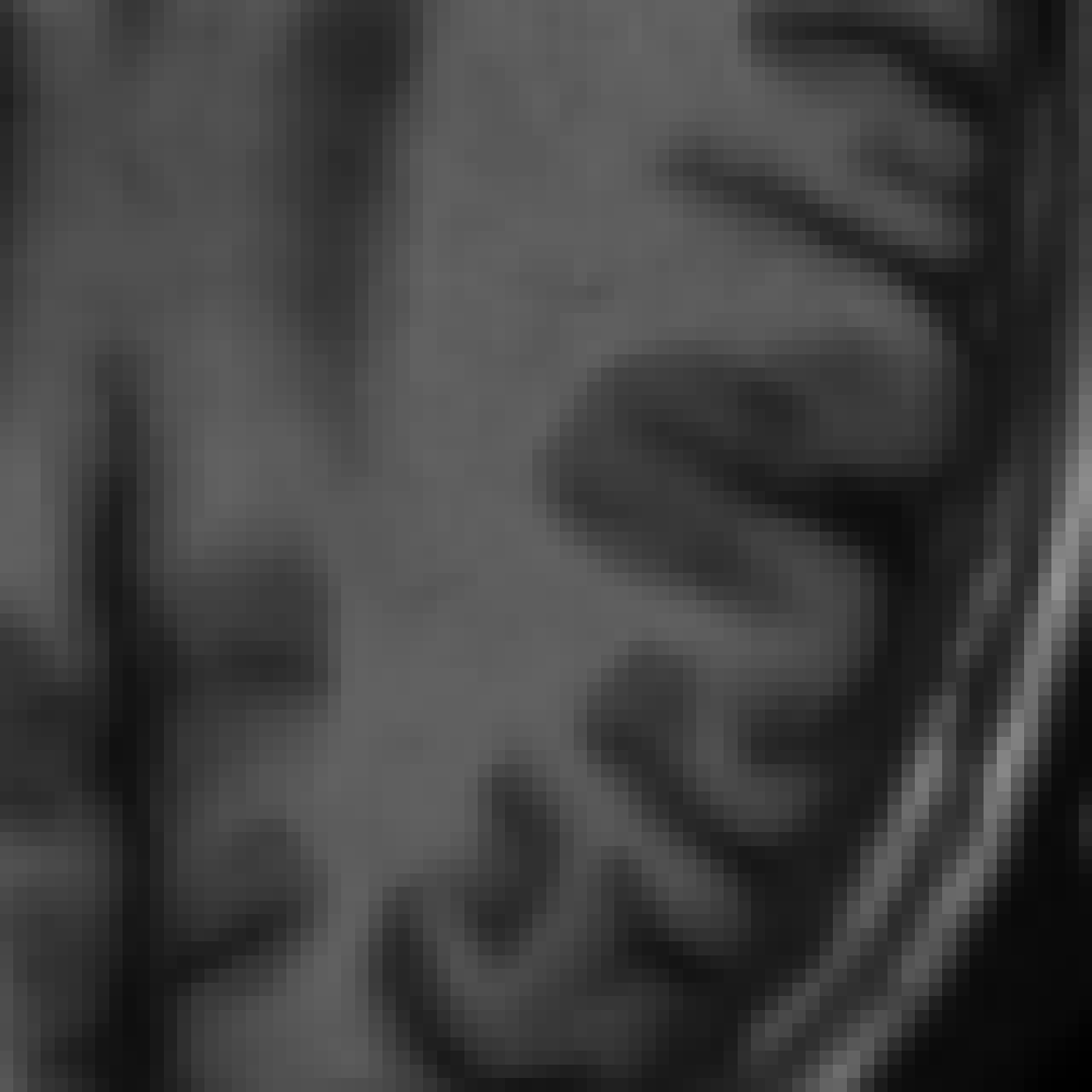} }; 
					% Create scope with normalized axes
					\begin{scope}[
					x={($0.1*(image.south east)$)},
					y={($0.1*(image.north west)$)}]
					% Grid
					% \draw[lightgray,step=1] (image.south west) grid (image.north east);
					\draw[thick,orange] (5,3) rectangle (9, 8) ;
					\end{scope}
					\end{tikzpicture} } &
				\subfloat[Neighboring slice 2] { \begin{tikzpicture}
					\node[above right, inner sep=0] (image) at (0,0)  { \includegraphics[width=.1\textwidth]{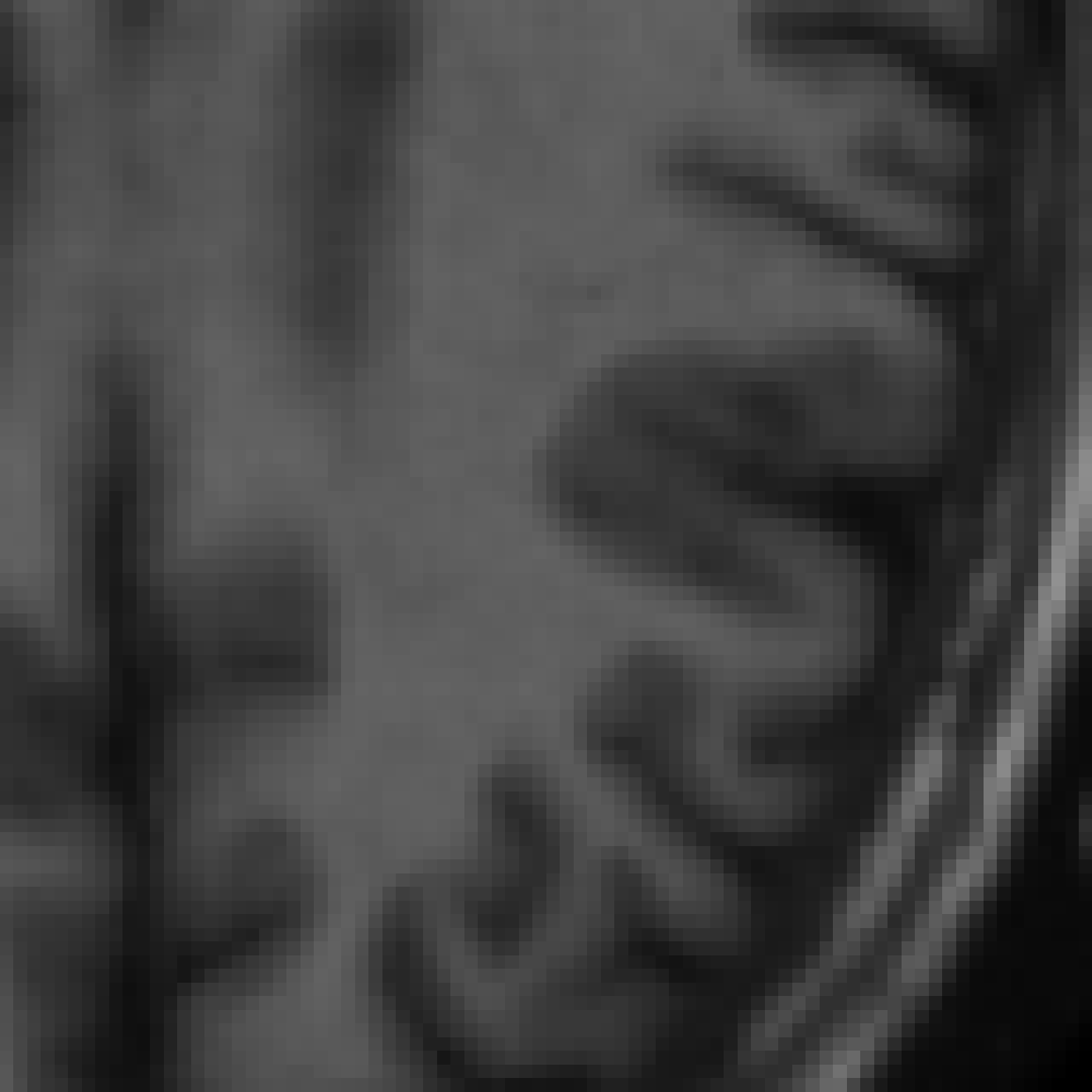} }; 
					% Create scope with normalized axes
					\begin{scope}[
					x={($0.1*(image.south east)$)},
					y={($0.1*(image.north west)$)}]
					% Grid
					% \draw[lightgray,step=1] (image.south west) grid (image.north east);
					\draw[thick,orange] (5,3) rectangle (9, 8) ;
					\end{scope}
					\end{tikzpicture} }  \\
				% row 2
				\subfloat[Neighboring slice 1]{ \begin{tikzpicture}
					\node[above right, inner sep=0] (image) at (0,0)  { \includegraphics[width=.1\textwidth]{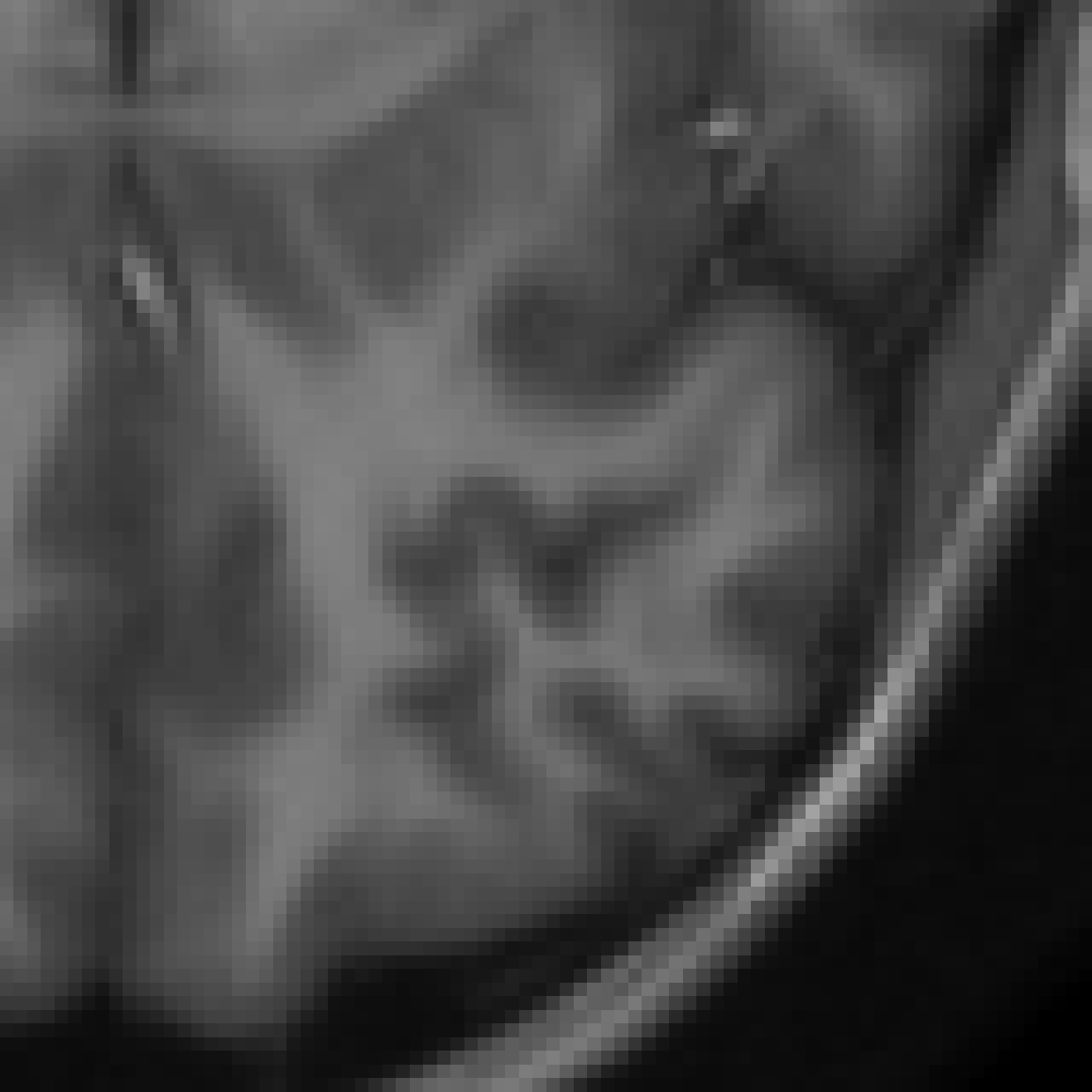} }; 
					% Create scope with normalized axes
					\begin{scope}[
					x={($0.1*(image.south east)$)},
					y={($0.1*(image.north west)$)}]
					% Grid
					% \draw[lightgray,step=1] (image.south west) grid (image.north east);
					\draw[thick,orange] (3,2) rectangle (8, 6) ;
					\end{scope}
					\end{tikzpicture} }
				& 
				\subfloat[$\alpha=$ \nicefrac{1}{7}]{ \begin{tikzpicture}
					\node[above right, inner sep=0] (image) at (0,0)  {\includegraphics[width=.1\textwidth]{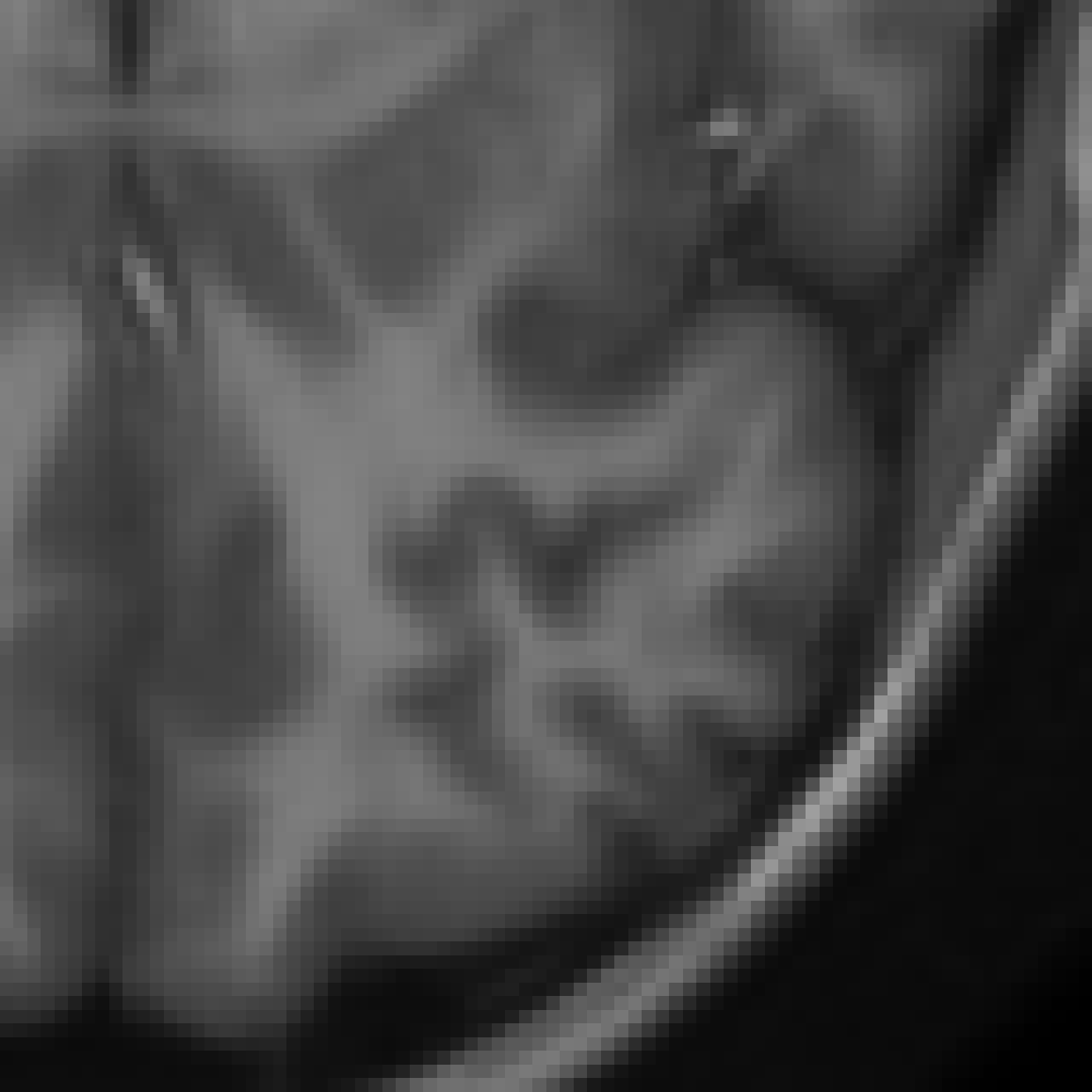}  }; 
					% Create scope with normalized axes
					\begin{scope}[
					x={($0.1*(image.south east)$)},
					y={($0.1*(image.north west)$)}]
					% Grid
					% \draw[lightgray,step=1] (image.south west) grid (image.north east);
					\draw[thick,orange] (3,2) rectangle (8, 6) ;
					\end{scope}
					\end{tikzpicture} } &
				\subfloat[$\alpha=$ \nicefrac{2}{7}]{ \begin{tikzpicture}
					\node[above right, inner sep=0] (image) at (0,0)  {\includegraphics[width=.1\textwidth]{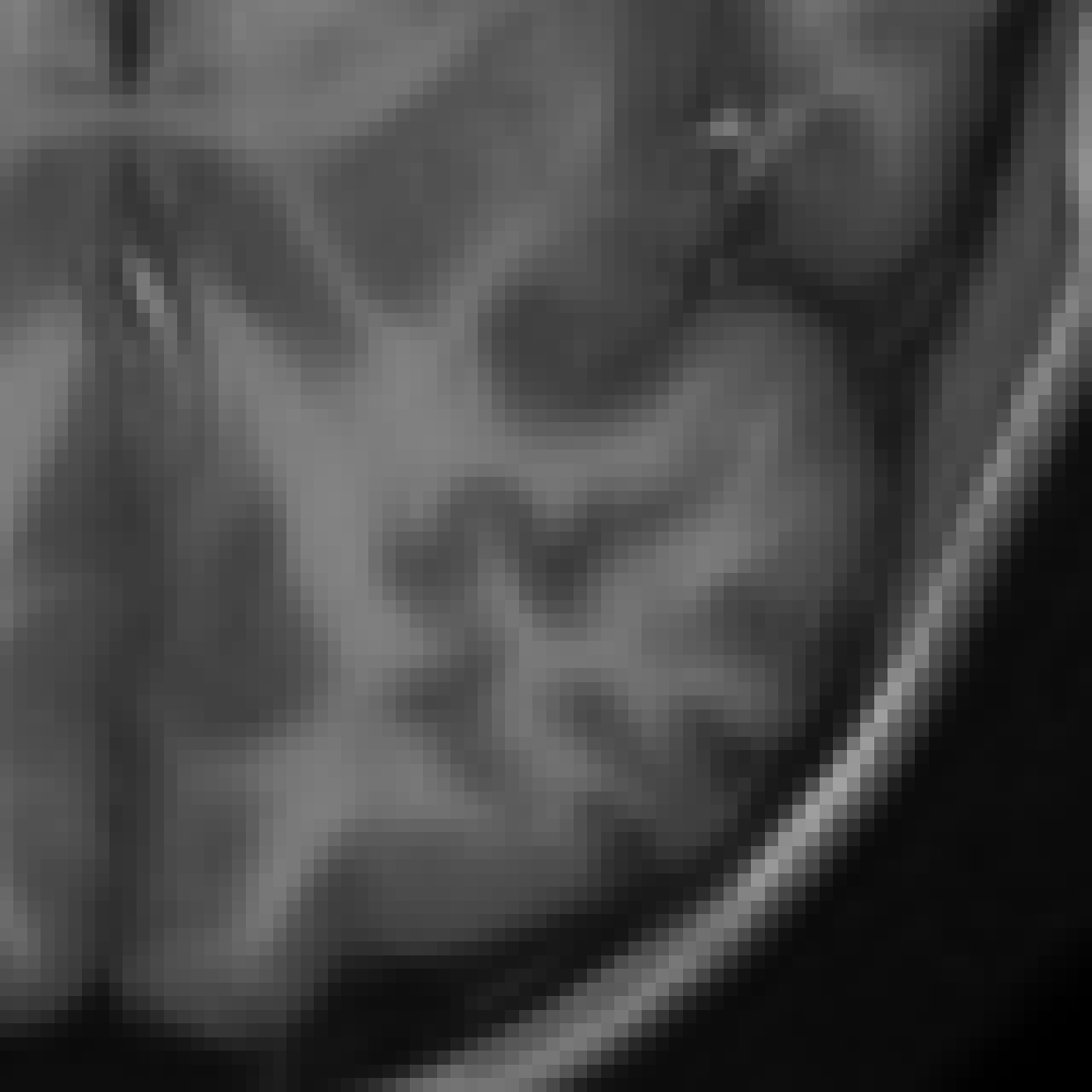}  }; 
					% Create scope with normalized axes
					\begin{scope}[
					x={($0.1*(image.south east)$)},
					y={($0.1*(image.north west)$)}]
					% Grid
					% \draw[lightgray,step=1] (image.south west) grid (image.north east);
					\draw[thick,orange] (3,2) rectangle (8, 6) ;
					\end{scope}
					\end{tikzpicture} } &
				\subfloat[$\alpha=$ \nicefrac{3}{7}]{ \begin{tikzpicture}
					\node[above right, inner sep=0] (image) at (0,0)  {\includegraphics[width=.1\textwidth]{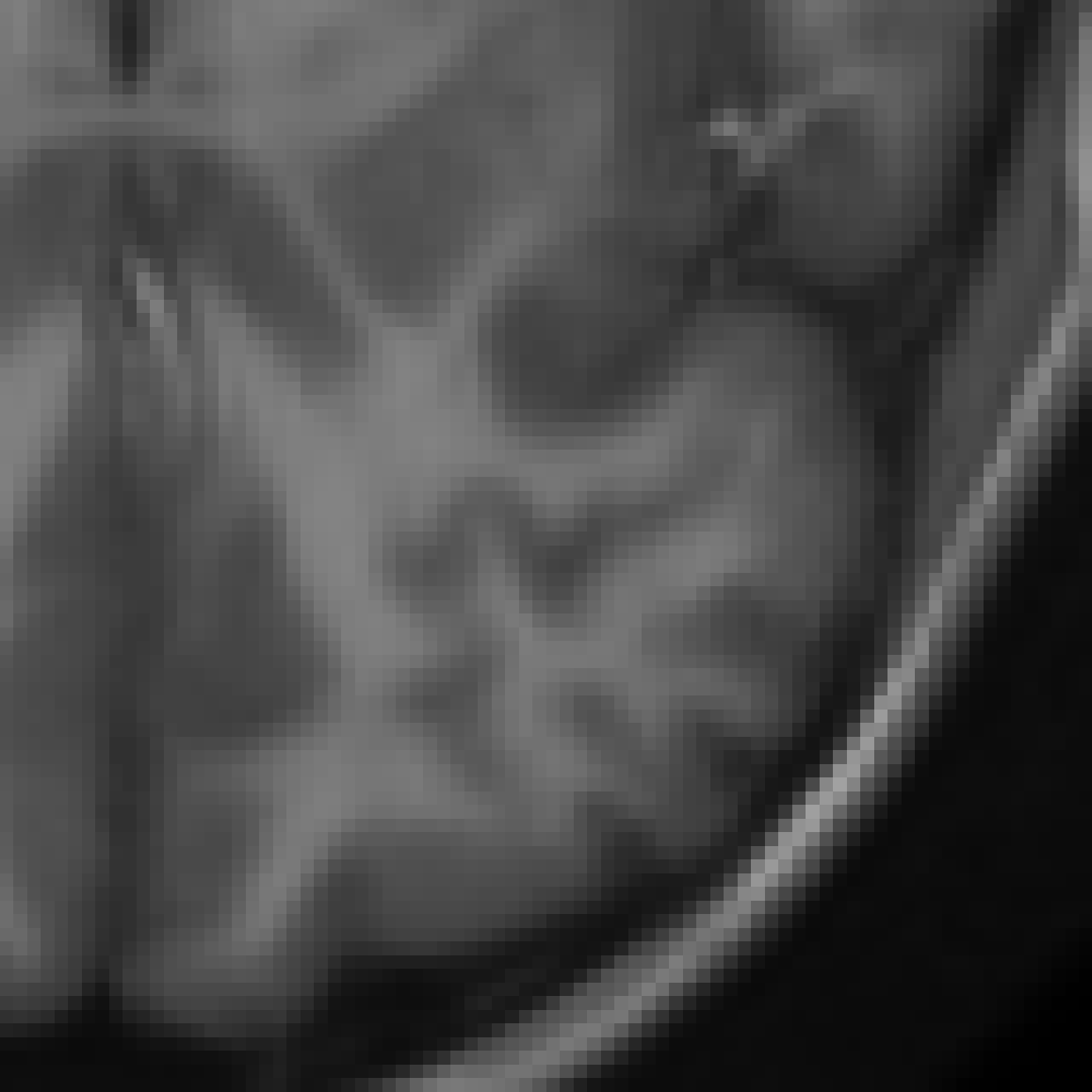}   }; 
					% Create scope with normalized axes
					\begin{scope}[
					x={($0.1*(image.south east)$)},
					y={($0.1*(image.north west)$)}]
					% Grid
					% \draw[lightgray,step=1] (image.south west) grid (image.north east);
					\draw[thick,orange] (3,2) rectangle (8, 6) ;
					\end{scope}
					\end{tikzpicture} } &
				\subfloat[$\alpha=$ \nicefrac{4}{7}]{ \begin{tikzpicture}
					\node[above right, inner sep=0] (image) at (0,0)  {\includegraphics[width=.1\textwidth]{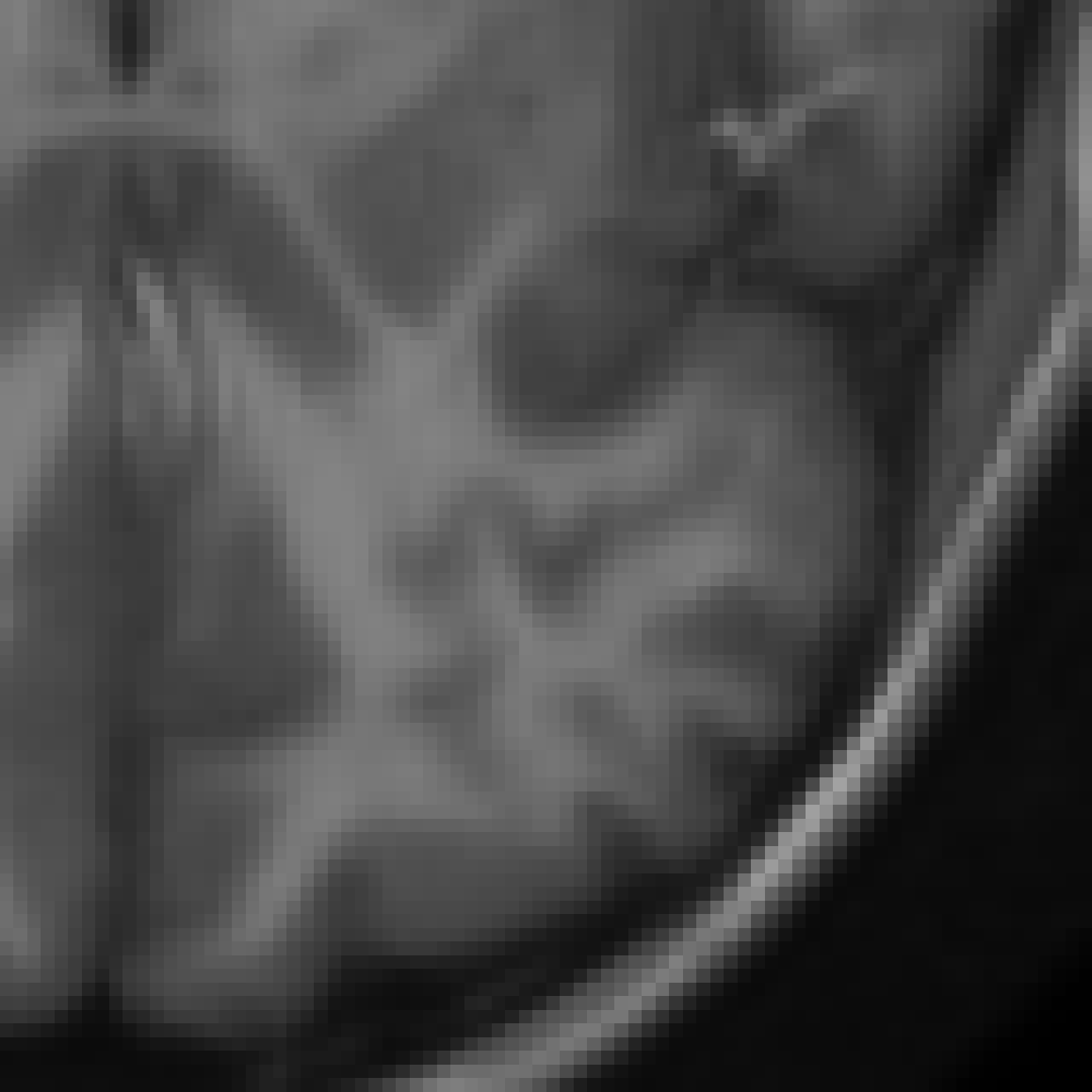}   }; 
					% Create scope with normalized axes
					\begin{scope}[
					x={($0.1*(image.south east)$)},
					y={($0.1*(image.north west)$)}]
					% Grid
					% \draw[lightgray,step=1] (image.south west) grid (image.north east);
					\draw[thick,orange] (3,2) rectangle (8, 6) ;
					\end{scope}
					\end{tikzpicture} } &
				\subfloat[$\alpha=$ \nicefrac{5}{7}]{ \begin{tikzpicture}
					\node[above right, inner sep=0] (image) at (0,0)  {\includegraphics[width=.1\textwidth]{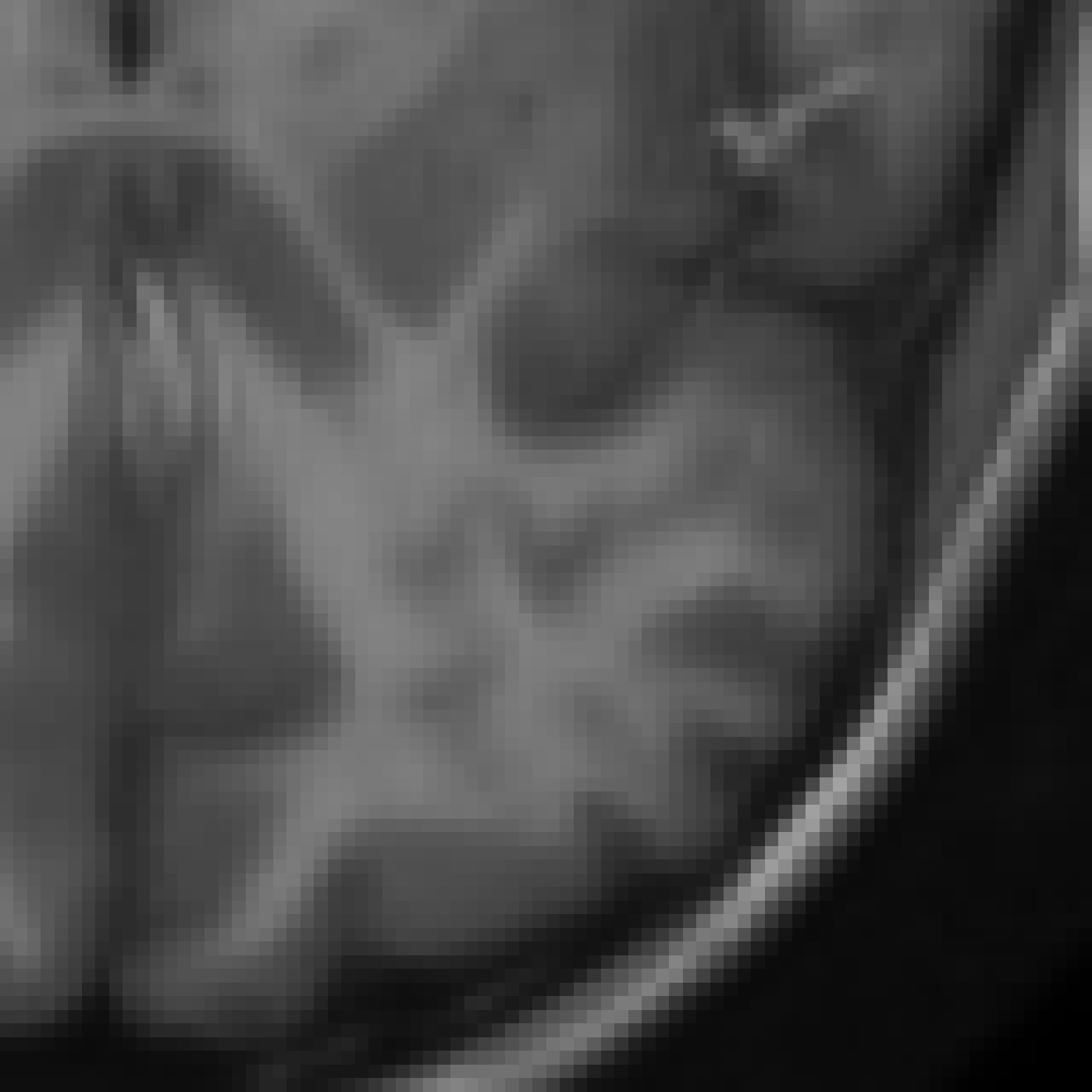}  }; 
					% Create scope with normalized axes
					\begin{scope}[
					x={($0.1*(image.south east)$)},
					y={($0.1*(image.north west)$)}]
					% Grid
					% \draw[lightgray,step=1] (image.south west) grid (image.north east);
					\draw[thick,orange] (3,2) rectangle (8, 6) ;
					\end{scope}
					\end{tikzpicture} } &
				\subfloat[$\alpha=$ \nicefrac{6}{7}]{ \begin{tikzpicture}
					\node[above right, inner sep=0] (image) at (0,0)  {\includegraphics[width=.1\textwidth]{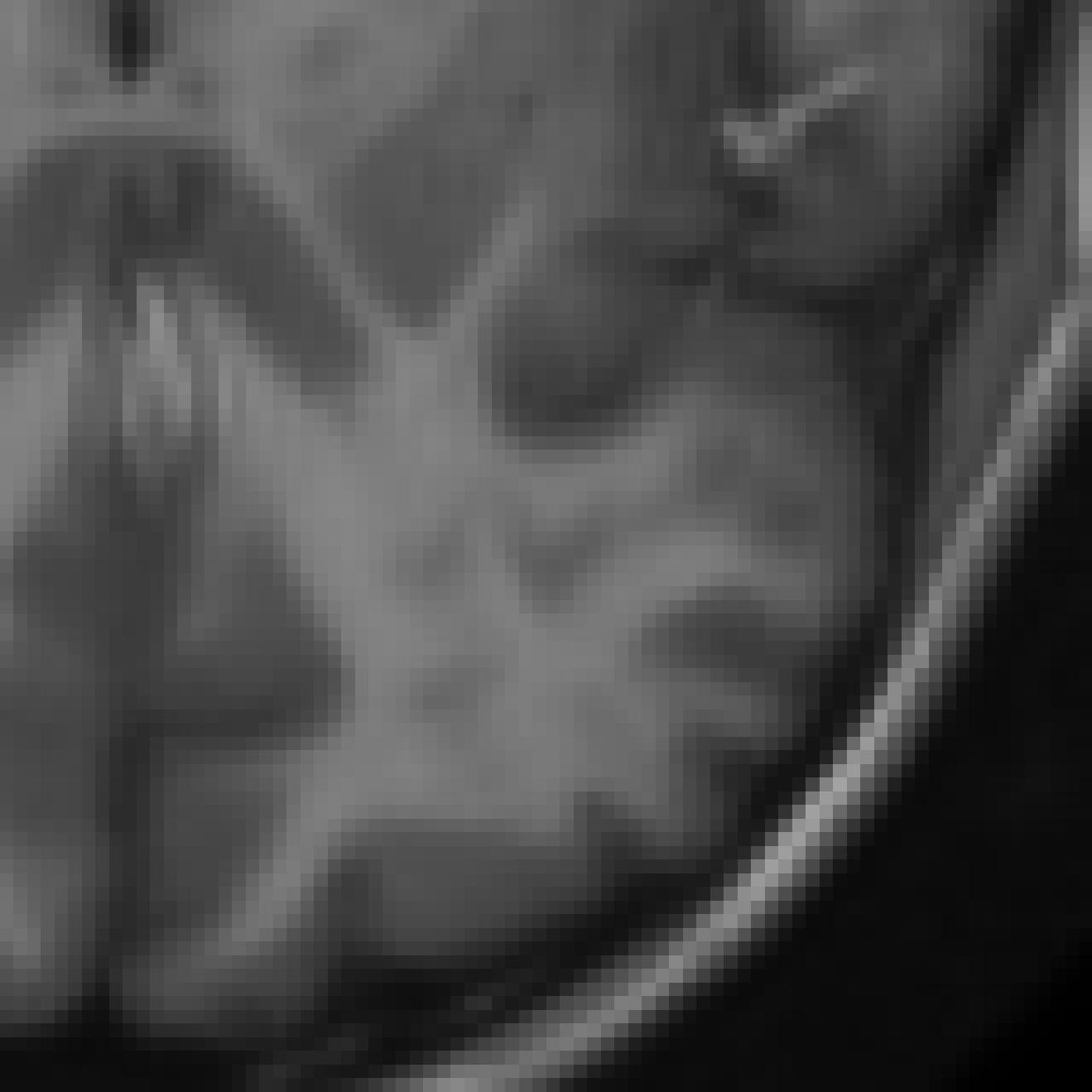}   }; 
					% Create scope with normalized axes
					\begin{scope}[
					x={($0.1*(image.south east)$)},
					y={($0.1*(image.north west)$)}]
					% Grid
					% \draw[lightgray,step=1] (image.south west) grid (image.north east);
					\draw[thick,orange] (3,2) rectangle (8, 6) ;
					\end{scope}
					\end{tikzpicture} } &
				\subfloat[Neighboring slice 2] { \begin{tikzpicture}
					\node[above right, inner sep=0] (image) at (0,0)  { \includegraphics[width=.1\textwidth]{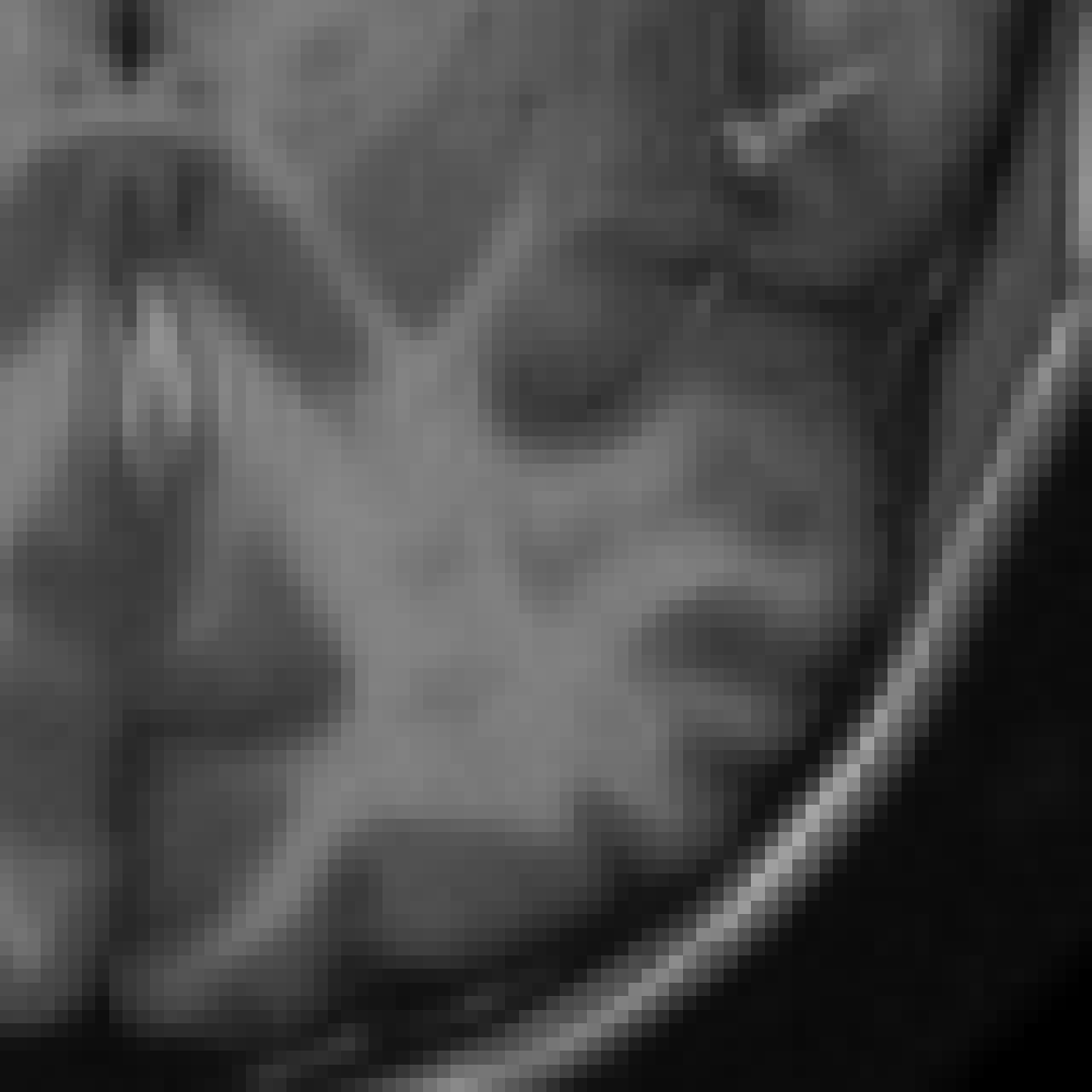}    }; 
					% Create scope with normalized axes
					\begin{scope}[
					x={($0.1*(image.south east)$)},
					y={($0.1*(image.north west)$)}]
					% Grid
					% \draw[lightgray,step=1] (image.south west) grid (image.north east);
					\draw[thick,orange] (3,2) rectangle (8, 6) ;
					\end{scope}
					\end{tikzpicture} }  
			\end{tabular}
			
		\end{center}
		\caption{Zoomed-in images of synthesis results in scans of adult brain MRIs of two different patients (one example per row) from the OASIS dataset. Slice spacing was improved from \num{4} to \SI{0.57}{\milli\meter} by synthesizing six intermediate slices (second to penultimate columns). For this, latent space encodings of the two neighboring slices (first and last column) were combined using their convex combination. $\alpha$ denotes the mixing coefficient as specified in Equation~\ref{eq_convex_combination}. Bounding boxes focus on anatomical variations between images in a row.}
		
		\label{fig_qualitative_synthesis_OASIS4x}
		
	\end{figure*}

	\subsection{Semantic Interpolation of Neonatal Brain MRI} \label{exp_neonatal_brain_dhcp}
	
	Acquisition of high-resolution neonatal brain MRIs is typically hampered by uncontrollable full-term infant motion and their small size of the brain \citep{dubois2021mri}. As a result, acquired neonatal brain MRIs are often anisotropic and poorly capture the \num{3}D brain structures \citep{gholipour2010robust}. Super-resolution of neonatal brain MRI may enhance the capacity of image analysis on the dynamics of brain maturation \citep{glenn2006magnetic} and brain development \citep{rutherford2008mr, moeskops2017prediction}. Therefore, our proposed approach was evaluated using \num{240} randomly selected T$_2$-weighted (T2w) neonatal brain MRIs from the developing Human Connectome Project (dHCP) (\cite{hughes2017developing}) hereafter referred to as dHCP dataset. 
	
	\subsubsection{Experimental Details}
	
	The dataset was randomly split into training (\num{200}), validation (\num{20}) and test set (\num{20}).  
	The images with isotropic resolution of \num{0.5}$\times$\num{0.5}$\times\SI{0.5}{\milli\meter}^3$ served as ground truth high-resolution (HR) images. To simulate low through-plane resolution (LR) images were downsampled by factor $K \in \{2,3,4,5,6\}$ in the z-axis. More specifically, low-resolution images were generated by using a Gaussian blur with the full-width-at-half-maximum (FWHM) set to the desired slice thickness (\cite{greenspan2009super}). Subsequently, volumes were downsampled with factor $K$ by including every $K^{th}$ slice in the test images to obtain \num{0.5}$\times$\num{0.5}$\times K * \SI{0.5}{\milli\meter}^3$ resolution. To assess upsampling performance of the proposed method downsampled test volumes were upsampled in through-plane direction by synthesizing $K-1$ new slices between each pair of neighboring slices using the set of mixing coefficients as defined in Equation~\ref{eq_alpha_set}. Resulting volumes were then compared with high-resolution ground truth data.
	%$\{$ \nicefrac{1}{3}, \nicefrac{2}{3} $\}$ as the set of mixing coefficients. Resulting volumes were then compared to high-resolution ground truth data.
	
	To train a model patches of \num{64}$\times$\num{64} voxels were randomly chosen from the training set using mini-batches of \num{8} randomly selected slice pairs (\num{16} slices) originating from the same volume as described in Section~\ref{section_implementation}. To balance loss terms in Equation~\ref{eq_combined_loss}, $\lambda$ was set to \num{0.001}. A model was trained in \num{1300} epochs and the best performing model on the validation set was selected for final evaluation on the test volumes.
	%  after performing a line search ($\lambda \in \{0.5, 0.1, 0.05, 0.01, 0.005, 0.001, 0.0005 \}$)

	%%%%%%% !!!!!!!!!!!!!!!!!!!!!!!!!!!!  ADULT BRAIN MRI - UPSAMPLING FACTOR 5x   !!!!!!!!!!!!!!!!
	% FIGURE 11
	\begin{figure}
		\captionsetup[subfigure]{justification=centering, labelformat=empty}
		\setlength{\tabcolsep}{1pt}
		\begin{center}
			\begin{tabular}{c c c}
				% row 1
				% \setcounter{subfigure}{0}
				\subfloat[Original axial slice to be synthesized]{\includegraphics[width=.15\textwidth]{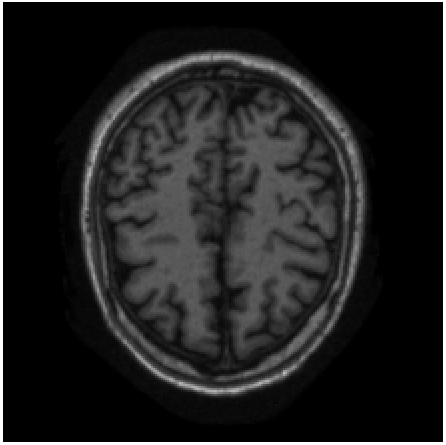} }  &
				\subfloat[B-spline ]{\includegraphics[width=.15\textwidth]{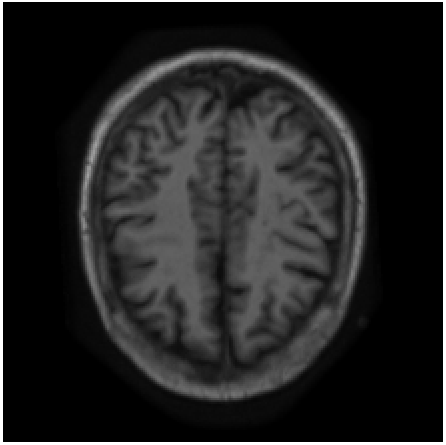} } &
				\subfloat[ours]{\includegraphics[width=.15\textwidth]{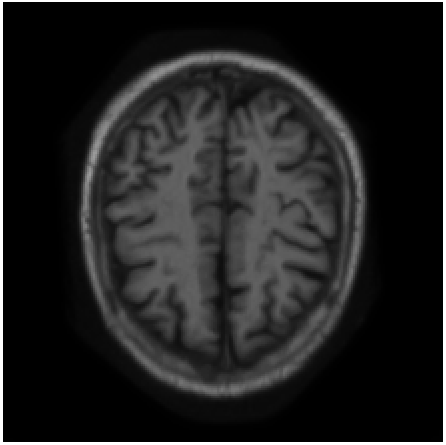} }	\\  
				% row 2
				&
				\subfloat{\includegraphics[width=.15\textwidth]{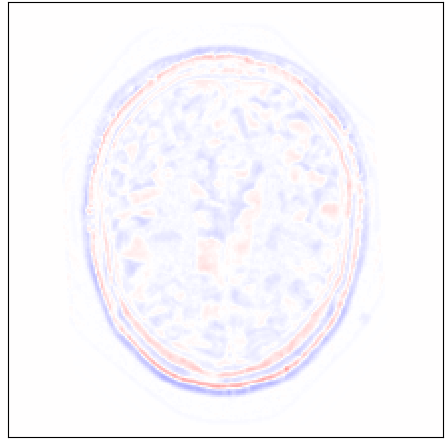}  } &
				\subfloat{\includegraphics[width=.15\textwidth]{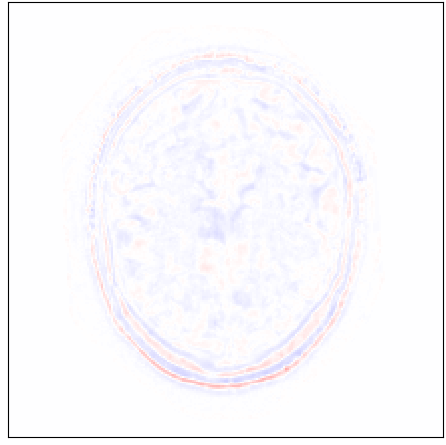}  }  \\
			\end{tabular}
			
		\end{center}
		% CORONAL SLICES
		\begin{center}
			\begin{tabular}{c c c}
				% row 2
				% \setcounter{subfigure}{0}
				\subfloat[Original coronal slice ]{\includegraphics[width=.15\textwidth]{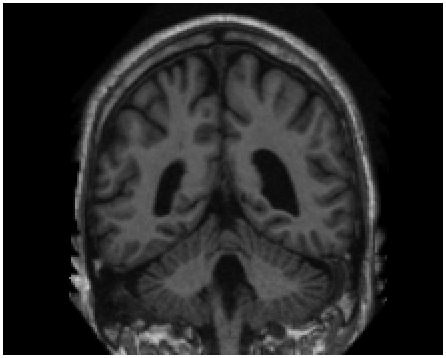} }  &
				\subfloat[B-spline ]{\includegraphics[width=.15\textwidth]{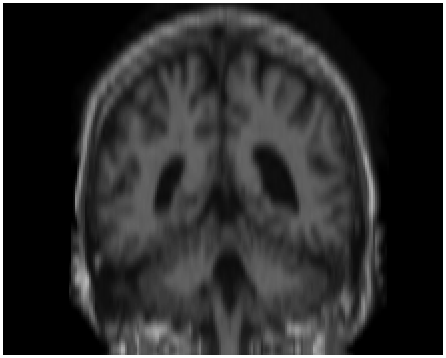} } &
				\subfloat[ours]{\includegraphics[width=.15\textwidth]{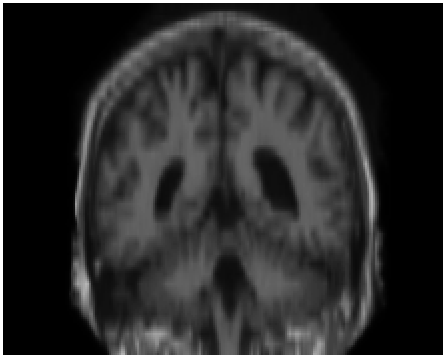} }	\\
				% row 2 
				&
				\subfloat{\includegraphics[width=.15\textwidth]{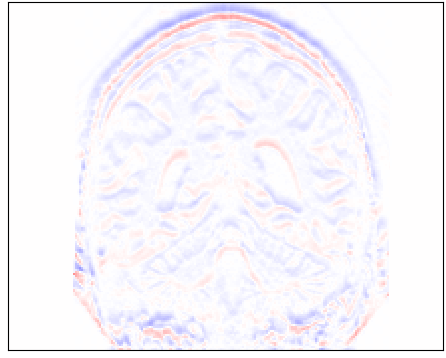}  } &
				\subfloat{\includegraphics[width=.15\textwidth]{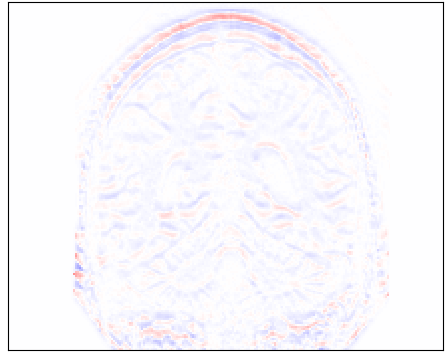}  }  \\
			\end{tabular}
			
		\end{center}
		% SAGITTAL SLICES
		\begin{center}
			\begin{tabular}{c c c}
				% row 1
				% \setcounter{subfigure}{0}
				\subfloat[Original sagittal slice ]{\includegraphics[width=.15\textwidth]{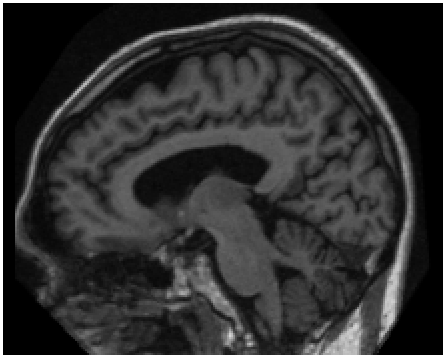} }  &
				\subfloat[B-spline ]{\includegraphics[width=.15\textwidth]{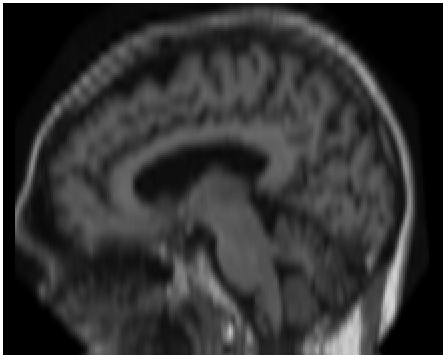} } &
				\subfloat[ours]{\includegraphics[width=.15\textwidth]{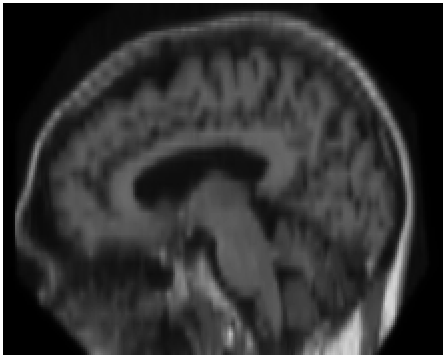} }	\\  
				% row 2
				&
				\subfloat{\includegraphics[width=.15\textwidth]{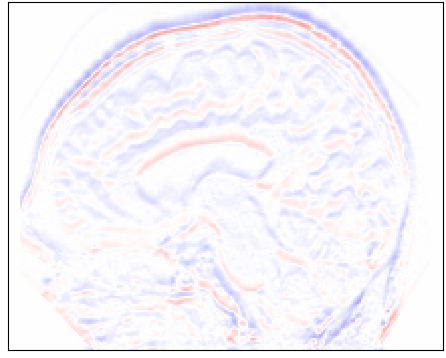}  } &
				\subfloat{\includegraphics[width=.15\textwidth]{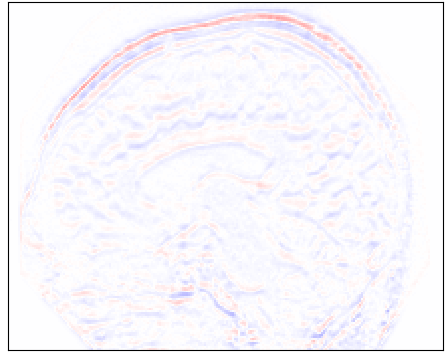}  }  \\
			\end{tabular}
			
		\end{center}
		\caption{Comparing upsampling performance between cubic B-spline interpolation and proposed approach using T$_1$-weighted adult brain MRI of the OASIS dataset. Original volumes with slice thickness and spacing of \SI{1}{\milli\meter} were downsampled to \SI{5}{\milli\meter} by applying a Gaussian blur before including every fifth slice in the test volume. Differences between reference (minuend) and synthesized slice (subtrahend). Blue corresponds to negative and red to positive differences. Image intensities are scaled to a $[0,1]$ range. All difference images use the same color scale $[-1, 1]$.}
		% First row: axial slices; third row: coronal slices; and fifth row: sagittal slices. First, third and fifth row: First column shows slice from reference volume. Second to fourth columns depict images generated by different conventional interpolation methods. Last column shows synthesized slice using proposed method. Second, fourth and sixth row:
		\label{fig_oasis_5x_qualitative_synthesis_compare_all_slices}
	\end{figure}
	
	\subsubsection{Results}
	
	\vspace{1ex}
	\textbf{Slice Synthesis: } Qualitative evaluation of the proposed approach on neonatal brain MRI with reveals that generated slices using a convex combination of neighboring slice encodings, comprise a smooth anatomical transition between adjacent slices. Examples depicted in Figure~\ref{fig_qualitative_synthesis_dhcp4x} show upsampling performance of proposed method for neighboring slices with large anatomical variations. In the depicted figures slice spacing was improved from \num{2} to $\SI{0.29}{\milli\meter}$ by synthesizing six intermediate slices using latent space encodings of the two neighboring slices.
	% the upsampling factor $K$ was set to \num{7} and $\mathcal{A}$, the set of mixing coefficients was equal to $\{\nicefrac{1}{7}, \nicefrac{2}{7}, \nicefrac{3}{7}, \nicefrac{4}{7}, \nicefrac{5}{7}, \nicefrac{6}{7}\}$.

	Visual inspection of Figure~\ref{fig_qualitative_synthesis_compare_dhcp4x} conveys that our proposed approach was able to synthesize excluded high-resolution axial slices more accurately than cubic B-spline interpolation. These results are corroborated by the coronal and sagittal views revealing that volumes generated by the proposed method are less blurry and contain smoother transitions between slices compared to volumes generated by the conventional interpolation method. 
	
	Moreover, quantitative comparison shown in Figure~\ref{fig_dhcp_compare_upsampling_factor} depicts that the proposed unsupervised method outperformed cubic B-spline interpolation in terms of SSIM and PSNR for all evaluated upsampling factors. Furthermore, the performance differences are statistically significant ($p<0.001$) using the one-sided Wilcoxon signed-rank test. 		
	
	\vspace{2ex}
	\textbf{Comparison With Supervised Super-Resolution Methods: } Supervised deep-learning super-resolution methods developed by \cite{pham2017brain, pham2019simultaneous} were evaluated on a subset of \num{20} neonatal brain MRIs from the dHCP dataset. Table~\ref{table_dHCP_other_methods} lists results as reported by \cite{pham2019simultaneous} together with quantitative evaluation of our proposed approach on the same dataset (see Section~\ref{exp_dHCP_quant_qual_evaluation}). Although methods of \cite{pham2017brain, pham2019simultaneous} used high-resolution ground-truth data for model training their results can be used to put results reported in this work into perspective. One may carefully conclude that our proposed unsupervised deep-learning approach is on par with supervised super-resolution approaches by \cite{pham2017brain, pham2019simultaneous} when evaluated on neonatal brain MRI.

	\subsection{Semantic Interpolation of Adult Brain MRI} \label{exp_adult_brain_mri}
	
	Upsampling performance of the proposed method was also evaluated using T$_1$-weighted (T1w) adult brain MRIs of the OASIS project (\cite{marcus2007open}). The images with isotropic resolution of \num{1.0}$\times$\num{1.0}$\times\SI{1.0}{\milli\meter}^3$ served as ground truth high-resolution (HR) images.
	
	\subsubsection{Experimental Details}
	
	Model was trained on the first \num{200} unique patients, validated on the subsequent \num{20} patients, and tested on \num{50} randomly selected scans from the remaining set. To ensure that test images were divisible by four, slices were zero-padded to \num{220}$\times$\num{220} voxels. Other experimental settings were identical to experiments performed on neonatal brain MRI described in section~\ref{exp_neonatal_brain_dhcp}. 
	
	\subsubsection{Results}
	
	\vspace{1ex}
	\textbf{Slice Synthesis: } Qualitative evaluation of proposed method on adult brain MRI with reveals that synthesized slices constitute a smooth anatomical transition between neighboring slices. The proposed method is able to bridge large anatomical variations between adjacent slices. These findings are depicted in Figure~\ref{fig_qualitative_synthesis_OASIS4x}. 

	\vspace{1ex}
	\textbf{Comparison With Conventional Interpolation Method: } Qualitative comparison of generated axial brain MRI slices between cubic B-spline interpolation and proposed approach shown in Figure~\ref{fig_oasis_5x_qualitative_synthesis_compare_all_slices} reveals that the proposed method can synthesize excluded axial slices with higher image quality than conventional interpolation method. 
	Moreover, visual inspection of coronal and sagittal slices shown in Figure~\ref{fig_oasis_5x_qualitative_synthesis_compare_all_slices} conveys that images generated by cubic B-spline interpolation more frequently suffer from aliasing artifacts than images generated by our proposed method. 
	
\begin{figure*}
		\captionsetup[subfigure]{justification=centering}
		\centering
		\includegraphics[width=.96\textwidth]{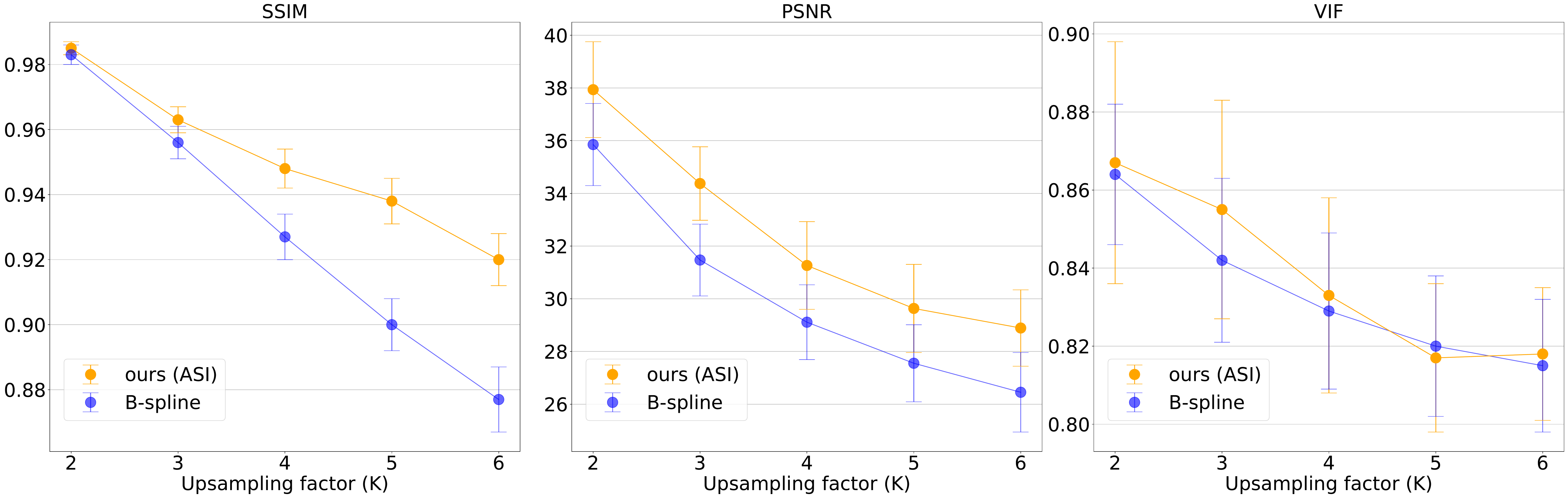} 
		\caption{Quantitative comparison of upsampling performance for proposed method (ASI) and cubic B-spline interpolation in terms of SSIM, PSNR, and VIF. T$_1$-weighted adult brain MRIs of \num{50} subjects from the OASIS dataset were upsampled with factor $K \in \{2, 3, 4, 5, 6 \}$. The differences between proposed and cubic B-spline interpolation approach in terms of SSIM and PSNR are statistically significant ($p < 0.001$) using the one-sided Wilcoxon signed-rank test.}		
		\label{fig_oasis_compare_upsampling_factor} 
	\end{figure*}

	In line with results reported for cardiac cine and neonatal brain MRI in Sections~\ref{exp_cardiac_acdc} and \ref{exp_neonatal_brain_dhcp}, respectively, quantitative evaluation in terms of SSIM, PSNR, and VIF depicted in Figure~\ref{fig_oasis_compare_upsampling_factor} corroborate the qualitative findings. Measures were computed on sagittal slices through volume. The proposed method outperformed cubic B-spline interpolation and the differences are statistically significant ($p < 0.0001$) in terms of SSIM and PSNR for all upsampling factors ($K \in \{2,3,4,5,6\}$).

	% Figure rebuttal: comparing SR performance for different upsampling factors OASIS 
	
	\vspace{1ex}
	\textbf{Comparison With Unsupervised Super-Resolution Methods: } Previously developed unsupervised super-resolution methods by \cite{jog2016self, zhao2018self} were evaluated on T$_1$-weighted adult brain MRIs from the \href{http://www.neuromorphometrics.com/}{Neuromorphometrics dataset}.
	The dataset is not publicly available. From the \num{114} brain scans in the Neuromorphometrics dataset a subset of \num{60} scans (\num{30} patients) was taken from the OASIS project. Therefore, reported results of previously developed methods for images from the Neuromorphometrics dataset can be used as estimates for a careful comparison. Quantitative results in terms of PSNR and SSIM listed in Table~\ref{table_oasis_comparison_with_other_methods} show that for upsampling factor \num{2} our proposed method is on par or better than previously developed super-resolution approaches.  For upsampling factor \num{3} the method of \cite{zhao2018self} achieved the best results and our proposed approach is on par with method of \cite{jog2016self}. 
	
	Generative image synthesis approach of \cite{dalca2018medical} was evaluated on \num{50} T$_1$-weighted brain MRIs of the ADNI dataset \citep{jack2008alzheimer}. To compare performances, our model was trained (\num{100} scans) and evaluated (\num{50} scans) on the ADNI dataset using identical experimental settings as described in section~\ref{exp_adult_brain_mri}. Table~\ref{table_adni_comparison_with_other_methods} lists quantitative results for both methods in terms of mean squared error (MSE). One can observe that our method achieved a lower mean squared error compared with approach of \cite{dalca2018medical}.

	% TABLE 3
	\begin{table}
		\centering
		\caption{Comparison between proposed method (ASI) and \textit{unsupervised} super-resolution approaches of \cite{jog2016self, zhao2018self} in terms of SSIM and PSNR. Approaches of \cite{jog2016self} and \cite{zhao2018self} were evaluated on T$_1$-weighted adult brain MRIs with $\SI{1.0}{\milli\meter}^3$ isotropic resolution from the \href{http://www.neuromorphometrics.com/}{Neuromorphometrics dataset}. Slice spacing was improved from \num{2} to \SI{1}{\milli\meter} (factor 2) and \num{3} to \SI{1}{\milli\meter} (factor 3). Results reported here are taken from the original work by \cite{jog2016self} and \cite{zhao2018self}.}
		\label{table_oasis_comparison_with_other_methods}
		% \begin{tabular}{l   C{1.3cm} C{1.cm} C{1.cm} C{2.5cm} }
		\begin{tabular}{L{2cm}  C{1.cm} C{1.cm} | C{1.cm} C{1.cm} }
			\toprule
			\multicolumn{1}{l}{\textbf{Method}} & \multicolumn{2}{c}{\textbf{Factor 2}} & \multicolumn{2}{c}{\textbf{Factor 3} }  \\
			& \multicolumn{1}{c}{SSIM} & \multicolumn{1}{c|}{PSNR} & \multicolumn{1}{c}{SSIM} & \multicolumn{1}{c}{PSNR} \\
			\hline
			% \vspace{0.2ex} \\
			\multicolumn{1}{l}{\cite{jog2016self}} & 0.983 & 37.98  & 0.968 & 33.49 \\
			\multicolumn{1}{l}{\cite{zhao2018self}} & 0.976 & 35.14 & \textbf{0.977} & \textbf{34.44} \\
			% \multicolumn{1}{l}{proposed AE} & 0.984 & 37.67   \\
			\multicolumn{1}{l}{ours (ASI)} & \textbf{0.985}  & \textbf{38.01} & 0.967 & 34.24 \\
			\bottomrule
		\end{tabular}
	\end{table}
	
	%---------------------------   A B L A T I O N      S T U D Y  -----------------------------
	
	\subsection{Ablation Study}
	
	To investigate the effect of the synthesis loss on upsampling performance, the proposed model was trained and evaluated on cardiac cine MRIs by minimizing the reconstruction loss only. For this, $\lambda$ in Equation~\ref{eq_combined_loss} is set to zero (referred to as ASI$_{\lambda=0}$). All other experimental conditions were held constant. 
	
	\begin{table}[ht]
		\centering
		\caption{Quantitative comparison of upsampling performance of proposed unsupervised method (ASI) compared with approach of \cite{dalca2018medical} in terms of mean squared error (MSE). Both approaches were evaluated on \num{50} randomly selected T$_1$-weighted adult brain MRIs with $\SI{1.0}{\milli\meter}^3$ isotropic resolution from the ADNI dataset. Slice spacing was improved from \num{6} to \SI{1}{\milli\meter}. Result listed here was reported in the work by \cite{dalca2018medical}. Values represent medians over \num{50} scans.}
		\label{table_adni_comparison_with_other_methods}
		\begin{tabular}{L{2cm}  C{1.5cm}   }
			\toprule
			\multicolumn{1}{l}{\textbf{Method}} & \multicolumn{1}{c}{\textbf{MSE}} \\
			\hline
			\multicolumn{1}{l}{\cite{dalca2018medical}} & \num{2.1} $\times 10^{-3}$ \\
			\multicolumn{1}{l}{ours (ASI)} &  \num{1.5} $\times 10^{-3}$ \\
			\bottomrule
		\end{tabular}
	\end{table}

\begin{figure*}
		\begin{center}
			\begin{tabular}{c c c}
				% row 1
				% \setcounter{subfigure}{0}
				\subfloat[Original slice to be synthesized]{\includegraphics[width=.18\textwidth]{fig_ACDC_synthesis_68_p-68_s1_reference_axis0.png} }  & 
				\subfloat[Synthesized using proposed ASI$_{\lambda=0.05}$]{\includegraphics[width=.18\textwidth]{fig_ACDC_synthesis_68_p-68_s1_aisr_axis0.png} } &
				\subfloat[Synthesized using ASI$_{\lambda=0}$ trained with reconstruction loss only ]{\includegraphics[width=.18\textwidth]{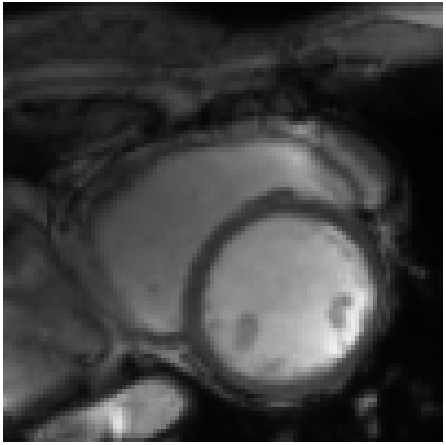} }	\\  
				&
				\subfloat{\begin{tikzpicture}
					\node[above right, inner sep=0] (image) at (0,0)  {\includegraphics[width=.18\textwidth]{fig_ACDC_synthesis_68_p-68_s1_aisr_diff_axis0.png}  }; 
					% Create scope with normalized axes
					\begin{scope}[
					x={($0.1*(image.south east)$)},
					y={($0.1*(image.north west)$)}]
					% Grid
					% \draw[lightgray,step=1] (image.south west) grid (image.north east);
					\draw[thick,darkgray] (6.5, 4.5) rectangle (9.5, 8) ;
					\draw[thick,darkgray] (0.5, 7.5) rectangle (6, 9.8) ;
					\end{scope}
					\end{tikzpicture} } & 
				\subfloat{\begin{tikzpicture}
					\node[above right, inner sep=0] (image) at (0,0)  {\includegraphics[width=.18\textwidth]{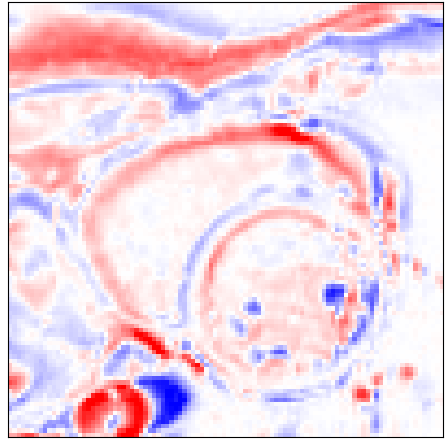}  } ;
					% Create scope with normalized axes
					\begin{scope}[
					x={($0.1*(image.south east)$)},
					y={($0.1*(image.north west)$)}]
					% Grid
					% \draw[lightgray,step=1] (image.south west) grid (image.north east);
					\draw[thick,darkgray] (6.5, 4.5) rectangle (9.5, 8) ;
					\draw[thick,darkgray] (0.5, 7.5) rectangle (6, 9.8) ;
					\end{scope}
					\end{tikzpicture} } 
				% row 3
			\end{tabular}
			
		\end{center}
		
		\caption{Comparison of synthesis performance between proposed model trained on cardiac MRI (ACDC dataset) using a combination of reconstruction and synthesis loss (b) compared to model trained with reconstruction loss only (denoted ASI$_{\lambda=0}$) (c). Bottom row: Differences between reference (minuend) and synthesized slice (subtrahend). Blue corresponds to negative and red to positive differences. Image intensities are scaled to a $[0,1]$ range. All difference images use the same color scale $[-1, 1]$.}
		\label{fig_qualitative_acdc_effect_of_synthesis_1}
	\end{figure*}

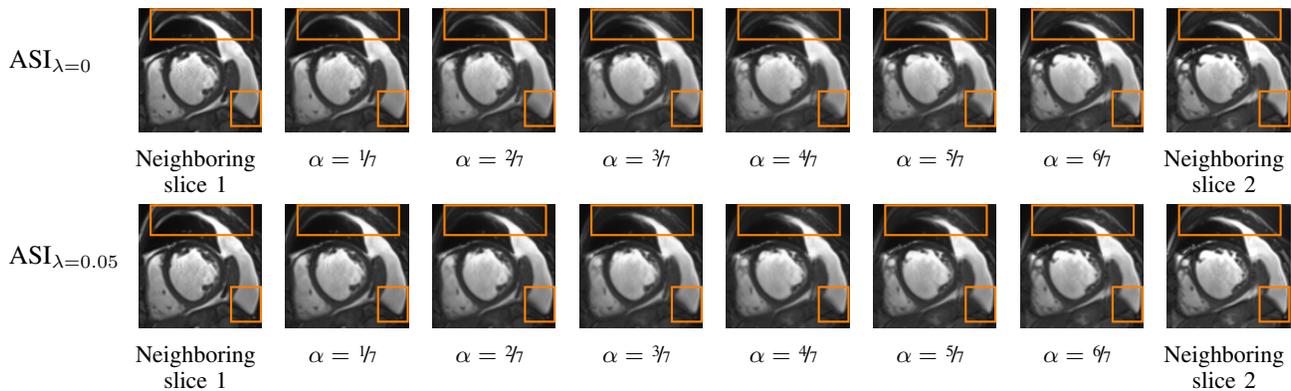
\begin{figure*}
		\captionsetup[subfigure]{justification=centering, labelformat=empty}
		\begin{center}
			\setlength{\tabcolsep}{1pt}
			\begin{tabular}{l c c c c c c c c}
				% row 1
				\begin{tabular}{@{}c@{}} ASI$_{\lambda=0}$  \\ \\ \\ \\ \\ \end{tabular} &
				\subfloat[Neighboring slice 1]{ \begin{tikzpicture}
					\node[above right, inner sep=0] (image) at (0,0)  { \includegraphics[width=.09\textwidth]{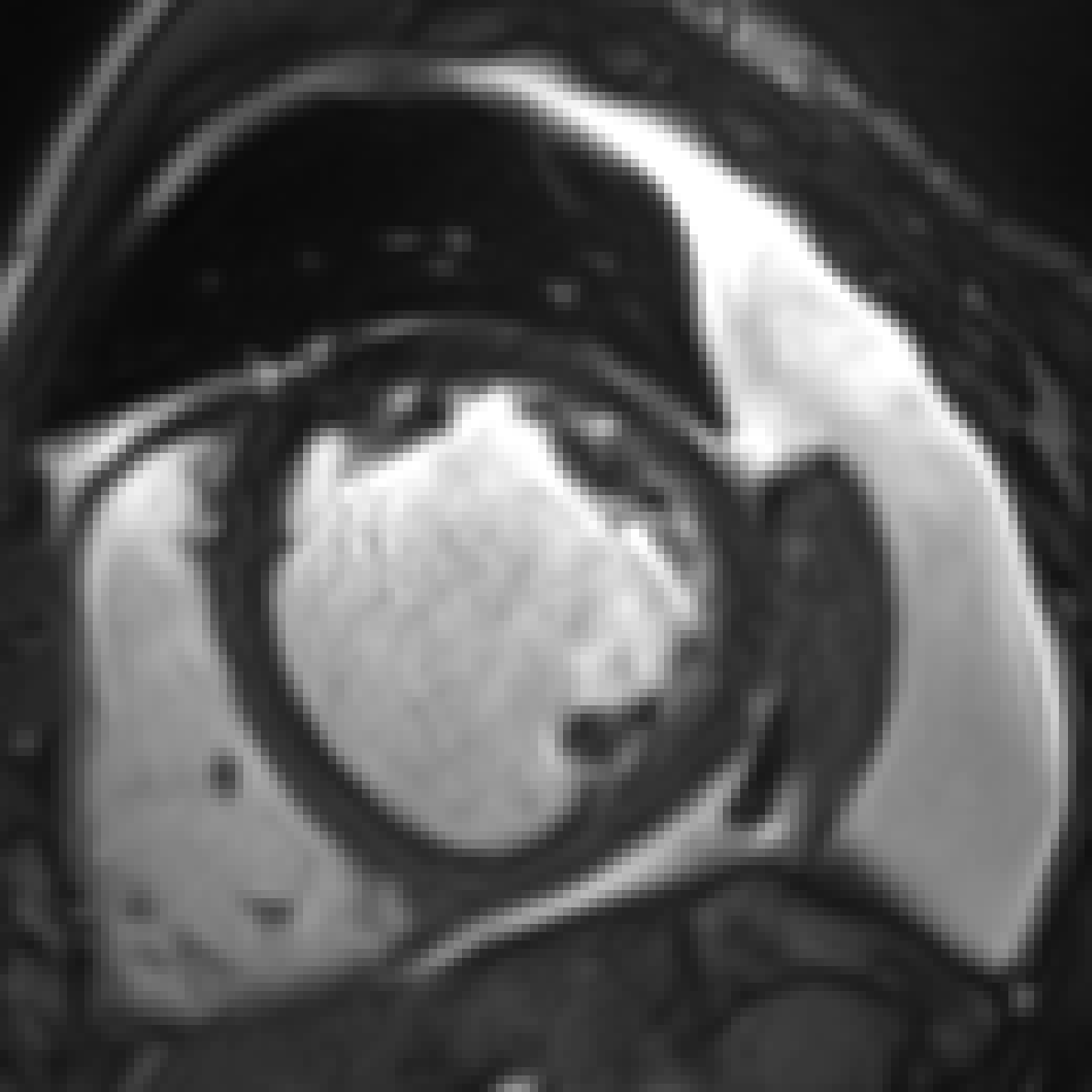} }; 
					% Create scope with normalized axes
					\begin{scope}[
					x={($0.1*(image.south east)$)},
					y={($0.1*(image.north west)$)}]
					% Grid
					% \draw[lightgray,step=1] (image.south west) grid (image.north east);
					\draw[thick,orange] (1,7.5) rectangle (9.2, 9.9) ;
					\draw[thick,orange] (7.5,0.5) rectangle (9.9, 3.3) ;
					\end{scope}
					\end{tikzpicture} } &
				\subfloat[$\alpha=$ \nicefrac{1}{7}]{ \begin{tikzpicture}
					\node[above right, inner sep=0] (image) at (0,0)  { \includegraphics[width=.09\textwidth]{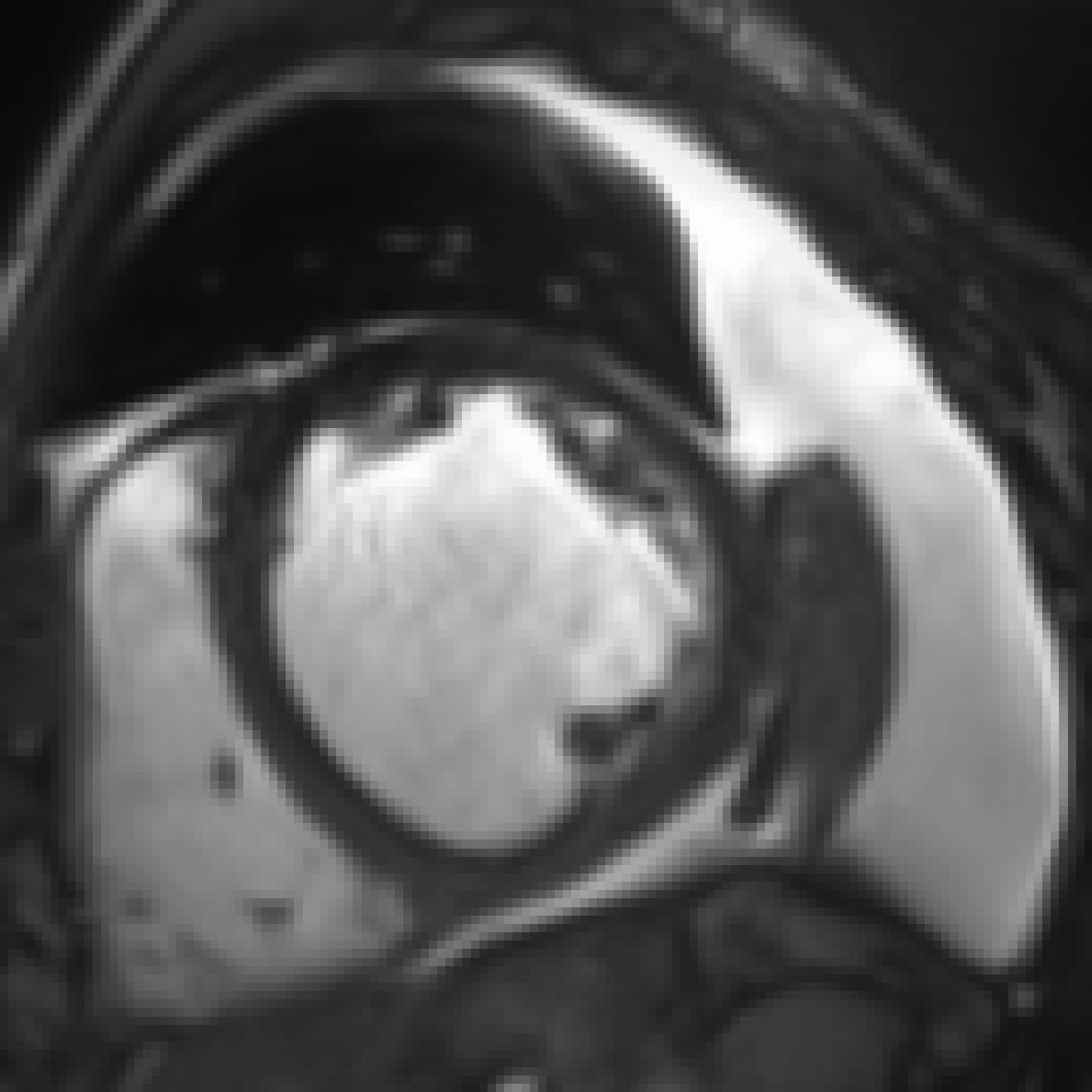} }; 
					% Create scope with normalized axes
					\begin{scope}[
					x={($0.1*(image.south east)$)},
					y={($0.1*(image.north west)$)}]
					% Grid
					% \draw[lightgray,step=1] (image.south west) grid (image.north east);
					\draw[thick,orange] (1,7.5) rectangle (9.2, 9.9) ;
					\draw[thick,orange] (7.5,0.5) rectangle (9.9, 3.3) ;
					\end{scope}
					\end{tikzpicture} } &			
				\subfloat[$\alpha=$ \nicefrac{2}{7}]{ \begin{tikzpicture}
					\node[above right, inner sep=0] (image) at (0,0)  { \includegraphics[width=.09\textwidth]{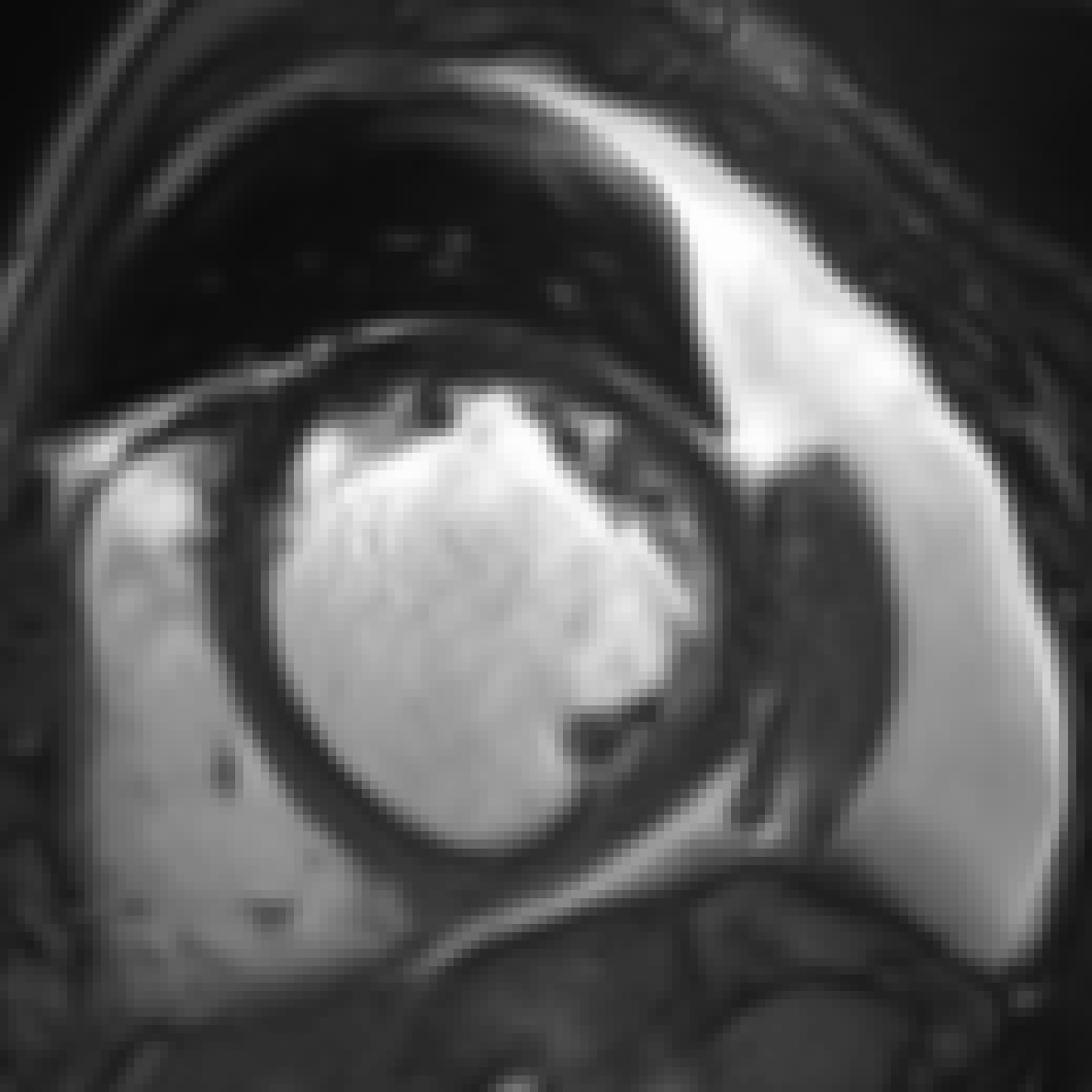} }; 
					% Create scope with normalized axes
					\begin{scope}[
					x={($0.1*(image.south east)$)},
					y={($0.1*(image.north west)$)}]
					% Grid
					% \draw[lightgray,step=1] (image.south west) grid (image.north east);
					\draw[thick,orange] (1,7.5) rectangle (9.2, 9.9) ;
					\draw[thick,orange] (7.5,0.5) rectangle (9.9, 3.3) ;
					\end{scope}
					\end{tikzpicture} } &
				\subfloat[$\alpha=$ \nicefrac{3}{7}]{ \begin{tikzpicture}
					\node[above right, inner sep=0] (image) at (0,0)  { \includegraphics[width=.09\textwidth]{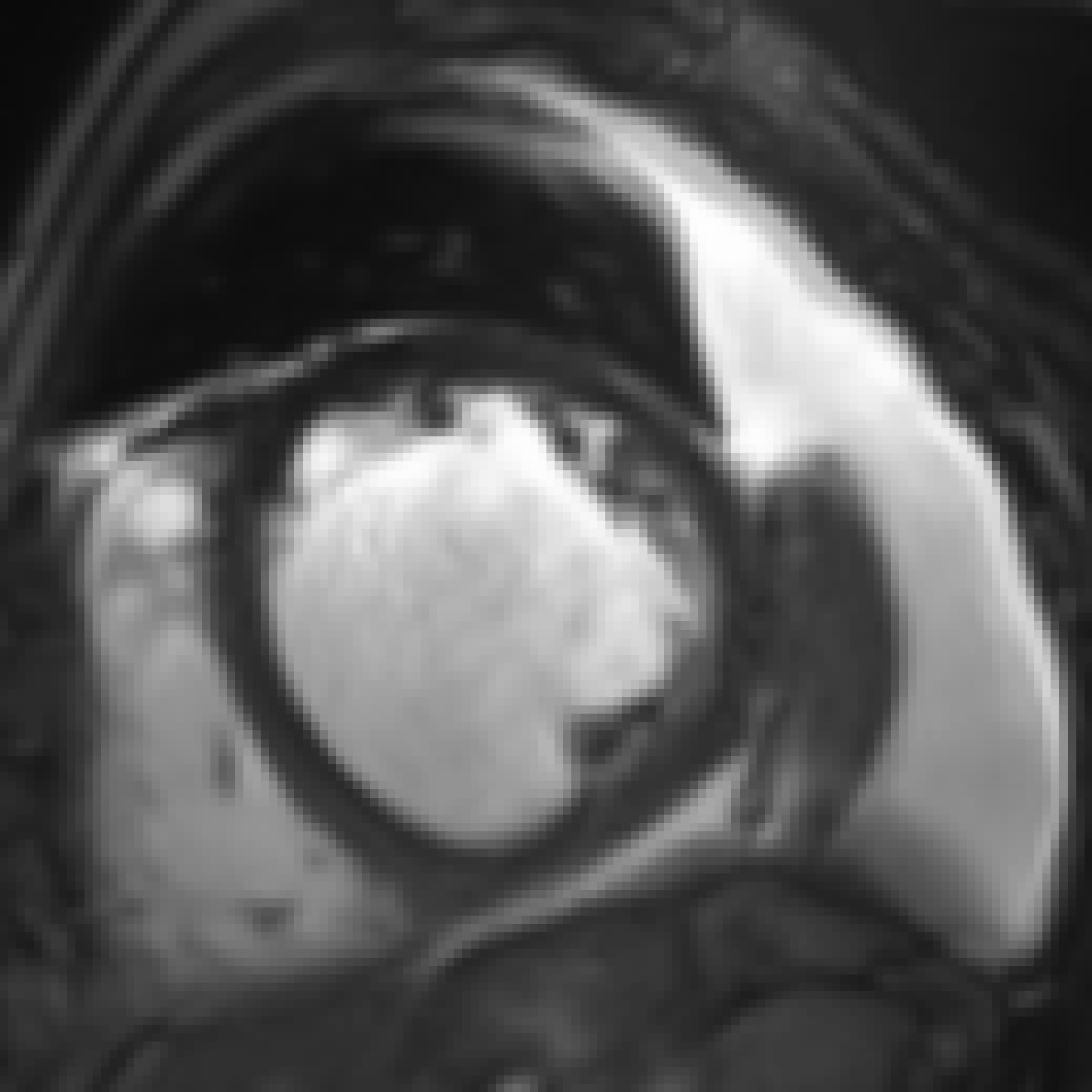} }; 
					% Create scope with normalized axes
					\begin{scope}[
					x={($0.1*(image.south east)$)},
					y={($0.1*(image.north west)$)}]
					% Grid
					% \draw[lightgray,step=1] (image.south west) grid (image.north east);
					\draw[thick,orange] (1,7.5) rectangle (9.2, 9.9) ;
					\draw[thick,orange] (7.5,0.5) rectangle (9.9, 3.3) ;
					\end{scope}
					\end{tikzpicture} } &
				\subfloat[$\alpha=$ \nicefrac{4}{7}]{ \begin{tikzpicture}
					\node[above right, inner sep=0] (image) at (0,0)  { \includegraphics[width=.09\textwidth]{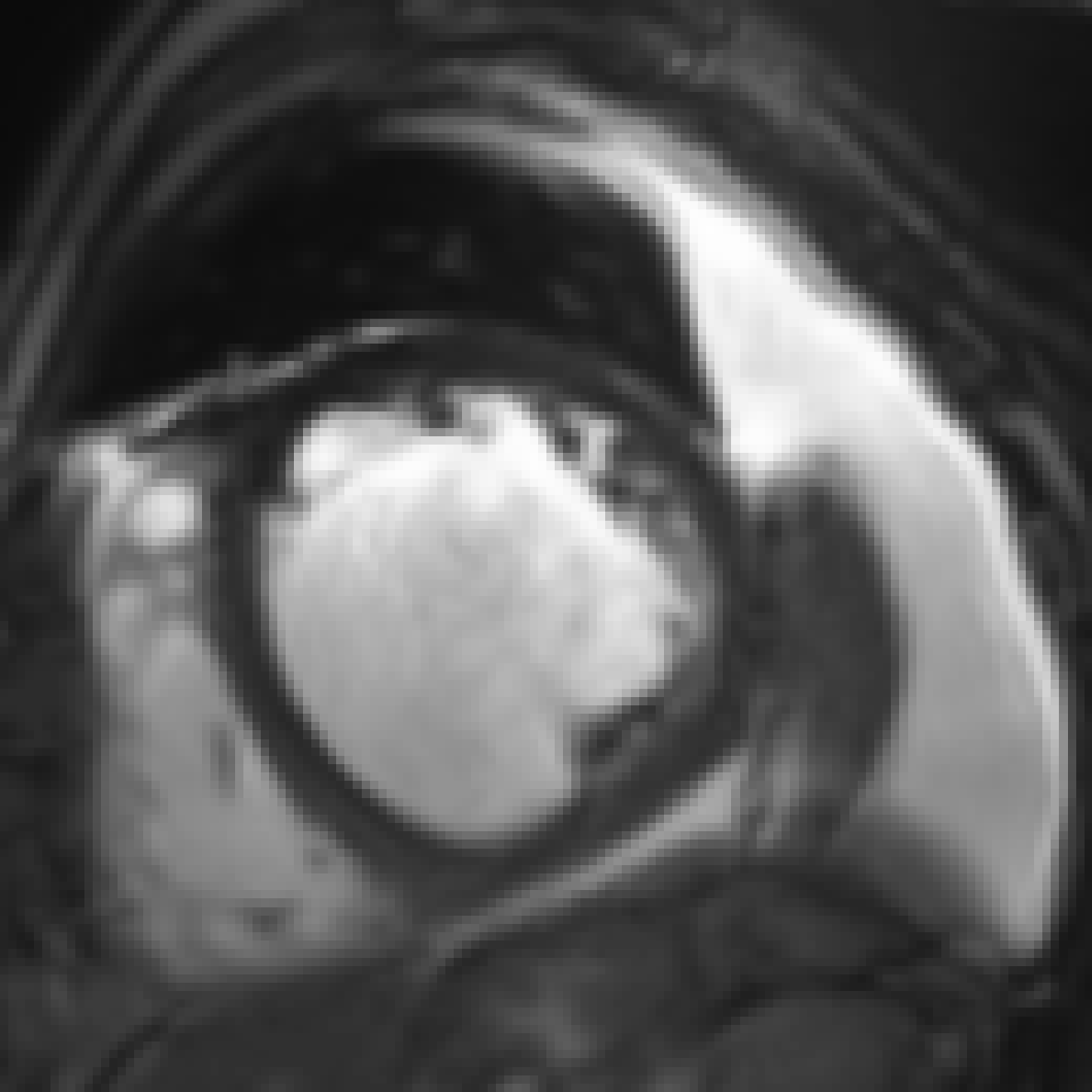} }; 
					% Create scope with normalized axes
					\begin{scope}[
					x={($0.1*(image.south east)$)},
					y={($0.1*(image.north west)$)}]
					% Grid
					% \draw[lightgray,step=1] (image.south west) grid (image.north east);
					\draw[thick,orange] (1,7.5) rectangle (9.2, 9.9) ;
					\draw[thick,orange] (7.5,0.5) rectangle (9.9, 3.3) ;
					\end{scope}
					\end{tikzpicture} } &
				\subfloat[$\alpha=$ \nicefrac{5}{7}]{ \begin{tikzpicture}
					\node[above right, inner sep=0] (image) at (0,0)  { \includegraphics[width=.09\textwidth]{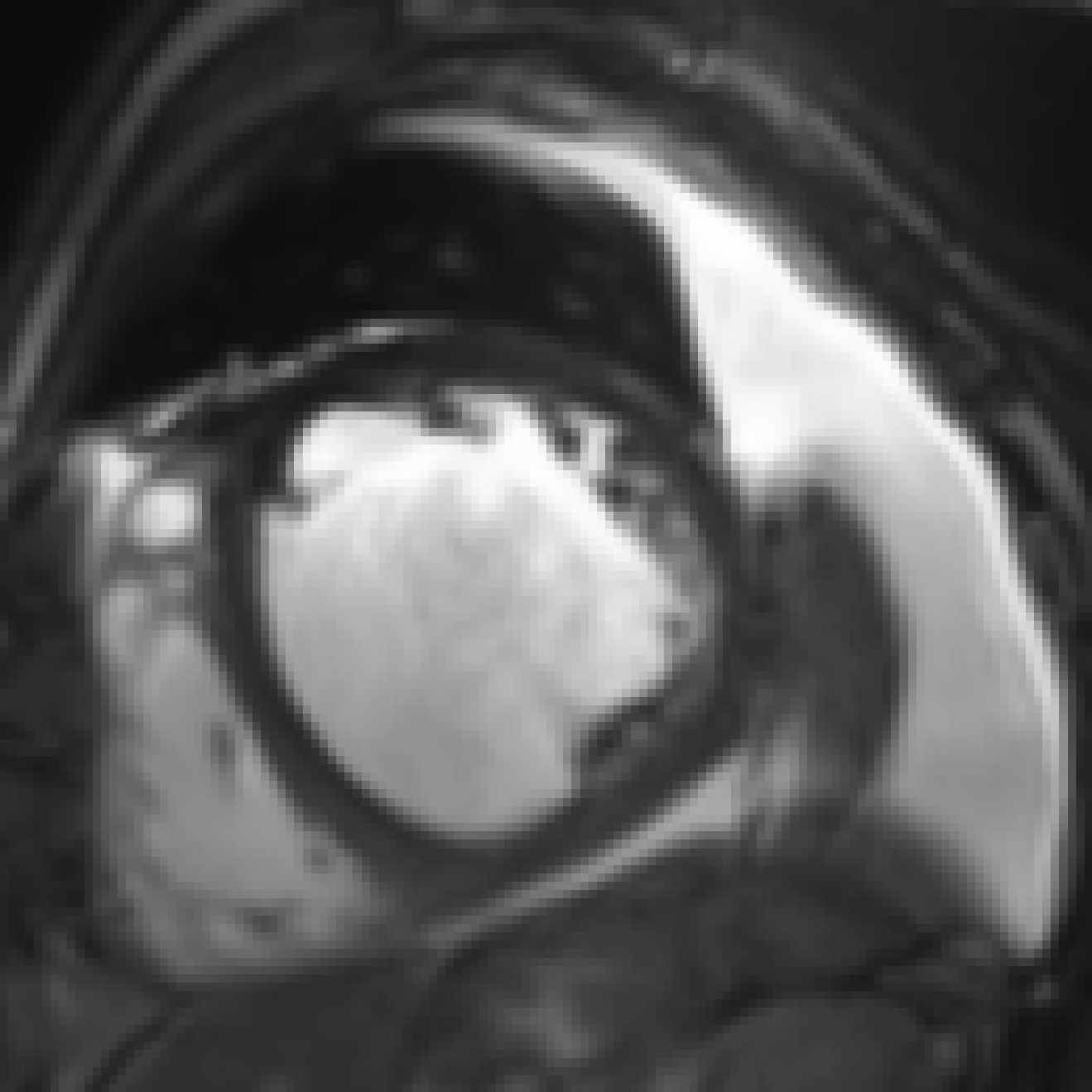} }; 
					% Create scope with normalized axes
					\begin{scope}[
					x={($0.1*(image.south east)$)},
					y={($0.1*(image.north west)$)}]
					% Grid
					% \draw[lightgray,step=1] (image.south west) grid (image.north east);
					\draw[thick,orange] (1,7.5) rectangle (9.2, 9.9) ;
					\draw[thick,orange] (7.5,0.5) rectangle (9.9, 3.3) ;
					\end{scope}
					\end{tikzpicture} } &
				\subfloat[$\alpha=$ \nicefrac{6}{7}]{ \begin{tikzpicture}
					\node[above right, inner sep=0] (image) at (0,0)  { \includegraphics[width=.09\textwidth]{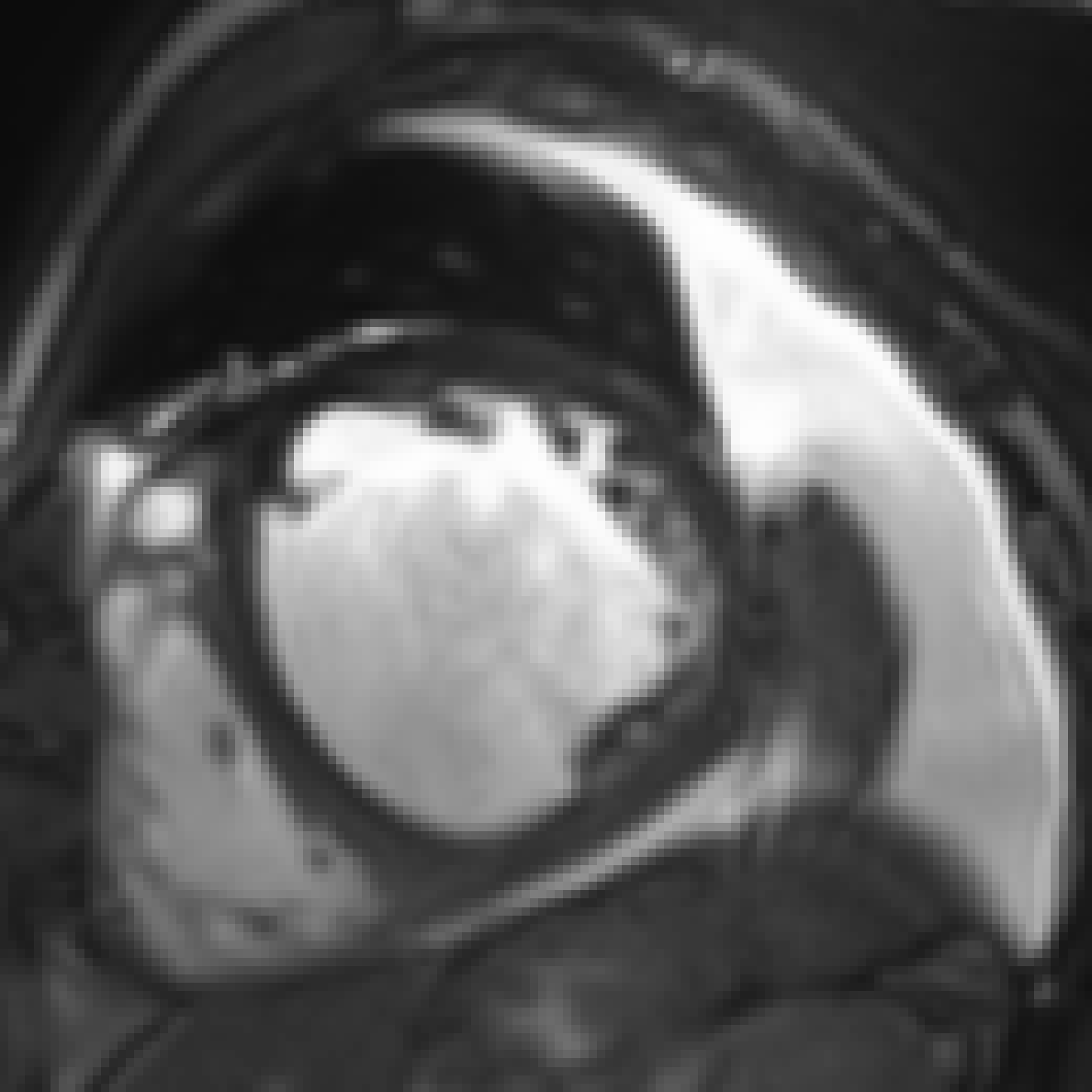} }; 
					% Create scope with normalized axes
					\begin{scope}[
					x={($0.1*(image.south east)$)},
					y={($0.1*(image.north west)$)}]
					% Grid
					% \draw[lightgray,step=1] (image.south west) grid (image.north east);
					\draw[thick,orange] (1,7.5) rectangle (9.2, 9.9) ;
					\draw[thick,orange] (7.5,0.5) rectangle (9.9, 3.3) ;
					\end{scope}
					\end{tikzpicture} } &
				\subfloat[Neighboring slice 2]{ \begin{tikzpicture}
					\node[above right, inner sep=0] (image) at (0,0)  { \includegraphics[width=.09\textwidth]{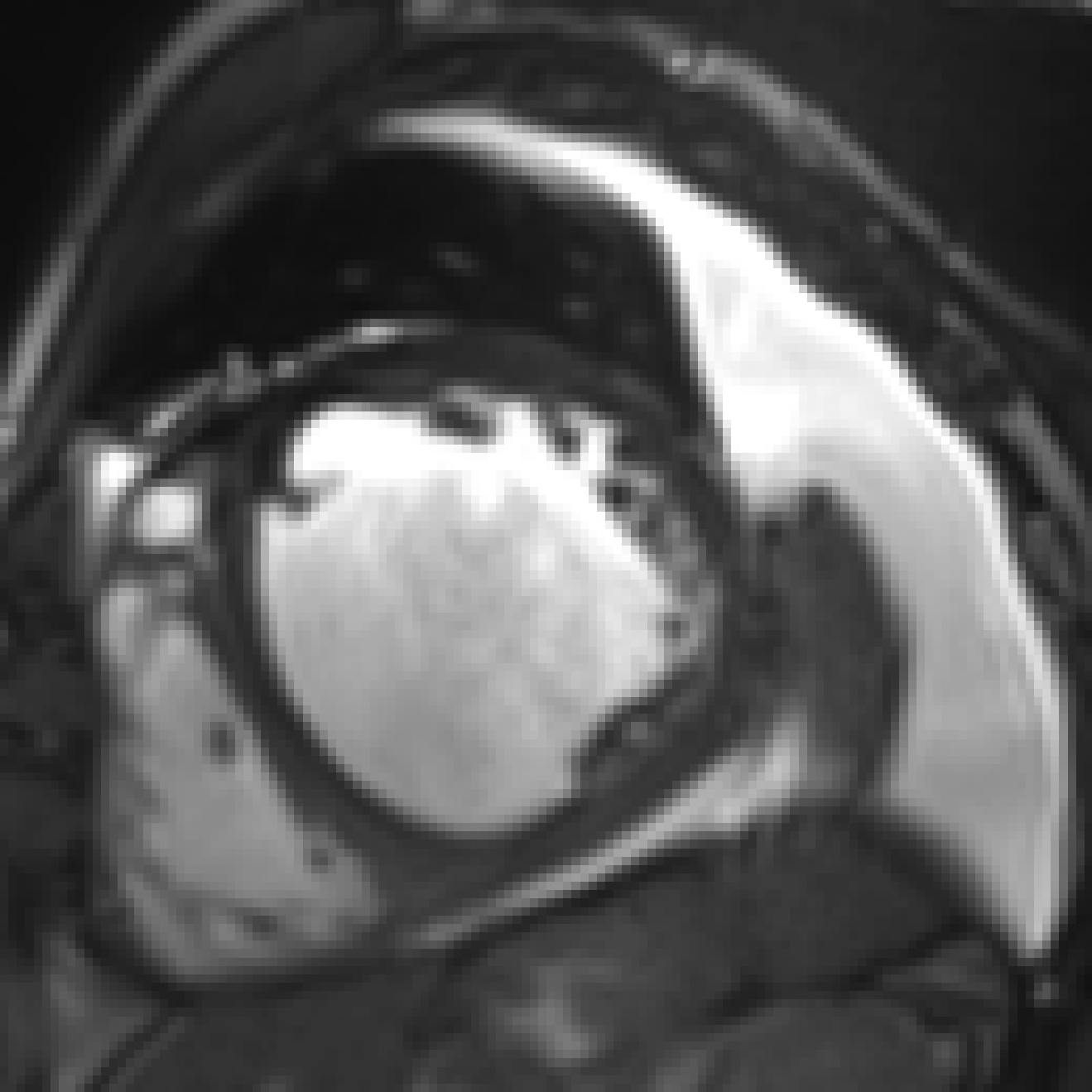} }; 
					% Create scope with normalized axes
					\begin{scope}[
					x={($0.1*(image.south east)$)},
					y={($0.1*(image.north west)$)}]
					% Grid
					% \draw[lightgray,step=1] (image.south west) grid (image.north east);
					\draw[thick,orange] (1,7.5) rectangle (9.2, 9.9) ;
					\draw[thick,orange] (7.5,0.5) rectangle (9.9, 3.3) ;
					\end{scope}
					\end{tikzpicture} } \\
				% row 2
				\begin{tabular}{@{}c@{}} ASI$_{\lambda=0.05}$  \\ \\ \\ \\ \\ \end{tabular} &
				\subfloat[Neighboring slice 1]{ \begin{tikzpicture}
					\node[above right, inner sep=0] (image) at (0,0)  { \includegraphics[width=.09\textwidth]{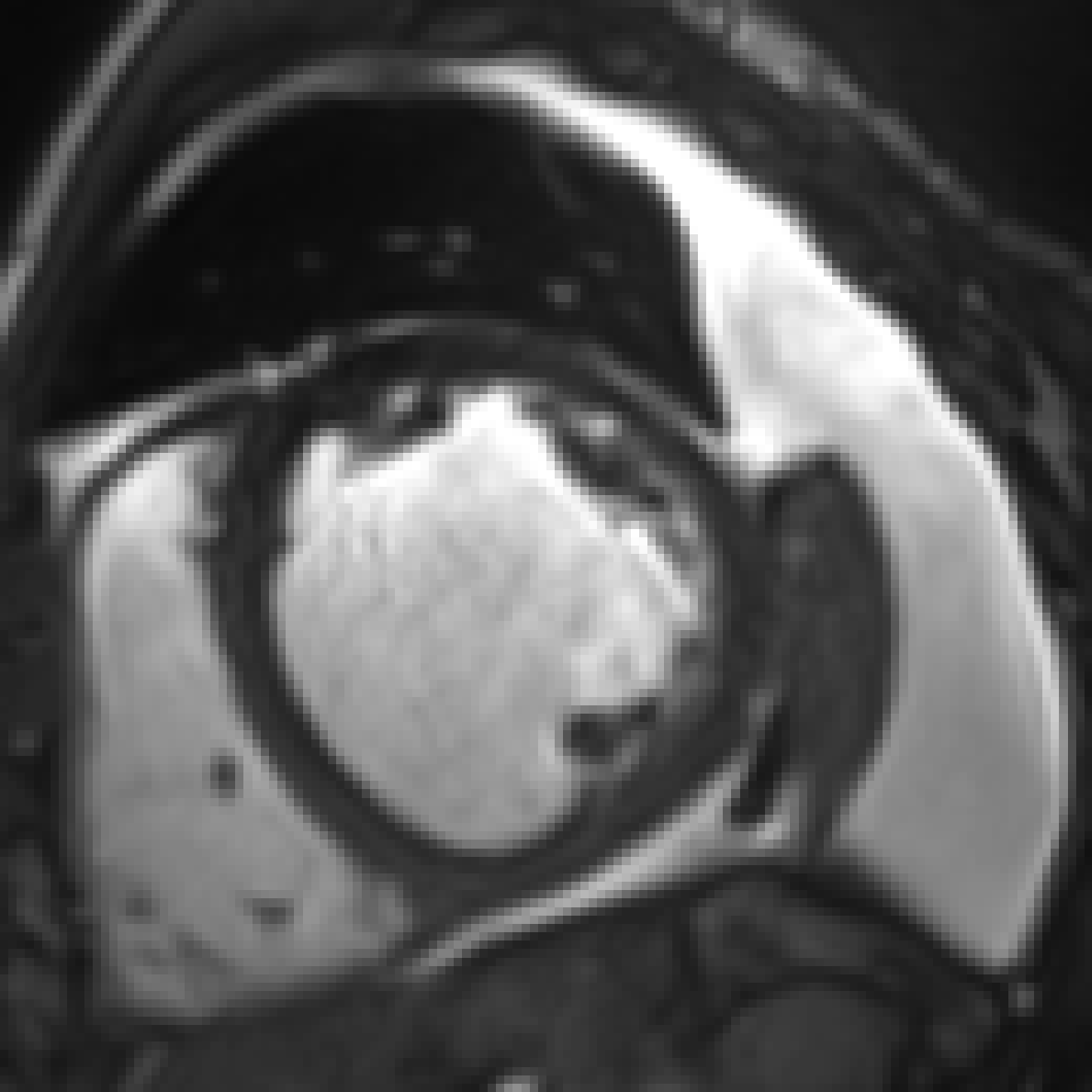} }; 
					% Create scope with normalized axes
					\begin{scope}[
					x={($0.1*(image.south east)$)},
					y={($0.1*(image.north west)$)}]
					% Grid
					% \draw[lightgray,step=1] (image.south west) grid (image.north east);
					\draw[thick,orange] (1,7.5) rectangle (9.2, 9.9) ;
					\draw[thick,orange] (7.5,0.5) rectangle (9.9, 3.3) ;
					\end{scope}
					\end{tikzpicture} } &
				\subfloat[$\alpha=$ \nicefrac{1}{7}]{ \begin{tikzpicture}
					\node[above right, inner sep=0] (image) at (0,0)  { \includegraphics[width=.09\textwidth]{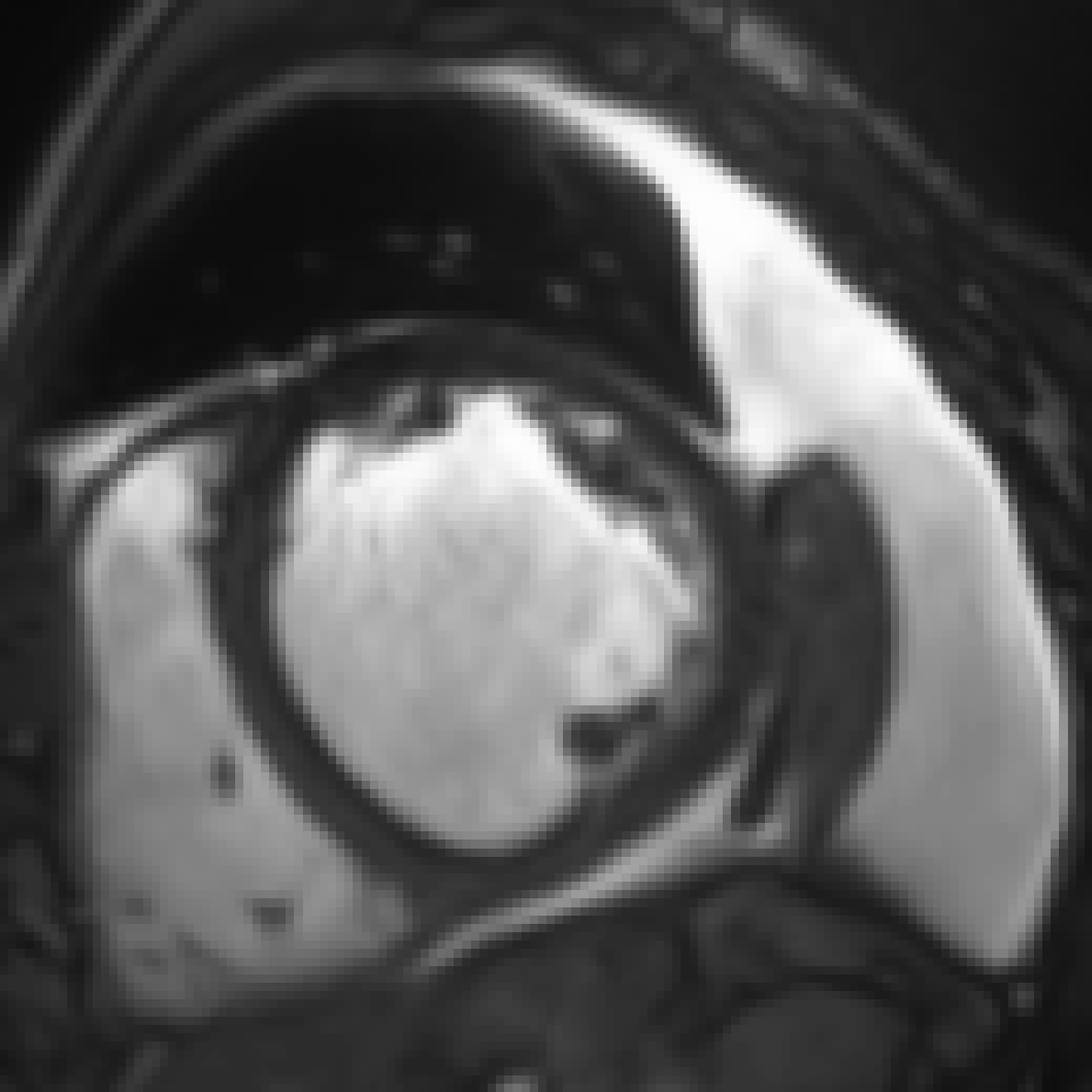} }; 
					% Create scope with normalized axes
					\begin{scope}[
					x={($0.1*(image.south east)$)},
					y={($0.1*(image.north west)$)}]
					% Grid
					% \draw[lightgray,step=1] (image.south west) grid (image.north east);
					\draw[thick,orange] (1,7.5) rectangle (9.2, 9.9) ;
					\draw[thick,orange] (7.5,0.5) rectangle (9.9, 3.3) ;
					\end{scope}
					\end{tikzpicture} } &
				\subfloat[$\alpha=$ \nicefrac{2}{7}]{ \begin{tikzpicture}
					\node[above right, inner sep=0] (image) at (0,0)  { \includegraphics[width=.09\textwidth]{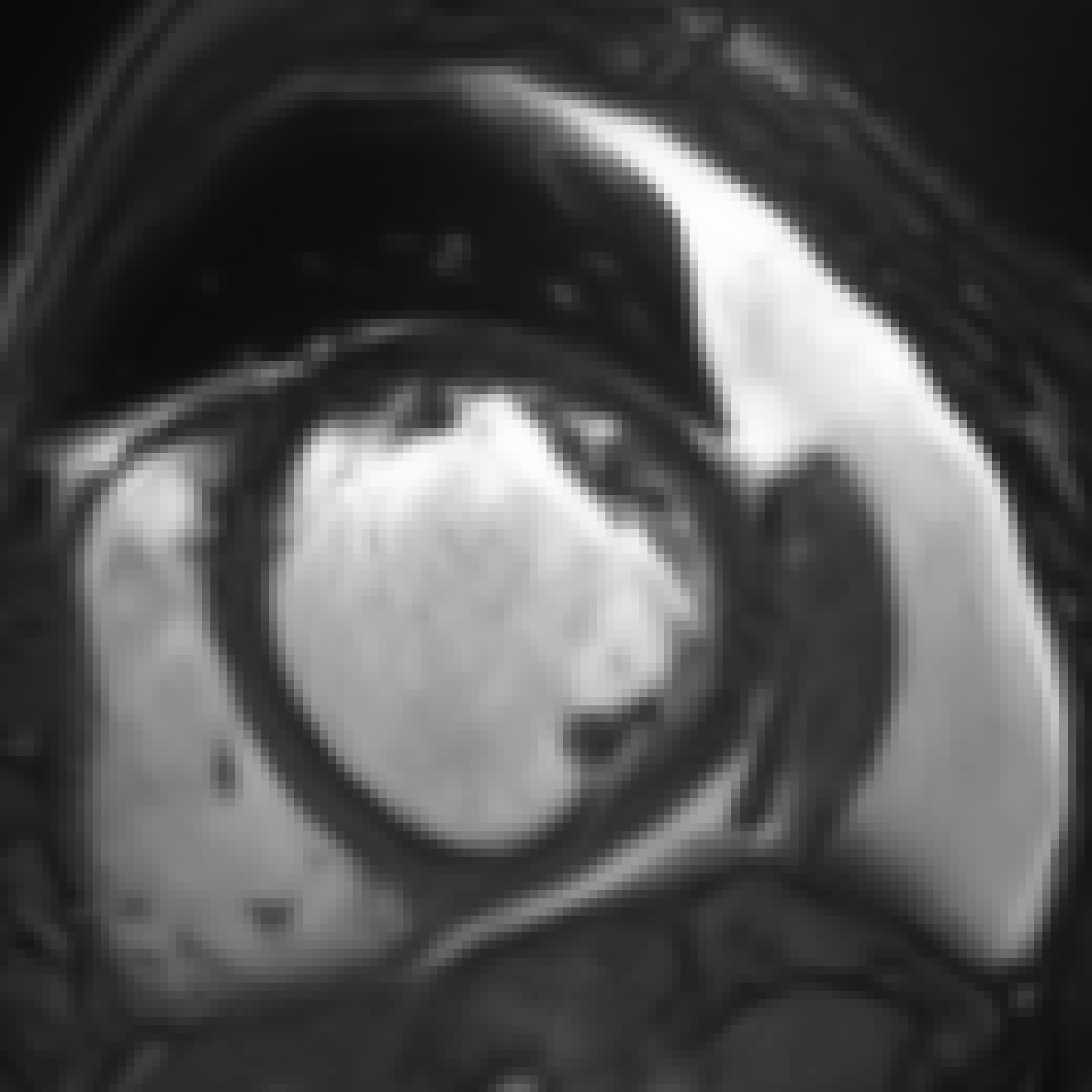} }; 
					% Create scope with normalized axes
					\begin{scope}[
					x={($0.1*(image.south east)$)},
					y={($0.1*(image.north west)$)}]
					% Grid
					% \draw[lightgray,step=1] (image.south west) grid (image.north east);
					\draw[thick,orange] (1,7.5) rectangle (9.2, 9.9) ;
					\draw[thick,orange] (7.5,0.5) rectangle (9.9, 3.3) ;
					\end{scope}
					\end{tikzpicture} } &
				\subfloat[$\alpha=$ \nicefrac{3}{7}]{ \begin{tikzpicture}
					\node[above right, inner sep=0] (image) at (0,0)  { \includegraphics[width=.09\textwidth]{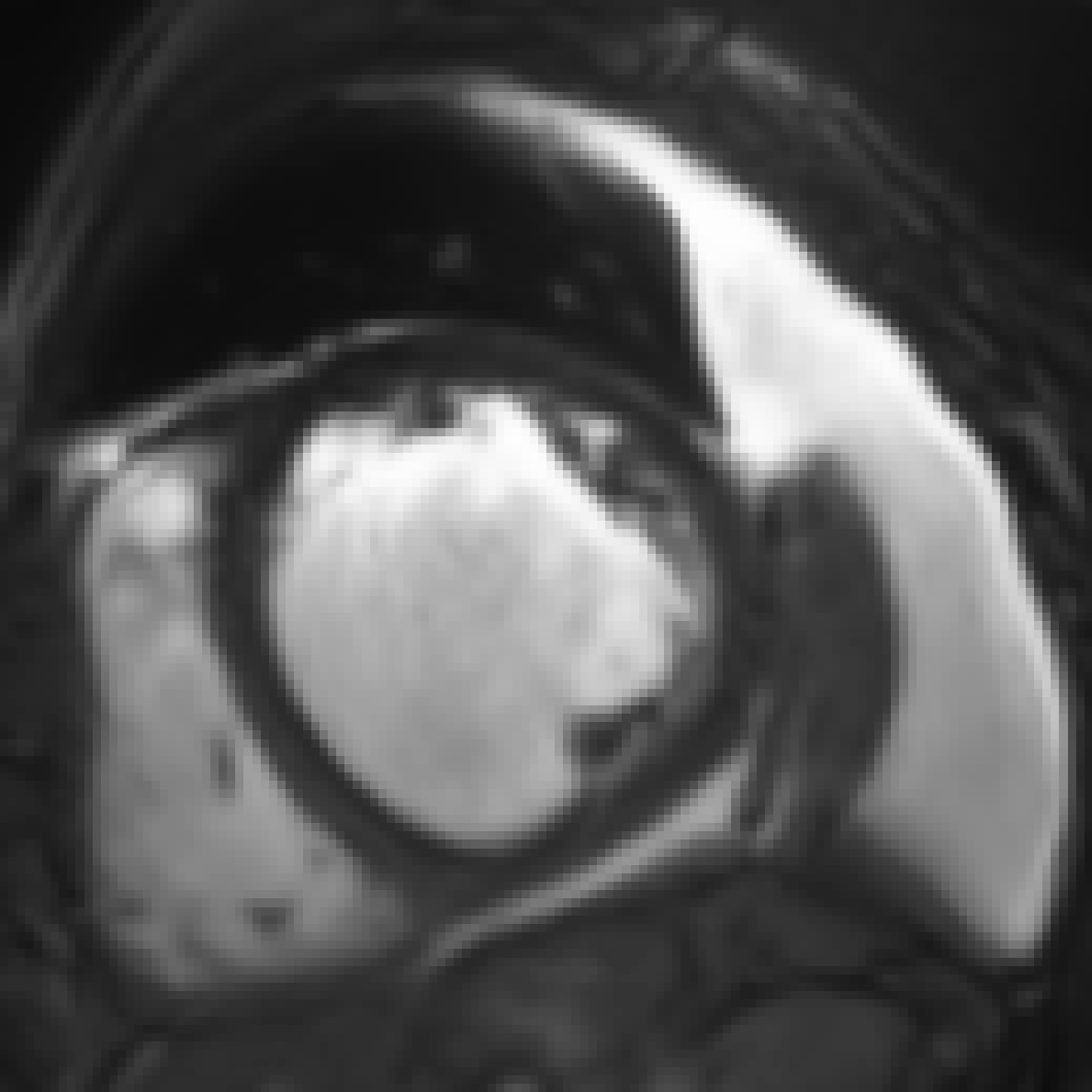} }; 
					% Create scope with normalized axes
					\begin{scope}[
					x={($0.1*(image.south east)$)},
					y={($0.1*(image.north west)$)}]
					% Grid
					% \draw[lightgray,step=1] (image.south west) grid (image.north east);
					\draw[thick,orange] (1,7.5) rectangle (9.2, 9.9) ;
					\draw[thick,orange] (7.5,0.5) rectangle (9.9, 3.3) ;
					\end{scope}
					\end{tikzpicture} } &
				\subfloat[$\alpha=$ \nicefrac{4}{7}]{ \begin{tikzpicture}
					\node[above right, inner sep=0] (image) at (0,0)  { \includegraphics[width=.09\textwidth]{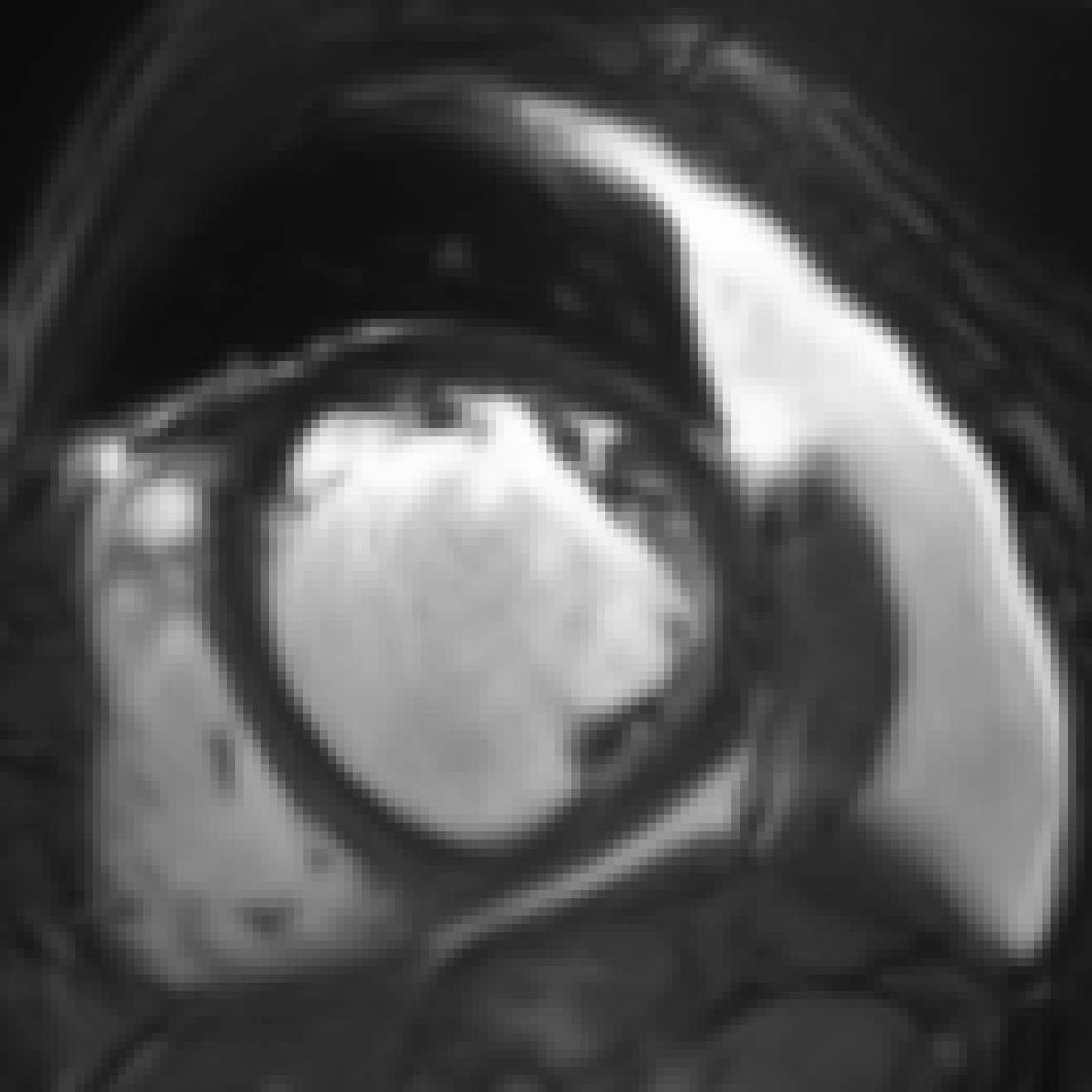} }; 
					% Create scope with normalized axes
					\begin{scope}[
					x={($0.1*(image.south east)$)},
					y={($0.1*(image.north west)$)}]
					% Grid
					% \draw[lightgray,step=1] (image.south west) grid (image.north east);
					\draw[thick,orange] (1,7.5) rectangle (9.2, 9.9) ;
					\draw[thick,orange] (7.5,0.5) rectangle (9.9, 3.3) ;
					\end{scope}
					\end{tikzpicture} } &
				\subfloat[$\alpha=$ \nicefrac{5}{7}]{ \begin{tikzpicture}
					\node[above right, inner sep=0] (image) at (0,0)  { \includegraphics[width=.09\textwidth]{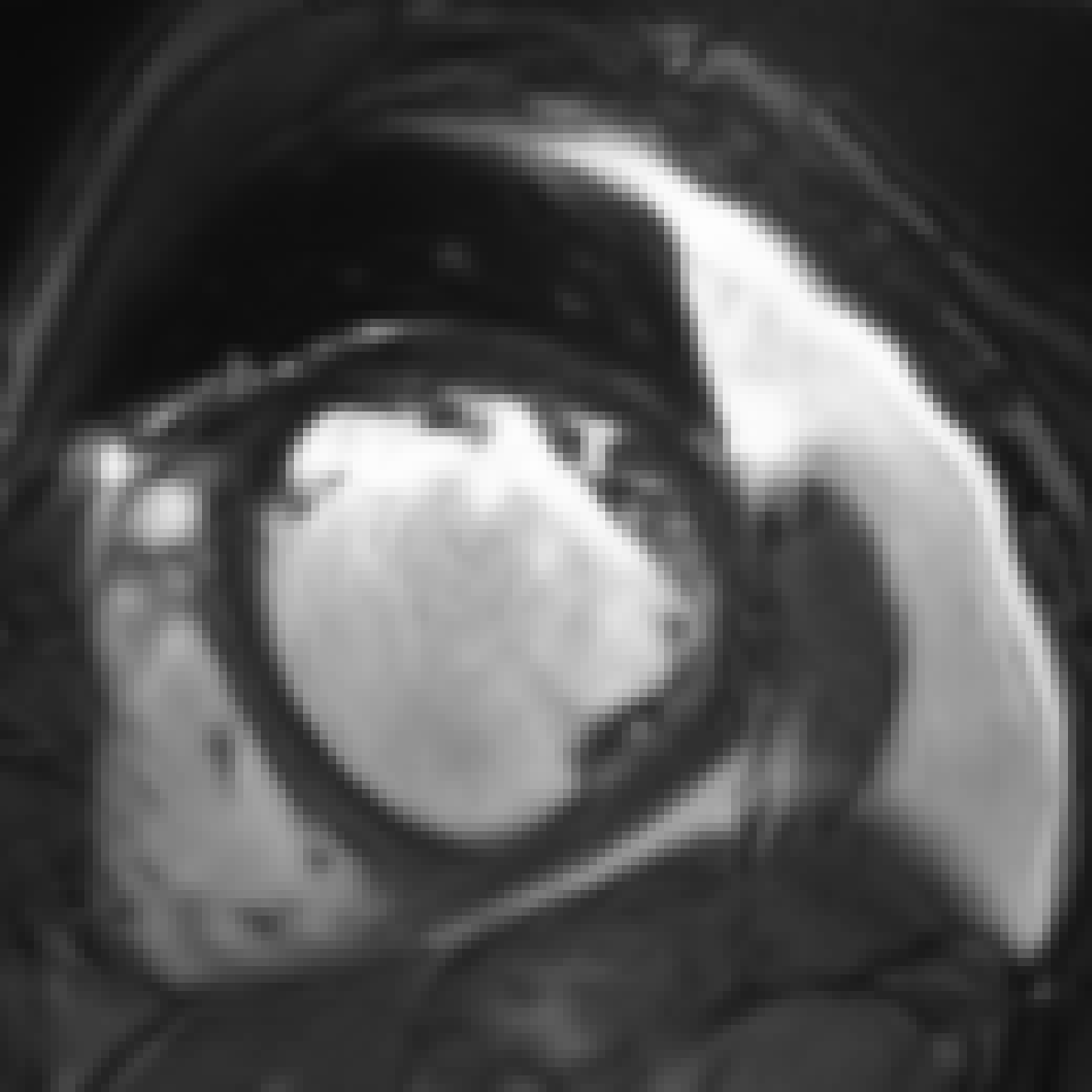} }; 
					% Create scope with normalized axes
					\begin{scope}[
					x={($0.1*(image.south east)$)},
					y={($0.1*(image.north west)$)}]
					% Grid
					% \draw[lightgray,step=1] (image.south west) grid (image.north east);
					\draw[thick,orange] (1,7.5) rectangle (9.2, 9.9) ;
					\draw[thick,orange] (7.5,0.5) rectangle (9.9, 3.3) ;
					\end{scope}
					\end{tikzpicture} } &
				\subfloat[$\alpha=$ \nicefrac{6}{7}]{ \begin{tikzpicture}
					\node[above right, inner sep=0] (image) at (0,0)  { \includegraphics[width=.09\textwidth]{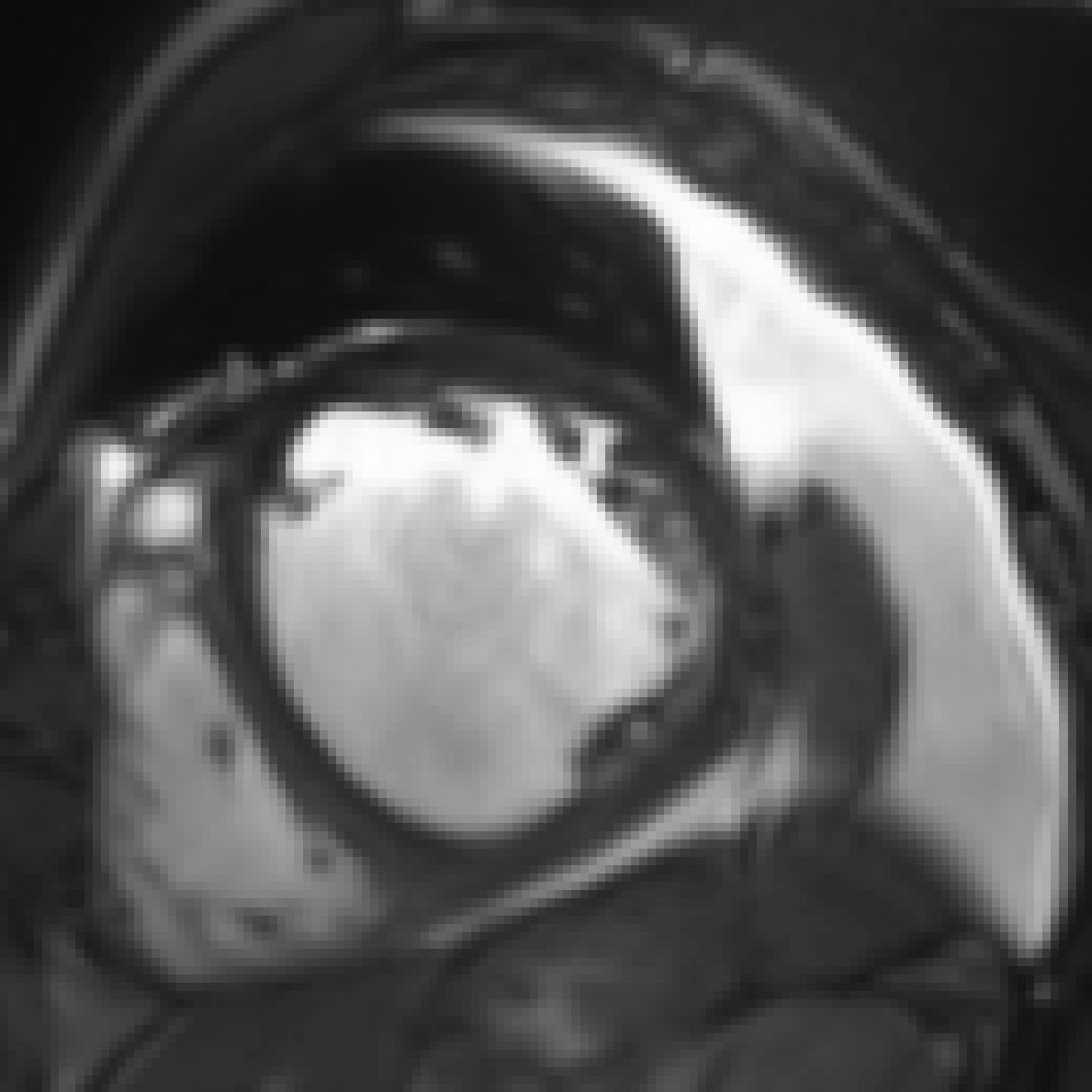} }; 
					% Create scope with normalized axes
					\begin{scope}[
					x={($0.1*(image.south east)$)},
					y={($0.1*(image.north west)$)}]
					% Grid
					% \draw[lightgray,step=1] (image.south west) grid (image.north east);
					\draw[thick,orange] (1,7.5) rectangle (9.2, 9.9) ;
					\draw[thick,orange] (7.5,0.5) rectangle (9.9, 3.3) ;
					\end{scope}
					\end{tikzpicture} } &
				\subfloat[Neighboring slice 2]{ \begin{tikzpicture}
					\node[above right, inner sep=0] (image) at (0,0)  { \includegraphics[width=.09\textwidth]{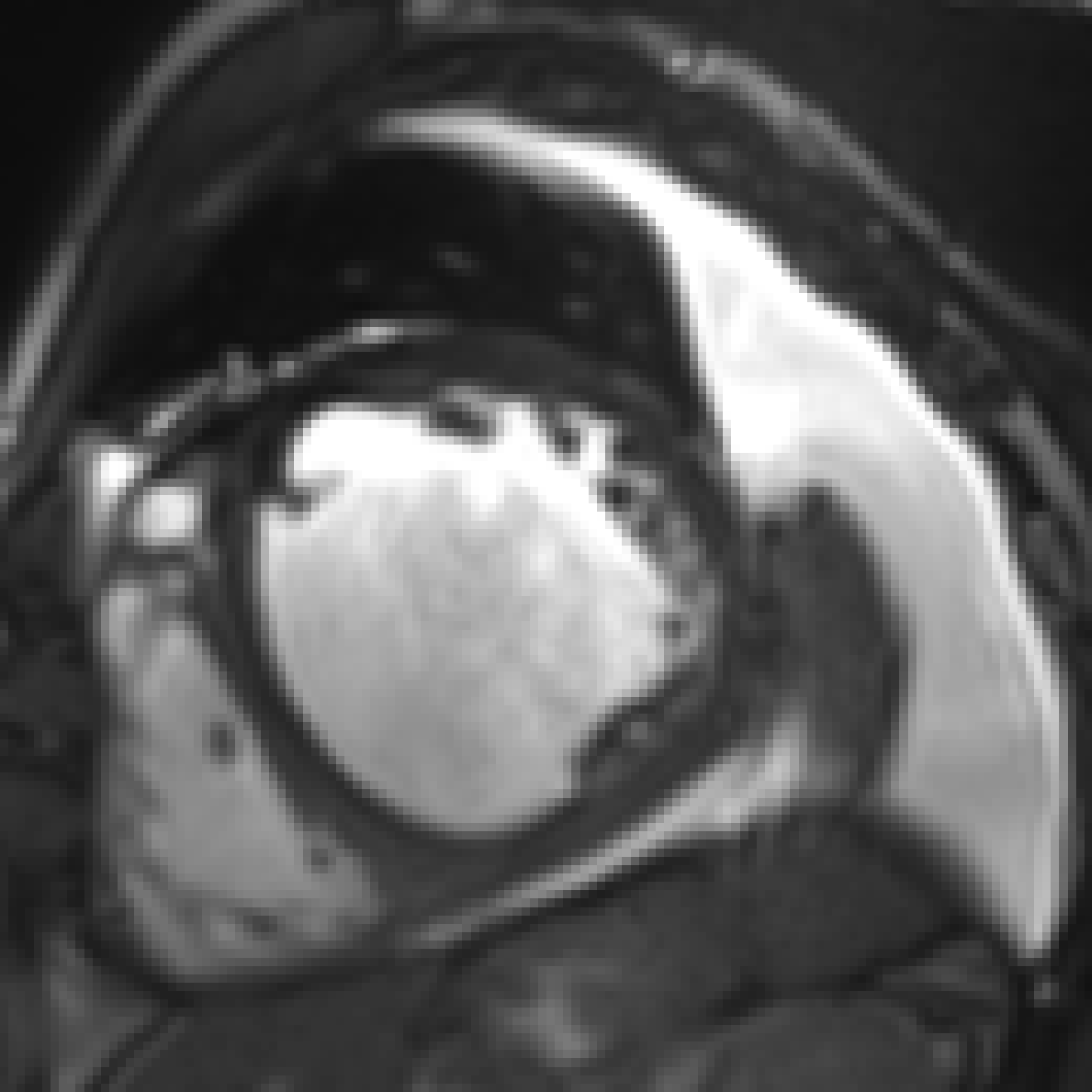} }; 
					% Create scope with normalized axes
					\begin{scope}[
					x={($0.1*(image.south east)$)},
					y={($0.1*(image.north west)$)}]
					% Grid
					% \draw[lightgray,step=1] (image.south west) grid (image.north east);
					\draw[thick,orange] (1,7.5) rectangle (9.2, 9.9) ;
					\draw[thick,orange] (7.5,0.5) rectangle (9.9, 3.3) ;
					\end{scope}
					\end{tikzpicture} } \\	
			\end{tabular}
			
		\end{center}
		
		\caption{Comparison of synthesis performance between proposed model trained with (top row) reconstruction loss only (denoted ASI$_{\lambda=0}$) and (bottom row) model trained with a combination of reconstruction and synthesis loss ASI$_{\lambda=0.05}$. Second to penultimate columns show six synthesized intermediate slices using latent space encodings of the two neighboring slices (first and last column). Model trained with just reconstruction loss (ASI$_{\lambda=0}$) generates \textit{cross-fade} artifacts (e.g. columns four to six) between the intensities of the two neighboring slices. Such artifacts are mostly suppressed by the ASI$_{\lambda=0.05}$ model. $\alpha$ denotes the mixing coefficient as specified in Equation~\ref{eq_convex_combination}.}
		\label{fig_qualitative_compare_ae_aisr_acdc_cross_fade}
		% Slice spacing of cardiac MRI (ACDC dataset) was reduced from \num{10} to \SI{1.43}{\milli\meter} by synthesizing six intermediate slices (second to penultimate columns) using latent space encodings of the two neighboring slices (first and last column).
	\end{figure*}

	This setting enables performance comparison between model ASI$_{\lambda=0}$ and a model trained with the combined reconstruction and synthesis loss (ASI$_{\lambda=0.05}$). Figure~\ref{fig_qualitative_acdc_effect_of_synthesis_1} depicts qualitative comparison between the two models for image synthesis of cardiac MRI. The results reveal that performance decreased for a model trained with the reconstruction loss only. The performance difference is more pronounced for larger anatomical variations between neighboring slices e.g. the shape of the right ventricle. Furthermore, Figure~\ref{fig_qualitative_compare_ae_aisr_acdc_cross_fade} demonstrates synthesis performance of the two models. The model trained with just the reconstruction loss (ASI$_{\lambda=0}$) generates \textit{cross-fade} artifacts between the intensities of the two neighboring slices \citep{arvanitidis2018, laine2018feature, oring21a} whereas the model trained with a combination of reconstruction and synthesis loss (ASI$_{\lambda=0.05}$) can substantially suppress such artifacts.
	% an autoencoder trained with reconstruction and synthesis loss (ASI$_{\lambda=0.05}$) compared with a model trained with reconstruction loss only (ASI$_{\lambda=0}$)
	
\begin{figure}[ht]
		\captionsetup[subfigure]{justification=centering}
		\centering
		\includegraphics[width=.48\textwidth]{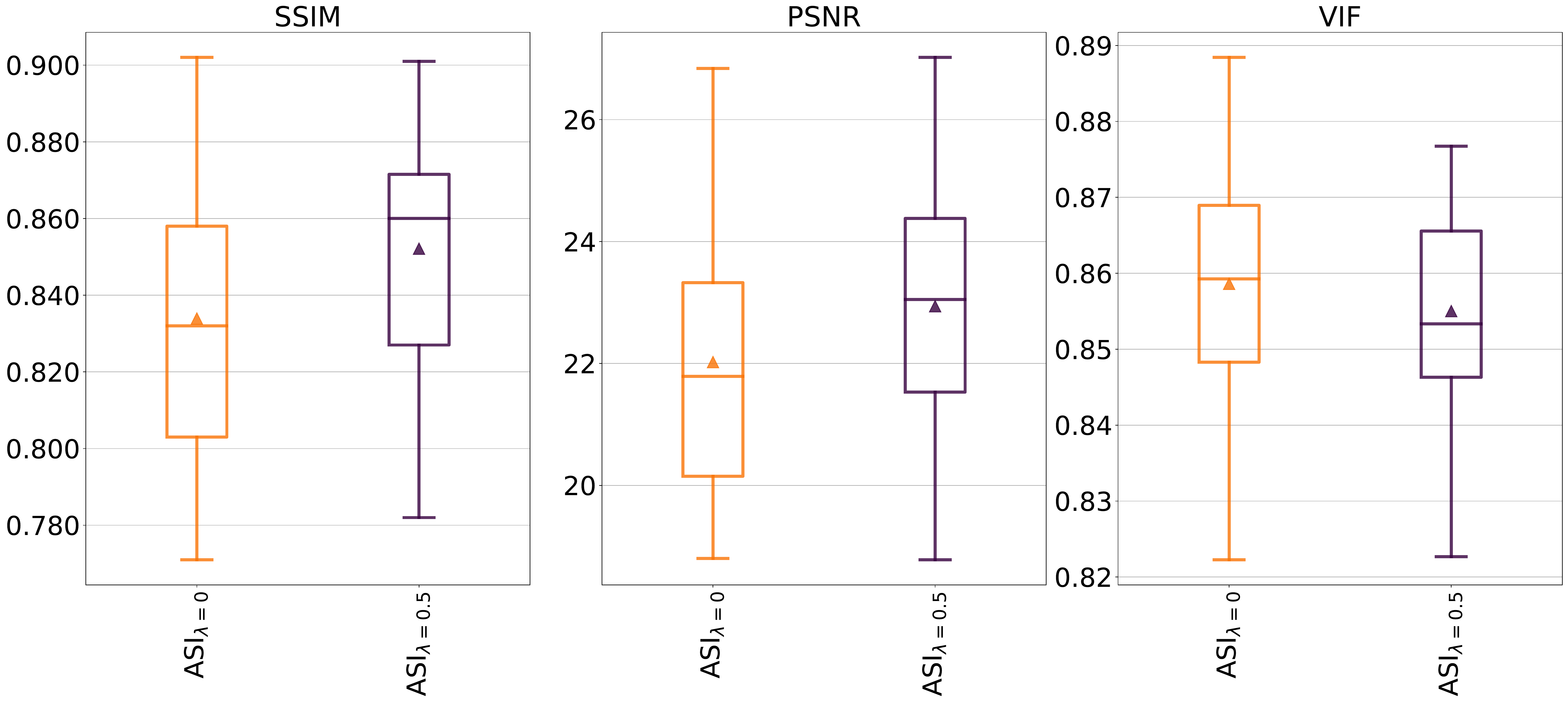} 
		\caption{Boxplots compare performance of the proposed model trained with (a) reconstruction loss only (denoted ASI$_{\lambda=0}$) compared with model trained with (b) reconstruction and synthesis loss ASI$_{\lambda=0.05}$ in terms of SSIM, PSNR, and VIF. Cardiac cine MRIs of \num{20} patients from the ACDC dataset were upsampled with factor \num{2} in through-plane direction. A higher score indicates better performance. Measures were computed on sagittal slices through volume. Triangle indicates mean value.}
		%Boxplots showing evaluation of performance with upsampling factor \num{2} in through-plane direction using cardiac cine MRIs of \num{20} patients from the ACDC dataset. Boxplots compare performance for proposed model trained with (a) reconstruction loss only (denoted ASI$_{\lambda=0}$) compared to model trained with (b) reconstruction and synthesis loss ASI$_{\lambda=0.05}$ in terms of SSIM, PSNR, and VIF. }		
		\label{fig_acdc_quantitative_results_effect_of_synthesis_1} 
	\end{figure}

	These results are corroborated by quantitative evaluation in terms of SSIM, PSNR, and VIF depicted in Figure~\ref{fig_acdc_quantitative_results_effect_of_synthesis_1}. One can observe that the model trained with the combined reconstruction and synthesis loss achieved better performance when evaluated by SSIM and PSNR compared to a model trained with the reconstruction loss only (ASI$_{\lambda=0}$). For the VIF measure the performance achievements are reversed. All differences are statistically significant ($p < 0.0001$) using the one-sided Wilcoxon signed-rank test.

		% FIGURE 14

	% FIGURE 18  TEMPORARY NOT SURE WHETHER IT SHOULD STAY
	
	To further investigate the effect of the synthesis loss on upsampling performance, separate quantitative evaluations were performed on reconstructed and synthesized short-axis slices, respectively. 
	The results listed in Table~\ref{table_acdc_ae_caisr_recon_versus_synth_1} reveal that a model trained with the reconstruction loss only (ASI$_{\lambda=0}$) achieved better performance for the reconstruction task compared to a model trained with the combined reconstruction and synthesis loss (ASI$_{\lambda=0.05}$). In contrast, the latter model performed better in terms of SSIM and PSNR when synthesizing the excluded slices. These results indicate that the additional synthesis loss resulted in increased interpolation performance but at the same time impacted reconstruction performance of the autoencoder. 
	
	% Figure rebuttal: comparing SR performance for different values of lambda. 
	\begin{figure*}
		\captionsetup[subfigure]{justification=centering}
		\centering
		\includegraphics[width=.96\textwidth]{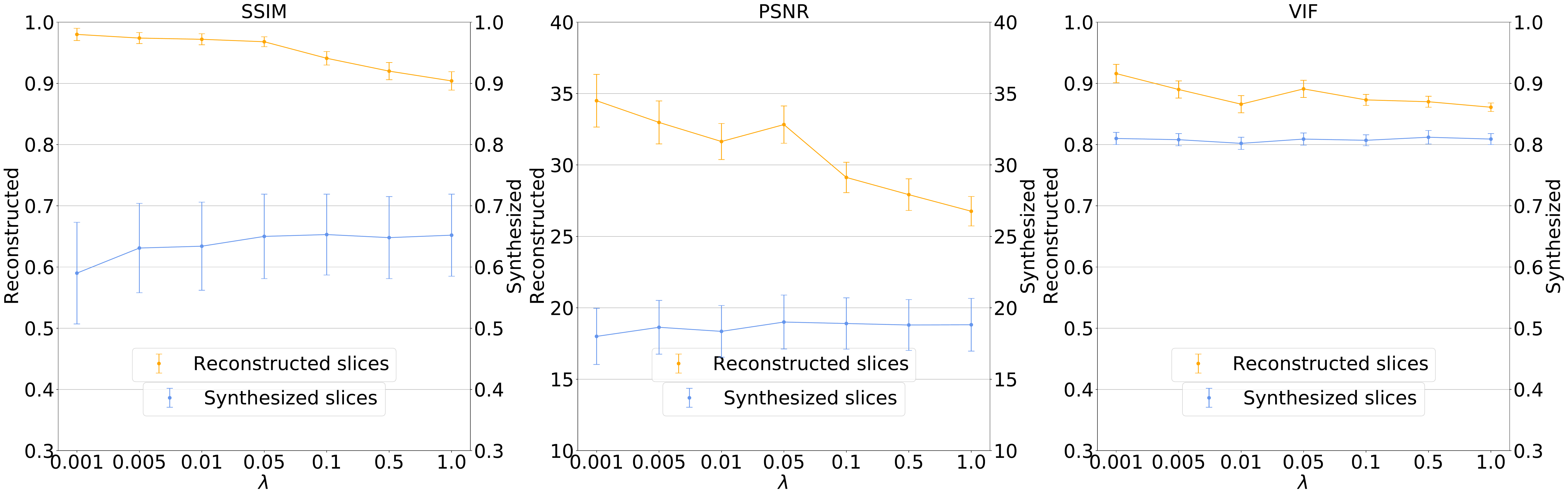} 
		\caption{Quantitative comparison of reconstruction and synthesis performance in terms of SSIM, PSNR, and VIF between proposed model trained on cardiac cine MRIs (ACDC dataset) using different values for the hyperparameter $\lambda \in \{0.001, 0.005, 0.01, 0.05, 0.1, 0.5, 1 \}$ as specified in Equation~\ref{eq_combined_loss}. y-axis shows performance for reconstructed and synthesized slices, respectively.}		
		\label{fig_acdc_compare_lambda} 
	\end{figure*}
	
	Finally, Figure~\ref{fig_acdc_compare_lambda} shows the effect of different values of $\lambda$ on reconstruction and synthesis performance of the proposed approach. Increasing the contribution of the synthesis loss, i.e. using larger values for $\lambda$, increases synthesis and lowers reconstruction performance of the model in terms of SSIM and PSNR. % This effect is not prominent in terms of VIF.
	
	\subsection{Evaluation on Cardiac Cine MRIs from Sunnybrook dataset} \label{exp_sunnybrook}
	
\begin{figure}
		\centering
		\includegraphics[width=.48\textwidth]{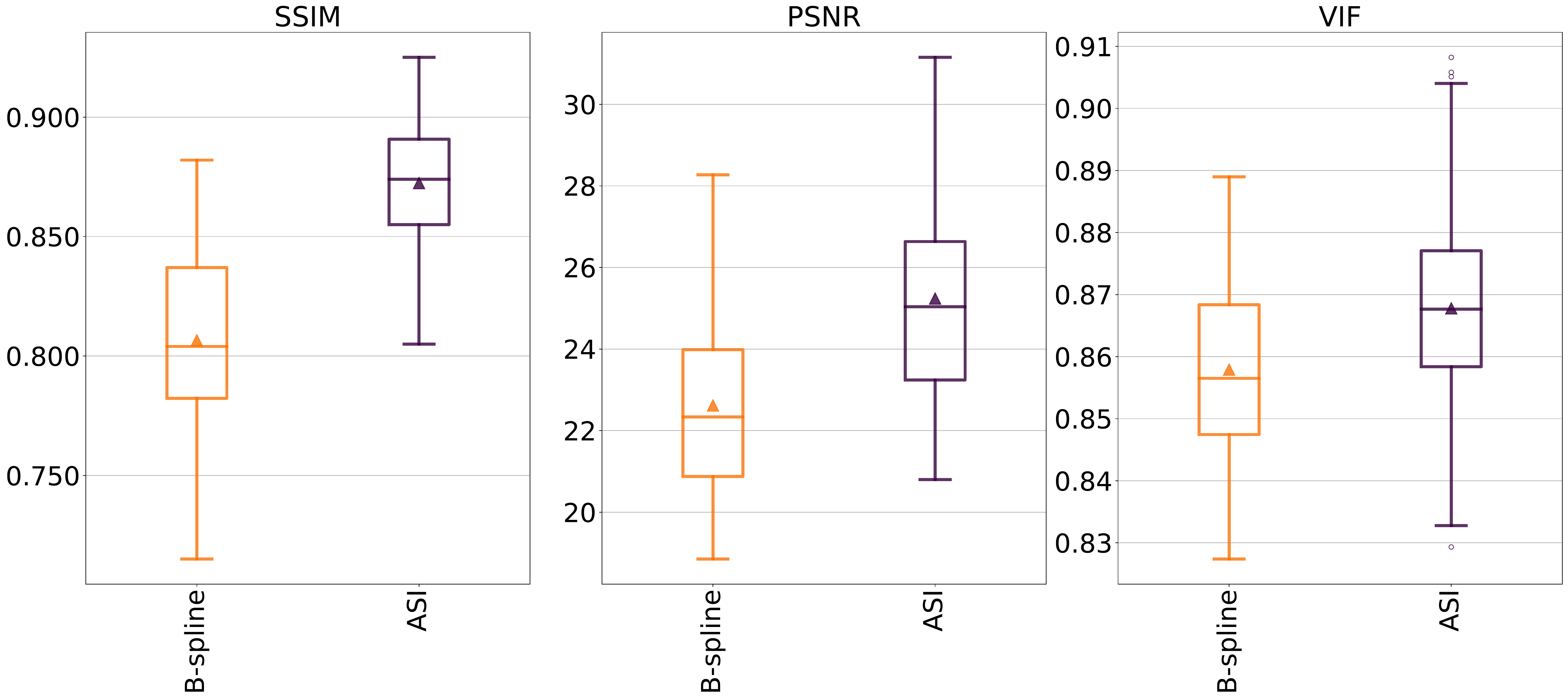} 
		\caption{Boxplots showing results for upsampling of \num{90} cardiac MRIs from \textit{Sunnybrook} dataset with upsampling factor \num{2} in through-plane direction for conventional interpolation method (cubic B-spline) compared to proposed method (ASI) in terms of SSIM, PSNR, and Visual Information Fidelity (VIF). Proposed approach was trained on CMR scans from \textit{ACDC} dataset. Hence, depicted results demonstrate generalization performance of proposed method. A higher score indicates better performance. Measures were computed on sagittal slices through short-axis volume. Performance differences are statistically significant ($p < 0.0001$) using the one-sided Wilcoxon signed-rank test. Triangle indicates mean value.}
		\label{fig_sunnybrook_quantitative_results} 
	\end{figure}

	To examine generalization performance of our proposed method a model trained on cardiac cine MRIs from the ACDC dataset was evaluated on cardiac MRI scans from the Sunnybrook dataset \citep{radau2009evaluation}. Figure~\ref{fig_sunnybrook_quantitative_results} shows quantitative comparison for cubic B-spline compared with proposed method (ASI) in terms of SSIM, PSNR, and VIF. One can observe that the proposed approach trained on ACDC images outperformed cubic B-spline interpolation for all measures on Sunnybrook dataset. The latter methods do not require training. These differences are statistically significant ($p<0.0001$) using the one-sided Wilcoxon signed-rank test. Furthermore, the results depict that relative performance differences between methods are nearly identical to those observed on ACDC dataset depicted in Figure~\ref{fig_acdc_quantitative_results}. Finally, achieved performance on cardiac MRI scans of Sunnybrook dataset is higher for all measures compared with performance reported for evaluation on ACDC dataset depicted in Figure~\ref{fig_acdc_quantitative_results}. This might have been caused by scans of Sunnybrook dataset having more consistent image quality, better alignment of adjacent slices, and higher bit depth compared with scans from ACDC dataset.
	
	% FIGURE 16

	\begin{figure*}
		\captionsetup[subfigure]{justification=centering, labelformat=empty}
		\begin{center}
			\begin{tabular}{c c c c}
				% row 1
				\subfloat[Neighboring slice 1]{\includegraphics[width=.18\textwidth]{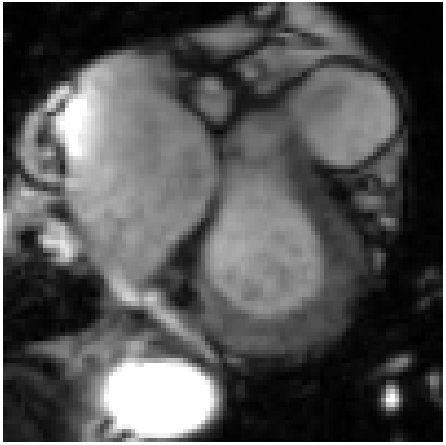} }  & 
				\subfloat[Slice \num{2}, located in-between slice \num{1} and \num{3}]{\includegraphics[width=.18\textwidth]{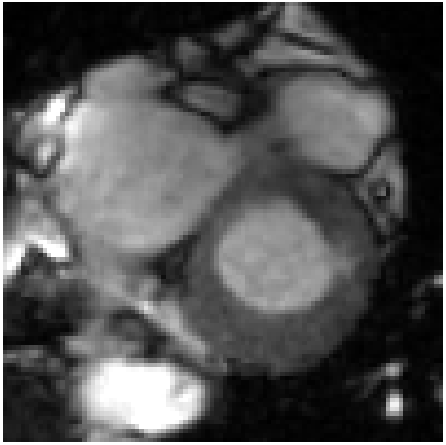} } &
				\subfloat[Synthesized slice \num{2} using proposed approach with $\alpha=0.5$ ]{\includegraphics[width=.18\textwidth]{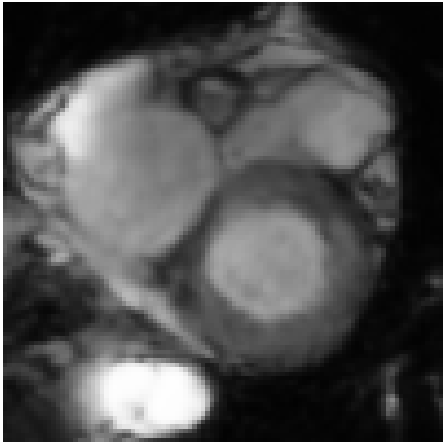} }  &
				\subfloat[Neighboring slice 3 ]{\includegraphics[width=.18\textwidth]{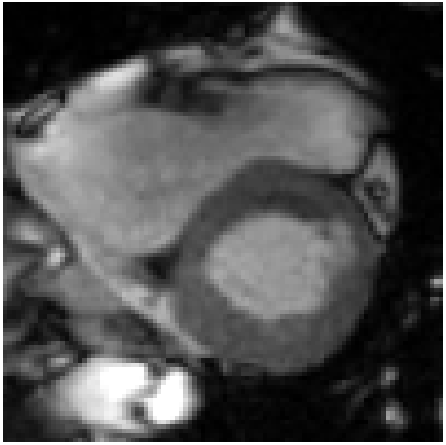} } \\
				% row 2
				\subfloat[Neighboring slice 1]{\includegraphics[width=.18\textwidth]{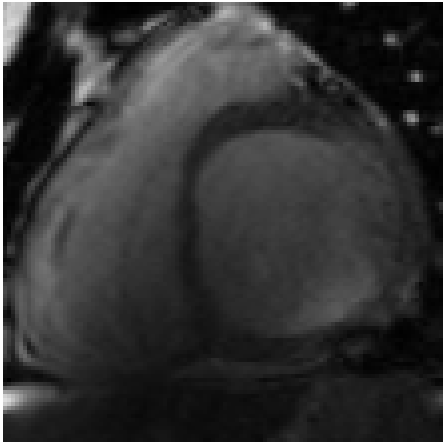}} &
				\subfloat[Slice \num{2}, located in-between slice \num{1} and \num{3}]{\includegraphics[width=.18\textwidth]{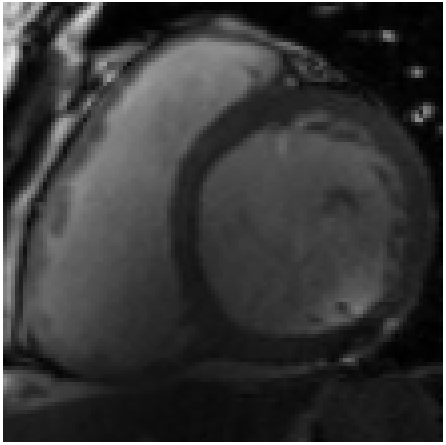}} &
				\subfloat[Synthesized slice \num{2} using proposed approach with $\alpha=0.5$]{\includegraphics[width=.18\textwidth]{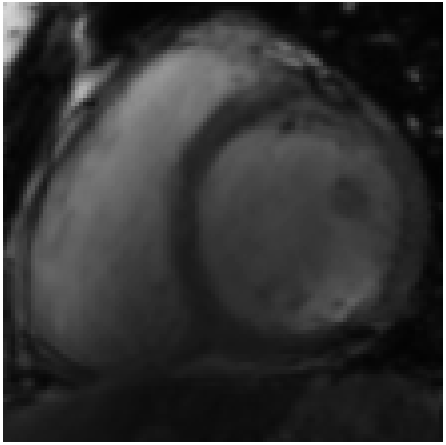}} &
				\subfloat[Neighboring slice 3]{\includegraphics[width=.18\textwidth]{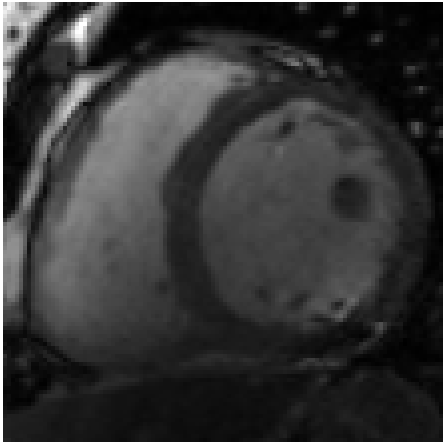}} 
			\end{tabular}
			
		\end{center}
		
		\caption{Qualitative comparison for upsampling factor \num{2} on cardiac MRI (ACDC dataset). Each row depicts an example of synthesizing intermediate slice \num{2} (second column) using latent space encodings of the two neighboring slices \num{1} and \num{3} (first and last column). The synthesized slice is shown in penultimate column. First row shows basal slices with large anatomical variations compared to second row that depicts mid-ventricular slices with mild anatomical variations. $\alpha$ denotes the mixing coefficient as specified in Equation \ref{eq_convex_combination}. Scans have a slice spacing of $\SI{10} {\milli\meter}$.}.
		\label{fig_qualitative_acdc_large_anatomical_variations}
	\end{figure*}

	\section{Discussion} \label{section_discussion}
	
	A method for unsupervised deep-learning image synthesis of medical images has been presented. To synthesize new intermediate slices and thereby recovering spatial information, the method exploits the latent space interpolation ability of autoencoders. New intermediate slices are generated by mixing the latent space encodings of two spatially adjacent slices. High-resolution ground-truth images are not required to train the approach. Results of our preliminary experiments using MNIST data demonstrated that our proposed approach outperformed a variational autoencoder (VAE) and Adversarially Constrained Autoencoder Interpolation (ACAI) approach \citep{berthelot2018understanding} for interpolating rotations of handwritten digits. Evaluation of the approach on cardiac and brain structures using four publicly available MRI datasets revealed that the method can outperform cubic B-spline interpolation. Performance differences between evaluated methods become more apparent when the upsampling task becomes more difficult i.e. for highly anisotropic volumes e.g. cardiac MRI or larger through-plane upsampling factors. This might indicate that the model can infer the missing information from contextual and anatomical information captured in the latent space. Furthermore, the experimental results revealed that our proposed approach can compete  with related unsupervised \citep{jog2016self, zhao2018self} and supervised \citep{pham2019simultaneous} super-resolution approaches. Compared with unsupervised super-resolution methods of \cite{jog2016self, zhao2018self,zhao2018deep, zhao2020smore, dalca2018medical, xia2021super} our approach can be applied with any desired upsampling factor and uses a single encoder-decoder structure. Moreover, applying methods of \cite{jog2016self} and \cite{zhao2018self,zhao2018deep, zhao2020smore} requires optimization during inference for every image at hand and therefore, inference requires several minutes of GPU processing time \citep{zhao2020smore}. In contrast, at test time our method can synthesize multiple intermediate slices between each pair of adjacent slices in an MRI scan in less than a second on a GPU. Furthermore, using the method of \cite{dalca2018medical} necessitates creation of a common atlas space to which each image must be transformed. Finally, it is fair to note that the approach of \cite{zhao2018deep, zhao2020smore} performs explicitly anti-aliasing by using an additional CNN.
	
	Even though autoencoders are designed to learn a lower-dimensional representation of the input while minimizing information loss, in this work, we used them to perform semantic interpolation and thereby recover spatial information in anisotropic medical images. Specifically, we used an over-complete autoencoder that can potentially retain all information contained in the input. Theoretically, such an approach could learn an identity to minimize the reconstruction loss. Nonetheless, in line with previous research (\cite{bengio2007greedy}), the results presented in this work seem to indicate that such a model can learn a useful representation of its input.
	% Why latent space interpolation for super-resolution
	Furthermore, the experimental results revealed that interpolation between image representations is feasible to approximate information orthogonal to the input images. This suggests that the proposed approach learns to extract contextual and high level conceptual information from the input images. Moreover, the results demonstrate that the decoder learns to exploit this information to instantiate semantically meaningful intermediate slices. While previously developed shape-based interpolation approaches of \cite{raya1990shape, grevera1996shape} exploit anatomical shape information to achieve high-order interpolation between cross sections of \num{3}D anatomical structures, we argue that our approach performs semantic interpolation between two spatially adjacent slices.

	% no guarantees synthesized slices are always semantically meaningful
	% synthesis loss in image space rather than latent space
	% cross-fade artifacts
	Synthesized images are not guaranteed to be semantically meaningful. However, the solution space of the proposed approach is constrained using encodings of two spatially adjacent slices. Moreover, compared with an adversarial training objective \citep{berthelot2018understanding} training the autoencoder with the proposed synthesis loss enforces an explicit constraint on the solution space because the synthesized image is evaluated against its reference. Nevertheless, image interpolation in latent space can still result in \textit{cross-fade} artifacts between the intensities of the two images i.e. neighboring slices \citep{arvanitidis2018, laine2018feature, oring21a}. Appearance of such an artifact can be observed in Figure~\ref{fig_qualitative_synthesis_dhcp4x} (second row, columns four to six). However, results presented in Figure~\ref{fig_qualitative_compare_ae_aisr_acdc_cross_fade} demonstrated that the proposed model can synthesize images with substantially less artifacts compared with a standard autoencoding approach. Although adversarial approaches are extremely difficult to train \citep{arjovsky2017towards}, adding a critic to our approach could further constrain the model and improve synthesis performance. Moreover, to further constrain the model an additional synthesis loss in latent space could have been proposed \citep{oring21a}. We deliberately leave the challenge of defining a useful distance metric in latent space for future work.

	% add paragraph to discuss assumption that physical slice spacing coincides with latent space spacing. -> rebuttal alpha \neq 0.5
	Our approach assumes that a linear spacing in latent space corresponds to slice spacing in image space. Although, alternatives such as spherical latent space interpolation \citep{roberts2018hierarchical, white2016sampling} or an enforced Riemannian latent space \citep{shao2018riemannian, arvanitidis2018} can be used, linear interpolation in the latent space of an autoencoder trained with the proposed synthesis loss showed excellent results. Nevertheless, human anatomy does not change linearly along spatial dimensions. We conjecture that the model can learn such a nonlinearity from the training data. Furthermore, we presume that the synthesis loss encourages the model to learn the nonlinear mapping between \textit{distances} in latent and image space. For example, our experiments on cardiac MRI revealed that the model has learned that structural changes at the base of the heart are substantially different than at the apex. Finally, our experiments on neonatal and adult brain MRI apply mixing coefficients ($\alpha$) unequal to \num{0.5}. Results of these experiments corroborate our assumption that linear steps taken in latent space can approximate anatomical distances in image space, however, the approach does not guarantee such a relationship to be exact.
	% Laine: ...we minimize the energy of an explicit metric that can be chosen to suit the task at hand.
	% Nevertheless, we conjecture that the proposed model can learn a shared structure of the heart encoded in the latent space. Furthermore, although in all experiments the synthesis loss is evaluated for slices halfway between ($\alpha = 0.5$) the two neighboring slices, the synthesis loss encourages the model to learn e.g. that cardiac anatomy changes more }
	
	%	Conducted experiments on neonatal and adult brain MRI for upsampling factors higher than \num{2}        apply mixing coefficients $\alpha$ unequal to \num{0.5}. Hence, we think that our experimental          results     reveal 
	% This could also be a possible explanation of why reconstruction performance of the model decreases when increasing the value of $\lambda$ (weighting of the synthesis loss in Equation~\ref{eq_combined_loss}).
	
	% limitation: large anatomical variations between slices
	Performance of the proposed method is affected by large anatomical variations between adjacent slices as shown in Figure~\ref{fig_qualitative_acdc_large_anatomical_variations}. Furthermore, quality of synthesized images is decreased when adjacent slices within the original volume are misaligned. This is a known problem in cardiac cine MR imaging caused by surrounding organ motion during breath-hold acquisition. In these cases intermediate points along the interpolation path spanned by two adjacent slice encodings result in anatomically implausible images i.e. cross-fades. This might indicate that the latent space between the two endpoints is too sparsely populated. For cardiac cine MRI this could be alleviated by choosing additional interpolation endpoints e.g. from other time frames. This direction will be investigated in future work.
	
	Training the autoencoder with a combination of reconstruction and synthesis loss slightly hampered reconstruction performance when compared to training with reconstruction loss only as shown in Table~\ref{table_acdc_ae_caisr_recon_versus_synth_1}. However, quality of synthesized slices considerably improved when adding the synthesis loss during training. % when compared to model training with reconstruction loss only. 
	To render \num{3}D high-resolution images the method only relies on the quality of newly synthesized slices. Hence, all existing slices do not need to be reconstructed but can be taken from the original anisotropic \num{3}D volumes.
	% the model i.e. encouraging the model to generate intermediate slices that are perceptually similar to the original slice 
	
	% Comparison with XIA 2021 Unsupervised SR
	The here presented approach was developed in parallel with the super-resolution method proposed by \cite{xia2021super}. In the current work quantitative results are presented in terms of SSIM, PSNR and VIF for all cardiac MRI slices of test patients from the ACDC dataset and \num{45} subjects from the Sunnybrook dataset. In comparison, work of \cite{xia2021super} depicts results in terms of PSNR and correlation coefficient for a limited selection of two cardiac MRI slices (mid-ventricular and basal) from the UK Biobank. Note that intensity statistics of images from different datasets may be very different and hence, PSNR measurements might be inaccurate \citep{oktay2017anatomically}. Therefore, thorough quantitative comparison of the two methods is hardly feasible.
	
	To conclude, we presented a method for unsupervised semantic interpolation of anisotropic \num{3}D medical images achieving anatomically smooth transitions in through-plane direction. New intermediate slices are generated by mixing the latent space encodings of two spatially adjacent slices. The experiments using cardiac cine and brain MRIs demonstrated that the proposed approach outperforms cubic B-spline interpolation on cardiac cine and brain MRIs. Given the unsupervised nature of the method, high-resolution training data is not required and hence, the method can be readily applied in clinical settings. 
	% The approach} exploits the ability of autoencoders to interpolate in latent space.
	
	\section*{Acknowledgment}
	This study was performed within the DLMedIA program (P15-26) funded by Dutch Technology Foundation with participation of Pie Medical Imaging.
	
	%%Harvard
	%\bibliographystyle{IEEEtran}
	% \balance
	%\bibliography{sr_bibliography}

	% \section*{Supplementary Material}

\end{document}